\documentclass[12pt,a4paper]{book}
\usepackage{bm}
\usepackage{amssymb}
\usepackage{amsmath}
\usepackage{amsfonts}
\usepackage[dvips]{graphicx}
\usepackage{psfrag}
\usepackage{rotating}
\usepackage[spanish]{babel}
\usepackage{multirow}
\usepackage{bbm}%

\newtheorem{theorem}{Teorema}
\newtheorem{definition}[theorem]{Definici\'{o}n}
\newtheorem{corollary}[theorem]{Corolario}
\newenvironment{proof}[1][Demostraci\'{o}n]{\noindent\textbf{#1.} }{\ \rule{0.5em}{0.5em}}

\usepackage[twoside,bindingoffset=5mm,a4paper]{geometry}

\begin{document}

\pagestyle{empty}

\begin{center}

{\large UNIVERSIDAD DE CONCEPCI\'{O}N} \\
{\large FACULTAD DE CIENCIAS F\'{I}SICAS Y MATEM\'{A}TICAS} \\
{\large DEPARTAMENTO DE F\'{I}SICA}

\vspace{.5cm}

\includegraphics[width=2.5cm]{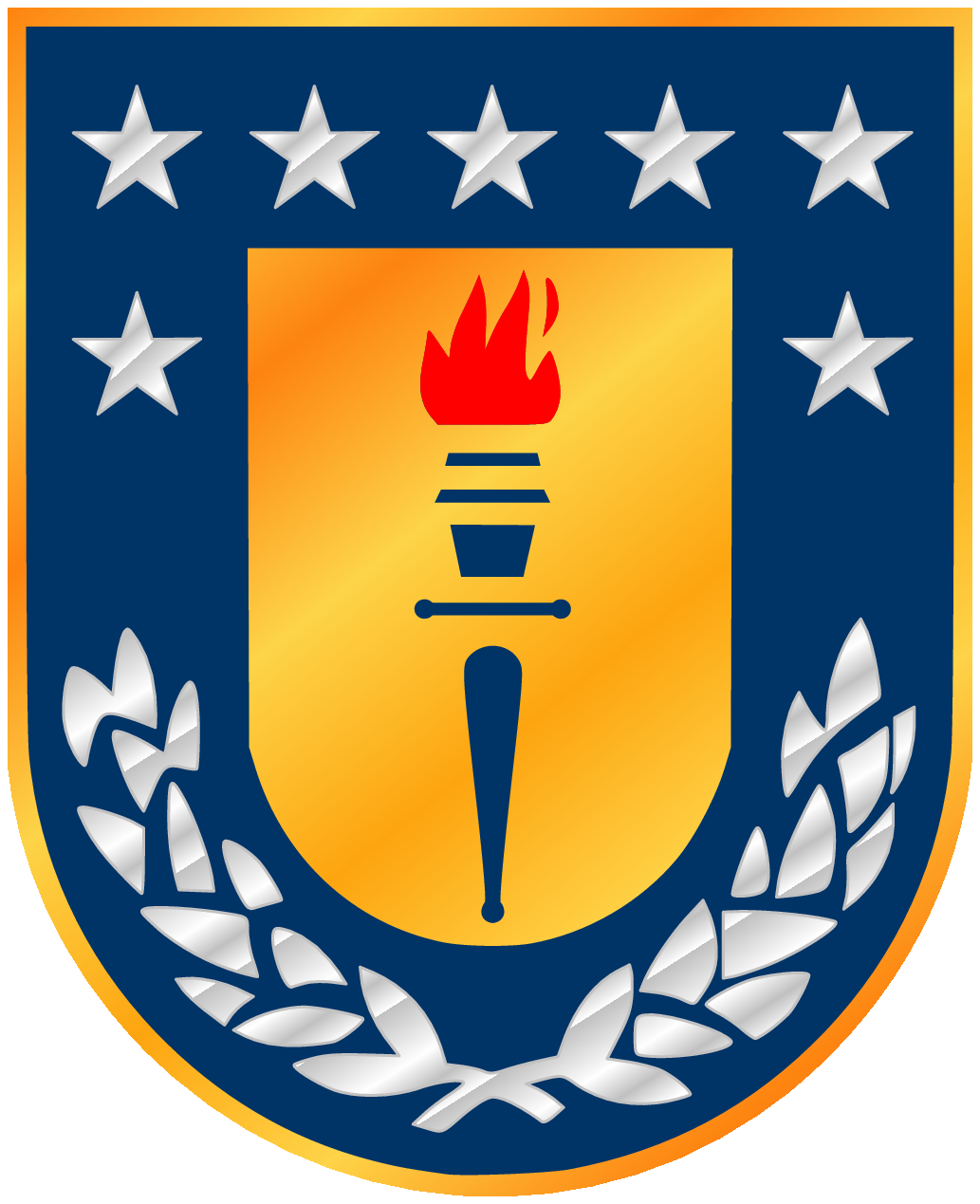}

\vspace{.5cm}

\rule{15cm}{0.1cm}

\vspace{1.0cm}

{\huge \textbf{Expansi\'{o}n en Semigrupos y}} \\
{\huge \textbf{M-Supergravedad en 11 dimensiones}}

\vspace{1.5cm}

Tesis para optar al grado acad\'emico

de Doctor en Ciencias F\'{\i}sicas

\vspace{0.5cm}

por

\vspace{0.5cm}

{\Large Fernando Esteban Izaurieta Aranda}
\rule{15cm}{0.1cm}
\vspace{1cm}

\begin{tabular}[c]{lll}
Director de Tesis & : & Dr. Patricio Salgado \\
& & \\
Comisi\'{o}n & : & Dr. Jos\'{e} A. de Azc\'{a}rraga\\
& & Dr. Sergio del Campo\\
& & Dr. Jorge Zanelli
\end{tabular}

\vfill

\vspace{0.7cm}

Concepci\'{o}n, Chile

Octubre 2006
\end{center}

\pagebreak

\ \\
\newpage

\ \\
\ \\
\ \\

\emph{Dedicado a la Mujer.}

\emph{A la Mujer Madre, que nos di\'{o} la vida.}

\emph{A la Mujer Hermana, que nos llen\'{o} de orgullo.}

\emph{A la Mujer Amiga, que nos di\'{o} su comprensi\'{o}n.}

\emph{A la Mujer Amante, que nos abras\'{o} de pasi\'{o}n.}

\emph{Gracias, por llenarlo todo con calor y vida.}
\newpage

\ \\
\ \\
\ \\
\ \\
\ \\
\textquotedblleft\textit{Pick a flower on Earth and you move the farthest
star.}\textquotedblright

(P.~A.~M.~Dirac)
\ \\
\newpage

\tableofcontents

\pagestyle{plain} \pagenumbering{Roman}

\chapter*{Agradecimientos}

\addcontentsline{toc}{chapter}{Agradecimientos}

Esta tesis constituye el fruto de largos a\~{n}os de investigaci\'{o}n en
varios pa\'{\i}ses, en los cuales much\'{\i}simas personas e instituciones me
han brindado su amistad, ense\~{n}anza y apoyo en las m\'{a}s diversas formas.
Este es un peque\~{n}o intento para agradecerles a los que m\'{a}s recuerdo;
s\'{e} que olvido a muchos (espero me perdonen), pero si los nombrase a todos,
esta tesis tendr\'{\i}a el doble de p\'{a}ginas.

Primero que nada quisiera agrader a los m\'{a}s cercanos en el coraz\'{o}n,
incluso cuando oc\'{e}anos nos separaban: mi familia. Doy gracias a mi madre,
mi padre y especialmente a mis hermanas, Pamela y Ana, por su amor y apoyo
durante todo este tiempo.

Por otra parte, en esta apasionada b\'{u}squeda del saber que constituye la
Ciencia, uno adquiere en el camino una deuda de gratitud inmensa hacia sus
Profesores. Ellos nos llevan pacientemente hasta los bordes del conocimiento,
para as\'{\i} darnos el privilegio de poder dar un paso m\'{a}s all\'{a}.
As\'{\i}, quisiera agradecer a algunos de los cuales han tenido incumbencia
directa con el desarrollo de esta tesis [orden cronol\'{o}gico]:

Deseo agradecer a mi tutor, Prof. Patricio Salgado de la Universidad de
Concepci\'{o}n, por su apoyo y ayuda como profesor y amigo. Sus sugerencias a
medida que se desarrollaba la investigaci\'{o}n resultaron ser de crucial
importancia y a\'{u}n m\'{a}s, su gu\'{\i}a desde el inicio a trav\'{e}s de
los m\'{a}s bellos parajes de la F\'{\i}sica ha dejado una profunda huella en
mi modo de ver la Ciencia. Tambi\'{e}n quisiera agradecerle el que siempre me
haya considerado no s\'{o}lo como estudiante, sino tambi\'{e}n como persona,
preocup\'{a}ndose no s\'{o}lo por mi desempe\~{n}o acad\'{e}mico sino que
tambi\'{e}n por mi situaci\'{o}n personal. En la Universidad de Concepci\'{o}n
quisiera agradecer tambi\'{e}n al Prof. Paul Minning, el cual me introdujo en
variados t\'{o}picos en matem\'{a}ticas simplemente por su elegancia, aunque
no pareciesen tener una utilidad inmediata. Entre estos t\'{o}picos, estuvo el
tema semigrupos, el cual result\'{o} ser fundamental para desarrollar la
presente tesis. Agradezco asimismo a los Profs. Jorge Zanelli y Ricardo
Troncoso, por su generosa hospitalidad en el Centro de Estudios
Cient\'{\i}ficos en Valdivia. En amenas e interesantes discusiones con ellos
tuve la oportunidad de aprender much\'{\i}simo, sobre los m\'{a}s diversos
t\'{o}picos en F\'{\i}sica, y en particular en el de teor\'{\i}as de
Chern--Simons y su profunda elegancia matem\'{a}tica. Debo destacar en
especial la extrema amabilidad del Prof. Zanelli el cual brind\'{o} su apoyo
para poder realizar la estad\'{\i}a en Alemania. En el extranjero, deseo
agradecer al Prof.~Dieter~L\"{u}st por su amable hospitalidad, primero en el
Lisa-Meitner Institut en Adlershof de la Humboldt-Universit\"{a}t zu Berlin y
luego en el Arnold Sommerfeld Zentrum de la Ludwig-Maximilians-Universit\"{a}t
M\"{u}nchen, en donde \'{e}l tuvo la gentileza de mostrarme diversos
t\'{o}picos relacionados con Teor\'{\i}a de Cuerdas, adem\'{a}s de siempre
preocuparse porque tuviera una grata estad\'{\i}a en Alemania y sobre todo en
Baviera. Tambi\'{e}n deseo agradecer la c\'{a}lida hospitalidad del
Prof.~Jos\'{e}~de~Azc\'{a}rraga y Prof.~Mar\'{\i}a~A.~Lled\'{o} de la
Universitat de Val\`{e}ncia. Gracias a su amable disposici\'{o}n, nuestra
estad\'{\i}a en Valencia fue especialmente agradable y productiva. Les debo
agradecer especialmente por sus explicaciones extremadamente detalladas y
pacientes en diversos t\'{o}picos relacionados con la expansi\'{o}n de
\'{a}lgebras, \'{a}lgebras diferenciales libres y teor\'{\i}as de
Chern--Simons, discusiones que resultaron ser finalmente el germen de donde
surgi\'{o} buena parte de los resultados de la presente tesis.

Tambi\'{e}n deseo recordar aqu\'{\i} a los Profs. Enrique Oelker y Juan Rivera
de la Universidad de Concepci\'{o}n. Tuve el gran honor de contarme entre sus
\'{u}ltimos estudiantes antes del lamentable fallecimiento de ambos, pero sus
ense\~{n}anzas y su modo de hacer F\'{\i}sica es algo que recordar\'{e}
siempre. Su filosof\'{\i}a de rigurosidad y claridad al ense\~{n}ar, junto con
su profunda conciencia de que la b\'{u}squeda del conocimiento es una empresa
heredada de generaci\'{o}n en generaci\'{o}n, fue y ser\'{a} siempre fuente de inspiraci\'{o}n.

La vida universitaria resulta especialmente estimulante debido al incesante
torbellino de intercambio de nuevas ideas en el cual uno se ve envuelto. En
este punto, el rol protag\'{o}nico lo juegan los compa\~{n}eros y amigos con
los cuales se conversa de los m\'{a}s variados t\'{o}picos, se r\'{\i}e y se
llora. Entre ellos debo agradecer en especial a mi amigo Eduardo
Rodr\'{\i}guez, con el cual desarrollamos este trabajo codo a codo,
comprobando siempre que el trabajo combinado de ambos es mucho m\'{a}s que la
simple suma del trabajo de ambos. Resulta d\'{\i}ficil encontrar en una sola
persona esa mezcla de inteligencia, paciencia, generosidad e integridad, que
me ha hecho considerarlo m\'{a}s que un amigo, un hermano a lo largo de estos
a\~{n}os. En la Universidad de Concepci\'{o}n tuve la fortuna de hacer
numerosos amigos en todo este tiempo (muchos est\'{a}n ahora esparcidos por el
mundo), a los que deseo agradecer por su compa\~{n}erismo, apoyo y ayuda:
Marcelo Alid, Leonardo B\'{a}ez, Carla Baldov\'{\i}n, Carlos Bascu\~{n}an,
Benjam\'{\i}n Burgos, Jorge Castillo, Antonella Cid,
Marcelo Contreras, Arturo Fern\'{a}ndez, Roberto Gaete, Marcela Godoy, Arturo
G\'{o}mez, Katherine Gonz\'{a}lez, Jorge Inostroza, Juan L\'{o}pez, Amerika
Manzanares (y a su familia, que me ayud\'{o} much\'{\i}simo en situaciones
dif\'{\i}ciles), Patricio Mella, Nelson Merino, Julio Oliva, Yazmina Olmos,
Jos\'{e} Luis Romero, Cristian Salas, Marcela Salazar, C\'{e}sar S\'{a}nchez, Ruth Sandoval,
Carolina Trichet, Gabriela Ulloa, Paola Utreras, Ana Vald\'{e}s, Claudia
Vald\'{e}s, Omar Valdivia, Pablo Viveros y Marisol Zambrano. Le deseo
agradecer tambi\'{e}n a Elzbieta Grosiak especialmente por su comprensi\'{o}n
y paciencia en los \'{u}ltimos meses de escritura de la tesis; de otra forma
este trabajo no hubiese sido posible.

Quisiera agradecer tambi\'{e}n grandemente la simpat\'{\i}a, cordialidad y
buena disposici\'{o}n de las secretarias y auxiliares del Departamento de
F\'{\i}sica de la Universidad de Concepci\'{o}n, Marta Astudillo, Patricia
Luarte, Marcela Sanhueza, Heraldo Manr\'{\i}quez y V\'{\i}ctor Mora.

La amistad tambi\'{e}n superan las barreras de los lenguajes y las distintas
culturas. En Alemania tuve la oportunidad de hacer buenos amigos, con los
cuales aprend\'{\i} mucho de F\'{\i}sica y sobre muchas otras cosas,
enriqueciendo mi visi\'{o}n del mundo grandemente.

Un abrazo para todos aquellos que me brindaron su amistad tanto en Berl\'{\i}n
como en M\'{u}nich, entre los cuales deseo nombrar a unos pocos, como Danilo
D\'{\i}az, Marija Dimitrijevi\'{c}, Viviane Gra\ss , Florian Koch, Daniel
Krefl, Frank Meyer, Johannes Oberreuter, Dan Oprisa, Susanne Reffert, Waldemar
Schulgin, Maren Stein y Prasanta Tripathy. Debo agradecer en especial a mis
compa\~{n}eros de oficina, Murad Alim, Rachid Benhamid, Matteo Cardella y
Enrico Pajer, con los cuales compartimos muchas horas de una mezcla de
F\'{\i}sica, buena (y mala) comida y buena cerveza. Asimismo un agradecimiento
especial a Natalia Borodatchenkova, por su simpat\'{\i}a, amenas
conversaciones y buen caf\'{e}, a Novara Sari Jambak, la cual enriqueci\'{o}
mi visi\'{o}n de mundo grandemente, ense\~{n}\'{a}ndome la incre\'{\i}ble
variedad cultural del planeta, a Branislav Jurco, el cual siempre estuvo
dispuesto a explicar en forma clara y amable las matem\'{a}ticas m\'{a}s
abstractas, y a Jan Perz, por su cordialidad y gran ayuda resolviendo
enrevesados problemas de traducci\'{o}n de gran importancia para m\'{\i}.

Tambi\'{e}n quisiera agradecer a Maria Hartmann y Sophie von Werder del
Servicio Alem\'{a}n de Intercambio Acad\'{e}mico (DAAD), por su amabilidad y
siempre buena disposici\'{o}n para resolver los m\'{a}s diversos problemas que
surgen en forma natural cuando se vive en un pa\'{\i}s extranjero.

Por \'{u}ltimo, quisiera agradecer a numerosas instituciones que hicieron esta
investigaci\'{o}n posible. En particular, quisiera agradecer el apoyo
econ\'{o}mico entregado a trav\'{e}s de becas por la Universidad de
Concepci\'{o}n (2001, 2006), la Comisi\'{o}n Nacional de Investigaci\'{o}n
Cient\'{\i}fica y Tecnol\'{o}gica CONICYT (2002--2003) y el Servicio
Alem\'{a}n de Intercambio Acad\'{e}mico DAAD (2003--2006). Agradezco
tambi\'{e}n el apoyo circunstancial prestado por el Ministerio de
Educaci\'{o}n a trav\'{e}s del Proyecto MECESUP UCO 0209, la
Ludwig-Maximilians-Universit\"{a}t M\"{u}nchen por medio del Arnold Sommerfeld
Center for Theoretical Physics y la Universitat de Val\`{e}ncia a trav\'{e}s
del Departament de F\'{\i}sica Te\`{o}rica.

\chapter*{Acknowledgements}

\addcontentsline{toc}{chapter}{Acknowledgements}

This Ph.D. Thesis is the final outcome of long years of research in several
countries, where a lot of people and institutions have given me their
friendship, teachings and support in different ways. Here I will try to thank
some of you; I will fail to mention all of you (I hope you will forgive me),
but anyway, if I could write the whole list, this Thesis would have twice as
many pages.

At first, I want to thank the ones closest to my heart, even when oceans were
separating us: my family. I thank my mother, my father and especially my
sisters, Pamela and Ana, for their love and support all along this time.

On the other hand, in this big adventure which is Science, our Professors play
a central r\^{o}le. They take us all the way to the cutting edge of knowledge,
and give us the privilege of taking one step further. Therefore, here I want
to thank some of you, who played a direct r\^{o}le in the development of the
present work [in chronological ordering]:

I want to thank my advisor, Prof. Patricio Salgado from the Universidad de
Concepci\'{o}n, for his support and help as teacher and friend. His
suggestions during researchtime were of crucial importance, and even more, his
guiding from the very beginning through Physics' most beautiful landscapes has
left a deep impact in my way of seeing Science. I want to thank him also for
considering me not only as a student but also as a person, taking care not
only of academic aspects but also of my personal situation. At the Universidad
de Concepci\'{o}n I also want to thank Prof. Paul Minning, who has introduced
me to several topics in mathematics just because of their elegance, even if
they were not of immediate usefulness. Among these topics was that of
semigroups, which was finally fundamental for the current research. Also I
want to thank Profs. Jorge Zanelli and Ricardo Troncoso, for their kind
hospitality at the Centro de Estudios Cient\'{\i}ficos, Valdivia. In lively
and interesting discussions with them I had the opportunity to learn a lot on
several topics in Physics and especially on that of Chern--Simons Theories and
their deep mathematical elegance. I must also thank Prof. Zanelli for his
strong support in order to make the stay in Germany. Abroad, I want to thank
Prof.~Dieter~L\"{u}st for his kind hospitality, first at the Lisa-Meitner
Institut in Adlershof of the Humboldt-Universit\"{a}t zu Berlin and afterwards
at the Arnold Sommerfeld Zentrum of the Ludwig-Maximilians-Universit\"{a}t
M\"{u}nchen. He had the kindness of showing me several interesting topics in
String Theory, besides of always taking care of making my stay in Germany and
particularly in Bavaria a grateful experience. I want to thank the warm
hospitality of Prof.~Jos\'{e}~de~Azc\'{a}rraga and
Prof.~Mar\'{\i}a~A.~Lled\'{o} from the Universitat de Val\`{e}ncia. Because of
their kindness, our stay in Valencia was specially nice and productive. I want
to thank them specially for their detailed and patient explanations on several
topics related with algebra expansion, free differential algebras and
Chern--Simons Theories. Those discussions were finally the seed for a big
portion of the current work.

I want to remember here the late Profs. Enrique Oelker and Juan Rivera from
the Universidad de Concepci\'{o}n. I had the big honour of being one of their
last students, but their teachings and their way of doing physics is something
that I will always remember. Their philosophy of rigour and clearness at
teaching, together with their deep awareness of the fact that the search for
knowledge is an enterprise inherited from generation to generation, was and
will always be a source of inspiration.

Universitary life is especially stimulating because of the big ideas
interchange medium where one is immersed. At this point, the central r\^{o}le
is played by the classmates and friends with whom you talk about everything,
laugh and cry. Among them I must especially thank my friend Eduardo
Rodr\'{\i}guez, with whom we developed this research side by side, proving
that joint work is a lot more than the simple addition of both individual
works. It is hard to find in only one person this intelligence, patience,
goodness and integrity, which made me consider him more than a friend, a
brother along these years. At the Universidad de Concepci\'{o}n I had the luck
of finding a lot of friends all this time (a lot of them are now spread along
the world), and I want to thank them for their friendship, support and help:
Marcelo Alid, Leonardo B\'{a}ez, Carla Baldov\'{\i}n, Carlos Bascu\~{n}an,
Benjam\'{\i}n Burgos, Jorge Castillo, Antonella Cid,
Marcelo Contreras, Arturo Fern\'{a}ndez, Roberto Gaete, Marcela Godoy, Arturo
G\'{o}mez, Katherine Gonz\'{a}lez, Jorge Inostroza, Juan L\'{o}pez, Amerika
Manzanares (and her family, who helped me a lot during hard times), Patricio
Mella, Nelson Merino, Julio Oliva, Yazmina Olmos, Jos\'{e} Luis Romero,
Cristian Salas, Marcela Salazar, C\'{e}sar S\'{a}nchez, Ruth Sandoval, Carolina Trichet,
Gabriela Ulloa, Paola Utreras, Ana Vald\'{e}s, Claudia Vald\'{e}s, Omar
Valdivia, Pablo Viveros and Marisol Zambrano. I also want to thank Elzbieta
Grosiak for her understanding and patience during the last months of thesis
work; this work would not have been possible otherwise.

I must also thank a lot the cheerfulness and kindness of the secretaries and
assistants from the Physics Department of the Universidad de Concepci\'{o}n,
Marta Astudillo, Patricia Luarte, Marcela Sanhueza, Heraldo Manr\'{\i}quez y
V\'{\i}ctor Mora.

Friendship breaks barriers of languages and cultures. In Germany I had the
chance of making good friends, with whom I have learned a lot of physics and
other things, making my vision of the world a lot wider.

A big hug for everyone who has offered his/her friendship in Berlin as well as
in Munich. Among them I want to mention some few of you; Danilo D\'{\i}az,
Marija Dimitrijevi\'{c}, Viviane Gra\ss , Florian Koch, Daniel Krefl, Frank
Meyer, Johannes Oberreuter, Dan Oprisa, Susanne Reffert, Waldemar Schulgin,
Maren Stein and Prasanta Tripathy. I must especially thank my officemates,
Murad Alim, Rachid Benhamid, Matteo Cardella and Enrico Pajer, with whom we
shared several hours of physics, good (and bad) food and very good beer. Big
thanks also to Natalia Borodatchenkova, for her cheerfulness, lively
discussions and good coffee; Novara Sari Jambak, for widening my vision of the
world, teaching me about the big cultural diversity of our planet; Branislav
Jurco, who was always so kind to nicely and clearly explain the most abstract
topics in mathematics, and Jan Perz for his kindness and big help solving some
hard translation problems which were of big importance to me.

Ich bedanke mich auch bei Frau Maria Hartmann und Frau Sophie von Werder vom
Deutschen Akademischen Austausch Dienst DAAD, f\"{u}r Ihre Freundlichkeit und
Fertigkeit um die komischsten Probleme zu l\"{o}sen, die normalerweise erlebt
werden m\"{u}ssen, wenn man im Ausland lebt.

Finally, I must thank several institutions which made possible this research.
Particularly, I thank the financial support from scholarships from the
Universidad de Concepci\'{o}n (2001, 2006), the Comisi\'{o}n Nacional de
Investigaci\'{o}n Cient\'{\i}fica y Tecnol\'{o}gica CONICYT (2002--2003) and
the Deutschen Akademischen Austausch Dienst DAAD (2003--2006). I thank also
the timely support given by the Ministerio de Educaci\'{o}n through Proyecto
MECESUP UCO 0209, by Ludwig-Maximilians-Universit\"{a}t M\"{u}nchen through
the Arnold-Sommerfeld-Zentrum f\"{u}r Theorethische Physik and the Universitat
de Val\`{e}ncia through the Departament de F\'{\i}sica Te\`{o}rica.


\chapter*{Resumen}

\addcontentsline{toc}{chapter}{Resumen}

Esta tesis trata sobre la construcci\'{o}n de una teor\'{\i}a de gauge
invariante off-shell para el \'{A}lgebra~M en 11 dimensiones, a trav\'{e}s del
uso de una forma de Transgresi\'{o}n como Lagrangeano.

Para realizar esto, primero analizamos la construcci\'{o}n general de
teor\'{\i}as de gauge a trav\'{e}s de formas de Transgresi\'{o}n para un grupo
de simetr\'{\i}a arbitrario (Cap\'{\i}tulo~\ref{SecTrans}). Algunos resultados
interesantes con respecto a este punto constituyen

\begin{enumerate}
\item el c\'{a}lculo de cargas de Noether conservadas \textit{off-shell},

\item la asociaci\'{o}n de la estructura de dos conexiones propia de una forma
de Transgresi\'{o}n con distintas orientaciones de la variedad base y

\item la construcci\'{o}n de un M\'{e}todo de Separaci\'{o}n en Subespacios,
el cual permite dividir la acci\'{o}n en un t\'{e}rmino de volumen
(\textit{bulk}) y uno de borde, y separar cada uno de ellos en trozos que
reflejen la f\'{\i}sica asociada con una cierta elecci\'{o}n de grupo de simetr\'{\i}a.
\end{enumerate}

Para llevar a cabo la construcci\'{o}n de la teor\'{\i}a de gauge, ser\'{a}
necesario crear una nueva herramienta matem\'{a}tica, llamada $S$-Expansiones,
para analizar la estructura del \'{A}lgebra~M y crear un tensor invariante
para ella (Cap\'{\i}tulo~\ref{SecS_Exp}). Este m\'{e}todo es desarrollado en
forma general, y permite, dada una cierta \'{a}lgebra de Lie y un semigrupo
abeliano discreto, crear nuevas \'{a}lgebras de Lie (\'{A}lgebras
$S$-Expandidas, Sub\'{a}lgebras resonantes, \'{A}lgebras reducidas en forma
resonante). Aplicando estas herramientas, se construye un tensor invariante
para el \'{A}lgebra~M, el cual ser\'{a} usado como fundamento para la
construcci\'{o}n de una teor\'{\i}a de gauge de Transgresi\'{o}n para el
\'{A}lgebra~M (Cap\'{\i}tulo~\ref{SecAcc_M_Alg}). La relaci\'{o}n entre la
din\'{a}mica cuadridimensional asociada a esta teor\'{\i}a y la torsi\'{o}n en
$D=11$ son asimismo considerados. Por \'{u}ltimo, concluimos con un
an\'{a}lisis de las posibles aplicaciones de las herramientas desarrolladas,
en el contexto de Cosmolog\'{\i}a, Supergravedad y Teor\'{\i}a de Cuerdas.


\chapter*{Abstract}

\addcontentsline{toc}{chapter}{Abstract}

This thesis deals with the construction of an eleven-dimensional gauge theory,
off-shell invariant, for the M~Algebra. The theory is buit using a
Transgression Form as a Lagrangian.

In order to accomplish this, one must first analyze the general construction
of Transgression Gauge Field Theories, for an arbitrary symmetry group
(Chapter~\ref{SecTrans}). Some interesting results regarding this point are

\begin{enumerate}
\item the calculation of Noether Charges which are off-shell conserved,

\item the association of the double connection structure of the Transgression
Form with both orientations of the basis manifold and

\item the Subspace Separation Method, which allows us to divide the action in
bulk and boundary terms, and to split them in terms which reflect the physics
corresponding to a symmetry group choice.
\end{enumerate}

To construct the gauge theory explicitly, it is necessary to buid a new
mathematical tool, called $S$-Expansion procedure. Analyzing the M~Algebra
under the light of this method, it is possible to construct an invariant
tensor for it. This method is developed in a general way and, given a Lie
algebra and an Abelian, finite semigroup, it allows us to generate new Lie
algebras ($S$-Expanded Algebras, Resonant Subalgebras, Resonant Forced Algebras).

Applying this tool, an invariant tensor for the M~Algebra is constructed,
which serves as the basis upon which a Transgression Gauge Field Theorie for
the M~Algebra (Chapter~\ref{SecAcc_M_Alg}) is constructed. The relationship
between the four-dimensional dynamics from this theory and the
eleven-dimensional torsion is also considered. Finally, we close with an
analysis of the possible applications of the developed tools, in the context
of Cosmology, Supergravity and String Theory.


\newpage

\chapter{\label{SecIntro}Introducci\'{o}n}

\pagenumbering{arabic} \setcounter{page}{1} \pagestyle{headings}

\begin{center}
\textquotedblleft\textit{Venient annis}

\textit{S\ae cula seris, quibus Oceanus}

\textit{Vincula rerum laxet \& ingens}

\textit{Pateat tellus, Typhi\'{s}que nouos}

\textit{Detegat Orbes,}

\textit{Nec sit terris vltima Thyle \footnote{\textquotedblleft\emph{A\~{n}os vendr\'{a}n}/
\emph{Al paso de los tiempos, en que suelte el Oc\'{e}ano}
/\emph{Las barreras del mundo y se abra la tierra}/
\emph{En toda su extensi\'{o}n, y Tetis nos descubra}/
\emph{Nuevos Orbes,}/
\emph{Y el conf\'{\i}n de la tierra ya no sea Tule.}\textquotedblright} }
\textquotedblright

(Abraham Ortelius, Theatrvm Orbis Terrarvm de 1570, considerado el primer
atlas moderno, citando la Medea de S\'{e}neca, Acto II, en relaci\'{o}n al
descubrimiento de nuevos continentes)
\end{center}

\section{Motivaci\'{o}n}

Los \'{u}ltimos cientos de a\~{n}os han sido una \'{e}poca de profundos
cambios para nuestro mundo. El planeta ha sido, y est\'{a} siendo,
transformado por nuestra especie a una escala quiz\'{a}s solamente equiparable
con la aparici\'{o}n de las primeras formas de vida vegetal en los
oc\'{e}anos, las cuales transformaron la atm\'{o}sfera primitiva en una de
ox\'{\i}geno. Esto no es un accidente, sino m\'{a}s bien se debe a la
aplicaci\'{o}n de la Ciencia y el M\'{e}todo Cient\'{\i}fico en la
exploraci\'{o}n de nuestro mundo. Ella nos ha provisto de la capacidad para
entender y predecir en forma precisa los fen\'{o}menos naturales. Como
consecuencia de esto, la humanidad ha sido capaz de desarrollar una
tecnolog\'{\i}a que le ha permitido modificar en forma radical nuestro medio ambiente.

\begin{sidewaysfigure}[ptb]
\includegraphics[width=\textheight]{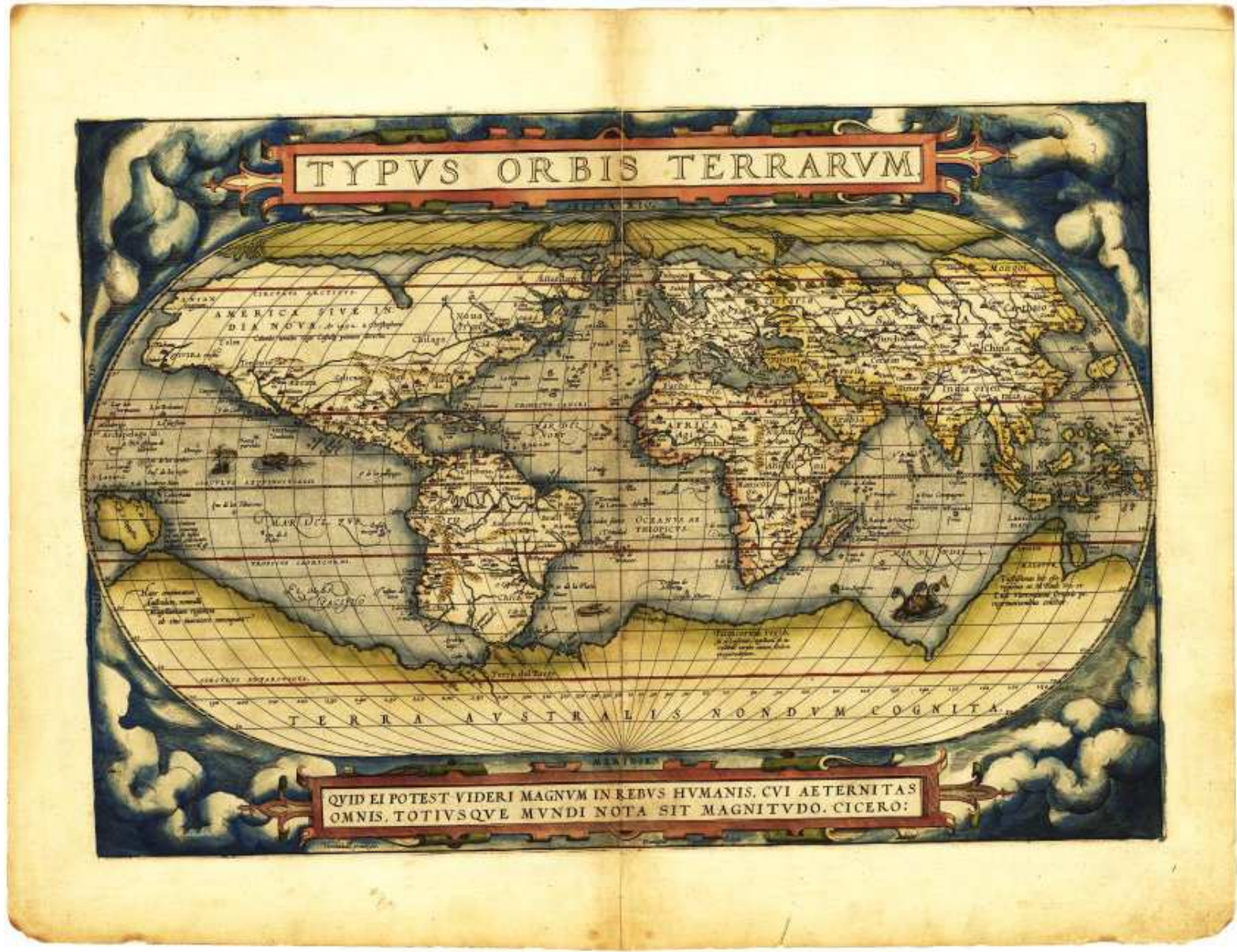}
\caption{Mapa del Mundo del primer atlas moderno, el Theatrvm Orbis Terrarvm,
realizado en 1570 por Abraham Ortelius. Obs\'{e}rvese en particular la
inexactitud de los conocimientos geogr\'{a}ficos (y de la topolog\'{\i}a
inclusive!) de regiones como Tierra del Fuego y Australia.}%
\label{FigOrbis}%
\end{sidewaysfigure}

Sin embargo, y pese a este avasallador \'{e}xito, la F\'{\i}sica dista mucho
de estar \textquotedblleft finalizada\textquotedblright. De hecho, si hay una
cosa que podemos decir con seguridad, es que nuestro conocimiento de las leyes
fundamentales de la f\'{\i}sica no es s\'{o}lo incompleto, sino que ni
siquiera forma un todo autoconsistente. Para entender mejor esta
afirmaci\'{o}n, es provechoso a modo de ilustraci\'{o}n hacer un paralelo
entre el estado de la f\'{\i}sica actual y el conocimiento geogr\'{a}fico del
mundo occidental durante los s.~XV y~XVI, \textquotedblleft la era de los
descubrimientos\textquotedblright. En aqu\'{e}llas \'{e}pocas, el mundo
occidental, reci\'{e}n salido de la oscuridad intelectual del medioevo, se
lanza a la exploraci\'{o}n de oc\'{e}anos desconocidos. Cuando se observa un
mapa de aquel entonces, es posible observar un hecho muy natural: las
\'{a}reas conocidas (las m\'{a}s accesibles) son dibujadas con gran detalle,
mientras que extensas zonas inexploradas s\'{o}lo aparecen esbozadas. En
particular, se volv\'{\i}a un gran problema el saber si dos regiones distintas
eran parte de un mismo continente, o si estaban separadas por un \'{o}ceano.
Las costas sol\'{\i}an ser exploradas antes que los interiores de las regiones
reci\'{e}n descubiertas, esto conllev\'{o} equ\'{\i}vocos famosos, como por
ejemplo, confundir bah\'{\i}as con desembocaduras de r\'{\i}os (ciudades como
Rio de Janeiro deben sus nombres a errores como aquellos). Poco a poco, los
viajeros llegaban m\'{a}s lejos, e islas separadas en los mapas se iban
fundiendo para llegar a ser continentes, mientras que nuevos \'{o}ceanos y
estrechos uni\'{e}ndolos fueron apareciendo en lugares inesperados (Ve\'{a}se
Fig.\ref{FigOrbis}).

\begin{figure}[ptb]
\begin{center}
\includegraphics[width=\textwidth]{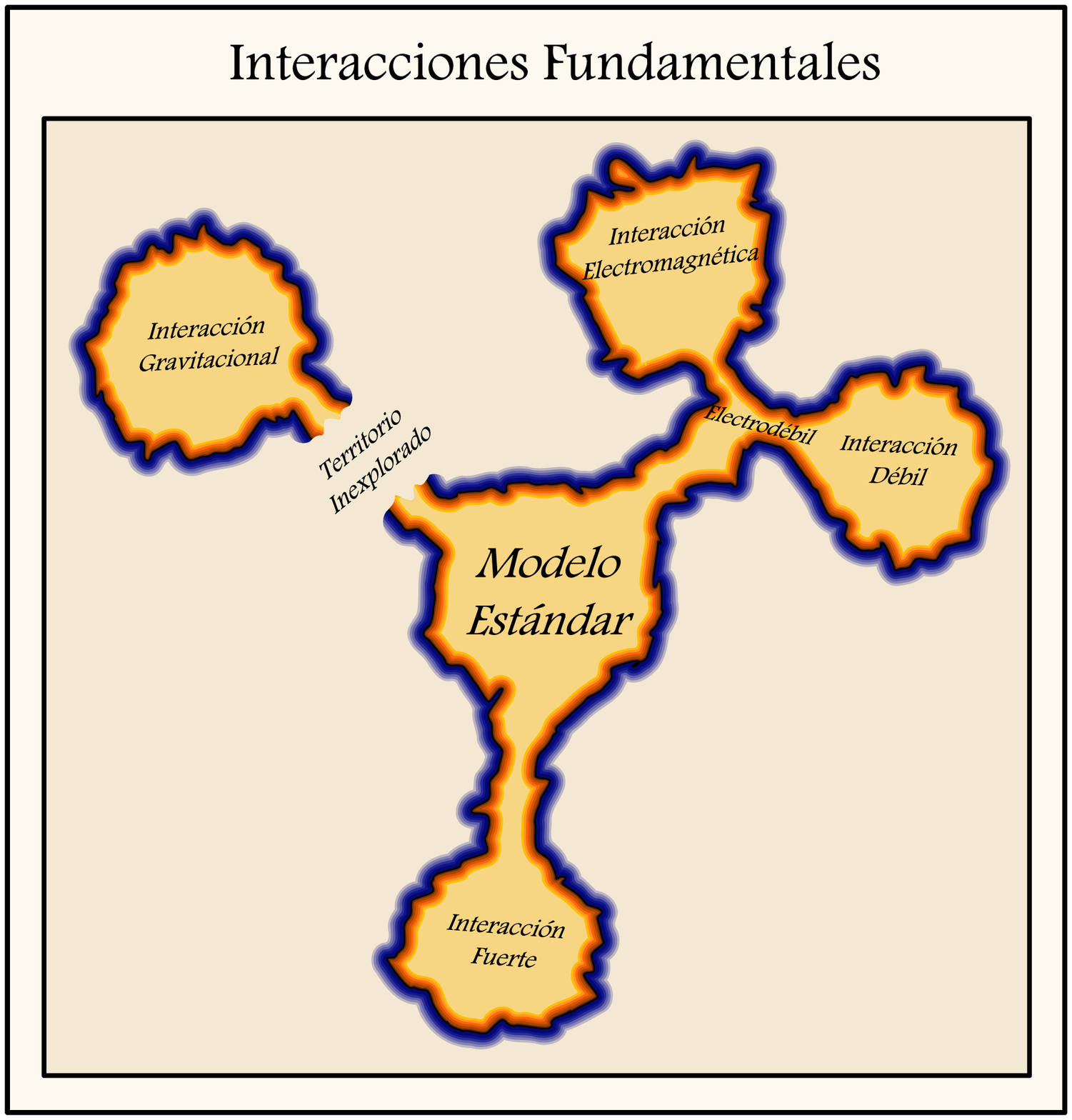}
\caption{Un \textquotedblleft mapa\textquotedblright de las cuatro interacciones fundamentales}
\label{FigCuatroInteracciones}
\end{center}
\end{figure}

En f\'{\i}sica, la situaci\'{o}n ha sido, y sigue siendo, bastante similar.
Usualmente, se empieza con fen\'{o}menos aislados, explicados a trav\'{e}s de
teor\'{\i}as \textit{ad hoc}, inconexas entre s\'{\i}. Con el tiempo, se
vuelve de manifiesto que en realidad estas teor\'{\i}as corresponden a
distintos aspectos de una \'{u}nica teor\'{\i}a, m\'{a}s sencilla y elegante,
la cual engloba las teor\'{\i}as m\'{a}s peque\~{n}as en \'{u}nico marco
unificador. Este ha sido precisamente el caso (a\'{u}n no completamente
resuelto) de las cuatro interacciones fundamentales: Electrom\'{a}gnetica,
D\'{e}bil, Fuerte y Gravitacional. As\'{\i} por ejemplo, la
Electrodin\'{a}mica Cu\'{a}ntica y la Interacci\'{o}n D\'{e}bil fueron las
primeras en llegar a ser unificadas, y posteriormente, la Cromodin\'{a}mica
cu\'{a}ntica se lig\'{o} a ellas en el contexto del Modelo Est\'{a}ndar. Por
otra parte, gravedad se ha resistido tercamente a los intentos de
unificaci\'{o}n con las otras interacciones (Ve\'{a}se Fig.~\ref{FigCuatroInteracciones}).

Quiz\'{a}s, los enfoques m\'{a}s prometedores con este objetivo en mira han
sido los entregados por Teor\'{\i}a de Cuerdas, Gravedad Cu\'{a}ntica de Lazos
y Geometr\'{\i}a No-Conmutativa. De entre estos enfoques, Teor\'{\i}a de
Cuerdas se muestra desde el inicio particularmente apropiada para unificar las
interacciones fundamentales. Esto se debe a que en este contexto, las cuatro
interacciones fundamentales corresponder\'{\i}an sencillamente a diferentes
modos de vibraci\'{o}n de un s\'{o}lo objeto fundamental, la cuerda. Como
consecuencia, la teor\'{\i}a de cuerdas ha arrojado alguna luz sobre qu\'{e}
podemos esperar de una teor\'{\i}a cu\'{a}ntica para la gravitaci\'{o}n.

Sin embargo, nuestra analog\'{\i}a con los exploradores del s.~XV sigue siendo
v\'{a}lida en este contexto. No existe s\'{o}lo una posible formulaci\'{o}n
para la Teor\'{\i}a de Cuerdas, sino cinco de ellas: Tipo~I, Tipo~IIA,
Tipo~IIB, Heter\'{o}tica~$\mathrm{SO}\left(  32\right)  $ y
Heter\'{o}tica~$\mathrm{E}_{8}\times\mathrm{E}_{8}$. Hasta antes de la
dec\'{a}da de 1990, estas teor\'{\i}as eran como \textquotedblleft islas en un
mapa\textquotedblright, completamente independientes entre s\'{\i}. Sin
embargo, a principios de esa dec\'{a}da se descubri\'{o} que estaban
estrechamente relacionadas a trav\'{e}s de una red de dualidades\footnote{Una
dualidad entre teor\'{\i}as es, en pocas palabras, un mapeo entre los estados
cu\'{a}nticos de ellas, tal que la descripci\'{o}n de estos estados es
preservada bajo este mapeo. As\'{\i} por ejemplo, es posible realizar obtener
ciertos resultados en el contexto de cierta teor\'{\i}a (amplitudes de
scattering por ejemplo), y estos pueden ser mapeados a la otra, sin perder su
validez.}. Las m\'{a}s importantes entre ellas son la Dualidad~T (dualidad
bajo el intercambio $R\rightleftarrows1/R$ del radio de compactificaci\'{o}n)
y la Dualidad~S (dualidad entre acoplamiento d\'{e}bil y fuerte). As\'{\i} por
ejemplo, Tipo~IIA y Tipo~IIB son T-duales, lo que significa que en realidad
ambas son dos descripciones equivalentes de la misma teor\'{\i}a. De la misma
forma, Heter\'{o}tica~$\mathrm{SO}\left(  32\right)  $ y
Heter\'{o}tica~$\mathrm{E}_{8}\times\mathrm{E}_{8}$ son tambi\'{e}n T-duales,
y Tipo~I y Heter\'{o}tica~$\mathrm{SO}\left(  32\right)  $ son S-duales.
As\'{\i}, tenemos simplemente dos grandes grupos de teor\'{\i}as de cuerdas:
teor\'{\i}as tipo~II y tipo~I-heter\'{o}tico.

Pero a su vez, estos dos grupos de teor\'{\i}as no est\'{a}n del todo
desconectados entre s\'{\i}. Sucede que las teor\'{\i}as de cuerdas Tipo~IIA y
Heter\'{o}tica~$\mathrm{SO}\left(  32\right)  $ tienen como l\'{\i}mite de
bajas energ\'{\i}as supergravedad~IIA en 10 dimensiones, mientras que las
teor\'{\i}as de cuerdas Tipo~IIB y Heter\'{o}tica~$\mathrm{E}_{8}%
\times\mathrm{E}_{8}$ tienen como l\'{\i}mite de bajas energ\'{\i}as
supergravedad~IIB en 10 dimensiones. En 1995, E.~Witten present\'{o} una
posible explicaci\'{o}n para esta situaci\'{o}n. Sucede que supergravedad~IIA
en 10 dimensiones tambi\'{e}n puede ser obtenida a partir de la
formulaci\'{o}n de supergravedad en 11 dimensiones debida a Cremer, Julia y
Scherk. Sin embargo, esta supergravedad no es completamente consistente por
s\'{\i} misma, y por lo tanto, parece bastante plausible la existencia de una
teor\'{\i}a m\'{a}s general en 11 dimensiones ---apodada Teor\'{\i}a~M por
Witten--- la cual tiene como l\'{\i}mite de bajas energ\'{\i}as la
supergravedad CJS y est\'{a} relacionada con teor\'{\i}a de cuerdas
\textit{via} reducci\'{o}n dimensional.

\begin{figure}[ptb]
\begin{center}
\includegraphics[width=\textwidth]{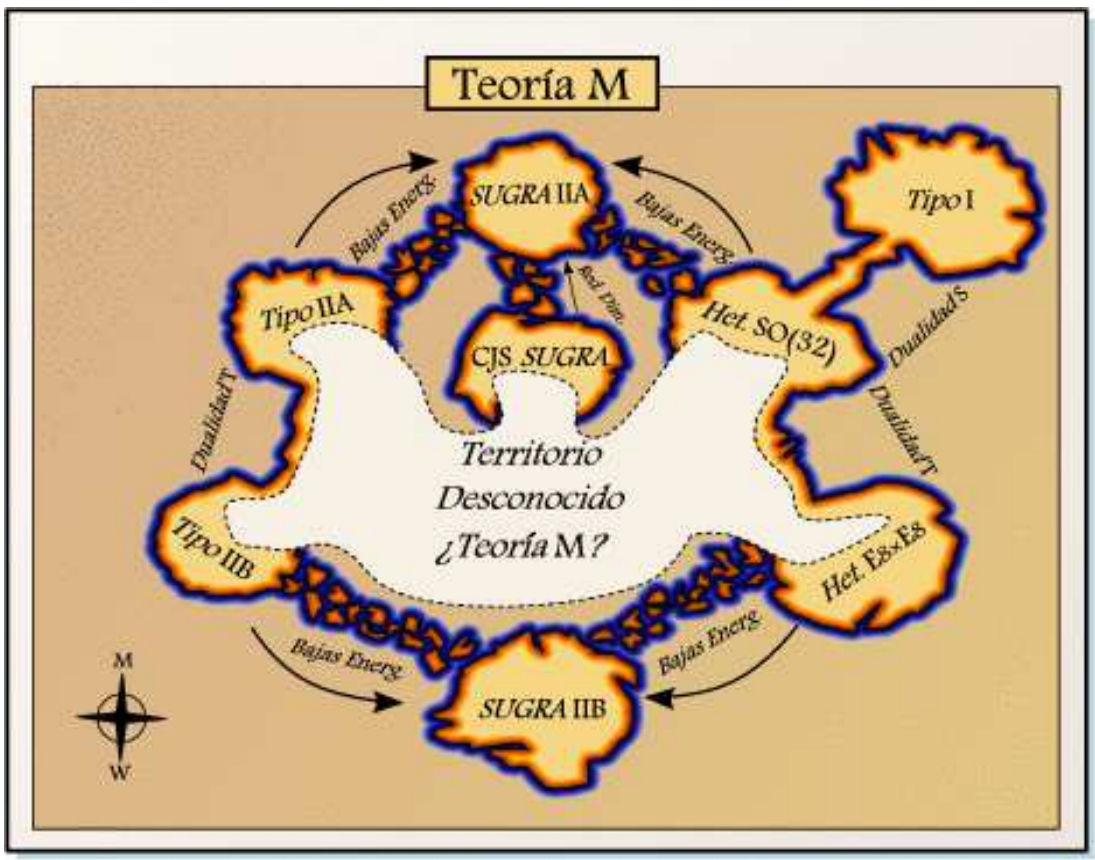}
\caption{Las Teor\'{\i}as de Cuerdas y Supergravedad est\'{a}n relacionadas a trav\'{e}s de dualidades, l\'{\i}mites de bajas energ\'{\i}as y reducci\'{o}n dimensional: esto sugiere que ellas no son mutuamente incompatibles, sino m\'{a}s bien distintos casos l\'{\i}mite de una teor\'{\i}a m\'{a}s fundamental.}
\label{FigTeoriaM}
\end{center}
\end{figure}

Curiosamente, si quisi\'{e}semos en este punto dibujar un mapa representando
este conjunto de teor\'{\i}as y sus relaciones, obtendr\'{\i}amos algo similar
a una de las ya mencionadas cartas de navegaci\'{o}n del s.~XV (Ve\'{a}se
Fig.\ref{FigTeoriaM}).

Hasta el momento s\'{o}lo se perfilan algunas regiones costeras, pero se
tienen buenas razones como para sospechar la existencia de un continente
desconocido en el interior.

Se conoce (o especula) sobre algunas de las caracter\'{\i}sticas que
deber\'{\i}a presentar la Teor\'{\i}a~M. En particular, se postula que su
acci\'{o}n deber\'{\i}a ser invariante bajo el \'{A}lgebra~M, la cual
corresponde a la extensi\'{o}n supersim\'{e}trica maximal del \'{a}lgebra de
Poincar\'{e} en 11 dimensiones, con dos cargas centrales bos\'{o}nicas,
$\boldsymbol{Z}_{ab}$ y $\boldsymbol{Z}_{a_{1}\cdots a_{5}}$. As\'{\i} su anticonmutador es de la forma%
\[
\left\{  \boldsymbol{Q},\bar{\boldsymbol{Q}}\right\}  =\frac{1}{8}\left(
\Gamma^{a}\boldsymbol{P}_{a}-\frac{1}{2}\Gamma^{ab}\boldsymbol{Z}_{ab}%
+\frac{1}{5!}\Gamma^{abcde}\boldsymbol{Z}_{abcde}\right)  .
\]

La raz\'{o}n para hacer esta elecci\'{o}n como simetr\'{\i}a fundamental es para incluir gravedad en forma natural en este contexto, para lo
cual debe asociarse el \textit{elfbein} $e^{a}$ con los generadores de
traslaciones de Poincar\'{e} $\boldsymbol{P}_{a},$ y la conexi\'{o}n de spin
$\omega^{ab}$ con los generadores de Lorentz, $\boldsymbol{J}_{ab}$.

Siguiendo esta l\'{\i}nea de pensamiento, el problema se vuelve
particularmente interesante. Cuando se parte del \textit{postulado} de que la
Teor\'{\i}a~M debe ser invariante bajo el \'{A}lgebra~M, resulta de especial
inter\'{e}s constru\'{\i}r una teor\'{\i}a de gauge para la citada
\'{a}lgebra. Esto significa que usar una 1-forma conexi\'{o}n $\boldsymbol{A}%
\left(  x\right)  =A_{\;\mu}^{A}\boldsymbol{T}_{A}\mathrm{d}x^{\mu}$ valuada
en el \'{A}lgebra~M como campo fundamental.

Dada una cierta simetr\'{\i}a, encontrar un principio de acci\'{o}n invariante
de gauge es un problema que puede llegar a ser altamente no trivial. El
procedimiento \textquotedblleft tradicional\textquotedblright\ para resolver
este problema (usado en el Modelo Est\'{a}ndar por ejemplo) es utilizar la
Acci\'{o}n de Yang--Mills,
\begin{equation}
\int\left\langle \boldsymbol{F}^{2}\right\rangle =\int\mathrm{d}x^{n}\sqrt
{-g}g^{\mu\rho}g^{\nu\sigma}\operatorname*{Tr}\left(  \boldsymbol{T}%
_{A}\boldsymbol{T}_{B}\right)  F_{\mu\nu}^{A}F_{\rho\sigma}^{B}%
.\label{F2 Yang Mills}%
\end{equation}
As\'{\i} es como se describen por ejemplo, la interacci\'{o}n electrod\'{e}bil
GWS o la Teor\'{\i}a de Gran Unificaci\'{o}n GUT (Modelo Est\'{a}ndar). Sin
embargo, existe una diferencia fundamental entre las interacciones del Modelo
Est\'{a}ndar y el \'{A}lgebra~M. Debemos observar que para escribir el
Lagrangeano de Yang--Mills, es necesaria la existencia de una m\'{e}trica
invertible. Las interacciones del Modelo Est\'{a}ndard se modelan sobre un
espacio-tiempo plano, sin din\'{a}mica, con una m\'{e}trica Minkowskiana de
fondo. Sin embargo, en el contexto de la Teor\'{\i}a~M, se debe de incluir la
interacci\'{o}n gravitacional y por lo tanto la m\'{e}trica pasa ser un campo
din\'{a}mico; as\'{\i} ya no puede constituir el fondo (\textit{background})
de la teor\'{\i}a, puesto que no pertenecer\'{\i}a a la estructura de gauge.

Sin embargo, tambi\'{e}n existe un procedimiento conocido (del cual
utlizaremos una generalizaci\'{o}n en la presente tesis) para construir
teor\'{\i}as de gauge sin utilizar una m\'{e}trica de fondo. En el a\~{n}o
1974, a partir del trabajo de Shiing-Shen Chern and James Harris Simons
(Ve\'{a}se Ref.~\cite{ChernSimons-Original}) se introdujo lo que pas\'{o} a
llamarse posteriormente Forma de Chern--Simons. Esta pas\'{o} a jugar un rol
fundamental en la construcci\'{o}n de un nuevo tipo de teor\'{\i}a de gauge a
a partir de los trabajos de E.~Witten y A.~Chammseddine en los 80 (V\'{e}ase
Refs.~\cite{Witten-(2+1)Grav,Witten-QFT Jones
Poly,Witten-TopQuantFieldTheory,Chamseddine-CS1,Chamseddine-CS2}).

Dada un \'{a}lgebra de Lie, una forma de Chern--Simons permite constru\'{\i}r
en cada dimensi\'{o}n impar con una teor\'{\i}a de gauge que no requiere una
estructura m\'{e}trica de fondo. La idea ha resultado particularmente
fruct\'{\i}fera en la construcci\'{o}n de gravedades y supergravedades como
teor\'{\i}as de gauge\footnote{En dimensiones impares, y siempre y cuando
\textit{no} se impongan constraints como la condici\'{o}n de torsi\'{o}n nula.
En dimensiones pares, el problema de escribir una teor\'{\i}a de gauge genuina
(\textit{i.e.}, sin imponer nulidad de la torsi\'{o}n) es un problema mucho
m\'{a}s d\'{\i}ficil de resolver. Algunos intentos en esta direcci\'{o}n han
sido los de MacDowell-Mansouri (Ve\'{a}se Ref.~\cite{McDowell-Mansouri}), (la
acci\'{o}n es un invariante topol\'{o}gico) y los de
Refs.~\cite{Nosotros2-GravAdS,NosotrosBrasil} en donde se utilizan
realizaciones no lineales del grupo de Lie. La relaci\'{o}n de este \'{u}ltimo
procedimiento y Chern--Simons a trav\'{e}s de reducci\'{o}n dimensional es
analizado en Refs.~\cite{Nosotros1-Lovelock,NosotrosSuperSim}.} (Ve\'{a}se por
ejemplo Refs.~\cite{Nosotros3-TransLargo,Nosotros6-SExp M,Witten-(2+1)Grav,
Chamseddine-CS1,Chamseddine-CS2, CECS CS 1ro,CECS-DimContBH,CECS Quant
Const,CECS-HighCSsugra, CECS NewSUGRA,CECS-SugraAllOdd,
CECS-HighDimGravPropgTorsion,CECS-BHScan, CECS-BHAdSAsymp,CECS-FiniteGrav,
CECS-VacuumOddDim,CECS-MAlgNoether-1, CECS-MAlgNoether-2,MoraNishino-BranaCS1,
Mora-BraneCS,CECS-Exp1, CECS-Exp2,Polaco1, Polaco2,Azcarraga-Expansion1}).

La posibilidad de utilizar formas de Chern--Simons para construir un principio
de acci\'{o}n para la Teor\'{\i}a M ha sido estudiada en Refs.~\cite{CECS
NewSUGRA}, y~\cite{Horava,Nastase}. En Ref.~\cite{CECS NewSUGRA}, la
construcci\'{o}n ha sido llevada a cabo utilizando para ello el \'{a}lgebra
$\mathfrak{osp}\left(  32|1\right)  ,$ y en las Refs.~\cite{Horava,Nastase}
usando el \'{a}lgebra $\mathfrak{osp}\left(  32|1\right)  \times
\mathfrak{osp}\left(  32|1\right)  .$

Recientemente ha sido postulada la posibilidad de explorar formas de
transgresi\'{o}n en lugar de formas de CS para constru\'{\i}r teor\'{\i}as de
gauge. En Refs.~\cite{CECS-FiniteGrav,CECS-VacuumOddDim} ha sido mostrado que
este punto de vista resulta particularmente ventajoso desde el punto
f\'{\i}sico, en particular en lo que respecta a condiciones de contorno y
definici\'{o}n de cargas conservadas.

La presente tesis muestra como construir una forma de Transgresi\'{o}n
invariante bajo el \'{A}lgebra~M. Esto significa analizar en forma exhaustiva
los v\'{\i}nculos entre el \'{a}lgebra $\mathfrak{osp}\left(  32|1\right)  $ y
el \'{A}lgebra~M. El procedimiento est\'{a}ndar para obtener el \'{A}lgebra~M
es a trav\'{e}s de un proceso de expansi\'{o}n en potencias de las formas de
Maurer--Cartan (Ve\'{a}se
Refs.~\cite{Azcarraga-Expansion1,Azcarraga-Expansion2}). Sin embargo, la
construcci\'{o}n expl\'{\i}cita de la forma de transgresi\'{o}n requiere de la
creaci\'{o}n nuevos m\'{e}todos que permitan generar nuevas \'{a}lgebras de
Lie. Estos m\'{e}todos, que abarcar\'{a}n buena parte de la presente tesis, se
basan en el uso de semigrupos abelianos, con los cuales se generan nuevas
\'{a}lgebras y sus correspondientes principios de acci\'{o}n.

\section{Plan de la Tesis}

Toda teor\'{\i}a de gauge se basa en el concepto matem\'{a}tico de fibrados.
Por ello, en el Cap\'{\i}tulo~\ref{Sec Fibrados} se hace una revisi\'{o}n de
algunos conceptos introductorios en fibrados, conexiones y grupos de Lie,
finalizando con el Teorema de Chern--Weil y las f\'{o}rmulas de Homotop\'{\i}a.

En el Cap\'{\i}tulo~\ref{SecTrans} se analizan aspectos generales de la
construcci\'{o}n de una teor\'{\i}a de campos usando formas de
transgresi\'{o}n como lagrangeano as\'{\i} como su relaci\'{o}n con una
teor\'{\i}a de Chern--Simons. A modo de ejemplo expl\'{\i}cito analizaremos la
construcci\'{o}n de gravedad como teor\'{\i}a de gauge usando formas de
transgresi\'{o}n y Chern--Simons, comparando ambas formulaciones.

En el Cap\'{\i}tulo~\ref{SecS_Exp} estudiaremos algunas caracter\'{\i}sticas
b\'{a}sicas de semigrupos y construiremos un nuevo m\'{e}todo ---al que
llamaremos $S$-Expansiones--- para generar \'{a}lgebras de Lie. En el contexto
de este m\'{e}todo, mostraremos como construir tensores invariantes, los
cuales son el ingrediente clave en la construcci\'{o}n de formas de
Chern--Simons y Transgresiones. Por \'{u}ltimo, analizaremos distintos
ejemplos expl\'{\i}citos del procedimiento, y como el \'{A}lgebra M cabe
dentro de este esquema.

En el Cap\'{\i}tulo~\ref{SecAcc_M_Alg}, se llevar\'{a} a cabo la
construcci\'{o}n de una teor\'{\i}a de gauge para el \'{A}lgebra~M a
trav\'{e}s de una forma de transgresi\'{o}n, utilizando las herramientas
matem\'{a}ticas desarrolladas en Cap\'{\i}tulos~\ref{SecTrans}
y~\ref{SecS_Exp}. A partir de este lagrangeano, analizaremos algunas
caracter\'{\i}sticas interesantes de la f\'{\i}sica cuadridimensional. Por
\'{u}ltimo, concluiremos en Cap\'{\i}tulo~\ref{SecConcl} con un an\'{a}lisis
de las posibles aplicaciones de los m\'{e}todos presentados.

\chapter{\label{Sec Fibrados}Fibrados y Conexiones}

\medskip

\begin{center}
\textquotedblleft\textit{To those who do not know mathematics it is difficult to get across a real feeling as to the beauty, the deepest beauty, of nature...}\\
\textit{If you want to learn about nature, to appreciate nature, it is
necessary to understand the language that she speaks in.}\textquotedblright

(R. Feynman\footnote{\textquotedblleft \textit{A aquellos que no conocen las matem\'{a}ticas, es d\'{\i}ficil transmitirles un sentimiento real de la belleza, la profunda belleza, de la naturaleza... Si quieres aprender sobre la naturaleza, si quieres apreciar la naturaleza, es necesario entender el lenguaje que ella habla}\textquotedblright})

\medskip

\textquotedblleft\textit{All these difficulties are but consequences of our
refusal to see that mathematics cannot be defined without acknowledging its
more obvious feature: namely, that it is interesting}\textquotedblright

(M. Polanyi, 1958, Personal Knowledge\footnote{\textquotedblleft \textit{Todas estas dificultades no son sino consecuencias de nuestro rechazo a ver que las matem\'{a}ticas no pueden ser definidas sin reconocer su caracter\'{\i}stica m\'{a}s obvia: a saber, que es interesante}\textquotedblright})

\medskip
\end{center}

Para describir una teor\'{\i}a de gauge en forma rigurosa, debemos recurrir al
concepto matem\'{a}tico de \textit{fibrado }o\textit{\ haz de fibras} (fiber
bundle). Un fibrado corresponde es una variedad que \emph{localmente}
corresponde al producto directo de dos variedades.

Este concepto matem\'{a}tico es fundamental para llevar a cabo la
construcci\'{o}n de cualquier teor\'{\i}a de gauge, y por lo tanto, se
presentar\'{a}n aqu\'{\i} en forma general pero sucinta, para aplicarlos
despu\'{e}s en la construcci\'{o}n de una teor\'{\i}a de gauge para el \'{A}lgebra~M.

Para un an\'{a}lisis detallado de estos conceptos, se recomienda el libro
Ref.~\cite{Nakahara} y especialmente el libro Ref.~\cite{Azcarraga-Libro}.

\section{Fibrados}

\begin{definition}
Un fibrado consiste de los elementos $\left\{  E,\pi,M,F\right\}  ,$ en donde
$E,M$ y $F$ corresponden a espacios topol\'{o}gicos y $\pi:E\rightarrow M$
corresponde a un mapeo continuo y sobreyectivo. Nos referiremos a $E$ como
espacio total, a $M$ como espacio base, a $F$ como fibra y a $\pi$ como proyecci\'{o}n.
\end{definition}

\begin{figure}[ptb]
\begin{center}
\includegraphics[width=.5\textwidth]{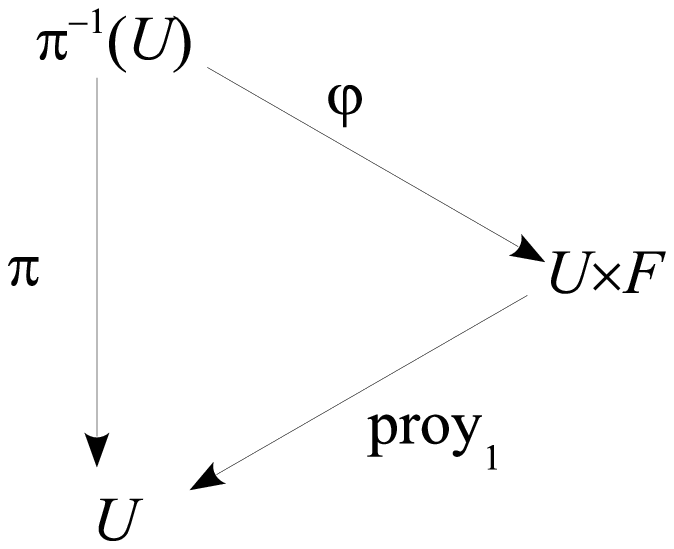}
\end{center}
\caption{Trivializaci\'{o}n Local.}%
\label{FigTrivLocal}%
\end{figure}

\begin{definition}
La condici\'{o}n de \emph{trivialidad local} consiste en requerir que para
todo $x\in M,$ exista una vecindad abierta $U$ de $x$ tal que $\pi^{-1}\left(
U\right)  $ es homeom\'{o}rfico al espacio producto $U\times F. $ El
homeomorfismo\footnote{Una funci\'{o}n $\varphi:X\rightarrow Y$ es llamada un
\textit{homeomorfismo} cuando $\varphi$ es uno a uno, continua y de inversa
continua. No confundir con un \textit{homomorfismo}.} asociado $\varphi
:\pi^{-1}\left(  U\right)  \rightarrow U\times F,$ es tal que el diagrama de
Fig.~\ref{FigTrivLocal} conmuta, en donde $\operatorname{proy}_{1}$
corresponde a la proyecci\'{o}n natural $\operatorname{proy}_{1}:U\times
F\rightarrow U.$ M\'{a}s expl\'{\i}citamente, exigiremos que para cada $x$ en
$U,$%
\begin{equation}
\operatorname{proy}_{1}\circ\varphi\circ\pi^{-1}\left(  x\right)
=x.\label{EqTrivLocal}%
\end{equation}

\end{definition}

Sea $\left\{  U_{\alpha}\right\}  $ un cubrimiento de abiertos para $M,$
$\cup_{\alpha}U_{\alpha}=M.$ Cada uno de llos tiene asociado un homeomorfismo
$\varphi_{\alpha}:\pi^{-1}\left(  U_{\alpha}\right)  \rightarrow U_{\alpha
}\times F.$ El conjunto de pares $\left\{  U_{\alpha},\varphi_{\alpha
}\right\}  $ forma una \textit{trivializaci\'{o}n local} del fibrado.As\'{\i},
para todo $x\in M,$ la preimagen $\pi^{-1}\left(  x\right)  $ es
homeom\'{o}rfica a $F$ y ser\'{a} llamada simplemente \textit{fibra en} $x$.
En lo que sigue, asumiremos que el espacio base es conexo, y que se satisface
la condici\'{o}n de trivialidad local.

\subsection{Funciones de Transici\'{o}n}

Para describir el fibrado completo en t\'{e}rminos de la trivializaci\'{o}n
local $\left\{  U_{\alpha},\varphi_{\alpha}\right\}  ,$ es necesario encontrar
condiciones de juntura en las intersecciones no vac\'{\i}as entre distintos
abiertos. Dados dos abiertos $U_{\alpha}$ y $U_{\beta},$ de intersecci\'{o}n
no nula, $U_{\alpha}\cap U_{\beta}\neq\varnothing,$ sus respectivos
homeomorfismos $\varphi_{\alpha}\ $y $\varphi_{\beta}$ en general mapear\'{a}n
en forma distinta $\pi^{-1}\left(  U_{i}\cap U_{j}\right)  $ en $\left(
U_{i}\cap U_{j}\right)  \times F,$
\begin{align*}
\varphi_{\alpha}  &  :\pi^{-1}\left(  U_{\alpha}\cap U_{\beta}\right)
\rightarrow\left(  U_{\alpha}\cap U_{\beta}\right)  \times F,\\
\varphi_{\beta}  &  :\pi^{-1}\left(  U_{\alpha}\cap U_{\beta}\right)
\rightarrow\left(  U_{\alpha}\cap U_{\beta}\right)  \times F.
\end{align*}
\'{o} m\'{a}s expl\'{\i}citamente,%
\begin{align*}
\varphi_{\alpha}\left(  p\right)   &  =\left(  \pi\left(  p\right)
,y_{\alpha}\left(  p\right)  \right)  =\left(  x,y_{\alpha}\left(  p\right)
\right) \\
\varphi_{\beta}\left(  p\right)   &  =\left(  \pi\left(  p\right)  ,y_{\beta
}\left(  p\right)  \right)  =\left(  x,y_{\beta}\left(  p\right)  \right)
\end{align*}
en donde $x\in U_{\alpha}\cap U_{\beta}$, $y_{\alpha}$ $\in F.$ Esto induce la siguiente

\begin{definition}
Componiendo los homoemorfismos $\varphi_{\alpha}$ y $\varphi_{\beta}^{-1}$,
tenemos que%
\[
\varphi_{\alpha}\circ\varphi_{\beta}^{-1}:\left(  U_{\alpha}\cap U_{\beta
}\right)  \times F\rightarrow\left(  U_{\alpha}\cap U_{\beta}\right)  \times
F.
\]
o m\'{a}s expl\'{\i}citamente,%
\begin{equation}
\varphi_{\alpha}\circ\varphi_{\beta}^{-1}\left(  x,y_{\beta}\right)  =\left(
x,\tau_{\alpha\beta}\left(  x\right)  y_{\beta}\right)  ,\label{EqFuncTrans}%
\end{equation}
en donde $\tau_{\alpha\beta}\left(  x\right)  $ corresponde a un operador
continuo actuando por la izquierda sobre los puntos de la fibra, $\tau
_{\alpha\beta}\left(  x\right)  :F\rightarrow F,$ al que llamaremos
\emph{funci\'{o}n de transici\'{o}n}. Las funciones de transici\'{o}n deben de
satisfacer las siguientes condiciones de consistencia:

\begin{enumerate}
\item $\tau_{\alpha\alpha}\left(  x\right)  =\boldsymbol{1}_{\text{{\tiny F}}%
}$ (operador identidad sobre $F$),

\item $\tau_{\alpha\beta}\left(  x\right)  =\tau_{\beta\alpha}^{-1}\left(
x\right)  ,$

\item $\tau_{\alpha\gamma}\left(  x\right)  =\tau_{\alpha\beta}\left(
x\right)  \tau_{\beta\gamma}\left(  x\right)  .$
\end{enumerate}

La \'{u}ltima condici\'{o}n debe cumplirse en el caso $U_{\alpha}\cap
U_{\beta}\cap U_{\gamma}\neq\varnothing,$ y es llamada \textit{condici\'{o}n
de cociclo}. Estas tres condiciones implican que estos operadores forman un
grupo, al que llamaremos $G$. As\'{\i}, tenemos que%
\[
\tau_{\alpha\beta}:U_{\alpha}\cap U_{\beta}\rightarrow G.
\]

\end{definition}

En el caso de un \textit{fibrado suave} (es decir, cuando $E,M$ y $F$
corresponden a variedades diferenciables, y los mapeos $\pi$ y $\varphi$ son
suaves)\textit{\ }$G$ corresponde a un grupo de Lie. De ahora en adelante,
consideraremos s\'{o}lo este tipo de fibrados. Hemos hecho actuar las
funciones de transici\'{o}n por la izquierda, pero por supuesto, esto es
s\'{o}lo materia de convenci\'{o}n. Es posible definir la acci\'{o}n de un
grupo de simetr\'{\i}a sobre la fibra por la derecha, lo cual induce en forma
\'{u}nivoca la \emph{acci\'{o}n derecha} de este grupo sobre el fibrado, tal
como en la siguiente

\begin{definition}
Sea $g$ un elemento de un grupo de simetr\'{\i}a $G$. Denotemos la acci\'{o}n
derecha de $g$ sobre sobre la fibra como%
\[
y_{\beta}^{\prime}\left(  p\right)  =y_{\beta}\left(  p\right)  g.
\]

Sean dos puntos del fibrado principal, $p$ y $p^{\prime}$ en $P.$ Diremos que%
\[
p^{\prime}=pg
\]
siempre y cuando se cumplan las condiciones%
\begin{align*}
\pi\left(  pg\right)   &  =\pi\left(  p\right)  ,\\
y_{\beta}\left(  pg\right)   &  =y_{\beta}\left(  p\right)  g.
\end{align*}

\end{definition}

Esta acci\'{o}n inducida sobre el fibrado es completamente independiente del
homeomorfismo $\varphi_{\beta}$ escogido. En efecto, de ec.~(\ref{EqFuncTrans}%
), tenemos que%
\[
y_{\alpha}\left(  p\right)  =\tau_{\alpha\beta}\left(  x\right)  y_{\beta
}\left(  p\right)
\]
y por lo tanto,%
\begin{align*}
y_{\alpha}\left(  pg\right)   &  =\tau_{\alpha\beta}\left(  x\right)
y_{\beta}\left(  pg\right)  ,\\
&  =\tau_{\alpha\beta}\left(  x\right)  y_{\beta}\left(  p\right)  g,\\
&  =y_{\alpha}\left(  p\right)  g.
\end{align*}

As\'{\i}, vemos que el concepto de acci\'{o}n derecha sobre el fibrado es
independiente de la elecci\'{o}n de $y_{\alpha}$.

Ambas operaciones, actuar por la izquierda y por la derecha sobre la fibra
ser\'{a}n importantes en lo que sigue. Usaremos la notaci\'{o}n $L_{g}$ para
denotar la acci\'{o}n izquierda de $g$ y $R_{g}$ para la derecha.

Por otra parte, la existencia de acciones por la izquierda y por la derecha
sugiere la elecci\'{o}n de una estructura particularmente elegante de fibrado:
el \textit{fibrado principal}.

\begin{definition}
Un \emph{fibrado principal} es aqu\'{e}l en donde

\begin{enumerate}
\item la fibra $F$,

\item el conjunto de funciones de transici\'{o}n $\left\{  \tau_{\alpha\beta
}\right\}  $ y

\item el grupo de simetr\'{\i}a $G$ que act\'{u}a por la derecha
\end{enumerate}

coinciden y corresponden al mismo grupo de Lie, al cual llamaremos $G$.
\end{definition}

Como veremos a continuaci\'{o}n, este tipo de fibrados representa el bloque
b\'{a}sico de construcci\'{o}n de las teor\'{\i}as de gauge. Para un fibrado
principal, la convenci\'{o}n usual en la literatura es llamar al espacio total
no $E,$ sino m\'{a}s bien $P.$ De ahora en adelante, todo el desarrollo
ser\'{a} llevado a cabo en el contexto de fibrados principales, \textit{i.e.},
con $F=G$.

\subsection{Secciones Locales}

\begin{definition}
Dado el cubrimiento de abiertos $\left\{  U_{\alpha}\right\}  $ para $M,$ es
posible definir una \textit{secci\'{o}n local }$\sigma_{\alpha}$, como el
mapeo%
\[
\sigma_{\alpha}:U_{\alpha}\rightarrow P
\]
tal que para todo $x$ en $U_{\alpha}$cumple%
\[
\pi\circ\sigma_{\alpha}\left(  x\right)  =x.
\]

\end{definition}

Una secci\'{o}n $\sigma_{\alpha}$ y un homeomorfismo local $\varphi_{\alpha
}:\pi^{-1}\left(  U_{\alpha}\right)  \rightarrow U_{\alpha}\times G$ est\'{a}n
int\'{\i}mamente relacionados. En efecto, dado un homeomorfismo $\varphi
_{\alpha}$ este induce una \textit{secci\'{o}n natural} como%
\[
\sigma_{\alpha}\left(  x\right)  =\varphi_{\alpha}^{-1}\left(  x,e\right)
\]
en donde $e$ es la identidad de $G.$ Lo inverso tambi\'{e}n es cierto: una
secci\'{o}n induce el homeomorfismo local rec\'{\i}proco. As\'{\i}, sea el
mapeo $y_{\alpha}:\pi^{-1}\left(  U_{\alpha}\right)  \rightarrow G$ tal que%
\begin{equation}
\sigma_{\alpha}\left(  x\right)  y_{\alpha}\left(  p\right)
=p,\label{EcSigmaY(p)=p}%
\end{equation}
en donde $y_{\alpha}\left(  p\right)  \in G$ est\'{a} actuando por la derecha
sobre el punto $\sigma_{\alpha}\left(  x\right)  .$ Es trivial demostrar que
esta definici\'{o}n satisface la condici\'{o}n $y_{\alpha}\left(  pg\right)
=y_{\alpha}\left(  p\right)  g$. Esta definici\'{o}n tambi\'{e}n es
autoconsistente, pues $p\left[  y_{\alpha}\left(  p\right)  \right]
^{-1}=\sigma_{\alpha}\left(  x\right)  $ sin depender de $p.$ En efecto, sea
$p^{\prime}=pg;$ entonces%
\begin{align*}
p^{\prime}\left[  y_{\alpha}\left(  p^{\prime}\right)  \right]  ^{-1}  &
=pg\left[  y_{\alpha}\left(  pg\right)  \right]  ^{-1},\\
&  =pgg^{-1}\left[  y_{\alpha}\left(  p\right)  \right]  ^{-1},\\
&  =p\left[  y_{\alpha}\left(  p\right)  \right]  ^{-1}.
\end{align*}

As\'{\i}, vemos que es posible definir%
\[
\varphi_{\alpha}\left(  p\right)  =\left(  x,y_{\alpha}\left(  p\right)
\right)  ,
\]
y en donde la condici\'{o}n~(\ref{EcSigmaY(p)=p}) implica que $\sigma_{\alpha
}\left(  x\right)  =\varphi_{\alpha}^{-1}\left(  x,e\right)  .$

Las secciones locales $\sigma_{\alpha}$ y $\sigma_{\beta}$ sobre $U_{\alpha
}\cap U_{\beta}$ est\'{a}n relacionadas entre s\'{\i} a trav\'{e}s de las
funciones de transici\'{o}n $\tau_{\alpha\beta}:U_{\alpha}\cap U_{\beta
}\rightarrow G$. Sean las secciones naturales $\sigma_{\alpha}$ y
$\sigma_{\beta};$ de ec.~(\ref{EcSigmaY(p)=p}) tenemos que%
\[
\sigma_{\alpha}\left(  x\right)  y_{\alpha}\left(  p\right)  =\sigma_{\beta
}\left(  x\right)  y_{\beta}\left(  p\right)
\]
y por lo tanto%
\begin{align}
\sigma_{\beta}\left(  x\right)   &  =\sigma_{\alpha}\left(  x\right)
y_{\alpha}\left(  p\right)  \left[  y_{\beta}\left(  p\right)  \right]
^{-1},\nonumber\\
&  =\sigma_{\alpha}\left(  x\right)  \tau_{\alpha\beta}\left(  x\right)
,\label{EcSigmaBeta=SigmaAlfaTauAlfaBeta}%
\end{align}
ya que%
\[
y_{\alpha}\left(  p\right)  =\tau_{\alpha\beta}\left(  x\right)  y_{\beta
}\left(  p\right)  .
\]

\subsection{Grupo de Simetr\'{\i}a}

Dado que la fibra en nuestro caso corresponde a la variedad del grupo $G,$ es
conveniente precisar algunas de sus caracter\'{\i}sticas fundamentales. En la
presente tesis trabajaremos en general con supergrupos y super\'{a}lgebras,
por lo tanto, usaremos indistintamente las palabras \textquotedblleft
grupo\textquotedblright\ y \textquotedblleft\'{a}lgebra\textquotedblright,
tambi\'{e}n para supergrupos y super\'{a}lgebras.

El espacio tangente $T_{e}G$ de $G$ sobre el elemento identidad $e$,
constituye el \'{a}lgebra de Lie $\mathfrak{g}$ asociada a $G$. Sea $\left\{
\boldsymbol{T}_{A}\right\}  $ una base de $T_{e}G=\mathfrak{g};$ entonces%
\begin{equation}
\left[  \boldsymbol{T}_{A},\boldsymbol{T}_{B}\right]  =C_{AB}^{\quad
\;C}\boldsymbol{T}_{C},\label{EcAlgebraLie}%
\end{equation}
en donde $C_{AB}^{\quad\;C}$ son las constantes de estructura, las cuales
satisfacen\footnote{Como es est\'{a}ndar, usaremos $\mathfrak{q}\left(
A\right)  =0$ para $\boldsymbol{T}_{A}$ bos\'{o}nico y $\mathfrak{q}\left(
A\right)  =1$ para $\boldsymbol{T}_{A}$ fermi\'{o}nico.}%
\[
C_{AB}^{\quad\;C}=\left(  -1\right)  ^{\mathfrak{q}\left(  A\right)
\mathfrak{q}\left(  B\right)  }C_{BA}^{\quad\;C},
\]
y la identidad de Jacobi,%
\[
\left(  -1\right)  ^{\mathfrak{q}\left(  A\right)  \mathfrak{q}\left(
D\right)  }C_{AB}^{\phantom{AB}C}C_{CD}^{\phantom{CD}E}+\left(  -1\right)
^{\mathfrak{q}\left(  D\right)  \mathfrak{q}\left(  B\right)  }C_{DA}%
^{\phantom{DA}C}C_{CB}^{\phantom{CB}E}+\left(  -1\right)  ^{\mathfrak{q}%
\left(  B\right)  \mathfrak{q}\left(  A\right)  }C_{BD}^{\phantom{BD}C}%
C_{CA}^{\phantom{CA}E}=0.
\]
. En general, usaremos tipograf\'{\i}a $\boldsymbol{negrita}$ o
$\mathbb{PIZARRA}$ para denotar objetos valuados en $\mathfrak{g}.$

El conmutador ec.~(\ref{EcAlgebraLie}) induce la noci\'{o}n de conmutador
sobre formas diferenciales valuadas en el \'{a}lgebra. Sean $\boldsymbol{M}$ y
$\boldsymbol{N}$ una $m$ y $n$-forma respectivamente sobre una variedad $V, $
valuadas en el \'{a}lgebra $\mathfrak{g}$, \textit{i.e.,}%
\begin{align*}
\boldsymbol{M}  &  =M^{A}\boldsymbol{T}_{A},\\
\boldsymbol{N}  &  =N^{A}\boldsymbol{T}_{A},
\end{align*}
en donde $M^{A}\in\Omega^{m}\left(  V\right)  $ y $N^{A}\in\Omega^{n}\left(
V\right)  $. Se define el conmutador entre $\boldsymbol{M}$ y $\boldsymbol{N}$
como%
\begin{align*}
\left[  \boldsymbol{M},\boldsymbol{N}\right]   &  =M^{A}N^{B}\left[
\boldsymbol{T}_{A},\boldsymbol{T}_{B}\right]  ,\\
&  =M^{A}N^{B}C_{AB}^{\quad\;C}\boldsymbol{T}_{C}.
\end{align*}

Obs\'{e}rvese que en la definici\'{o}n de $\left[  \boldsymbol{M}%
,\boldsymbol{N}\right]  $ el producto \textquotedblleft$\wedge$%
\textquotedblright\ est\'{a} impl\'{\i}cito; $M^{A}N^{B}=M^{A}\wedge N^{B}.$

Cuando el conmutador del \'{a}lgebra viene dado por la representaci\'{o}n
inducida por el \'{a}lgebra cobertora universal,%
\[
\left[  \boldsymbol{T}_{A},\boldsymbol{T}_{B}\right]  =\boldsymbol{T}%
_{A}\boldsymbol{T}_{B}-\left(  -1\right)  ^{\mathfrak{q}\left(  A\right)
\mathfrak{q}\left(  B\right)  }\boldsymbol{T}_{B}\boldsymbol{T}_{A},
\]
entonces\footnote{Cuando escribimos $\boldsymbol{M}=m^{a}\boldsymbol{B}%
_{a}+\mu^{\alpha}\boldsymbol{F}_{\alpha},$ siendo $\boldsymbol{B}_{a}$ un
generador bos\'{o}nico y $\boldsymbol{F}_{\alpha}$ uno fermi\'{o}nico, $m^{a}$
corresponder\'{a} a una forma real, y $\mu^{\alpha}$ a una de Grassmann, por
lo que $\boldsymbol{M}$ siempre ser\'{a} una forma \textquotedblleft
bos\'{o}nica\textquotedblright.}%
\begin{equation}
\left[  \boldsymbol{M},\boldsymbol{N}\right]  =\boldsymbol{MN}-\left(
-1\right)  ^{mn}\boldsymbol{NM}.\label{EcConmFormas}%
\end{equation}

Debido a que $T_{e}G=\mathfrak{g},$ resulta interesante definir un mapeo desde
vectores en un espacio tangente $T_{g}G$ hacia el espacio tangente $T_{e}G;$
para ello debemos definir cierta notaci\'{o}n, como sigue a continuaci\'{o}n.

Sean $V$ y $W$ dos variedades y sea $f:V\rightarrow W$ un mapeo entre ellas.
Denotaremos con $f^{\ast}$ la \emph{imagen rec\'{\i}proca} (\textit{pullback})
inducida por $f:V\rightarrow W$ sobre una forma en $W$ hacia $V,$ y con
$f_{\ast}$ la \emph{imagen directa} (\textit{pushforward}) inducida sobre un
vector en $V$ hacia $W$.

De esta forma, dado un vector $X\left(  g\right)  \in T_{g}G,$ el
correspondiente elemento del \'{a}lgebra de Lie, $\boldsymbol{X}%
\in\mathfrak{g}=T_{e}G$ est\'{a} dado por%
\[
\boldsymbol{X}=L_{g^{-1}\ast}\left(  X\left(  g\right)  \right)  .
\]

Esta operaci\'{o}n induce la definici\'{o}n de la \textit{forma can\'{o}nica}
de un grupo de Lie o \textit{forma de Maurer--Cartan}, $\boldsymbol{a}%
_{+}\left(  g\right)  \in\Omega^{1}\left(  G\right)  \otimes\mathfrak{g},$
como la forma que satisface%
\[
\boldsymbol{a}_{+}\left(  g\right)  \left(  X\left(  g\right)  \right)
=\boldsymbol{X}.
\]

Dada una representaci\'{o}n matricial de $G$\textbf{,} es directo demostrar
que la condici\'{o}n%
\[
\boldsymbol{a}_{+}\left(  g\right)  \left(  X\left(  g\right)  \right)
=L_{g^{-1}\ast}\left(  X\left(  g\right)  \right)
\]
implica que%
\[
\boldsymbol{a}_{+}\left(  g\right)  =g^{-1}\mathrm{d}_{\text{{\tiny G}}}g
\]
en donde $\mathrm{d}_{\text{{\tiny G}}}$ corresponde a la derivada exterior
sobre $G.$

Las componentes $a^{A}$ de $\boldsymbol{a}_{+}=a_{+}^{A}\boldsymbol{T}_{A}$
constituyen una representaci\'{o}n dual del \'{a}lgebra de Lie, y satisfacen
las \textit{ecuaciones de estructura de Maurer--Cartan},%
\begin{equation}
\mathrm{d}_{\text{{\tiny G}}}a_{+}^{C}+\frac{1}{2}C_{AB}^{\quad\;C}a_{+}%
^{A}a_{+}^{B}=0,\label{EcStructuraMC}%
\end{equation}
o en t\'{e}rminos de $\boldsymbol{a}_{+},$%
\begin{equation}
\mathrm{d}_{\text{{\tiny G}}}\boldsymbol{a}_{+}+\frac{1}{2}\left[
\boldsymbol{a}_{+},\boldsymbol{a}_{+}\right]  =0.\label{EcMC F=0}%
\end{equation}

Las ecuaciones~(\ref{EcAlgebraLie}) y~(\ref{EcStructuraMC}) son duales una con
respecto a la otra; contienen la misma informaci\'{o}n. As\'{\i} mismo, el
dual de la identidad de Jacobi es simplemente la derivada exterior de
ecuaci\'{o}n~(\ref{EcStructuraMC}),%
\[
\frac{1}{2}a_{+}^{D}a_{+}^{A}a_{+}^{B}C_{AB}^{\quad\;C}C_{DC}^{\quad\;E}=0,
\]
o en t\'{e}rminos de $\boldsymbol{a}_{+},$%
\[
\frac{1}{2}\left[  \boldsymbol{a}_{+},\left[  \boldsymbol{a}_{+}%
,\boldsymbol{a}_{+}\right]  \right]  =0.
\]

\section{Conexi\'{o}n sobre Fibrados Principales}

Dado que localmente el fibrado tiene una estructura del tipo $U\times G,$
parece razonable esperar que el espacio tangente al fibrado pueda
descomponerse en una estructura de suma directa. La descomposici\'{o}n se
efect\'{u}a en un subespacio tangente \textquotedblleft
vertical\textquotedblright, tangente a la fibra $G$, y uno horizontal,
\textquotedblleft ortogonal\textquotedblright\ a \'{e}l. Esta operaci\'{o}n se
lleva a cabo a trav\'{e}s de la llamada \textit{Conexi\'{o}n de Ehresmann}.

Denotaremos con $\mathrm{d}_{\text{{\tiny P}}}$ la derivada exterior sobre el
fibrado $P$ y con $\mathrm{d}_{\text{{\tiny M}}}$ la derivada exterior sobre
la variedad base $M.$

Sea $T_{p}P$ el espacio tangente del fibrado principal en $p$. Lo
descompondremos como $T_{p}P=V_{p}P\oplus H_{p}P$ en donde $V_{p}P$
corresponde al espacio tangente a la fibra (subespacio vertical) y $H_{p}P$ a
su complemento (subespacio horizontal), tal como se muestra en la siguiente

\begin{definition}
El subespacio vertical $V_{p}P$ de $T_{p}P$ estar\'{a} definido como el kernel
de $\pi_{\ast},$ \textit{i.e.},%
\begin{equation}
V_{p}P=\left\{  Y\in T_{p}P\text{ tal que }\pi_{\ast}\left(  Y\right)
=0\right\}  .\label{EcVp=KernelPi*}%
\end{equation}

\end{definition}

Por otra parte, para definir el subespacio horizontal en forma un\'{\i}voca,
se debe definir primero una \textit{conexi\'{o}n sobre el fibrado} como en la siguiente

\begin{definition}
Sea $\mathbb{A}\in\Omega^{1}\left(  P\right)  \otimes\mathfrak{g}$ una 1-forma
sobre $P$ valuada en el \'{a}lgebra de Lie, la cual satisface las siguientes condiciones:

\begin{enumerate}
\item La forma $\mathbb{A}$ es continua y suave sobre $P,$

\item para todo $Y\in V_{p}P$ se cumple que
\begin{equation}
\mathbb{A}\left(  Y\right)  =\boldsymbol{a}_{+}\left(  y_{\alpha}\left(
p\right)  \right)  \left(  y_{\alpha\ast}Y\right)  =\boldsymbol{Y}%
,\label{EcA(Y)=Y_Fibrado}%
\end{equation}

\item y la acci\'{o}n derecha del grupo viene dada por%
\begin{equation}
R_{g}^{\ast}\left(  \mathbb{A}\left(  pg\right)  \right)  =g^{-1}%
\mathbb{A}\left(  p\right)  g,\label{EcRgA=gAq-1_Fibrado}%
\end{equation}
en donde $R_{g}^{\ast}$ es la imagen rec\'{\i}proca inducida por la acci\'{o}n
derecha, $p^{\prime}=pg$.
\end{enumerate}

Entonces, $\mathbb{A}$ ser\'{a} llamada una \emph{conexi\'{o}n sobre el
fibrado}.
\end{definition}

Es interesante observar que de la primera condici\'{o}n se tiene que
$\mathbb{A}$ debe ser definida globalmente sobre todo el fibrado. La segunda
condici\'{o}n implica que $\mathbb{A}$ asocia a cada vector del subespacio
vertical su correspondiente elemento del \'{a}lgebra de Lie. Aqu\'{\i}
$y_{\alpha\ast}$ representa a la imagen directa $y_{\alpha\ast}:T_{p}%
P\rightarrow G$ inducida por el mapeo $y_{\alpha}:P\rightarrow G$.

Una definici\'{o}n de conexi\'{o}n $\mathbb{A}$ induce una definici\'{o}n de
subespacio horizontal, tal como en la siguiente

\begin{definition}
El subespacio horizontal ser\'{a} definido como el kernel de $\mathbb{A}$,
\textit{i.e.,}%
\[
H_{p}P\equiv\left\{  X\in T_{p}P\text{ tal que }\mathbb{A}\left(  X\right)
=0\right\}  .
\]

\end{definition}

As\'{\i}, dada una conexi\'{o}n $\mathbb{A}$, tenemos una definici\'{o}n
un\'{\i}voca para $H_{p}P$. Esta definici\'{o}n cumple con la condici\'{o}n de
consistencia%
\begin{equation}
H_{pg}P=R_{g\ast}H_{p}P,\label{EcHpgP=RgHpP}%
\end{equation}
\textit{i.e.}, la distribuci\'{o}n de $H_{p}P$ es invariante bajo la
acci\'{o}n de $G.$ En efecto, sea $X\in H_{p}P.$ Entonces,%
\[
\mathbb{A}\left(  R_{g\ast}\left(  X\right)  \right)  =R_{g}^{\ast}\left(
\mathbb{A}\left(  X\right)  \right)  =0,
\]
y por lo tanto, $R_{g\ast}X\in H_{pg}P$.

\subsection{Transformaciones de Gauge}

La conexi\'{o}n sobre el fibrado $\mathbb{A}$ est\'{a} \'{\i}ntimamente ligada
con el concepto de conexi\'{o}n encontrado en teor\'{\i}as de gauge. Definamos
sobre cada abierto $U_{\alpha}$ de $M$ la \textit{conexi\'{o}n de gauge}
$\boldsymbol{A}_{\alpha}\in\Omega^{1}\left(  U_{\alpha}\right)  \otimes
\mathfrak{g}$ como%
\begin{equation}
\boldsymbol{A}_{\alpha}=\sigma_{\alpha}^{\ast}\left(  \mathbb{A}\right)
.\label{EcAalfa=SigmaAlfaA}%
\end{equation}

Por lo tanto, dados dos abiertos $U_{\alpha}$ y $U_{\beta}$ de
intersecci\'{o}n no nula, $U_{\alpha}\cap U_{\beta}\neq\varnothing,$ tenemos
sobre $U_{\alpha}\cap U_{\beta}$ dos conexiones de gauge, $\boldsymbol{A}%
_{\alpha}=\sigma_{\alpha}^{\ast}\left(  \mathbb{A}\right)  $ y $\boldsymbol{A}%
_{\beta}=\sigma_{\beta}^{\ast}\left(  \mathbb{A}\right)  .$ Dado que las
secciones $\sigma_{\alpha}$ y $\sigma_{\beta}$ est\'{a}n relacionadas a
trav\'{e}s de las funciones de transici\'{o}n $\tau_{\alpha\beta}$[v\'{e}ase
ec.$~$(\ref{EcSigmaBeta=SigmaAlfaTauAlfaBeta})], entonces $\boldsymbol{A}%
_{\alpha}$ y $\boldsymbol{A}_{\beta}$ tambi\'{e}n deben de estarlo. Para
encontrar la relaci\'{o}n entre $\boldsymbol{A}_{\alpha}$ y $\boldsymbol{A}%
_{\beta}$, debemos de considerar cuidadosamente ec.$~$%
(\ref{EcSigmaBeta=SigmaAlfaTauAlfaBeta}),%
\[
\sigma_{\beta}\left(  x\right)  =\sigma_{\alpha}\left(  x\right)  \tau
_{\alpha\beta}\left(  x\right)  .
\]

Aqu\'{\i}, $\tau_{\alpha\beta}\left(  x\right)  $ act\'{u}a como un elemento
del grupo por la derecha sobre el punto $\sigma_{\alpha}\left(  x\right)  ,$
\[
\tau_{\alpha\beta}\left(  x\right)  :\sigma_{\alpha}\left(  x\right)
\rightarrow\sigma_{\beta}\left(  x\right)  .
\]

Por otra parte, $\tau_{\alpha\beta}$ le asigna a cada punto de $U_{\alpha}\cap
U_{\beta}$ un elemento del grupo,
\[
\tau_{\alpha\beta}:U_{\alpha}\cap U_{\beta}\rightarrow G.
\]
De la misma forma, $\sigma_{\alpha}\left(  x\right)  $ es por definici\'{o}n
un mapeo
\[
\sigma_{\alpha}:U_{\alpha}\rightarrow P,
\]
pero tambi\'{e}n act\'{u}a como un mapeo
\begin{align*}
\sigma_{\alpha}\left(  x\right)   &  :G\rightarrow P\\
\sigma_{\alpha}\left(  x\right)   &  :\tau_{\alpha\beta}\left(  x\right)
\rightarrow\sigma_{\alpha}\left(  x\right)  \tau_{\alpha\beta}\left(
x\right)
\end{align*}
Este doble estatus de $\tau_{\alpha\beta}\ $y $\sigma_{\alpha}$ hace que no
sea trivial expresar la imagen rec\'{\i}proca $\sigma_{\beta}^{\ast}$ en
t\'{e}rminos de $\sigma_{\alpha}^{\ast}$, pues debemos considerar cada una de
estos roles al escribir la imagen rec\'{\i}proca.

En efecto, consideremos un vector $X\in T_{x}\left(  U_{\alpha}\cap U_{\beta
}\right)  .$ Entonces, la imagen directa $\sigma_{\beta\ast}:T_{x}\left(
U_{\alpha}\cap U_{\beta}\right)  \rightarrow T_{\sigma_{\beta}\left(
x\right)  }P$ puede expresarse como%
\begin{align*}
\sigma_{\beta\ast}X  &  =\left[  \sigma_{\alpha}\left(  x\right)  \tau
_{\alpha\beta}\left(  x\right)  \right]  _{\ast}\left(  X\right) \\
&  =R_{\tau_{\alpha\beta}\ast}\circ\sigma_{\alpha\ast}\left(  X\right)
+\sigma_{\alpha}\left(  x\right)  _{\ast}\circ\tau_{\alpha\beta\ast}\left(
X\right)
\end{align*}
en donde tenemos las im\'{a}genes directas respectivas,%
\begin{align*}
\tau_{\alpha\beta\ast}  &  :T_{x}\left(  U_{\alpha}\cap U_{\beta}\right)
\rightarrow T_{\tau_{\alpha\beta}\left(  x\right)  }G,\\
\sigma_{\alpha}\left(  x\right)  _{\ast}  &  :T_{\tau_{\alpha\beta}\left(
x\right)  }G\rightarrow T_{\sigma_{\alpha}\left(  x\right)  \tau_{\alpha\beta
}\left(  x\right)  }P,\\
\sigma_{\alpha\ast}  &  :T_{x}\left(  U_{\alpha}\cap U_{\beta}\right)
\rightarrow T_{\sigma_{\alpha}\left(  x\right)  }P,\\
R_{\tau_{\alpha\beta}\ast}  &  :T_{\sigma_{\alpha}\left(  x\right)
}P\rightarrow T_{\sigma_{\alpha}\left(  x\right)  \tau_{\alpha\beta}\left(
x\right)  }P.
\end{align*}

Por lo tanto, como $\boldsymbol{A}_{\beta}\left(  X\right)  =\sigma_{\beta
}^{\ast}\mathbb{A}\left(  X\right)  ,$ tenemos
\begin{align*}
\boldsymbol{A}_{\beta}\left(  X\right)   &  =\sigma_{\beta}^{\ast}%
\mathbb{A}\left(  X\right) \\
&  =\mathbb{A}\left(  \sigma_{\beta\ast}\left(  X\right)  \right) \\
&  =\mathbb{A}\left(  R_{\tau_{\alpha\beta}\ast}\circ\sigma_{\alpha\ast
}\left(  X\right)  \right)  +\mathbb{A}\left(  \sigma_{\alpha}\left(
x\right)  _{\ast}\circ\tau_{\alpha\beta\ast}\left(  X\right)  \right)
\end{align*}

Ahora bien, usando ec.~(\ref{EcA(Y)=Y_Fibrado}), tenemos que%
\[
\mathbb{A}\left(  \sigma_{\alpha}\left(  x\right)  _{\ast}\circ\tau
_{\alpha\beta\ast}\left(  X\right)  \right)  =\boldsymbol{a}_{+}\left(
\tau_{\alpha\beta}\right)  \left(  \tau_{\alpha\beta\ast}\left(  X\right)
\right)  ,
\]
y por lo tanto,%
\[
\boldsymbol{A}_{\beta}\left(  X\right)  =\sigma_{\alpha}^{\ast}\left(
R_{\tau_{\alpha\beta}}^{\ast}\left(  \mathbb{A}\left(  X\right)  \right)
\right)  +\boldsymbol{a}_{+}\left(  \tau_{\alpha\beta}\right)  \left(
\tau_{\alpha\beta\ast}\left(  X\right)  \right)  .
\]

Por otra parte, usando $\boldsymbol{a}_{+}\left(  g\right)  =g^{-1}%
\mathrm{d}_{\text{{\tiny G}}}g,$ tenemos%
\begin{align*}
\boldsymbol{A}_{\beta}\left(  X\right)   &  =\sigma_{\alpha}^{\ast}\left(
\tau_{\alpha\beta}^{-1}\mathbb{A}\tau_{\alpha\beta}\right)  \left(  X\right)
+\tau_{\alpha\beta}^{-1}\mathrm{d}_{\text{{\tiny G}}}\tau_{\alpha\beta}\left(
\tau_{\alpha\beta\ast}\left(  X\right)  \right)  ,\\
&  =\tau_{\alpha\beta}^{-1}\sigma_{\alpha}^{\ast}\left(  \mathbb{A}\right)
\tau_{\alpha\beta}\left(  X\right)  +\tau_{\alpha\beta}^{\ast}\left(
\tau_{\alpha\beta}^{-1}\mathrm{d}_{\text{{\tiny G}}}\tau_{\alpha\beta}\right)
\left(  X\right)  ,\\
&  =\left(  \tau_{\alpha\beta}^{-1}\boldsymbol{A}_{\alpha}\tau_{\alpha\beta
}+\tau_{\alpha\beta}^{-1}\mathrm{d}_{\text{{\tiny M}}}\tau_{\alpha\beta
}\right)  \left(  X\right)  ,
\end{align*}
y por lo tanto, arribamos a%
\begin{equation}
\boldsymbol{A}_{\beta}=\tau_{\alpha\beta}^{-1}\boldsymbol{A}_{\alpha}%
\tau_{\alpha\beta}+\tau_{\alpha\beta}^{-1}\mathrm{d}_{\text{{\tiny M}}}%
\tau_{\alpha\beta}.\label{EcTransfGauge}%
\end{equation}

La ecuaci\'{o}n~(\ref{EcTransfGauge}) corresponde precisamente a lo que en
f\'{\i}sica se conoce como \textit{transformaciones de gauge}. Desde el punto
de vista del fibrado, debemos observar que siempre existe $\tau_{\alpha\beta
}\left(  x\right)  \in G$ tal que la conexi\'{o}n $\boldsymbol{A}_{\alpha}$
sobre $U_{\alpha}$ y la conexi\'{o}n $\boldsymbol{A}_{\beta}$ sobre $U_{\beta
}$ est\'{e}n relacionadas por \ref{EcTransfGauge} sobre $U_{\alpha}\cap
U_{\beta}.$ Esto es simplemente un reflejo de que siempre es posible
descomponer el espacio tangente como $T_{p}P=V_{p}P\oplus H_{p}P$ en cada
punto del fibrado.

Para no recargar la notaci\'{o}n, en el cap\'{\i}tulo siguiente escribiremos
simplemente $g=\tau_{\alpha\beta}$ y ec.~(\ref{EcTransfGauge}) simplemente
como%
\begin{equation}
\boldsymbol{A}\rightarrow\boldsymbol{A}^{\prime}=g^{-1}\boldsymbol{A}%
g+\boldsymbol{a}_{+},\label{Ec TransfGauge a+}%
\end{equation}
con\footnote{N\'{o}tese que hay un peque\~{n}o abuso de notaci\'{o}n; se
est\'{a} usando $\boldsymbol{a}_{+}$ para denotar tanto a la forma canonica
del \'{a}lgebra como a su retroceso sobre el espacio base. Sin embargo,
resultar\'{a} siempre claro a partir del contexto a qu\'{e} forma nos estamos
refiriendo} $\boldsymbol{a}_{+}=g^{-1}\mathrm{d}_{\text{{\tiny M}}}g.$
Llamaremos as\'{\i} mismo $\boldsymbol{a}_{-}=g\mathrm{d}_{\text{{\tiny M}}%
}g^{-1};$ as\'{\i} se tiene que $\boldsymbol{a}_{+}$ y $\boldsymbol{a}_{-}%
$est\'{a}n relacionados a trav\'{e}s de%
\[
\boldsymbol{a}_{+}=-g^{-1}\boldsymbol{a}_{-}g.
\]

\subsection{Derivada Covariante y Curvatura.}

Es interesante observar que por definici\'{o}n, la conexi\'{o}n del fibrado
$\mathbb{A}$ proyecta cualquier vector de $T_{p}P$ s\'{o}lo en su componente
vertical (V\'{e}ase la condici\'{o}n 2 en la definici\'{o}n de conexi\'{o}n y
la definici\'{o}n de subespacio horizontal). As\'{\i}, parece natural
considerar una $b$-forma $\mathbb{B}$ sobre el bundle, $\mathbb{B}\in
\Omega^{b}\left(  P\right)  \otimes\mathfrak{g}$ que haga precisamente lo
contrario, \textit{i.e.}, que considere la componente horizontal de los
vectores en $T_{p}P$.

Para ello, definamos el mapeo%
\[
h:T_{p}P\rightarrow H_{p}P
\]
que a cada vector $Z\in T_{p}P$ le asocia su proyecci\'{o}n sobre el
subespacio horizontal $Z^{h}=h\left(  Z\right)  .$ Con la ayuda de esta
funci\'{o}n, podemos definir formas pseudo tensoriales y tensoriales, tal como
en la siguiente

\begin{definition}
Una $b$-forma $\mathbb{B}\ $sobre el fibrado $P$ ser\'{a} llamada
\emph{pseudotensorial}\textit{\ }cuando satisface

\begin{enumerate}
\item $R_{g}^{\ast}\left(  \mathbb{B}\left(  pg\right)  \right)
=g^{-1}\mathbb{B}\left(  p\right)  g,$ en donde $R_{g}^{\ast}$ es la imagen
rec\'{\i}proca inducida por la acci\'{o}n derecha, $p^{\prime}=pg$. Diremos
que $\mathbb{B}$ es \emph{tensorial} cuando \textit{adem\'{a}s} satisface la condici\'{o}n

\item $\mathbb{B}\left(  Z_{1},\ldots,Z_{b}\right)  =\mathbb{B}\left(
Z_{1}^{h},\ldots,Z_{b}^{h}\right)  $, en donde $Z_{i}^{h}=h\left(
Z_{i}\right)  $, $i=1,\ldots,b$.
\end{enumerate}
\end{definition}

El nombre de forma tensorial tiene su origen en el siguiente

\begin{theorem}
Sea $\mathbb{B}$ una $b$-forma tensorial sobre el fibrado $P.$ Entonces, sobre
la intesecci\'{o}n $U_{\alpha}\cap U_{\beta}\neq\varnothing$ se tieneque%
\begin{equation}
\boldsymbol{B}_{\beta}=\tau_{\alpha\beta}^{-1}\boldsymbol{B}_{\alpha}%
\tau_{\alpha\beta},
\end{equation}
en donde la $b$-forma $\boldsymbol{B}_{\alpha}$ sobre el espacio base $M$
corresponde a $\boldsymbol{B}_{\alpha}=\sigma_{\alpha}^{\ast}\mathbb{B}$.
\end{theorem}

\begin{proof}
En efecto, tenemos que%
\begin{align*}
R_{g}^{\ast}\mathbb{B}\left(  p\right)  \left(  Z_{1}\left(  p\right)
,\ldots,Z_{b}\left(  p\right)  \right)   &  =R_{g}^{\ast}\mathbb{B}\left(
p\right)  \left(  Z_{1}^{h}\left(  p\right)  ,\ldots,Z_{b}^{h}\left(
p\right)  \right)  ,\\
&  =\mathbb{B}\left(  pg\right)  \left(  R_{g\ast}Z_{1}^{h}\left(  p\right)
,\ldots,R_{g\ast}Z_{b}^{h}\left(  p\right)  \right)  .
\end{align*}
Usando ec.~(\ref{EcHpgP=RgHpP}) concluimos que%
\begin{align*}
R_{g}^{\ast}\mathbb{B}\left(  p\right)  \left(  Z_{1}\left(  p\right)
,\ldots,Z_{b}\left(  p\right)  \right)   &  =\mathbb{B}\left(  pg\right)
\left(  Z_{1}^{h}\left(  pg\right)  ,\ldots,Z_{b}^{h}\left(  pg\right)
\right)  ,\\
&  =\mathbb{B}\left(  pg\right)  \left(  Z_{1}\left(  pg\right)  ,\ldots
,Z_{b}\left(  pg\right)  \right)  .
\end{align*}
y as\'{\i}, cuando $\mathbb{B}$ es tensorial,%
\begin{equation}
R_{g}^{\ast}\mathbb{B}\left(  p\right)  \left(  Z_{1}\left(  p\right)
,\ldots,Z_{b}\left(  p\right)  \right)  =\mathbb{B}\left(  pg\right)  \left(
Z_{1}\left(  pg\right)  ,\ldots,Z_{b}\left(  pg\right)  \right)
.\label{EcRgB=B(pg)}%
\end{equation}

Repitiendo ahora el procedimiento utilizado para deducir la transformaci\'{o}n
de gauge ec.(\ref{EcTransfGauge}), concluimos que sobre la intesecci\'{o}n
$U_{\alpha}\cap U_{\beta}$
\begin{equation}
\boldsymbol{B}_{\beta}=\tau_{\alpha\beta}^{-1}\boldsymbol{B}_{\alpha}%
\tau_{\alpha\beta},\label{EcBbeta=Tau-1BalfaTau}%
\end{equation}

\end{proof}

Esto nos lleva naturalmente a otra definici\'{o}n, la del operador de
\textit{derivada exterior covariante }$\mathcal{D}$, tal como en la siguiente

\begin{definition}
Sea $\mathbb{B}$ una forma pseudotensorial sobre el fibrado. Entonces,
definiremos su derivada exterior covariante como%
\[
\mathcal{D}\mathbb{B}=\mathrm{d}_{\text{{\tiny P}}}\mathbb{B}\circ h
\]
en donde $\mathrm{d}_{\text{{\tiny P}}}$ es la derivada exterior sobre $P$.
Usando un razonamiento an\'{a}logo al usado para demostrar
ec.~(\ref{EcRgB=B(pg)}), es posible demostrar que cuando $\mathbb{B}$ es una
$b$-forma \textit{pseudotensorial}, entonces $\mathcal{D}\mathbb{B}$ es es una
$\left(  b+1\right)  $-forma \textit{tensorial}.
\end{definition}

Debemos observar que la conexi\'{o}n sobre el fibrado $\mathbb{A}$ es
pseudotensorial. Por lo tanto, a partir de ella es posible definir la forma
tensorial%
\[
\mathbb{F}=\mathcal{D}\mathbb{A},
\]
a la cual llamaremos curvatura del fibrado. Es posible demostrar en forma
directa (Ve\'{a}se teorema 2.13 Ref.~\cite{Azcarraga-Libro}) que%
\[
\mathbb{F}=\mathrm{d}_{\text{{\tiny P}}}\mathbb{A}+\frac{1}{2}\left[
\mathbb{A},\mathbb{A}\right]
\]
o usando ec.~(\ref{EcConmFormas}),%
\[
\mathbb{F}=\mathrm{d}_{\text{{\tiny P}}}\mathbb{A}+\mathbb{A}^{2}.
\]

Dada la curvatura $\mathbb{F}$ sobre el fibrado, es directo definir a partir
de ella la curvatura de gauge (o intesidad de campo) sobre un abierto
$U_{\alpha}$ como $\boldsymbol{F}_{\alpha}=\sigma_{\alpha}^{\ast}\mathbb{F}$.
En t\'{e}rminos de $\boldsymbol{A}_{\alpha}=\sigma_{\alpha}^{\ast}\mathbb{A},
$ \'{e}sta corresponde a
\[
\boldsymbol{F}_{\alpha}=\mathrm{d}_{\text{{\tiny M}}}\boldsymbol{A}_{\alpha
}+\boldsymbol{A}_{\alpha}^{2}.
\]

Ahora bien, por construcci\'{o}n $\mathbb{F}$ es tensorial, y por lo tanto
satisface la ec.~(\ref{EcBbeta=Tau-1BalfaTau}),%
\[
\boldsymbol{F}_{\beta}=\tau_{\alpha\beta}^{-1}\boldsymbol{F}_{\alpha}%
\tau_{\alpha\beta}.
\]

En efecto, es directo verificar la consistencia de esta ecuaci\'{o}n con
ec.~(\ref{EcTransfGauge}).

Sobre una $b$-forma tensorial $\mathbb{B}\in\Omega^{b}\left(  P\right)
\otimes\mathfrak{g},$ es posible demostrar (V\'{e}ase p\'{a}g.~94
Ref.~\cite{Azcarraga-Libro}) que la derivada covariante $\mathcal{D}%
\mathbb{B}$ corresponde a%
\begin{equation}
\mathcal{D}\mathbb{B}=\mathrm{d}_{\text{{\tiny P}}}\mathbb{B}+\left[
\mathbb{A},\mathbb{B}\right]  .\label{Ec DerCovTens Fibrado}%
\end{equation}

Esta derivada satisface las identidades de Bianchi,%
\begin{align}
\mathcal{DD}\mathbb{B}  &  =\left[  \mathbb{F},\mathbb{B}\right]
,\label{EcDDB=[F,B]_Fibrado}\\
\mathcal{D}\mathbb{F}  &  =\mathcal{DD}\mathbb{A}=0.\label{EcDF=0_Fibrado}%
\end{align}

A partir de esta derivada, resulta natural definir la derivada covariante en
el espacio base,%
\begin{align}
\mathrm{D}_{\alpha}\boldsymbol{B}_{\alpha}  &  =\sigma_{\alpha}^{\ast}\left(
\mathcal{D}\mathbb{B}\right) \nonumber\\
&  =\mathrm{d}_{\text{{\tiny M}}}\boldsymbol{B}_{\alpha}+\left[
\boldsymbol{A}_{\alpha},\boldsymbol{B}_{\alpha}\right]
\label{Ec DerCovTens EspBase}%
\end{align}
la cual satisface las correspondientes identidades de Bianchi sobre el espacio
base,%
\begin{align*}
\mathrm{D}_{\alpha}\mathrm{D}_{\alpha}\boldsymbol{B}_{\alpha}  &  =\left[
\boldsymbol{F}_{\alpha},\boldsymbol{B}_{\alpha}\right]  ,\\
\mathrm{D}_{\alpha}\boldsymbol{F}_{\alpha}  &  =0.
\end{align*}

Dado que por definici\'{o}n $\mathcal{D}\mathbb{B}$ es una forma tensorial,
tenemos que satisface la ec.~(\ref{EcBbeta=Tau-1BalfaTau}),%
\[
\mathrm{D}_{\beta}\boldsymbol{B}_{\beta}=\tau_{\alpha\beta}^{-1}%
\mathrm{D}_{\alpha}\boldsymbol{B}_{\alpha}\tau_{\alpha\beta},
\]
y nuevamente, es directo verificar la consistencia de esta ecuaci\'{o}n con la
transformaci\'{o}n de gauge ec.~(\ref{EcTransfGauge}).

\section{Formas de Transgresi\'{o}n y el Car\'{a}cter de Chern.}

Una de las propiedades m\'{a}s interesantes de un fibrado principal (y la cual
ser\'{a} la base de todo el trabajo desarrollado en la presente tesis) es la
existencia de cantidades \textit{caracter\'{\i}sticas }del fibrado en s\'{\i}
mismo. Estas cantidades, pese a estar definidas a partir de la conexi\'{o}n,
son independientes de la elecci\'{o}n de $\mathbb{A}$ que se haga. As\'{\i},
\'{e}stas definen invariantes topol\'{o}gicos que miden la obstrucci\'{o}n a
la trivializaci\'{o}n del fibrado (\textit{i.e., }a considerarlo como el
producto $M\times G$ en forma global). Primero consideraremos algunas
definiciones previas, y posteriormente usaremos el teorema de Chern--Weil para
definir \textit{Formas de Transgresi\'{o}n }y el \textit{Car\'{a}cter de
Chern}.

\subsection{Polinomios Invariantes}

\begin{definition}
Definiremos un \emph{polinomio invariante} de grado $n$, como el mapeo
$n$-lineal%
\[
\left\vert \cdots\right\vert :\text{ }\underset{n\text{ veces}}{\underbrace
{\mathfrak{g}\times\cdots\times\mathfrak{g}}}\rightarrow\mathbb{R}%
\]
que satisface la condici\'{o}n%
\begin{equation}
\left\vert \left(  g^{-1}\boldsymbol{Z}_{1}g\right)  \cdots\left(
g^{-1}\boldsymbol{Z}_{n}g\right)  \right\vert =\left\vert \boldsymbol{Z}%
_{1}\cdots\boldsymbol{Z}_{n}\right\vert \label{EcCondInvGrupo}%
\end{equation}
en donde $\boldsymbol{Z}_{i}\in\Omega^{z_{i}}\otimes\mathfrak{g}$,
$i=1,\ldots,n$ y $g\in G.$ Cuando adem\'{a}s satisface la condici\'{o}n%
\begin{equation}
\left\vert \boldsymbol{Z}_{1}\cdots\boldsymbol{Z}_{i}\cdots\boldsymbol{Z}%
_{j}\cdots\boldsymbol{Z}_{n}\right\vert =\left(  -1\right)  ^{z_{i}z_{j}%
}\left\vert \boldsymbol{Z}_{1}\cdots\boldsymbol{Z}_{j}\cdots\boldsymbol{Z}%
_{i}\cdots\boldsymbol{Z}_{n}\right\vert \label{Ec TensInv Sim}%
\end{equation}
para todo $\boldsymbol{Z}_{i},\boldsymbol{Z}_{j},$ entonces diremos que es un
\emph{polinomio invariante sim\'{e}trico} y lo denotaremos por $\left\langle
\cdots\right\rangle $. Cuando se satisface la condici\'{o}n%
\begin{equation}
\left\vert \boldsymbol{Z}_{1}\cdots\boldsymbol{Z}_{i}\cdots\boldsymbol{Z}%
_{j}\cdots\boldsymbol{Z}_{n}\right\vert =-\left(  -1\right)  ^{z_{i}z_{j}%
}\left\vert \boldsymbol{Z}_{1}\cdots\boldsymbol{Z}_{j}\cdots\boldsymbol{Z}%
_{i}\cdots\boldsymbol{Z}_{n}\right\vert \label{Ec TensInv Antisim}%
\end{equation}
diremos que es un \emph{polinomio invariante antisim\'{e}trico} y lo
denotaremos por $\left\rangle \cdots\right\langle .$
\end{definition}

Dado que $\boldsymbol{Z}_{i}=Z_{i}^{A}\boldsymbol{T}_{A},$ es posible escribir%
\[
\left\vert \boldsymbol{Z}_{1}\cdots\boldsymbol{Z}_{n}\right\vert =Z_{1}%
^{A_{1}}\cdots Z_{n}^{A_{n}}\left\vert \boldsymbol{T}_{A_{1}}\cdots
\boldsymbol{T}_{A_{n}}\right\vert ,
\]
en donde llamaremos a $\left\vert \boldsymbol{T}_{A_{1}}\cdots\boldsymbol{T}%
_{A_{n}}\right\vert $ \textit{tensor invariante.}

Muchas veces, resulta pr\'{a}ctico expresar la condici\'{o}n de invariancia en
t\'{e}rminos de las constantes de estructura del \'{a}lgebra. Para hacerlo,
basta con considerar un elemento del grupo de la forma%
\[
g=\exp\left(  g^{A}\boldsymbol{T}_{A}\right)
\]
y considerarlo infinitesimalmente cercano a la identidad. En este caso, la
condici\'{o}n de invariancia se expresa en t\'{e}rminos del tensor $\left\vert
\boldsymbol{T}_{A_{1}}\cdots\boldsymbol{T}_{A_{n}}\right\vert $ como%
\begin{equation}
\sum_{p=1}^{n}X_{A_{0}\cdots A_{n}}^{\left(  p\right)  }%
=0,\label{EcCondInvCtesStruc}
\end{equation}
en donde\footnote{Una forma particularmente elegante de obtener la condici\'{o}n de invariancia ec.~(\ref{EcCondInvCtesStruc}) es a trav\'{e}s de la derivada de Lie; ve\'{a}se por ejemplo Refs.~\cite{Azcarraga-Libro,Azcarraga-InvTens}}
\begin{equation}
X_{A_{0}\cdots A_{n}}^{\left(  p\right)  }=\left(  -1\right)  ^{\mathfrak{q}%
\left(  A_{0}\right)  \left(  \mathfrak{q}\left(  A_{1}\right)  +\cdots
+\mathfrak{q}\left(  A_{p-1}\right)  \right)  }C_{A_{0}A_{p}}^{\text{\qquad}%
B}\left\vert \boldsymbol{T}_{A_{1}}\cdots\boldsymbol{T}_{A_{p-1}%
}\boldsymbol{T}_{B}\boldsymbol{T}_{A_{p+1}}\cdots\boldsymbol{T}_{A_{n}%
}\right\vert .\label{Ec X(p) = C x TensInv}%
\end{equation}

Es posible escribir esta condici\'{o}n de invariancia en forma compacta,
extendiendo la definici\'{o}n de conmutador%
\[
\left[  \;,\;\right]  :\mathfrak{g}\times\mathfrak{g}\rightarrow\mathfrak{g},
\]
a una definici\'{o}n de la forma%
\[
\left[  \;,\;\right]  :\underset{n+1\text{ veces}}{\underbrace{\mathfrak{g}%
\times\cdots\times\mathfrak{g}}}\rightarrow\underset{n\text{ veces}%
}{\underbrace{\mathfrak{g}\times\cdots\times\mathfrak{g}}}.
\]
que debe satisfacer%
\begin{equation}
\left[  \boldsymbol{M},\boldsymbol{N}_{1}\cdots\boldsymbol{N}_{n}\right]
=\left[  \boldsymbol{M},\boldsymbol{N}_{1}\right]  \boldsymbol{N}_{2}%
\cdots\boldsymbol{N}_{n}+\left(  -1\right)  ^{mn_{1}}\boldsymbol{N}_{1}\left[
\boldsymbol{M},\boldsymbol{N}_{2}\cdots\boldsymbol{N}_{n}\right]
\label{Ec GeneralizConm}%
\end{equation}
\'{o} m\'{a}s expl\'{\i}citamente,%
\begin{equation}
\left[  \boldsymbol{M},\boldsymbol{N}_{1}\cdots\boldsymbol{N}_{n}\right]
=\sum_{i=1}^{n}\left(  -1\right)  ^{m\left(  n_{1}+\cdots+n_{i-1}\right)
}\boldsymbol{N}_{1}\cdots\boldsymbol{N}_{i-1}\left[  \boldsymbol{M}%
,\boldsymbol{N}_{i}\right]  \boldsymbol{N}_{i+1}\cdots\boldsymbol{N}%
_{n}\label{Ec ConmGenExplicito}%
\end{equation}
para cualquier $\boldsymbol{M}\in\Omega^{m}\otimes\mathfrak{g},$
$\boldsymbol{N}_{i}\in\Omega^{n_{i}}\otimes\mathfrak{g}.$ En t\'{e}rminos de
este conmutador generalizado, la invariancia de $\left\vert \boldsymbol{N}%
_{1}\cdots\boldsymbol{N}_{n}\right\vert $ puede expresarse como%
\begin{equation}
\left\vert \left[  \boldsymbol{M},\boldsymbol{N}_{1}\cdots\boldsymbol{N}%
_{n}\right]  \right\vert =0.\label{EcCondInvConm}%
\end{equation}

Es importante observar que usando la definici\'{o}n generalizada de conmutador
ec.~(\ref{Ec GeneralizConm}), es posible generalizar as\'{\i} mismo la
definici\'{o}n de derivada covariante, ecs.~(\ref{Ec DerCovTens Fibrado}%
,~\ref{Ec DerCovTens EspBase}). Esta nueva derivada covariante act\'{u}a sobre
formas valuadas en varias copias del \'{a}lgebra, y satisface la regla de
Leibniz de la misma forma que la derivada exterior ordinaria.

Cuando existe una representaci\'{o}n matricial de un (super)grupo, la
(super)traza provee en forma directa de un tensor invariante, debido a que es
c\'{\i}clica. Debido a su sencilla construcci\'{o}n, este es el tensor
invariante m\'{a}s utilizado en la literatura. Sin embargo, debemos recordar
que no es la \'{u}nica alternativa, tal como veremos en
secci\'{o}n~\ref{Sec TensInv S-Exp}.

\subsubsection{Tensores Invariantes y Operadores de Casimir}

Los tensores invariantes est\'{a}n int\'{\i}mamente ligados con los
\textit{operadores de Casimir} de un \'{a}lgebra de Lie. En pocas palabras, un
Casimir es un operador que conmuta con todos los elementos del \'{a}lgebra,
sin estar valuado en el \'{a}lgebra. Por supuesto, la definici\'{o}n s\'{o}lo
tiene sentido cuando se usa una noci\'{o}n generalizada de conmutador, como la
que hemos utilizado en ec.~(\ref{Ec GeneralizConm}). En efecto, haremos la siguiente

\begin{definition}
Sea $\boldsymbol{C}=C^{A_{1}\cdots A_{N}}\boldsymbol{T}_{A_{1}}\cdots
\boldsymbol{T}_{A_{N}}$ tal que para todo generador $\boldsymbol{T}_{A}$ de
$\mathfrak{g}$ se satisface%
\[
\left[  \boldsymbol{T}_{A},\boldsymbol{C}\right]  =0.
\]

Entonces se dice que $\boldsymbol{C}$ es un operador de Casimir de
$\mathfrak{g}$.
\end{definition}

Utilizando ec.~(\ref{Ec GeneralizConm}), resulta directo expresar la
condici\'{o}n de Casimir $\left[  \boldsymbol{T}_{A},\boldsymbol{C}\right]
=0$ como una condici\'{o}n sobre $C^{A_{1}\cdots A_{N}},$ como%
\[
\sum_{p=1}^{N}\left[  Y_{\left(  p\right)  }\right]  _{A_{0}}^{\quad
A_{1}\cdots A_{N}}=0
\]
en donde%
\begin{equation}
\left[  Y_{\left(  p\right)  }\right]  _{A_{0}}^{\quad A_{1}\cdots A_{N}%
}=\left(  -1\right)  ^{\mathfrak{q}\left(  A_{0}\right)  \left(
\mathfrak{q}\left(  A_{1}\right)  +\cdots+\mathfrak{q}\left(  A_{p-1}\right)
\right)  }C_{A_{0}B}^{\text{\qquad}A_{p}}C^{A_{1}\cdots A_{p-1}BA_{p+1}\cdots
A_{N}}.\label{Ec Def Y Casimir}%
\end{equation}

Esta condici\'{o}n parece muy similar a la condici\'{o}n de invariancia ec.~(\ref{EcCondInvCtesStruc}). Esto no es una coincidencia; como veremos a continuaci\'{o}n, los tensores invariantes sim\'{e}tricos y los operadores de Casimir son mutuamente duales bajo una m\'{e}trica invertible la cual sea por s\'{\i} misma un tensor invariante sim\'{e}trico tambi\'{e}n (este es el caso por ejemplo, de la m\'{e}trica de Killing para \'{a}lgebras semisimples).

En efecto, sea $k_{AB}$ una m\'{e}trica invertible sobre $\mathfrak{g},$%
\[
k_{BC}k^{CA}=\delta_{B}^{A},
\]
la cual es adem\'{a}s un tensor invariante [ec.~(\ref{EcCondInvCtesStruc})],%
\begin{equation}
C_{A_{0}A_{1}}^{\text{\qquad}B}k_{BA_{2}}+\left(  -1\right)  ^{\mathfrak{q}%
\left(  A_{0}\right)  \mathfrak{q}\left(  A_{1}\right)  }C_{A_{0}A_{2}%
}^{\text{\qquad}B}k_{A_{1}B}=0\label{Ec Cond Inv Kab}%
\end{equation}
sim\'{e}trico [ec.~(\ref{Ec TensInv Sim})],
\[
k_{AB}=\left(  -1\right)  ^{\mathfrak{q}\left(  A\right)  \mathfrak{q}\left(
B\right)  }k_{BA}.
\]

Este es el caso por ejemplo de la m\'{e}trica de Killing, $k_{AB}%
=\operatorname*{STr}\left(  \boldsymbol{T}_{A}\boldsymbol{T}_{B}\right)  $
para \'{a}lgebras semisimples, pues se cumple que $\operatorname*{STr}\left(
\boldsymbol{T}_{A}\boldsymbol{T}_{B}\right)  $ es invertible.

Ahora bien, usando $k_{AB}$ para bajar los \'{\i}ndices de
ec.~(\ref{Ec Def Y Casimir}), y teniendo en cuenta ec.~(\ref{Ec Cond Inv Kab}%
), es directo demostrar que $C_{A_{1}\cdots A_{N}}=k_{A_{1}B_{1}}\cdots
k_{A_{N}B_{N}}C^{B_{1}\cdots B_{N}}$ es un tensor invariante [satisface
ec.~(\ref{EcCondInvCtesStruc})]. De la misma forma, subiendo los \'{\i}ndices
de un tensor invariante con el inverso de la m\'{e}trica $k^{AB}$ obtendremos
un operador de Casimir. As\'{\i}, ambos entes son duales.

Los operadores de Casimir y los tensores invariantes no son s\'{o}lo duales
uno de los otros, sino que adem\'{a}s se pueden contraer entre ellos, dando
origen a nuevos tensores invariantes o Casimirs. En efecto, dado un Casimir
$\boldsymbol{C}=C^{B_{1}\cdots B_{N}}\boldsymbol{T}_{B_{1}}\cdots
\boldsymbol{T}_{B_{N}}$ y un tensor invariante $\left\vert \boldsymbol{T}%
_{B_{1}}\cdots\boldsymbol{T}_{B_{N}}\boldsymbol{T}_{A_{1}}\cdots
\boldsymbol{T}_{A_{n}}\right\vert $ con $N<n,$ es directo demostrar que%
\begin{equation}
\left\vert \boldsymbol{CT}_{A_{1}}\cdots\boldsymbol{T}_{A_{n}}\right\vert
=C^{B_{1}\cdots B_{N}}\left\vert \boldsymbol{T}_{B_{1}}\cdots\boldsymbol{T}%
_{B_{N}}\boldsymbol{T}_{A_{1}}\cdots\boldsymbol{T}_{A_{n}}\right\vert
\label{Ec TensInv con Casimir}%
\end{equation}
es tambi\'{e}n un tensor invariante. De la misma forma,
\[
\bar{C}^{A_{1}\cdots A_{N}}=C^{A_{1}\cdots A_{N}B_{1}\cdots B_{n}}\left\vert
\boldsymbol{T}_{B_{1}}\cdots\boldsymbol{T}_{B_{n}}\right\vert
\]
corresponde a un operador de Casimir.

Esta proliferaci\'{o}n de distintos tipos de tensores invariantes hace
importante una clasificaci\'{o}n para ellos, que ser\'{a} el tema que
trataremos a continuaci\'{o}n.

\subsubsection{Clasificaci\'{o}n de Tensores Invariantes}

\label{Sec TensInv Prim-NoPrim}

Un detalle trivial pero importante es que dados dos tensores invariantes
$\left\vert \boldsymbol{T}_{A_{1}}\cdots\boldsymbol{T}_{A_{m}}\right\vert $ y
$\left\vert \boldsymbol{T}_{A_{1}}\cdots\boldsymbol{T}_{A_{n}}\right\vert ,$
entonces su producto
\begin{equation}
\left\vert \boldsymbol{T}_{A_{1}}\cdots\boldsymbol{T}_{A_{m+n}}\right\vert
=\left\vert \boldsymbol{T}_{A_{1}}\cdots\boldsymbol{T}_{A_{m}}\right\vert
\left\vert \boldsymbol{T}_{A_{1}}\cdots\boldsymbol{T}_{A_{n}}\right\vert
\label{Ec Prod InvTensor = InvTensor}%
\end{equation}
tambi\'{e}n es un tensor invariante. Este sencillo hecho induce una
clasificaci\'{o}n en el espacio de tensores invariantes.

Un tensor invariante que no puede expresarse como el producto de otros es
llamado \textit{primitivo}, y uno que corresponde al producto de otros es
llamado \textit{no-primitivo}. De la misma forma, hablaremos de Casimirs
primitivos y no-primitivos. En general, dado un Casimir o un tensor
invariante, resulta no trivial determinar cu\'{a}l es su componente primitiva.
Este problema se puede resolver, como veremos a continuaci\'{o}n, usando la
forma de Maurer--Cartan $\boldsymbol{a}_{+}$ (Ver Ref.~\cite{Azcarraga-InvTens}. para un
an\'{a}lisis detallado de esta clasificaci\'{o}n).

Sea $\left\vert \boldsymbol{T}_{A_{1}}\cdots\boldsymbol{T}_{A_{q}}\right\vert
$ un tensor invariante c\'{\i}clico (como la traza) y $\boldsymbol{a}_{+}$ la
forma de Maurer-Cartan de $G$. A partir de ellos se define el $q$%
-\textit{cociclo de Chevalley--Eilenberg de} $G$ como%
\begin{align*}
\Omega^{\left(  q\right)  }  &  =\frac{1}{q!}\left\vert \boldsymbol{a}_{+}%
^{q}\right\vert \\
&  =\frac{1}{q!}\underset{q\text{-veces}}{\underbrace{\left\vert
\boldsymbol{a}_{+}\cdots\boldsymbol{a}_{+}\right\vert }}.
\end{align*}

Resulta directo demostrar que $\Omega^{\left(  q\right)  }$ satisface las
siguientes propiedades:

\begin{enumerate}
\item El cociclo se anula para $q$ par, $\Omega^{\left(  2n\right)  }=0$.

\item El cociclo $\Omega^{\left(  2n+1\right)  }$ es una forma cerrada
\textrm{d}$_{{\tiny G}}\Omega^{\left(  2n+1\right)  }=0,$ pero no exacta
(utilizar ec.~(\ref{EcMC F=0}) y $\Omega^{\left(  2n+2\right)  }=0$)

\item Debido a que $\boldsymbol{a}_{+}$ es una $1$-forma, el cociclo
$\Omega^{\left(  2n+1\right)  }$ depende s\'{o}lo de la componente
\textit{antisim\'{e}trica} del tensor invariante $\left\vert \boldsymbol{T}%
_{A_{1}}\cdots\boldsymbol{T}_{A_{2n+1}}\right\vert $,%
\[
\Omega^{\left(  2n+1\right)  }=\frac{1}{\left(  2n+1\right)  !}\left\rangle
\boldsymbol{a}_{+}^{2n+1}\right\langle .
\]

\end{enumerate}

Cuando el conmutador del \'{a}lgebra viene dado por la representaci\'{o}n
inducida por el \'{a}lgebra cobertora universal, $\boldsymbol{a}_{+}^{2}%
=\frac{1}{2}\left[  \boldsymbol{a}_{+},\boldsymbol{a}_{+}\right]  $, entonces
el cociclo $\Omega^{\left(  2n+1\right)  }$ depende en forma indirecta de un
tensor invariante $\left\langle \boldsymbol{T}_{A_{1}}\cdots\boldsymbol{T}%
_{A_{n+1}}\right\rangle $ \textit{sim\'{e}trico}. En efecto, tenemos que%
\begin{align*}
\Omega^{\left(  2n+1\right)  }  &  =\frac{1}{\left(  2n+1\right)  !}\left\vert
\boldsymbol{a}_{+}^{2n+1}\right\vert ,\\
&  =\frac{1}{\left(  2n+1\right)  !}\frac{1}{2^{n}}|\underset{n\text{ veces}%
}{\underbrace{\left[  \boldsymbol{a}_{+},\boldsymbol{a}_{+}\right]
\cdots\left[  \boldsymbol{a}_{+},\boldsymbol{a}_{+}\right]  }}\boldsymbol{a}%
_{+}|,\\
&  =\frac{1}{\left(  2n+1\right)  !}\frac{1}{2^{n}}a_{+}^{A_{1}}a_{+}^{B_{1}%
}\cdots a_{+}^{A_{n}}a_{+}^{B_{n}}a_{+}^{C_{n+1}}C_{A_{1}B_{1}}%
^{\phantom{A_{1}B_{1}}C_{1}}\cdots C_{A_{n}B_{n}}^{\phantom{A_{n}B_{n}}C_{n}%
}\left\vert \boldsymbol{T}_{C_{1}}\cdots\boldsymbol{T}_{C_{n}}\boldsymbol{T}%
_{C_{n+1}}\right\vert ,
\end{align*}

y como $\frac{1}{2}a_{+}^{A}a_{+}^{B}C_{AB}^{\phantom{AB}C}\boldsymbol{T}_{C}
$ es una $2$-forma,%
\begin{equation}
\Omega^{\left(  2n+1\right)  }=\frac{1}{\left(  2n+1\right)  !}\frac{1}{2^{n}%
}a_{+}^{A_{1}}a_{+}^{B_{1}}\cdots a_{+}^{A_{n}}a_{+}^{B_{n}}a_{+}^{C_{n+1}%
}C_{A_{1}B_{1}}^{\phantom{A_{1}B_{1}}C_{1}}\cdots C_{A_{n}B_{n}}%
^{\phantom{A_{n}B_{n}}C_{n}}\left\langle \boldsymbol{T}_{C_{1}}\cdots
\boldsymbol{T}_{C_{n}}\boldsymbol{T}_{C_{n+1}}\right\rangle
.\label{Ec Cociclo = CtesStruct x TensInv}%
\end{equation}

De la misma forma, es posible escribir la identidad $\Omega^{\left(
2n\right)  }=0$ como
\begin{equation}
a_{+}^{A_{1}}a_{+}^{B_{1}}\cdots a_{+}^{A_{n}}a_{+}^{B_{n}}C_{A_{1}B_{1}%
}^{\phantom{A_{1}B_{1}}C_{1}}\cdots C_{A_{n}B_{n}}^{\phantom{A_{n}B_{n}}C_{n}%
}\left\vert \boldsymbol{T}_{C_{1}}\cdots\boldsymbol{T}_{C_{n}}\right\vert
=0.\label{Ec Identidad Cociclo Par=0}%
\end{equation}

En ec.~(\ref{Ec Cociclo = CtesStruct x TensInv}), diremos que $\Omega^{\left(
2n+1\right)  }$ es el \textit{cociclo asociado a} $\left\langle \boldsymbol{T}%
_{C_{1}}\cdots\boldsymbol{T}_{C_{n}}\boldsymbol{T}_{C_{n+1}}\right\rangle $.
Esto nos permite probar el siguiente teorema:

\begin{theorem}
El cociclo asociado a un tensor invariante no primitivo es cero.
\end{theorem}

\begin{proof}
Sea el tensor invariante no primitivo $\left\vert \boldsymbol{T}_{C_{1}}%
\cdots\boldsymbol{T}_{C_{p}}\right\vert \left\vert \boldsymbol{T}_{C_{p+1}%
}\cdots\boldsymbol{T}_{C_{p+q+1}}\right\vert .$ Entonces, el cociclo asociado
con \'{e}l corresponde a%
\begin{align*}
\Omega^{\left(  2\left(  p+q\right)  +1\right)  }  &  =\frac{1}{\left(
2n+1\right)  !}\frac{1}{2^{p+q}}a_{+}^{A_{1}}a_{+}^{B_{1}}\cdots a_{+}^{A_{p}%
}a_{+}^{B_{p}}a_{+}^{A_{p+1}}a_{+}^{B_{p+1}}\cdots a_{+}^{A_{p+q}}%
a_{+}^{B_{p+q}}a_{+}^{C_{p+q+1}}\times\\
&  \times C_{A_{1}B_{1}}^{\phantom{A_{1}B_{1}}C_{1}}\cdots C_{A_{p}B_{p}%
}^{\phantom{A_{p}B_{p}}C_{p}}C_{A_{p+1}B_{p+1}}%
^{\phantom{A_{p+1}B_{p+1}}C_{p+1}}\cdots C_{A_{p+q}B_{p+q}}%
^{\phantom{A_{p+q}B_{p+q}}C_{p+q}}\left\vert \boldsymbol{T}_{C_{1}}%
\cdots\boldsymbol{T}_{C_{p}}\right\vert \left\vert \boldsymbol{T}_{C_{p+1}%
}\cdots\boldsymbol{T}_{C_{p+q+1}}\right\vert .
\end{align*}

Usando ec.~(\ref{Ec Identidad Cociclo Par=0}) para $\left\vert \boldsymbol{T}%
_{C_{1}}\cdots\boldsymbol{T}_{C_{p}}\right\vert ,$ obtenemos%
\[
\Omega^{\left(  2\left(  p+q\right)  +1\right)  }=0.
\]

\end{proof}

Debido a esto, es directo ver que el cociclo de Chevalley-Eilenberg puede ser
usado justamente para construir una base de tensores invariantes primitivos. A
partir de un tensor invariante $\left\vert \boldsymbol{T}_{A_{1}}%
\cdots\boldsymbol{T}_{A_{n}}\boldsymbol{T}_{A_{n+1}}\right\vert $ es posible
construir el cociclo asociado $\Omega^{\left(  2n+1\right)  }=\frac{1}{\left(
2n+1\right)  !}\left[  \Omega^{\left(  2n+1\right)  }\right]  _{A_{1}%
A_{2}\cdots A_{2n-1}A_{2n}B}a_{+}^{A_{1}}\cdots a_{+}^{A_{2n}}a_{+}^{B}$, y a
partir de \'{e}ste, construir un nuevo tensor invariante, $P_{A_{1}\cdots
A_{n+1}}$ el cual ser\'{a} primitivo. En efecto, es posible demostrar
(Ve\'{a}se Ref.~\cite{Azcarraga-InvTens}) que dada una m\'{e}trica invariante invertible, entonces%

\begin{equation}
P_{C_{1}\cdots C_{n}B}=2^{n}\left[  \Omega^{\left(  2n+1\right)  }\right]
_{A_{1}A_{2}\cdots A_{2n-1}A_{2n}B}C_{\qquad C_{1}}^{A_{2}A_{1}}\cdots
C_{\qquad\qquad C_{n}}^{A_{2n}A_{2n-1}}\label{Ec TensInv Primitivo}%
\end{equation}
corresponde a un tensor invariante primitivo y sim\'{e}trico.

Cuando se utiliza la supertraza simetrizada como tensor invariante, entonces
los tensores invariantes ec.~(\ref{Ec TensInv Primitivo}) forman una base, en
el sentido de que todo tensor invariante constru\'{\i}do a partir de la
supertraza puede escribirse como productos de tensores invariantes del tipo
ec.~(\ref{Ec TensInv Primitivo}). A\'{u}n m\'{a}s, a partir de la identidad
ec.~(\ref{Ec Identidad Cociclo Par=0}) se tiene que este tipo de tensores
invariantes se anulan id\'{e}nticamente al realizar una contracci\'{o}n con un
Casimir [ec.~(\ref{Ec TensInv con Casimir})],%
\[
C^{A_{1}\cdots A_{p}}P_{A_{1}\cdots A_{p}A_{p+1}\cdots A_{q}}=0\text{ }p<q,
\]
por lo que se dice que en este sentido son \textquotedblleft
sin-traza\textquotedblright.

De esta forma, el cociclo de Chevalley--Eilenberg nos provee de una poderosa
herramienta, la cual permite disponer de una base en el espacio de tensores
invariantes, y por ende, de una clasificaci\'{o}n en tensores primitivos y no
primitivos. Por otra parte, el t\'{e}rmino de Chevalley--Eilenberg
aparecer\'{a} otra vez, en un contexto m\'{a}s f\'{\i}sico, al analizar las
transformaciones de gauge de un lagrangeano de Chern--Simons.

\subsection{Proyecci\'{o}n de Formas}

Hasta el momento, hemos definido formas sobre el espacio base $M$ a partir de
formas sobre el fibrado $P.$ Para ello hemos usado el imagen rec\'{\i}proca
inducida por las secciones locales $\sigma_{\alpha}:U_{\alpha}\subset
M\rightarrow P$. Parece natural preguntar sobre la posibilidad de realizar el
procedimiento contrario, \textit{i.e., }definir una forma sobre el fibrado $P$
a partir de una forma sobre el espacio base $M.$ En efecto, esto es posible
utilizando el imagen rec\'{\i}proca inducida por la proyecci\'{o}n
$\pi:P\rightarrow M.$ As\'{\i}, dada una forma $\boldsymbol{H}$ sobre $M,$ es
posible definir una forma $\mathbb{H}$ sobre $P$ como%
\[
\mathbb{H}=\pi^{\ast}\boldsymbol{H}.
\]

Ahora bien, dados dos puntos sobre una misma fibra, $p$ y $p^{\prime}=pg,$ y
una $q$-forma $\mathbb{H}$ que corresponde a la imagen rec\'{\i}proca de
alguna forma en $M,$ en general se cumplir\'{a} que $\mathbb{H}\left(
p\right)  $ y $\mathbb{H}\left(  pg\right)  $ corresponder\'{a}n a las
im\'{a}genes rec\'{\i}procas de formas diferentes de $M$%
\begin{align*}
\mathbb{H}\left(  p\right)   &  =\pi^{\ast}\boldsymbol{H}\left(  x\right)  ,\\
\mathbb{H}\left(  pg\right)   &  =\pi^{\ast}\boldsymbol{H}^{\prime}\left(
x\right)  ,
\end{align*}
con $\boldsymbol{H}\left(  x\right)  \neq\boldsymbol{H}^{\prime}\left(
x\right)  $ en general. Cuando $\mathbb{H}$ sobre toda la fibra $\pi
^{-1}\left(  x\right)  ,$ corresponde a la imagen rec\'{\i}proca de una
\textit{\'{u}nica} forma $\boldsymbol{H}\left(  x\right)  ,$ diremos que
$\mathbb{H}$ es \textit{proyectable} a $\boldsymbol{H}\left(  x\right)  $.

Es posible definir formas proyectables en forma sistem\'{a}tica a trav\'{e}s
del siguiente

\begin{theorem}
Sea una forma $\mathbb{H}$ sobre el fibrado $P$ que satisface las condiciones:

\begin{enumerate}
\item Ser invariante bajo la acci\'{o}n derecha del grupo,
\begin{equation}
R_{g}^{\ast}\mathbb{H}=\mathbb{H}\label{EcCondProy1}%
\end{equation}

\item su acci\'{o}n sobre vectores de $T_{p}P$ es de la forma
\begin{equation}
\mathbb{H}\left(  X_{1}\cdots X_{q}\right)  =\mathbb{H}\left(  X_{1}^{h}\cdots
X_{q}^{h}\right)  .\label{EcCondProy2}%
\end{equation}

\end{enumerate}

Entonces, existe una \'{u}nica forma $\boldsymbol{H}\left(  x\right)  $ sobre
el espacio base $M$ que corresponde a la proyecci\'{o}n de $\mathbb{H},$
\textit{i.e.}, $\mathbb{H}=\pi^{\ast}\boldsymbol{H}$, y por lo tanto,
$\mathbb{H}$ es \emph{proyectable}.
\end{theorem}

\begin{proof}
En efecto, sea el empuje inducido por $\pi$ sobre los vectores $X_{i}\left(
p\right)  \in T_{p}P,$ $i=1,\ldots,q$,%
\begin{align*}
\pi_{\ast}  &  :T_{p}P\rightarrow T_{x}M,\\
&  :X_{i}\left(  p\right)  \rightarrow\check{X}_{i}\left(  x\right)
=\pi_{\ast}X_{i}\left(  p\right)  .
\end{align*}

Sea $\boldsymbol{H}\left(  x\right)  $ tal que%
\[
\mathbb{H}\left(  p\right)  =\pi^{\ast}\boldsymbol{H}\left(  x\right)  .
\]

Entonces, tenemos que%
\begin{align*}
\boldsymbol{H}\left(  x\right)  \left(  \check{X}_{1}\left(  x\right)
,\ldots,\check{X}_{q}\left(  x\right)  \right)   &  =\boldsymbol{H}\left(
x\right)  \left(  \pi_{\ast}X_{1}\left(  p\right)  ,\ldots,\pi_{\ast}%
X_{q}\left(  p\right)  \right)  ,\\
&  =\pi^{\ast}\boldsymbol{H}\left(  x\right)  \left(  X_{1}\left(  p\right)
,\ldots,X_{q}\left(  p\right)  \right)  ,\\
&  =\mathbb{H}\left(  p\right)  \left(  X_{1}\left(  p\right)  ,\ldots
,X_{q}\left(  p\right)  \right)  ,
\end{align*}
y usando ec.~(\ref{EcCondProy2}),%
\begin{equation}
\pi^{\ast}\boldsymbol{H}\left(  x\right)  \left(  X_{1}\left(  p\right)
,\ldots,X_{q}\left(  p\right)  \right)  =\mathbb{H}\left(  p\right)  \left(
X_{1}^{h}\left(  p\right)  ,\ldots,X_{q}^{h}\left(  p\right)  \right)
.\label{EcWea1}%
\end{equation}

Ahora bien, usando ec.~(\ref{EcHpgP=RgHpP}) tenemos que%
\begin{align*}
\mathbb{H}\left(  pg\right)  \left(  X_{1}^{h}\left(  pg\right)  ,\ldots
,X_{q}^{h}\left(  pg\right)  \right)   &  =\mathbb{H}\left(  pg\right)
\left(  R_{g\ast}X_{1}^{h}\left(  p\right)  ,\ldots,R_{g\ast}X_{q}^{h}\left(
p\right)  \right)  ,\\
&  =R_{g}^{\ast}\mathbb{H}\left(  p\right)  \left(  X_{1}^{h}\left(  p\right)
,\ldots,X_{q}^{h}\left(  p\right)  \right)  ,
\end{align*}
y usando ec.~(\ref{EcCondProy1}),%
\begin{equation}
\mathbb{H}\left(  pg\right)  \left(  X_{1}^{h}\left(  pg\right)  ,\ldots
,X_{q}^{h}\left(  pg\right)  \right)  =\mathbb{H}\left(  p\right)  \left(
X_{1}^{h}\left(  p\right)  ,\ldots,X_{q}^{h}\left(  p\right)  \right)
.\label{EcWea2}%
\end{equation}

As\'{\i} comparando ecs.~(\ref{EcWea1}) y~(\ref{EcWea2}), tenemos que
$\mathbb{H}\left(  pg\right)  $ y $\mathbb{H}\left(  p\right)  $ corresponden
a las im\'{a}genes rec\'{\i}procas de la misma forma $\boldsymbol{H}\left(
x\right)  ,$%
\[
\mathbb{H}\left(  pg\right)  \left(  X_{1}^{h}\left(  pg\right)  ,\ldots
,X_{q}^{h}\left(  pg\right)  \right)  =\pi^{\ast}\boldsymbol{H}\left(
x\right)  \left(  X_{1}\left(  p\right)  ,\ldots,X_{q}\left(  p\right)
\right)  ,
\]
y por lo tanto, $\boldsymbol{H}$ corresponde a la proyecci\'{o}n de
$\mathbb{H}$.
\end{proof}

La importancia de la operaci\'{o}n de proyecci\'{o}n consiste en que dada una
forma suave $\mathbb{H}$ definida sobre todo el fibrado $P$, su proyecci\'{o}n
$\boldsymbol{H}$ ser\'{a} tambi\'{e}n una forma suave \textit{globalmente}
definida sobre todo el espacio base $M$. Esto contrasta fuertemente con las
formas construidas utilizando la imagen rec\'{\i}proca inducida por una
secci\'{o}n $\sigma_{\alpha},$ ya que dada una forma globalmente definida
sobre el fibrado (como la conexi\'{o}n $\mathbb{A}$ por ejemplo) en general
s\'{o}lo es posible obtener formas \textit{localmente} definidas sobre los
abiertos $U_{\alpha}\subset M.$

Sin embargo, es posible encontrar relaciones entre ambos procedimientos. Es
necesario observar que dada una secci\'{o}n local arbitraria $\sigma_{\alpha
},$ siempre es posible encontrar un elemento del grupo $g_{\alpha}$ tal que%
\[
\sigma_{\alpha}\circ\pi=g_{\alpha}.
\]

Cuando $\mathbb{H}$ es \textit{proyectable}, tenemos que%
\begin{align*}
\mathbb{H}  &  =R_{g_{\alpha}}^{\ast}\mathbb{H}\\
&  =R_{\sigma_{\alpha}\circ\pi}^{\ast}\mathbb{H}\\
&  =\pi^{\ast}\sigma_{\alpha}^{\ast}\mathbb{H}%
\end{align*}
y escribiendo $\boldsymbol{H}_{\alpha}=\sigma_{\alpha}^{\ast}\mathbb{H},$
tenemos%
\begin{equation}
\mathbb{H}=\pi^{\ast}\boldsymbol{H}_{\alpha}.\label{Ec H=pi-backH}%
\end{equation}

Dado que cuando $\mathbb{H}$ es proyectable, su proyecci\'{o}n $\boldsymbol{H}%
$ tal que $\mathbb{H}=\pi^{\ast}\boldsymbol{H}$ es \'{u}nica, esto implica que
en la intersecci\'{o}n $U_{\alpha}\cap U_{\beta},$%
\begin{equation}
\boldsymbol{H}_{\alpha}=\boldsymbol{H}_{\beta}=\boldsymbol{H}%
\label{EcHalfa=Hbeta=H}%
\end{equation}
o sea, que en este caso $\boldsymbol{H}$ est\'{a} globalmente definida sobre
$M.$

Una propiedad importante del proceso de proyecci\'{o}n es su relaci\'{o}n con
la derivada covariante sobre el fibrado, tal como se indica en el siguiente

\begin{theorem}
Cuando una forma $\mathbb{H}$ sobre el fibrado es proyectable, entonces%
\begin{equation}
\mathcal{D}\mathbb{H}=\mathrm{d}_{\text{{\tiny P}}}\mathbb{H}.\label{EcDH=dH}%
\end{equation}

\end{theorem}

\begin{proof}
En efecto,%
\begin{align*}
\mathrm{d}_{\text{{\tiny P}}}\mathbb{H}\left(  X_{1},\ldots,X_{q}\right)   &
=\mathrm{d}_{\text{{\tiny P}}}\pi^{\ast}\boldsymbol{H}\left(  X_{1}%
,\ldots,X_{q}\right)  ,\\
&  =\pi^{\ast}\left(  \mathrm{d}_{\text{{\tiny M}}}\boldsymbol{H}\right)
\left(  X_{1},\ldots,X_{q}\right)  ,\\
&  =\mathrm{d}_{\text{{\tiny M}}}\boldsymbol{H}\left(  \pi_{\ast}X_{1}%
,\ldots,\pi_{\ast}X_{q}\right)  ,
\end{align*}
y ya que por definici\'{o}n de subespacio vertical, $\pi_{\ast}X_{i}=\pi
_{\ast}X_{i}^{h}$ (ver ec.~(\ref{EcVp=KernelPi*})), tenemos%
\begin{align*}
\mathrm{d}_{\text{{\tiny P}}}\mathbb{H}\left(  X_{1},\ldots,X_{q}\right)   &
=\mathrm{d}_{\text{{\tiny M}}}\boldsymbol{H}\left(  \pi_{\ast}X_{1}^{h}%
,\ldots,\pi_{\ast}X_{q}^{h}\right)  ,\\
&  =\pi^{\ast}\left(  \mathrm{d}_{\text{{\tiny M}}}\boldsymbol{H}\right)
\left(  X_{1}^{h},\ldots,X_{q}^{h}\right)  ,\\
&  =\mathrm{d}_{\text{{\tiny P}}}\pi^{\ast}\boldsymbol{H}\left(  X_{1}%
^{h},\ldots,X_{q}^{h}\right)  ,\\
&  =\mathrm{d}_{\text{{\tiny P}}}\mathbb{H}\left(  X_{1}^{h},\ldots,X_{q}%
^{h}\right)  ,\\
&  =\mathrm{d}_{\text{{\tiny P}}}\mathbb{H}\circ h\left(  X_{1},\ldots
,X_{q}\right)  ,\\
&  =\mathcal{D}\mathbb{H}\left(  X_{1},\ldots,X_{q}\right)  .
\end{align*}

Por lo tanto,%
\[
\mathrm{d}_{\text{{\tiny P}}}\mathbb{H}=\mathcal{D}\mathbb{H}.
\]

\end{proof}

As\'{\i}, provistos de la definici\'{o}n de polinomio invariante y
proyecci\'{o}n de formas, consideraremos ahora un importante resultado
conocido como Teorema de Chern--Weil.

\subsection{Teorema de Chern--Weil}

\begin{theorem}
[Chern--Weil]Sea $\pi:P\rightarrow M$ un fibrado principal de grupo de
simetr\'{\i}a $G$, provisto de una conexi\'{o}n $\mathbb{A}\in\Omega
^{1}\left(  P\right)  \otimes\mathfrak{g}$ y su correspondiente curvatura
$\mathbb{F}\in\Omega^{2}\left(  P\right)  \otimes\mathfrak{g}$, siendo
$\mathfrak{g}$ el \'{a}lgebra de Lie asociada a $G$. Sea $\left\langle
\boldsymbol{T}_{A_{1}}\cdots\boldsymbol{T}_{A_{n+1}}\right\rangle $ un tensor
invariante sim\'{e}trico bajo $G,$ y sea la $2\left(  n+1\right)  $-forma
$\left\langle \mathbb{F}^{n+1}\right\rangle ,$%
\[
\left\langle \mathbb{F}^{n+1}\right\rangle =\text{ }\underset{n+1\text{
veces}}{\underbrace{\left\langle \mathbb{F}\cdots\mathbb{F}\right\rangle }}.
\]
Entonces, se cumple que:

\begin{enumerate}
\item $\left\langle \mathbb{F}^{n+1}\right\rangle $ es una forma cerrada,
$\mathrm{d}_{\text{{\tiny P}}}\left\langle \mathbb{F}^{n+1}\right\rangle =0,$
y proyectable,%
\[
\left\langle \mathbb{F}^{n+1}\right\rangle =\pi^{\ast}\left\langle
\boldsymbol{F}^{n+1}\right\rangle ,
\]
siendo su proyecci\'{o}n $\left\langle \boldsymbol{F}^{n+1}\right\rangle $ una
forma cerrada, $\mathrm{d}_{\text{{\tiny M}}}\left\langle \boldsymbol{F}%
^{n+1}\right\rangle =0$.

\item Dadas dos conexiones sobre el fibrado, $\mathbb{A}$ y $\bar{\mathbb{A}%
},$ siendo sus respectivas curvaturas $\mathbb{F}$ y $\bar{\mathbb{F}},$ la
diferencia $\left\langle \mathbb{F}^{n+1}\right\rangle -\left\langle
\bar{\mathbb{F}}^{n+1}\right\rangle $ es una forma exacta y proyectable.
\end{enumerate}
\end{theorem}

\begin{proof}
(1) La forma $\left\langle \mathbb{F}^{n+1}\right\rangle $ es proyectable, ya
que es invariante bajo la acci\'{o}n derecha del grupo. En efecto, ya que
$\mathbb{F}$ es una forma tensorial,%
\[
R_{g}^{\ast}\mathbb{F}=g^{-1}\mathbb{F}g,
\]
y ya que $\left\langle \mathbb{F}^{n+1}\right\rangle $ es un polinomio
invariante [ver ec.~(\ref{EcCondInvGrupo})], tenemos que $\left\langle
\mathbb{F}^{n+1}\right\rangle $ satisface la condici\'{o}n~(\ref{EcCondProy1}%
)
\[
R_{g}^{\ast}\left\langle \mathbb{F}^{n+1}\right\rangle =\left\langle
\mathbb{F}^{n+1}\right\rangle .
\]
Por otra parte, dado que $\mathbb{F}$ es tensorial, se tiene que
$\mathbb{F}\left(  X_{1},X_{2}\right)  =\mathbb{F}\left(  X_{1}^{h},X_{2}%
^{h}\right)  $ y por lo tanto, $\left\langle \mathbb{F}^{n+1}\right\rangle $
satisface la condici\'{o}n~(\ref{EcCondProy2}),%
\[
\left\langle \mathbb{F}^{n+1}\right\rangle \left(  X_{1},\ldots,X_{2n+2}%
\right)  =\left\langle \mathbb{F}^{n+1}\right\rangle \left(  X_{1}^{h}%
,\ldots,X_{2n+2}^{h}\right)  .
\]

As\'{\i}, dado que $\left\langle \mathbb{F}^{n+1}\right\rangle $ es
proyectable, se satisface la ec.~(\ref{EcDH=dH}),%
\[
\mathrm{d}_{\text{{\tiny P}}}\left\langle \mathbb{F}^{n+1}\right\rangle
=\mathcal{D}\left\langle \mathbb{F}^{n+1}\right\rangle ,
\]
y utilizando la identidad de Bianchi~(\ref{EcDF=0_Fibrado}), tenemos que en
efecto, \'{e}sta es cerrada,%
\[
\mathrm{d}_{\text{{\tiny P}}}\left\langle \mathbb{F}^{n+1}\right\rangle =0.
\]
Utilizando ec.~(\ref{Ec H=pi-backH}) tenemos que el polinomio%
\[
\left\langle \boldsymbol{F}_{\alpha}^{n+1}\right\rangle =\sigma_{\alpha}%
^{\ast}\left\langle \mathbb{F}^{n+1}\right\rangle
\]
corresponde a la proyecci\'{o}n de $\left\langle \mathbb{F}^{n+1}\right\rangle
$ sobre $M$ y est\'{a} globalmente definido (v\'{e}ase
ec.~(\ref{EcHalfa=Hbeta=H})). Una manera interesante de demostrar que
$\left\langle \boldsymbol{F}_{\alpha}^{n+1}\right\rangle $ es cerrada es
usando la identidad de Bianchi para escribir%
\[
\mathrm{D}_{\alpha}\left\langle \boldsymbol{F}_{\alpha}^{n+1}\right\rangle =0.
\]

Haciendo uso de la condici\'{o}n de invariancia del polinomio,
ec.~(\ref{EcCondInvConm}), tenemos%
\begin{align*}
\mathrm{D}_{\alpha}\left\langle \boldsymbol{F}_{\alpha}^{n+1}\right\rangle  &
=\mathrm{d}_{\text{{\tiny M}}}\left\langle \boldsymbol{F}_{\alpha}%
^{n+1}\right\rangle +\left\langle \left[  \boldsymbol{A}_{\alpha
},\boldsymbol{F}_{\alpha}^{n+1}\right]  \right\rangle \\
&  =\mathrm{d}_{\text{{\tiny M}}}\left\langle \boldsymbol{F}_{\alpha}%
^{n+1}\right\rangle ,
\end{align*}
y por lo tanto,
\[
\mathrm{d}\left\langle \boldsymbol{F}_{\alpha}^{n+1}\right\rangle =0.
\]

Dado que $\left\langle \mathbb{F}^{n+1}\right\rangle $ es proyectable, tenemos
que $\left\langle \boldsymbol{F}_{\alpha}^{n+1}\right\rangle $ est\'{a}
globalmente definido sobre $M,$ y por lo tanto, lo denotaremos sencillamente
por $\left\langle \boldsymbol{F}^{n+1}\right\rangle .$ Esta forma, con la
debida normalizaci\'{o}n, corresponde al $\left(  n+1\right)  $%
-\textit{\'{e}simo car\'{a}cter de Chern},%
\[
\operatorname{ch}_{n+1}\left(  \boldsymbol{F}\right)  =\frac{1}{\left(
n+1\right)  !}\left(  \frac{i}{2\pi}\right)  ^{n+1}\left\langle \boldsymbol{F}%
^{n+1}\right\rangle .
\]

(2) Consideremos dos conexiones $\mathbb{A}$ y $\bar{\mathbb{A}}.$ Su
diferencia%
\[
\mathbb{O}=\mathbb{A}-\bar{\mathbb{A}}%
\]
ser\'{a} tensorial, pues cuando $Y$ es un vector vertical,%
\[
\mathbb{O}\left(  Y\right)  =\boldsymbol{Y}-\boldsymbol{Y}=0.
\]
En t\'{e}rminos de ellas es posible definir otra conexi\'{o}n, que interpola
entre $\bar{\mathbb{A}}$ y $\mathbb{A}$,%
\[
\mathbb{A}_{t}=\bar{\mathbb{A}}+t\mathbb{O},
\]
con su correspondiente curvatura%
\begin{align*}
\mathbb{F}_{t}  &  =\mathcal{D}\mathbb{A}_{t}\\
&  =\mathrm{d}_{\text{{\tiny P}}}\mathbb{A}_{t}+\mathbb{A}_{t}^{2},\\
&  =\bar{\mathbb{F}}+t\mathcal{\bar{D}}\mathbb{O}+t^{2}\mathbb{O}^{2},
\end{align*}
en donde%
\begin{align*}
\mathcal{\bar{D}}\mathbb{O}  &  =\mathrm{d}_{\text{{\tiny P}}}\mathbb{O}%
+\left[  \bar{\mathbb{A}},\mathbb{O}\right]  ,\\
\mathbb{O}^{2}  &  =\frac{1}{2}\left[  \mathbb{O},\mathbb{O}\right]  .
\end{align*}
Ahora bien, resulta interesante observar que la derivada con respecto a $t$ de
la curvatura $\mathbb{F}_{t}$ corresponde a%
\begin{align}
\frac{\mathrm{d}\mathbb{F}_{t}}{\mathrm{d}t}  &  =\mathcal{\bar{D}}%
\mathbb{O}+t\left[  \mathbb{O},\mathbb{O}\right] \nonumber\\
&  =\mathcal{D}_{t}\mathbb{O}\label{EcdFt/dt=DtO_Fibrado}%
\end{align}
en donde%
\[
\mathcal{D}_{t}\mathbb{O}=\mathrm{d}_{\text{{\tiny P}}}\mathbb{O}+\left[
\mathbb{A}_{t},\mathbb{O}\right]  .
\]
Dado que $\left.  \mathbb{F}_{t}\right\vert _{t=0}=\bar{\mathbb{F}}$ y
$\left.  \mathbb{F}_{t}\right\vert _{t=1}=\mathbb{F},$ es posible expresar la
diferencia $\left\langle \mathbb{F}^{n+1}\right\rangle -\left\langle
\bar{\mathbb{F}}^{n+1}\right\rangle $ como%
\[
\left\langle \mathbb{F}^{n+1}\right\rangle -\left\langle \bar{\mathbb{F}%
}^{n+1}\right\rangle =\int_{t=0}^{t=1}\mathrm{d}t\frac{\mathrm{d}}%
{\mathrm{d}t}\left\langle \mathbb{F}_{t}^{n+1}\right\rangle .
\]
Dado que el polinomio $\left\langle \mathbb{F}_{t}^{n+1}\right\rangle $ es
sim\'{e}trico, y $\mathbb{F}_{t}$ es una $2$-forma, tenemos que%
\[
\left\langle \mathbb{F}^{n+1}\right\rangle -\left\langle \bar{\mathbb{F}%
}^{n+1}\right\rangle =\left(  n+1\right)  \int_{t=0}^{t=1}\mathrm{d}%
t~\left\langle \frac{\mathrm{d}\mathbb{F}_{t}}{\mathrm{d}t}\mathbb{F}_{t}%
^{n}\right\rangle
\]
y usando ec.~(\ref{EcdFt/dt=DtO_Fibrado}), tenemos%
\[
\left\langle \mathbb{F}^{n+1}\right\rangle -\left\langle \bar{\mathbb{F}%
}^{n+1}\right\rangle =\left(  n+1\right)  \int_{t=0}^{t=1}\mathrm{d}%
t~\left\langle \mathcal{D}_{t}\mathbb{OF}_{t}^{n}\right\rangle .
\]
Usando la identidad de Bianchi $\mathcal{D}_{t}\mathbb{F}_{t}=0,$ tenemos que%
\[
\left\langle \mathbb{F}^{n+1}\right\rangle -\left\langle \bar{\mathbb{F}%
}^{n+1}\right\rangle =\left(  n+1\right)  \int_{t=0}^{t=1}\mathrm{d}%
t~\mathcal{D}_{t}\left\langle \mathbb{OF}_{t}^{n}\right\rangle .
\]
La forma $\mathbb{O}$ es tensorial, y dado que $\left\langle \mathbb{OF}%
_{t}^{n}\right\rangle $ es un tensor invariante, tenemos que $\left\langle
\mathbb{OF}_{t}^{n}\right\rangle $ es una forma proyectable. Por lo
tanto\footnote{Una forma alternativa de probar $\mathcal{D}_{t}\left\langle
\mathbb{OF}_{t}^{n}\right\rangle =\mathrm{d}_{\text{{\tiny P}}}\left\langle
\mathbb{OF}_{t}^{n}\right\rangle $ es haciendo uso de la condici\'{o}n de
invariancia, ec.~(\ref{EcCondInvConm}).}, $\mathcal{D}_{t}\left\langle
\mathbb{OF}_{t}^{n}\right\rangle =\mathrm{d}_{\text{{\tiny P}}}\left\langle
\mathbb{OF}_{t}^{n}\right\rangle ,$ y as\'{\i},%
\[
\left\langle \mathbb{F}^{n+1}\right\rangle -\left\langle \bar{\mathbb{F}%
}^{n+1}\right\rangle =\mathrm{d}_{\text{{\tiny P}}}\mathbb{T}%
_{\mathbb{A\leftarrow}\bar{\mathbb{A}}}^{\left(  2n+1\right)  }%
\]
con%
\begin{equation}
\mathbb{T}_{\mathbb{A\leftarrow}\bar{\mathbb{A}}}^{\left(  2n+1\right)
}=\left(  n+1\right)  \int_{t=0}^{t=1}\mathrm{d}t~\left\langle \mathbb{OF}%
_{t}^{n}\right\rangle \label{Ec Transg Fibrado}%
\end{equation}
en donde llamaremos a la $\left(  2n+1\right)  $-forma $\mathbb{T}%
_{\mathbb{A\leftarrow}\bar{\mathbb{A}}}^{\left(  2n+1\right)  }$
\textit{transgresi\'{o}n sobre el fibrado}. Dado que $\mathbb{T}%
_{\mathbb{A\leftarrow}\bar{\mathbb{A}}}^{\left(  2n+1\right)  }$ es
proyectable, es posible escribir su proyecci\'{o}n sobre el espacio base $M$
como (v\'{e}ase ec.~(\ref{EcHalfa=Hbeta=H}))%
\begin{align}
T_{\boldsymbol{A}_{\alpha}\mathbb{\leftarrow}\bar{\boldsymbol{A}}_{\alpha}%
}^{\left(  2n+1\right)  }  &  =\sigma_{\alpha}^{\ast}\mathbb{T}%
_{\mathbb{A\leftarrow}\bar{\mathbb{A}}}^{\left(  2n+1\right)  }%
,\label{Ec Transg EspBase 1}\\
&  =\left(  n+1\right)  \int_{t=0}^{t=1}\mathrm{d}t\left\langle
\boldsymbol{\Theta}_{\alpha}\left[  \boldsymbol{F}_{t}\right]  _{\alpha}%
^{n}\right\rangle ,\label{Ec Transg EspBase 2}%
\end{align}
con%
\begin{align*}
\boldsymbol{\Theta}_{\alpha}  &  =\sigma_{\alpha}^{\ast}\mathbb{O},\\
\left[  \boldsymbol{F}_{t}\right]  _{\alpha}  &  =\sigma_{\alpha}^{\ast
}\mathbb{F}_{t}.
\end{align*}
Llamaremos a $T_{\boldsymbol{A}_{\alpha}\mathbb{\leftarrow}\bar{\boldsymbol{A}%
}_{\alpha}}^{\left(  2n+1\right)  }$ \textit{transgresi\'{o}n sobre el espacio
base}, o simplemente \textit{transgresi\'{o}n}, cuando no haya lugar a
equ\'{\i}vocos. Dado que, como en el caso de cualquier forma proyectable,
$T_{\boldsymbol{A}_{\alpha}\mathbb{\leftarrow}\bar{\boldsymbol{A}}_{\alpha}%
}^{\left(  2n+1\right)  }$ est\'{a} globalmente definida sobre $M,$ la
denotaremos sencillamente como $T_{\boldsymbol{A}\mathbb{\leftarrow}%
\bar{\boldsymbol{A}}}^{\left(  2n+1\right)  }.$ \'{E}sta satisface sobre el
espacio base la relaci\'{o}n%
\begin{equation}
\left\langle \boldsymbol{F}^{n+1}\right\rangle -\left\langle \bar
{\boldsymbol{F}}^{n+1}\right\rangle =\mathrm{d}_{\text{{\tiny M}}%
}T_{\boldsymbol{A}\mathbb{\leftarrow}\bar{\boldsymbol{A}}}^{\left(
2n+1\right)  }.\label{EcTeoChern-WeilSobreM}%
\end{equation}

\end{proof}

\subsection{Forma de Chern--Simons}

\label{Sec Def Forma CS}

Es importante se\~{n}alar que \textit{sobre el fibrado} la forma $\left\langle
\mathbb{F}^{n+1}\right\rangle $ no s\'{o}lo es cerrada, sino que tambi\'{e}n
exacta. Es importante recalcar que esto se cumple s\'{o}lo para $\left\langle
\mathbb{F}^{n+1}\right\rangle $ y \textit{no} para $\left\langle
\boldsymbol{F}^{n+1}\right\rangle ,$ tal como veremos a continuaci\'{o}n.

En la definici\'{o}n de forma de transgresi\'{o}n, entraron en juego dos
conexiones, $\mathbb{A}$ y $\bar{\mathbb{A}}$, y con ellas se construy\'{o}
una tercera conexi\'{o}n sobre el fibrado,
\[
\mathbb{A}_{t}=\bar{\mathbb{A}}+t\left(  \mathbb{A}-\bar{\mathbb{A}}\right)  .
\]

Es interesante observar que una conexi\'{o}n sobre el fibrado \textit{no}
puede ser cero, pues no cumplir\'{\i}a con la condici\'{o}n dada en
ec.~(\ref{EcA(Y)=Y_Fibrado}). Por lo tanto, al imponer $\bar{\mathbb{A}}=0,$ y
escribir%
\begin{align*}
\mathbb{A}_{t}  &  =t\mathbb{A},\\
\mathbb{F}_{t}  &  =t\mathrm{d}_{\text{{\tiny P}}}\mathbb{A}+t^{2}%
\mathbb{A}^{2},
\end{align*}
ellos no corresponden ni a una conexi\'{o}n ni a una curvatura sobre el
fibrado. Sin embargo, el desarrollo usado en la segunda parte del teorema de
Chern--Weil sigue siendo v\'{a}lido y por lo tanto, es posible escribir%
\[
\left\langle \mathbb{F}^{n+1}\right\rangle =\mathrm{d}_{\text{{\tiny P}}%
}\mathbb{Q}^{\left(  2n+1\right)  }\left(  \mathbb{A}\right)  ,
\]
en donde%
\begin{align}
\mathbb{Q}^{\left(  2n+1\right)  }\left(  \mathbb{A}\right)   &
=\mathbb{T}_{\mathbb{A\leftarrow}0}^{\left(  2n+1\right)  }%
\label{Ec Def CS Fibrado 1}\\
&  =\left(  n+1\right)  \int_{t=0}^{t=1}\mathrm{d}t\left\langle \mathbb{A}%
\left(  t\mathrm{d}_{\text{{\tiny P}}}\mathbb{A}+t^{2}\mathbb{A}^{2}\right)
^{n}\right\rangle \label{Ec Def CS Fibrado 2}%
\end{align}
ser\'{a} llamado \textit{forma de Chern--Simons sobre el fibrado}. Es
importante observar, que ya que $\mathbb{A}_{t}=t\mathbb{A},$ no es una
conexi\'{o}n y $\mathbb{F}_{t}=t\mathrm{d}_{\text{{\tiny P}}}\mathbb{A}%
+t^{2}\mathbb{A}^{2}$ no es una curvatura, la forma de Chern--Simons no es
invariante y por lo tanto, \textit{no es proyectable} sobre el espacio base.

Por lo tanto, dadas dos secciones $\sigma_{\alpha}$ y $\sigma_{\beta},$ con
sus respectivos abiertos $U_{\alpha}$ y $U_{\beta},$ tendremos en general que%
\[
\sigma_{\alpha}^{\ast}\mathbb{Q}^{\left(  2n+1\right)  }\neq\sigma_{\beta
}^{\ast}\mathbb{Q}^{\left(  2n+1\right)  }.
\]

De esta forma, es posible definir una forma de Chern--Simons sobre el espacio
base,%
\begin{align}
Q_{\alpha}^{\left(  2n+1\right)  }\left(  \boldsymbol{A}_{\alpha}\right)   &
=\sigma_{\alpha}^{\ast}\mathbb{Q}^{\left(  2n+1\right)  }\left(
\mathbb{A}\right) \label{Ec Def CS Esp Base 1}\\
&  =\left(  n+1\right)  \int_{t=0}^{t=1}\mathrm{d}t~\left\langle
\boldsymbol{A}_{\alpha}\left(  t\mathrm{d}_{\text{{\tiny M}}}\boldsymbol{A}%
_{\alpha}+t^{2}\boldsymbol{A}_{\alpha}^{2}\right)  ^{n}\right\rangle
\label{Ec Def CS Esp Base 2}%
\end{align}
s\'{o}lo l\textit{ocalmente}, sobre un abierto particular de $M$. Por lo
tanto, $\left\langle \boldsymbol{F}^{n+1}\right\rangle $ puede escribirse como
la derivada exterior de la forma de Chern--Simons $\boldsymbol{Q}_{\alpha
}^{\left(  2n+1\right)  }$ s\'{o}lo sobre el abierto $U_{\alpha}$,%
\begin{equation}
\left.  \left\langle \boldsymbol{F}^{n+1}\right\rangle \right\vert
_{U_{\alpha}}=\mathrm{d}_{\text{{\tiny M}}}Q_{\alpha}^{\left(  2n+1\right)
},\label{Ec <F^n+1>|Ua = d C-S}%
\end{equation}
pero \textit{no} globalmente sobre $M$. Para no recargar la notaci\'{o}n,
posteriormente simplemente omitiremos el sub\'{\i}ndice griego en la forma de
Chern--Simons, pero es importante recordar que ella est\'{a} definida s\'{o}lo
en forma local, \textit{i.e., }s\'{o}lo sobre cada abierto.

\section{Homotop\'{\i}a}

Es interesante observar que cuando el espacio base $M$ es de dimensi\'{o}n
par, $d=2n$ y carece de bordes, la integral del $n$-\'{e}simo car\'{a}cter de
Chern sobre $M$ es un invariante topol\'{o}gico. En efecto, integrando
ecuaci\'{o}n~(\ref{EcTeoChern-WeilSobreM}) sobre $M,$ y considerando $\partial
M=\varnothing,$ tenemos que
\[
\int_{M}\left\langle \boldsymbol{F}^{n}\right\rangle =\int_{M}\left\langle
\bar{\boldsymbol{F}}^{n}\right\rangle =\text{cte.}%
\]
donde $\boldsymbol{F}$ y $\bar{\boldsymbol{F}}$ son dos curvatura en dos
conexiones $\boldsymbol{A}$ y $\bar{\boldsymbol{A}}$ completamente
independientes. Sin embargo, ser\'{\i}a extremadamente interesante extender
esta definici\'{o}n de invariante topol\'{o}gico para variedades con borde.

Otra propiedad interesante relacionada con el car\'{a}cter de Chern es que la
siguiente suma de derivadas de formas de Transgresi\'{o}n se anula
id\'{e}nticamente\footnote{Escribimos esta ecuaci\'{o}n sobre el espacio base,
pero por supuesto, la misma relaci\'{o}n se satisface sobre el fibrado. En
general, toda la discusi\'{o}n que se llevar\'{a} a cabo en esta secci\'{o}n
es v\'{a}lida tanto sobre el espacio base como sobre el fibrado; escogeremos
escribir nuestras ecuaciones sobre el espacio base simplemente para no
recargar la notaci\'{o}n.}:%
\begin{eqnarray*}
\textrm{d}T_{\boldsymbol{A}\leftarrow\tilde{\boldsymbol{A}}}^{\left(
2n+1\right)  }+\textrm{d}T_{\tilde{\boldsymbol{A}}\leftarrow
\bar{\boldsymbol{A}}}^{\left(  2n+1\right)  }+\textrm{d}T_{\bar
{\boldsymbol{A}}\leftarrow\boldsymbol{A}}^{\left(  2n+1\right)  }
& = & \left\langle \boldsymbol{F}^{n+1}\right\rangle
-\left\langle \tilde{\boldsymbol{F}}^{n+1}\right\rangle
+\left\langle \tilde{\boldsymbol{F}}^{n+1}\right\rangle + \\
& & -\left\langle \bar{\boldsymbol{F}}^{n+1}\right\rangle
+\left\langle \bar{\boldsymbol{F}}^{n+1}\right\rangle
-\left\langle \boldsymbol{F}^{n+1}\right\rangle \\
& = & 0.
\end{eqnarray*}

Dado que $T_{\bar{\boldsymbol{A}}\leftarrow\boldsymbol{A}}^{\left(
2n+1\right)  }=-T_{\boldsymbol{A}\leftarrow\bar{\boldsymbol{A}}}^{\left(
2n+1\right)  },$ esto significa que%
\[
\text{\textrm{d}}T_{\boldsymbol{A}\leftarrow\bar{\boldsymbol{A}}}^{\left(
2n+1\right)  }=\text{\textrm{d}}T_{\boldsymbol{A}\leftarrow\tilde
{\boldsymbol{A}}}^{\left(  2n+1\right)  }+\text{\textrm{d}}T_{\tilde
{\boldsymbol{A}}\leftarrow\bar{\boldsymbol{A}}}^{\left(  2n+1\right)  }.
\]

Por lo tanto, siempre es posible descomponer una transgresi\'{o}n como la suma
de otras dos, m\'{a}s una forma cerrada $\alpha$, usando una conexi\'{o}n
intermedia,%
\[
T_{\boldsymbol{A}\leftarrow\bar{\boldsymbol{A}}}^{\left(  2n+1\right)
}=T_{\boldsymbol{A}\leftarrow\tilde{\boldsymbol{A}}}^{\left(  2n+1\right)
}+T_{\tilde{\boldsymbol{A}}\leftarrow\bar{\boldsymbol{A}}}^{\left(
2n+1\right)  }+\alpha.
\]

Resulta interesante el descubrir que en realidad ambos problemas est\'{a}n
\'{\i}ntimamente relacionados. Esto se logra analizando el teorema de
Chern--Weil a la luz de una poderosa herramienta, la \textit{F\'{o}rmula
Extendida de Homotop\'{\i}a de Cartan}.

\subsection{F\'{o}rmula Extendida de Homotop\'{\i}a de Cartan.}

Sea el conjunto de $r+2$ conexiones independientes $\left\{  \boldsymbol{A}%
_{i},i=0,\ldots,r+1\right\}  ,$ $\boldsymbol{A}_{i}\in\Omega\left(  M\right)
^{1}\otimes\mathfrak{g}$ y un simplex $\left(  r+1\right)  $-dimensional
$T_{r+1}.$ Es posible embeber el simplex $T_{r+1}$ en $\mathbb{R}^{r+2}$ como%
\[
T_{r+1}=\left\{  \left(  t^{0},t^{1},\ldots,t^{r+1}\right)  \in\mathbb{R}%
^{r+2}\text{ tal que: }%
\begin{array}
[c]{l}%
t^{i}\geq0,\text{ con }i=0,\ldots,r+1\text{ y}\\
\sum_{i=0}^{r+1}t^{i}=1,
\end{array}
\right\}
\]

La relaci\'{o}n entre el conjunto de conexiones y $T_{r+1}$ viene dada por el
hecho de que%
\begin{equation}
\boldsymbol{A}_{t}=\sum_{i=0}^{r+1}t^{i}\boldsymbol{A}_{i}%
\label{EcAt=SUM(tiAi)}%
\end{equation}
es una conexi\'{o}n \textit{cuando} $\left(  t^{0},t^{1},\ldots,t^{r+1}%
\right)  \in T_{r+1}.$ En efecto, es directo verificar que al transformar cada
$\boldsymbol{A}_{i}$ como una conexi\'{o}n bajo una transformaci\'{o}n de
gauge, entonces $\boldsymbol{A}_{t}$ tambi\'{e}n lo hace. As\'{\i} mismo, es
posible escribir una curvatura para esta conexi\'{o}n, $\boldsymbol{F}%
_{t}=\mathrm{d}\boldsymbol{A}_{t}+\boldsymbol{A}_{t}^{2}$. Por lo tanto, a
cada punto del simplex se le est\'{a} asociando una conexi\'{o}n; en
particular, el $i$-\'{e}simo v\'{e}rtice de $T_{r+1}$ est\'{a} asociado con la
$i$-\'{e}sima conexi\'{o}n $\boldsymbol{A}_{i}$.

Denotaremos el simplex de las conexiones asociadas como%
\[
\boldsymbol{T}_{r+1}=\left(  \boldsymbol{A}_{0}\boldsymbol{A}_{1}%
\cdots\boldsymbol{A}_{r+1}\right)  .
\]

Ahora, consideremos un polinomio $\Pi$ en las formas $\left\{  \boldsymbol{A}%
_{t},\boldsymbol{F}_{t},\mathrm{d}_{t}\boldsymbol{A}_{t},\mathrm{d}%
_{t}\boldsymbol{F}_{t}\right\}  $, el cual es tambi\'{e}n una $\left(
q+m\right)  $-forma sobre $T_{r+1}\times M,$ con el car\'{a}cter de $m$-forma
sobre $M$ y de $q$-forma sobre $T_{r+1}.$ Denotaremos las derivadas exteriores
sobre $M$ y $T_{r+1}$ como $\mathrm{d}$ y $\mathrm{d}_{t}$ respectivamente, y
definiremos el operador de \textit{derivaci\'{o}n homot\'{o}pica}, $l_{t}$
como%
\[
l_{t}:\Omega^{a}\left(  M\right)  \times\Omega^{b}\left(  T_{r+1}\right)
\rightarrow\Omega^{a-1}\left(  M\right)  \times\Omega^{b+1}\left(
T_{r+1}\right)  .
\]

Este operador satisface la regla de Leibniz, y forma la siguiente \'{a}lgebra
gradada con $\mathrm{d}$ y $\mathrm{d}_{t}:$%
\begin{align}
\mathrm{d}^{2}  &  =0,\label{EcAlgFEHC 1}\\
\mathrm{d}_{t}^{2}  &  =0,\label{EcAlgFEHC 2}\\
\left[  l_{t},\mathrm{d}\right]   &  =\mathrm{d}_{t},\label{EcAlgFEHC 3}\\
\left[  l_{t},\mathrm{d}_{t}\right]   &  =0,\label{EcAlgFEHC 4}\\
\left\{  \mathrm{d},\mathrm{d}_{t}\right\}   &  =0.\label{EcAlgFEHC 5}%
\end{align}

La \'{u}nica forma consistente de definir la acci\'{o}n de $l_{t}$ sobre
$\boldsymbol{A}_{t}$ y $\boldsymbol{F}_{t}$ es de la forma%
\begin{align}
l_{t}\boldsymbol{F}_{t}  &  =\mathrm{d}_{t}\boldsymbol{A}_{t}%
,\label{Ec ltFt=dtAt}\\
l_{t}\boldsymbol{A}_{t}  &  =0.\label{Ec ltAt=0}%
\end{align}

Usando el \'{a}lgebra~(\ref{EcAlgFEHC 1})-(\ref{EcAlgFEHC 5}), es posible
demostrar que%
\[
\left[  l_{t}^{p+1},\mathrm{d}\right]  \Pi=\left(  p+1\right)  \mathrm{d}%
_{t}l_{t}^{p}\Pi,
\]
en donde $p\leq m$ (debido a que $\Pi$ es una $m$-forma sobre $M$).

Ahora bien, integrando esta ecuaci\'{o}n sobre el simplex $T_{r+1},$ con
$p+q=r,$ tenemos que%
\[
\int_{T_{r+1}}l_{t}^{p+1}\mathrm{d}\Pi-\int_{T_{r+1}}\mathrm{d}l_{t}^{p+1}%
\Pi=\left(  p+1\right)  \int_{\partial T_{r+1}}l_{t}^{p}\Pi.
\]

Debido a que $l_{t}^{p+1}\Pi$ es una $\left(  r+1\right)  $-forma sobre
$T_{r+1},$ se cumple que%
\[
\int_{T_{r+1}}\mathrm{d}l_{t}^{p+1}\Pi=\left(  -1\right)  ^{r+1}\mathrm{d}%
\int_{T_{r+1}}l_{t}^{p+1}\Pi.
\]

Por lo tanto, tenemos que se cumple la identidad%
\[
\int_{T_{r+1}}l_{t}^{p+1}\mathrm{d}\Pi+\left(  -1\right)  ^{r}\mathrm{d}%
\int_{T_{r+1}}l_{t}^{p+1}\Pi=\left(  p+1\right)  \int_{\partial T_{r+1}}%
l_{t}^{p}\Pi,
\]
la cual puede ser escrita en forma m\'{a}s conveniente como%
\begin{equation}
\int_{\partial T_{r+1}}\frac{1}{p!}l_{t}^{p}\Pi=\int_{T_{r+1}}\frac{1}{\left(
p+1\right)  !}l_{t}^{p+1}\text{\textrm{d}}\Pi+\left(  -1\right)
^{r}\text{\textrm{d}}\int_{T_{r+1}}\frac{1}{\left(  p+1\right)  !}l_{t}%
^{p+1}\Pi.\label{EcFEHC}%
\end{equation}

Este importante resultado es conocido en la literatura como F\'{o}rmula
Extendida de Homotop\'{\i}a de Cartan (Ve\'{a}se Ref. \cite{Zumino-ECHF},
\cite{Nosotros3-TransLargo}, \cite{Nosotros4-TransCorto}).

Consideremos ahora el polinomio%
\[
\Pi=\left\langle \boldsymbol{F}_{t}^{n+1}\right\rangle ,
\]
en donde $\boldsymbol{F}_{t}$ corresponde a la curvatura en la conexi\'{o}n
dada en ec.~(\ref{EcAt=SUM(tiAi)}).

Este polinomio es cerrado sobre $M,$%
\[
\mathrm{d}\Pi=0,
\]
es una $0$-forma sobre $T_{r+1}$ y una $\left(  2n+2\right)  $-forma sobre $M$
($q=0,m=2n+2$). Por lo tanto, tenemos que $0\leq p\leq2n+2.$

Para este polinomio en particular, ec.~(\ref{EcFEHC}) se reduce a%
\begin{equation}
\int_{\partial T_{p+1}}\frac{l_{t}^{p}}{p!}\left\langle \boldsymbol{F}%
_{t}^{n+1}\right\rangle =\left(  -1\right)  ^{p}\text{\textrm{d}}\int
_{T_{p+1}}\frac{l_{t}^{p+1}}{\left(  p+1\right)  !}\left\langle \boldsymbol{F}%
_{t}^{n+1}\right\rangle .\label{EcFEHC<F^n+1>}%
\end{equation}

Revisaremos los casos $p=0,$ $p=1$ y $p=2,$ los cuales se mostrar\'{a}n
especialmente \'{u}tiles en el contexto de las formas de transgresi\'{o}n.

\subsubsection{Caso $p=0:$Teorema de Chern--Weil.}

En el caso $p=0,$ ec.~(\ref{EcFEHC<F^n+1>}) se reduce a%
\begin{equation}
\int_{\partial T_{1}}\left\langle \boldsymbol{F}_{t}^{n+1}\right\rangle
=\text{\textrm{d}}\int_{T_{1}}l_{t}\left\langle \boldsymbol{F}_{t}%
^{n+1}\right\rangle ,\label{EcCaso p=0}%
\end{equation}
en donde $\boldsymbol{F}_{t}$ corresponde a la curvatura en la conexi\'{o}n%
\[
\boldsymbol{A}_{t}=t^{0}\boldsymbol{A}_{0}+t^{1}\boldsymbol{A}_{1},
\]
con%
\[
t^{0}+t^{1}=1.
\]

El borde del simplex $\boldsymbol{T}_{1}=\left(  \boldsymbol{A}_{0}%
\boldsymbol{A}_{1}\right)  $ corresponde simplemente a%
\[
\partial\left(  \boldsymbol{A}_{0}\boldsymbol{A}_{1}\right)  =\left(
\boldsymbol{A}_{1}\right)  -\left(  \boldsymbol{A}_{0}\right)  ,
\]
y as\'{\i} el lado izquierdo de ec.~(\ref{EcCaso p=0}) corresponde a%
\[
\int_{\partial T_{1}}\left\langle \boldsymbol{F}_{t}^{n+1}\right\rangle
=\left\langle \boldsymbol{F}_{1}^{n+1}\right\rangle -\left\langle
\boldsymbol{F}_{0}^{n+1}\right\rangle .
\]

Dado que $\left\langle \boldsymbol{F}_{t}^{n+1}\right\rangle $ es un polinomio
sim\'{e}trico, al lado derecho de ec.~(\ref{EcCaso p=0}) tenemos que%
\[
l_{t}\left\langle \boldsymbol{F}_{t}^{n+1}\right\rangle =\left(  n+1\right)
\left\langle l_{t}\boldsymbol{F}_{t}\boldsymbol{F}_{t}^{n}\right\rangle
\]
y como%
\begin{align*}
l_{t}\boldsymbol{F}_{t}  &  =\mathrm{d}_{t}\boldsymbol{A}_{t},\\
&  =\mathrm{d}t^{0}\boldsymbol{A}_{0}+\mathrm{d}t^{1}\boldsymbol{A}_{1},\\
&  =\mathrm{d}t^{1}\left(  \boldsymbol{A}_{1}-\boldsymbol{A}_{0}\right)  ,
\end{align*}
entonces%
\[
l_{t}\left\langle \boldsymbol{F}_{t}^{n+1}\right\rangle =\left(  n+1\right)
\mathrm{d}t^{1}\left\langle \left(  \boldsymbol{A}_{1}-\boldsymbol{A}%
_{0}\right)  \boldsymbol{F}_{t}^{n}\right\rangle .
\]

As\'{\i}, hemos recuperado el teorema de Chern--Weil a trav\'{e}s de la
f\'{o}rmula de homotop\'{\i}a de Cartan para $p=0$,%
\[
\left\langle \boldsymbol{F}_{1}^{n+1}\right\rangle -\left\langle
\boldsymbol{F}_{0}^{n+1}\right\rangle =\left(  n+1\right)  \text{\textrm{d}%
}\int_{t=0}^{t=1}\mathrm{d}t\left\langle \left(  \boldsymbol{A}_{1}%
-\boldsymbol{A}_{0}\right)  \boldsymbol{F}_{t}^{n}\right\rangle .
\]

\subsubsection{Caso $p=1:$ Ecuaci\'{o}n Triangular}

Para $p=1,$ ec.~(\ref{EcFEHC<F^n+1>}) tenemos%
\begin{equation}
\int_{\partial T_{2}}l_{t}\left\langle \boldsymbol{F}_{t}^{n+1}\right\rangle
=-\frac{1}{2}\text{\textrm{d}}\int_{T_{2}}l_{t}^{2}\left\langle \boldsymbol{F}%
_{t}^{n+1}\right\rangle ,\label{EcCaso p=1}%
\end{equation}
en donde esta vez $\boldsymbol{F}_{t}$ corresponde a la curvatura en la
conexi\'{o}n%
\[
\boldsymbol{A}_{t}=t^{0}\boldsymbol{A}_{0}+t^{1}\boldsymbol{A}_{1}%
+t^{2}\boldsymbol{A}_{2},
\]
con%
\[
t^{0}+t^{1}+t^{2}=1.
\]

El simplex asociado es $\boldsymbol{T}_{2}=\left(  \boldsymbol{A}%
_{0}\boldsymbol{A}_{1}\boldsymbol{A}_{2}\right)  ,$ de borde%
\[
\partial\left(  \boldsymbol{A}_{0}\boldsymbol{A}_{1}\boldsymbol{A}_{2}\right)
=\left(  \boldsymbol{A}_{1}\boldsymbol{A}_{2}\right)  -\left(  \boldsymbol{A}%
_{0}\boldsymbol{A}_{2}\right)  +\left(  \boldsymbol{A}_{0}\boldsymbol{A}%
_{1}\right)  .
\]

Por lo tanto, el lado izquierdo de ec.~(\ref{EcCaso p=1}) corresponde a%
\[
\int_{\partial T_{2}}l_{t}\left\langle \boldsymbol{F}_{t}^{n+1}\right\rangle
=\int_{\left(  \boldsymbol{A}_{1}\boldsymbol{A}_{2}\right)  }l_{t}\left\langle
\boldsymbol{F}_{t}^{n+1}\right\rangle -\int_{\left(  \boldsymbol{A}%
_{0}\boldsymbol{A}_{2}\right)  }l_{t}\left\langle \boldsymbol{F}_{t}%
^{n+1}\right\rangle +\int_{\left(  \boldsymbol{A}_{0}\boldsymbol{A}%
_{1}\right)  }l_{t}\left\langle \boldsymbol{F}_{t}^{n+1}\right\rangle .
\]

Pero como ya hemos visto en la secci\'{o}n anterior,%
\[
\int_{\left(  \bar{\boldsymbol{A}A}\right)  }l_{t}\left\langle \boldsymbol{F}%
_{t}^{n+1}\right\rangle =T_{\boldsymbol{A}\leftarrow\bar{\boldsymbol{A}}%
}^{\left(  2n+1\right)  },
\]
y por lo tanto,%
\[
\int_{\partial T_{2}}l_{t}\left\langle \boldsymbol{F}_{t}^{n+1}\right\rangle
=T_{\boldsymbol{A}_{2}\leftarrow\boldsymbol{A}_{1}}^{\left(  2n+1\right)
}-T_{\boldsymbol{A}_{2}\leftarrow\boldsymbol{A}_{0}}^{\left(  2n+1\right)
}+T_{\boldsymbol{A}_{1}\leftarrow\boldsymbol{A}_{0}}^{\left(  2n+1\right)  }.
\]

As\'{\i}, es posible escribir ec.~(\ref{EcCaso p=1}) como%
\begin{equation}
T_{\boldsymbol{A}_{2}\leftarrow\boldsymbol{A}_{0}}^{\left(  2n+1\right)
}=T_{\boldsymbol{A}_{2}\leftarrow\boldsymbol{A}_{1}}^{\left(  2n+1\right)
}+T_{\boldsymbol{A}_{1}\leftarrow\boldsymbol{A}_{0}}^{\left(  2n+1\right)
}+\frac{1}{2}\text{\textrm{d}}\int_{T_{2}}l_{t}^{2}\left\langle \boldsymbol{F}%
_{t}^{n+1}\right\rangle .\label{EcYaNoSe}%
\end{equation}

Por otra parte, usando el hecho de que el polinomio es sim\'{e}trico y las
identidades~(\ref{Ec ltFt=dtAt}-\ref{Ec ltAt=0}), tenemos que%
\[
\frac{1}{2}l_{t}^{2}\left\langle \boldsymbol{F}_{t}^{n+1}\right\rangle
=\frac{1}{2}n\left(  n+1\right)  \left\langle \left(  \text{\textrm{d}}%
_{t}\boldsymbol{A}_{t}\right)  ^{2}\boldsymbol{F}_{t}^{n}\right\rangle
\]

Ahora bien, en el presente caso%
\[
\text{\textrm{d}}_{t}\boldsymbol{A}_{t}=\text{\textrm{d}}t^{0}\boldsymbol{A}%
_{0}+\text{\textrm{d}}t^{1}\boldsymbol{A}_{1}+\text{\textrm{d}}t^{2}%
\boldsymbol{A}_{2},
\]
y dado que \textrm{d}$t^{0}+$\textrm{d}$t^{1}+$\textrm{d}$t^{2}=0,$%
\[
\text{\textrm{d}}_{t}\boldsymbol{A}_{t}=\text{\textrm{d}}t^{0}\left(
\boldsymbol{A}_{0}-\boldsymbol{A}_{1}\right)  +\text{\textrm{d}}t^{2}\left(
\boldsymbol{A}_{2}-\boldsymbol{A}_{1}\right)  .
\]

Ya que el polinomio es sim\'{e}trico, tenemos que%
\[
\frac{1}{2}l_{t}^{2}\left\langle \boldsymbol{F}_{t}^{n+1}\right\rangle
=-n\left(  n+1\right)  \text{\textrm{d}}t^{0}\text{\textrm{d}}t^{2}%
\left\langle \left(  \boldsymbol{A}_{2}-\boldsymbol{A}_{1}\right)  \left(
\boldsymbol{A}_{1}-\boldsymbol{A}_{0}\right)  \boldsymbol{F}_{t}%
^{n}\right\rangle .
\]

Por conveniencia, definimos los par\'{a}metros de integraci\'{o}n%
\begin{align*}
t  &  =1-t^{0},\\
s  &  =t^{2},
\end{align*}
y realizamos en forma expl\'{\i}cita la integraci\'{o}n sobre $T_{2}.$
As\'{\i}, llegamos a%
\begin{equation}
\int_{T_{2}}\frac{l_{t}^{2}}{2}\left\langle \boldsymbol{F}_{t}^{n+1}%
\right\rangle =Q_{\boldsymbol{A}_{2}\leftarrow\boldsymbol{A}_{1}%
\leftarrow\boldsymbol{A}_{0}}^{\left(  2n\right)  },\label{EcQ(2n)=lt2<Ft>}%
\end{equation}
en donde $Q_{\boldsymbol{A}_{2}\leftarrow\boldsymbol{A}_{1}\leftarrow
\boldsymbol{A}_{0}}^{\left(  2n\right)  }$ est\'{a} definido como%
\begin{equation}
Q_{\boldsymbol{A}_{2}\leftarrow\boldsymbol{A}_{1}\leftarrow\boldsymbol{A}_{0}%
}^{\left(  2n\right)  }\equiv n\left(  n+1\right)  \int_{0}^{1}\mathrm{d}%
t\int_{0}^{t}\mathrm{d}s\left\langle \left(  \boldsymbol{A}_{2}-\boldsymbol{A}%
_{1}\right)  \left(  \boldsymbol{A}_{1}-\boldsymbol{A}_{0}\right)
\boldsymbol{F}_{st}^{n-1}\right\rangle ,\label{EcQ(2n)=Intgrl_Doble}%
\end{equation}
correspondiendo $\boldsymbol{F}_{st}$ a la curvatura en la conexi\'{o}n%
\[
\boldsymbol{A}_{st}=\boldsymbol{A}_{0}+t\left(  \boldsymbol{A}_{1}%
-\boldsymbol{A}_{0}\right)  +s\left(  \boldsymbol{A}_{2}-\boldsymbol{A}%
_{1}\right)  .
\]

Reemplazando en ec.~(\ref{EcYaNoSe}), arribamos finalmente a la identidad
conocida como \textquotedblleft Ecuaci\'{o}n Triangular\textquotedblright\ en
la literatura,%
\begin{equation}
T_{\boldsymbol{A}_{2}\leftarrow\boldsymbol{A}_{0}}^{\left(  2n+1\right)
}=T_{\boldsymbol{A}_{2}\leftarrow\boldsymbol{A}_{1}}^{\left(  2n+1\right)
}+T_{\boldsymbol{A}_{1}\leftarrow\boldsymbol{A}_{0}}^{\left(  2n+1\right)
}+\text{\textrm{d}}Q_{\boldsymbol{A}_{2}\leftarrow\boldsymbol{A}_{1}%
\leftarrow\boldsymbol{A}_{0}}^{\left(  2n\right)  }.\label{EcTriangular}%
\end{equation}

\subsubsection{Caso $p=2:$ Descomposici\'{o}n de t\'{e}rminos de borde.}

Considerar $p=2$ nos permitir\'{a} descomponer t\'{e}rminos del tipo
$Q_{\boldsymbol{A}_{2}\leftarrow\boldsymbol{A}_{1}\leftarrow\boldsymbol{A}%
_{0}}^{\left(  2n\right)  }$ a trav\'{e}s de conexiones intermedias. Para
$p=2, $ ec.~(\ref{EcFEHC<F^n+1>}) corresponde a%
\begin{equation}
\int_{\partial T_{3}}\frac{l_{t}^{2}}{2!}\left\langle \boldsymbol{F}_{t}%
^{n+1}\right\rangle =\text{\textrm{d}}\int_{T_{3}}\frac{l_{t}^{3}}%
{3!}\left\langle \boldsymbol{F}_{t}^{n+1}\right\rangle ,\label{EcCaso p=2}%
\end{equation}
con%
\[
\boldsymbol{T}_{3}=\left(  \boldsymbol{A}_{0}\boldsymbol{A}_{1}\boldsymbol{A}%
_{2}\boldsymbol{A}_{3}\right)
\]
y borde%
\[
\partial\left(  \boldsymbol{A}_{0}\boldsymbol{A}_{1}\boldsymbol{A}%
_{2}\boldsymbol{A}_{3}\right)  =\left(  \boldsymbol{A}_{1}\boldsymbol{A}%
_{2}\boldsymbol{A}_{3}\right)  -\left(  \boldsymbol{A}_{0}\boldsymbol{A}%
_{2}\boldsymbol{A}_{3}\right)  +\left(  \boldsymbol{A}_{0}\boldsymbol{A}%
_{1}\boldsymbol{A}_{3}\right)  -\left(  \boldsymbol{A}_{0}\boldsymbol{A}%
_{1}\boldsymbol{A}_{2}\right)  .
\]

Llamando%
\[
Q_{\boldsymbol{A}_{3}\leftarrow\boldsymbol{A}_{2}\leftarrow\boldsymbol{A}%
_{1}\leftarrow\boldsymbol{A}_{0}}^{\left(  2n-1\right)  }\equiv\int_{T_{3}%
}\frac{l_{t}^{3}}{3!}\left\langle \boldsymbol{F}_{t}^{n+1}\right\rangle ,
\]
y usando la definici\'{o}n dada en ec.~(\ref{EcQ(2n)=lt2<Ft>}) tenemos que
ec.~(\ref{EcCaso p=2}) toma la forma%
\[
Q_{\boldsymbol{A}_{3}\leftarrow\boldsymbol{A}_{2}\leftarrow\boldsymbol{A}_{1}%
}^{\left(  2n\right)  }-Q_{\boldsymbol{A}_{3}\leftarrow\boldsymbol{A}%
_{2}\leftarrow\boldsymbol{A}_{0}}^{\left(  2n\right)  }+Q_{\boldsymbol{A}%
_{3}\leftarrow\boldsymbol{A}_{1}\leftarrow\boldsymbol{A}_{0}}^{\left(
2n\right)  }-Q_{\boldsymbol{A}_{2}\leftarrow\boldsymbol{A}_{1}\leftarrow
\boldsymbol{A}_{0}}^{\left(  2n\right)  }=\text{\textrm{d}}Q_{\boldsymbol{A}%
_{3}\leftarrow\boldsymbol{A}_{2}\leftarrow\boldsymbol{A}_{1}\leftarrow
\boldsymbol{A}_{0}}^{\left(  2n-1\right)  }.
\]

As\'{\i}, sus derivadas est\'{a}n relacionadas a trav\'{e}s de%
\begin{equation}
\text{\textrm{d}}Q_{\boldsymbol{A}_{2}\leftarrow\boldsymbol{A}_{1}%
\leftarrow\boldsymbol{A}_{0}}^{\left(  2n\right)  }=\text{\textrm{d}%
}Q_{\boldsymbol{A}_{3}\leftarrow\boldsymbol{A}_{2}\leftarrow\boldsymbol{A}%
_{1}}^{\left(  2n\right)  }-\text{\textrm{d}}Q_{\boldsymbol{A}_{3}%
\leftarrow\boldsymbol{A}_{2}\leftarrow\boldsymbol{A}_{0}}^{\left(  2n\right)
}+\text{\textrm{d}}Q_{\boldsymbol{A}_{3}\leftarrow\boldsymbol{A}_{1}%
\leftarrow\boldsymbol{A}_{0}}^{\left(  2n\right)  }.\label{EcTetraedral}%
\end{equation}

\subsection{Aplicaciones de la F\'{o}rmula de Homotop\'{\i}a.}

Tal como se mencionaba al principio de esta secci\'{o}n, la F\'{o}rmula de
Homotop\'{\i}a de Cartan permite descomponer formas de transgresi\'{o}n y
as\'{\i}, entre otras cosas, obtener expresiones para invariantes
topol\'{o}gicos incluso en variedades con borde.

En efecto, usando ecs.~(\ref{EcTeoChern-WeilSobreM}) y~(\ref{EcTriangular}) en
forma conjunta, e integrando sobre una variedad $M$ $\left(  2n+2\right)
$-dimensional,%
\[
\int_{M}\left\langle \boldsymbol{F}^{n+1}\right\rangle -\int_{\partial
M}T_{\boldsymbol{A}\mathbb{\leftarrow}\tilde{\boldsymbol{A}}}^{\left(
2n+1\right)  }=\int_{M}\left\langle \bar{\boldsymbol{F}}^{n+1}\right\rangle
-\int_{\partial M}\mathrm{d}T_{\bar{\boldsymbol{A}}\mathbb{\leftarrow}%
\tilde{\boldsymbol{A}}}^{\left(  2n+1\right)  }\text{.}%
\]

As\'{\i}, vemos que una vez fijada una conexi\'{o}n $\tilde{\boldsymbol{A}}$
sobre $\partial M,$ la cantidad $\int_{M}\left\langle \boldsymbol{F}%
^{n+1}\right\rangle -\int_{\partial M}T_{\boldsymbol{A}\mathbb{\leftarrow
}\tilde{\boldsymbol{A}}}^{\left(  2n+1\right)  }$ corresponde a un invariante
topol\'{o}gico, el cual tendr\'{a} siempre el mismo valor sin importar que
conexi\'{o}n $\boldsymbol{A}$ se est\'{e} usando.

M\'{a}s all\'{a} de esta sencilla observaci\'{o}n, la ec.~(\ref{EcTriangular})
es especialmente \'{u}til en s\'{\i} misma. Esto, debido a que usaremos una
forma de transgresi\'{o}n como Lagrangeano, y por lo tanto, se vuelve de vital
importancia separar la transgresi\'{o}n en distintas partes para las distintas
interacciones presentes en la teor\'{\i}a. Por otra parte,
ec.~(\ref{EcTriangular}) es vital para entender la relaci\'{o}n entre
teor\'{\i}as de Chern--Simons y Transgresiones, en donde el t\'{e}rmino de
borde \textrm{d}$Q_{\boldsymbol{A}_{2}\leftarrow\boldsymbol{A}_{1}%
\leftarrow\boldsymbol{A}_{0}}^{\left(  2n\right)  }$ juega un rol
regularizador. A su vez, es posible descomponer f\'{a}cilmente este
t\'{e}rmino de borde usando ec.~(\ref{EcTetraedral}). Todos estos temas
ser\'{a}n analizados con mayor profundidad en el siguiente cap\'{\i}tulo.

\chapter{\label{SecTrans}Formas de Transgresi\'{o}n y Teor\'{\i}as de Gauge}

\begin{center}
\textquotedblleft\ldots\textit{a una Diana glacial en un nicho l\'{o}brego
correspond\'{\i}a en un segundo nicho otra Diana; un balc\'{o}n se reflejaba
en otro balc\'{o}n; dobles escalinatas se abr\'{\i}an en doble balaustrada}%
.\textquotedblright

(Jorge Luis Borges, La Muerte y la Br\'{u}jula)
\end{center}

Dada una cierta simetr\'{\i}a, la forma tradicional de construir una
teor\'{\i}a de gauge asociada a ella es a trav\'{e}s de la acci\'{o}n de
Yang-Mills. Este es por ejemplo, el caso de las interacciones del Modelo
Est\'{a}ndard. Sin embargo, una teor\'{\i}a de Yang--Mills necesita de una
estructura m\'{e}trica de fondo sobre el espacio base $M$. Por lo tanto,
cuando trabajamos sobre un espacio curvo, en donde la m\'{e}trica pasa a ser
un campo din\'{a}mico, una teor\'{\i}a de Yang--Mills ya no puede ser
considerada una teor\'{\i}a de gauge en el estricto sentido de la palabra, ya
que incorpora un campo din\'{a}mico que no es parte de la conexi\'{o}n.

Por esta raz\'{o}n, parece natural pensar en la posibilidad de usar la forma
de transgresi\'{o}n [ec.~(\ref{Ec Transg EspBase 2})] como Lagrangeano para
una teor\'{\i}a de gauge, dado que esta no necesita de una estructura
m\'{e}trica de fondo para ser definida.

Sin embargo, debemos observar que para implementar la transgresi\'{o}n como
lagrangeano, el espacio base debe ser de dimensi\'{o}n impar, $D=2n+1.$

Un lagrangeano transgresor satisface autom\'{a}ticamente las siguientes condiciones:

\begin{enumerate}
\item Es completamente invariante de gauge.

\item Est\'{a} globalmente definido sobre $M$ (Y por lo tanto, puede
integrarse sobre $M$).

\item No precisa de una m\'{e}trica de fondo

\item Todos los campos din\'{a}micos son parte de $1$-formas conexi\'{o}n
valuadas en el \'{a}lgebra.
\end{enumerate}

Sin embargo, es necesario observar que para definir la forma de
transgresi\'{o}n, es necesario utilizar dos conexiones, $\boldsymbol{A}$ y
$\bar{\boldsymbol{A}}$. Desde un punto de vista tradicional, esta duplicidad
en los campos din\'{a}micos pudiera parecer extra\~{n}a, principalmente debido
al hecho de que las teor\'{\i}as de gauge del Modelo Est\'{a}ndar est\'{a}n
contru\'{\i}das s\'{o}lo con una conexi\'{o}n y no con dos.

Es posible considerar varias configuraciones posibles para el par de
conexiones, las cuales desembocan en diversos tipos de teor\'{\i}as. El
espectro va desde trabajar con ambas conexiones sin imponer ninguna
condici\'{o}n sobre ellas, hasta imponer $\bar{\boldsymbol{A}}=0$ y dejar
$\boldsymbol{A}$ como \'{u}nico campo din\'{a}mico.

En la presente secci\'{o}n, consideraremos algunas de las consecuencias de
estas posibles elecciones. Empezaremos considerando el caso m\'{a}s general,
sin imponer restricciones sobre las conexiones, luego la configuraci\'{o}n de
variedades cobordantes, en donde se impone la nulidad de las conexiones en
distintos sectores del espacio base, y finalmente, las teor\'{\i}as de
Chern--Simons, en donde se impone la condici\'{o}n $\bar{\boldsymbol{A}}=0.$

Parte de los resultados presentados en esta secci\'{o}n fueron descubiertos en
forma independiente, por A.~Borowiec, M.~Ferraris y M.~Francaviglia en
Refs.~\cite{Polaco1} y~\cite{Polaco2}, y por P.~Mora en Ref.~\cite{Mora-Tesis}%
. Sin embargo, el punto de vista utilizado en nuestra investigaci\'{o}n es
parcialmente diferente; en particular la idea de asociar conexiones con
orientaciones distintas del espacio base (Sec.~\ref{Sec Accion T General}), la
conservaci\'{o}n \textit{off-shell} de las cargas de Noether
(Sec.~\ref{Sec Corrientes Noether}) y el M\'{e}todo de Separaci\'{o}n en
Subespacios proveen de un enfoque distinto del problema.

\section[Forma de Transgresi\'{o}n y Chern--Simons como Lagrangeano]%
{\label{Sec Trans y CS como Lagrang}Forma de Transgresi\'{o}n y Chern--Simons
como Lagrangeano \sectionmark{Lagrangeano Transgresi\'{o}n/Chern--Simons}}

\sectionmark{Lagrangeano Transgresi\'{o}n/Chern--Simons}

\subsection{Definiciones}

Sea un $\pi:P\rightarrow M$ un fibrado principal con un espacio base $M$
\textit{orientable}, $\left(  2n+1\right)  $-dimensional. Sean dos conexiones,
$\mathbb{A}$ y $\bar{\mathbb{A}}$ sobre $P,$ y consideremos su correspondiente
$\left(  2n+1\right)  $-forma transgresi\'{o}n [ec.~(\ref{Ec Transg Fibrado})]
sobre $P$,%
\[
\mathbb{T}_{\mathbb{A\leftarrow}\bar{\mathbb{A}}}^{\left(  2n+1\right)
}=\left(  n+1\right)  \int_{t=0}^{t=1}\mathrm{d}t~\left\langle \mathbb{OF}%
_{t}^{n}\right\rangle .
\]

Siendo $T_{\boldsymbol{A}\mathbb{\leftarrow}\bar{\boldsymbol{A}}}^{\left(
2n+1\right)  }$ la proyecci\'{o}n de $\mathbb{T}_{\mathbb{A\leftarrow}%
\bar{\mathbb{A}}}^{\left(  2n+1\right)  }$ sobre $M$ [v\'{e}ase
ecs.~(\ref{Ec Transg EspBase 1},~\ref{Ec Transg EspBase 2})],
\[
\mathbb{T}_{\mathbb{A\leftarrow}\bar{\mathbb{A}}}^{\left(  2n+1\right)  }%
=\pi^{\ast}T_{\boldsymbol{A}\mathbb{\leftarrow}\bar{\boldsymbol{A}}}^{\left(
2n+1\right)  },
\]
definimos como Lagrangeano Transgresor sobre $M$ a la $\left(  2n+1\right)
$-forma%
\begin{align}
L_{\mathrm{T}}^{\left(  2n+1\right)  }\left(  \boldsymbol{A},\bar
{\boldsymbol{A}}\right)   &  =kT_{\boldsymbol{A}\mathbb{\leftarrow}%
\bar{\boldsymbol{A}}}^{\left(  2n+1\right)  }\label{Ec LagrangT 1}\\
&  =k\left(  n+1\right)  \int_{t=0}^{t=1}\mathrm{d}t~\left\langle
\boldsymbol{\Theta F}_{t}^{n}\right\rangle \label{Ec LagrangT 2}%
\end{align}
siendo $k$ una constante, $\boldsymbol{\Theta}=\boldsymbol{A}-\bar
{\boldsymbol{A}}$ y $\boldsymbol{F}_{t}=\mathrm{d}\boldsymbol{A}%
_{t}+\boldsymbol{A}_{t}^{2}$ la curvatura en $\boldsymbol{A}_{t}%
=\bar{\boldsymbol{A}}+t\boldsymbol{\Theta}.$

Como ya ha sido mencionado, dado que $\mathbb{T}_{\mathbb{A\leftarrow}%
\bar{\mathbb{A}}}^{\left(  2n+1\right)  }$ es proyectable, $L_{\mathrm{T}%
}^{\left(  2n+1\right)  }\left(  \boldsymbol{A},\bar{\boldsymbol{A}}\right)  $
es invariante de gauge. En efecto, es directo demostrar que bajo las
transformaciones de gauge ec.~(\ref{Ec TransfGauge a+}),
\begin{align*}
\boldsymbol{A}  &  \rightarrow\boldsymbol{A}^{\prime}=g^{-1}\boldsymbol{A}%
g+\boldsymbol{a}_{+},\\
\bar{\boldsymbol{A}}  &  \rightarrow\bar{\boldsymbol{A}}^{\prime}=g^{-1}%
\bar{\boldsymbol{A}}g+\boldsymbol{a}_{+},
\end{align*}
se tiene%
\begin{align*}
\boldsymbol{\Theta}  &  \rightarrow\boldsymbol{\Theta}^{\prime}=g^{-1}%
\boldsymbol{\Theta}g,\\
\boldsymbol{F}_{t}  &  \rightarrow\boldsymbol{F}_{t}{}^{\prime}=g^{-1}%
\boldsymbol{F}_{t}g,
\end{align*}
y por lo tanto, dado que $\left\langle \boldsymbol{\Theta F}_{t}%
^{n}\right\rangle $ es un polinomio invariante [ver ec.~(\ref{EcCondInvGrupo}%
)],%
\[
L_{\mathrm{T}}^{\left(  2n+1\right)  }\left(  \boldsymbol{A}^{\prime}%
,\bar{\boldsymbol{A}}^{\prime}\right)  =L_{\mathrm{T}}^{\left(  2n+1\right)
}\left(  \boldsymbol{A},\bar{\boldsymbol{A}}\right)  .
\]

En el lagrangeano $L_{\mathrm{T}}^{\left(  2n+1\right)  }\left(
\boldsymbol{A},\bar{\boldsymbol{A}}\right)  ,$ las conexiones $\boldsymbol{A}$
y $\bar{\boldsymbol{A}}$ juegan roles sim\'{e}tricos; su intercambio s\'{o}lo
involucra un cambio global de signo,%
\[
T_{\boldsymbol{A}\mathbb{\leftarrow}\bar{\boldsymbol{A}}}^{\left(
2n+1\right)  }=-T_{\bar{\boldsymbol{A}}\mathbb{\leftarrow}\boldsymbol{A}%
}^{\left(  2n+1\right)  }.
\]

Por esta raz\'{o}n, consideraremos por ahora el caso m\'{a}s general, en el
cual $\boldsymbol{A}$ y $\bar{\boldsymbol{A}}$ corresponden a dos conexiones
completamente independientes.

Es conveniente observar que es posible utilizar ec.~(\ref{EcTriangular}) para
descomponer el lagrangeano como%
\begin{equation}
L_{\mathrm{T}}^{\left(  2n+1\right)  }\left(  \boldsymbol{A},\bar
{\boldsymbol{A}}\right)  =kT_{\boldsymbol{A}\leftarrow\tilde{\boldsymbol{A}}%
}^{\left(  2n+1\right)  }-kT_{\bar{\boldsymbol{A}}\leftarrow\tilde
{\boldsymbol{A}}}^{\left(  2n+1\right)  }+k\text{\textrm{d}}Q_{\boldsymbol{A}%
\leftarrow\tilde{\boldsymbol{A}}\leftarrow\bar{\boldsymbol{A}}}^{\left(
2n\right)  }\label{Ec LagrangTransConNhafle}%
\end{equation}
e imponiendo $\tilde{\boldsymbol{A}}=0,$
\[
L_{\mathrm{T}}^{\left(  2n+1\right)  }\left(  \boldsymbol{A},\bar
{\boldsymbol{A}}\right)  =kT_{\boldsymbol{A}\leftarrow\boldsymbol{0}}^{\left(
2n+1\right)  }-kT_{\bar{\boldsymbol{A}}\leftarrow\boldsymbol{0}}^{\left(
2n+1\right)  }+k\text{\textrm{d}}Q_{\boldsymbol{A}\leftarrow\boldsymbol{0}%
\leftarrow\bar{\boldsymbol{A}}}^{\left(  2n\right)  }.
\]

Es interesante notar que la imposici\'{o}n hecha $\tilde{\boldsymbol{A}}=0$ en
ec.~(\ref{Ec LagrangTransConNhafle}) no afecta la globalidad de $L_{\mathrm{T}%
}^{\left(  2n+1\right)  }\left(  \boldsymbol{A},\bar{\boldsymbol{A}}\right)
.$ A\'{u}n m\'{a}s, $L_{\mathrm{T}}^{\left(  2n+1\right)  }\left(
\boldsymbol{A},\bar{\boldsymbol{A}}\right)  $ no depende de la elecci\'{o}n de
$\tilde{\boldsymbol{A}};$ dicho en otras palabras, $L_{\mathrm{T}}^{\left(
2n+1\right)  }\left(  \boldsymbol{A},\bar{\boldsymbol{A}}\right)  $ es
invariante bajo una transformaci\'{o}n arbitraria $\tilde{\boldsymbol{A}%
}\rightarrow\tilde{\boldsymbol{A}}^{\prime}.$ Esto se debe a que en la
deducci\'{o}n de ec.~(\ref{EcTriangular}) nunca se utiliz\'{o}
expl\'{\i}citamente la hip\'{o}tesis de que $\tilde{\boldsymbol{A}}$ fuese una
conexi\'{o}n; basta con que sea una $1$-forma arbitraria valuada en el \'{a}lgebra.

Por otra parte dado que $Q^{\left(  2n+1\right)  }\left(  \boldsymbol{A}%
\right)  =T_{\boldsymbol{A}\leftarrow\boldsymbol{0}}^{\left(  2n+1\right)  }$
corresponde a la forma de Chern--Simons definida \textit{localmente} sobre el
espacio base, se define naturalmente el lagrangeano de Chern--Simons como%
\begin{align*}
L_{\mathrm{CS}}^{\left(  2n+1\right)  }\left(  \boldsymbol{A}\right)   &
=kT_{\boldsymbol{A}\leftarrow\boldsymbol{0}}^{\left(  2n+1\right)  },\\
&  =k\left(  n+1\right)  \int_{t=0}^{t=1}\mathrm{d}t~\left\langle
\boldsymbol{A}\left(  t\mathrm{d}\boldsymbol{A}+t^{2}\boldsymbol{A}%
^{2}\right)  ^{n}\right\rangle .
\end{align*}

As\'{\i}, es posible escribir el Lagrangeano transgresor como la diferencia de
dos Lagrangeanos de CS, m\'{a}s un t\'{e}rmino de borde,%
\begin{equation}
L_{\mathrm{T}}^{\left(  2n+1\right)  }\left(  \boldsymbol{A},\bar
{\boldsymbol{A}}\right)  =L_{\mathrm{CS}}^{\left(  2n+1\right)  }\left(
\boldsymbol{A}\right)  -L_{\mathrm{CS}}^{\left(  2n+1\right)  }\left(
\bar{\boldsymbol{A}}\right)  +k\text{\textrm{d}}Q_{\boldsymbol{A}%
\leftarrow\boldsymbol{0}\leftarrow\bar{\boldsymbol{A}}}^{\left(  2n\right)
}.\label{Ec Lt=Lcs-Lcs+dQ}%
\end{equation}

Es interesante observar que pese a que el lagrangeano de CS est\'{a} s\'{o}lo
\textit{localmente} definido, el Lagrangeano transgresor est\'{a}
\textit{globalmente} definido. En este sentido, el t\'{e}rmino $k$%
\textrm{d}$Q_{\boldsymbol{A}\leftarrow\boldsymbol{0}\leftarrow\bar
{\boldsymbol{A}}}^{\left(  2n\right)  }$ pese a no contribuir a las ecuaciones
de movimiento para $\boldsymbol{A}$ y $\bar{\boldsymbol{A}},$ juega el rol de
un t\'{e}rmino regularizador, que asegura la invariancia completa del
Lagrangeano bajo transformaciones de gauge. Por otra parte, debe de observarse
que este t\'{e}rmino s\'{\i} juega un rol fundamental en la existencia de
condiciones de borde y cargas conservadas de Noether, como veremos en las
secciones siguientes.

\section{\label{Sec Accion T General}Caso General}

\subsection{Ecuaciones del Movimiento y Condiciones de Borde para el
Lagrangeano Transgresor}

Para obtener las ecuaciones de movimiento y condiciones de borde, consideremos
la variaci\'{o}n del lagrangiano bajo las variaciones infinitesimales
independientes%
\begin{align*}
\boldsymbol{A}  &  \rightarrow\boldsymbol{A}^{\prime}=\boldsymbol{A}%
+\delta\boldsymbol{A},\\
\bar{\boldsymbol{A}}  &  \rightarrow\bar{\boldsymbol{A}}^{\prime}%
=\bar{\boldsymbol{A}}+\delta\bar{\boldsymbol{A}}.
\end{align*}

Bajo ellas, el Lagrangeano $L_{\mathrm{T}}^{\left(  2n+1\right)  }$ var\'{\i}a
de la forma (ve\'{a}se Ap\'{e}ndice~\ref{Apend VarTransg})%
\begin{equation}
\delta L_{\mathrm{T}}^{\left(  2n+1\right)  }=k\left(  n+1\right)  \left(
\left\langle \delta\boldsymbol{AF}^{n}\right\rangle -\left\langle \delta
\bar{\boldsymbol{A}}\bar{\boldsymbol{F}}^{n}\right\rangle \right)  +kn\left(
n+1\right)  \mathrm{d}\int_{0}^{1}\mathrm{d}t~\left\langle \delta
\boldsymbol{A}_{t}\boldsymbol{\Theta F}_{t}^{n-1}\right\rangle
.\label{Ec VarLagrangT}%
\end{equation}

Como ya se explic\'{o} en la introducci\'{o}n de este cap\'{\i}tulo, existen
varias maneras de manejar la existencia de dos conexiones,
$\boldsymbol{A}$ y $\bar{\boldsymbol{A}}$. Consideremos por ahora la
alternativa m\'{a}s general, en donde $\boldsymbol{A}$ y $\bar{\boldsymbol{A}%
}$ est\'{a}n definidas en forma independiente como campos din\'{a}micos sobre
la variedad $\left(  2n+1\right)  $-dimensional $M$.

En este caso, utilizando ec.~(\ref{Ec Lt=Lcs-Lcs+dQ}) la acci\'{o}n
corresponde a%
\begin{align}
S_{\mathrm{T}}^{\left(  2n+1\right)  }\left[  \boldsymbol{A},\bar
{\boldsymbol{A}}\right]   &  =\int_{M}L_{\mathrm{T}}^{\left(  2n+1\right)
}\left(  \boldsymbol{A},\bar{\boldsymbol{A}}\right)
,\label{Ec Def Accion M 1}\\
&  =\int_{M}L_{\mathrm{CS}}^{\left(  2n+1\right)  }\left(  \boldsymbol{A}%
\right)  -\int_{M}L_{\mathrm{CS}}^{\left(  2n+1\right)  }\left(
\bar{\boldsymbol{A}}\right)  +k\int_{\partial M}Q_{\boldsymbol{A}%
\leftarrow\boldsymbol{0}\leftarrow\bar{\boldsymbol{A}}}^{\left(  2n\right)
}.\label{Ec Def Accion M 2}%
\end{align}

En este contexto, estamos en presencia de una teor\'{\i}a con dos conexiones,
$\boldsymbol{A}$ y $\bar{\boldsymbol{A}}$ las cuales son auto-interactuantes,
pero que entre ellas s\'{o}lo interact\'{u}an en el borde de $M$. Sin embargo,
debe observarse que el segundo t\'{e}rmino del lado derecho tiene un
t\'{e}rmino cin\'{e}tico con el signo \textquotedblleft
equivocado\textquotedblright\ y por lo tanto, en prinicipio el campo
$\bar{\boldsymbol{A}}$ corresponder\'{\i}a a un \textquotedblleft
fantasma\textquotedblright\ el cual estar\'{\i}a asociado a estados que violan
conservaci\'{o}n de la probabilidad o con densidad de energ\'{\i}a negativa.
Este tipo de campos han sido utilizados \'{u}ltimamente en el contexto de
teor\'{\i}as efectivas en cosmolog\'{\i}a inflacionaria, debido a ciertas
indicaciones (basadas en datos de supernovas) sobre la posibilidad de violar
la condici\'{o}n d\'{e}bil de energ\'{\i}a. Pese a ello, este tipo de
teor\'{\i}as parecen no ser viables desde un punto de vista cu\'{a}ntico, ya
que se obtendr\'{\i}an teor\'{\i}as no unitarias o bien, estados de
energ\'{\i}a negativa sin un vac\'{\i}o bien determinado (Ve\'{a}se por
ejemplo, Ref.~\cite{GhostBusters}).

\begin{figure}[ptb]
\begin{center}
\includegraphics[width=.7\textwidth]{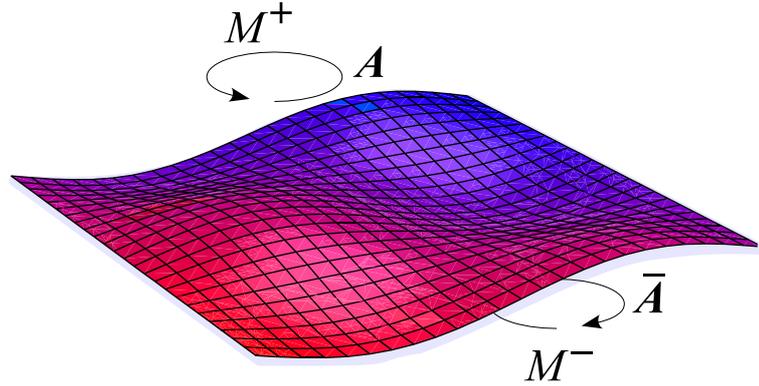}
\end{center}
\caption{Asociaci\'{o}n de cada conexi\'{o}n con una orientaci\'{o}n de la
variedad base.}%
\label{FigOrientacion}%
\end{figure}

Una manera de resolver el problema del fantasma es asociando cada una de las
conexiones con una orientaci\'{o}n distinta de $M.$ Sea $M^{+}$ la variedad
$M$ dotada de una cierta orientaci\'{o}n y $M^{-}$ la variedad $M$ dotada de
la orientaci\'{o}n contraria. Asociamos la conexi\'{o}n $\boldsymbol{A}$ con
$M^{+}$ y $\bar{\boldsymbol{A}}$ con $M^{-}$ [ve\'{a}se
Fig.~\ref{FigOrientacion}].

As\'{\i} la diferencia de signo en ec.~(\ref{Ec Def Accion M 2}) puede ser
interpretada como debida a la integraci\'{o}n sobre $M$ utilizando distintas
orientaciones,%
\begin{equation}
S_{\mathrm{T}}^{\left(  2n+1\right)  }\left[  \boldsymbol{A},\bar
{\boldsymbol{A}}\right]  =\int_{M^{+}}L_{\mathrm{CS}}^{\left(  2n+1\right)
}\left(  \boldsymbol{A}\right)  +\int_{M^{-}}L_{\mathrm{CS}}^{\left(
2n+1\right)  }\left(  \bar{\boldsymbol{A}}\right)  +k\int_{\partial M^{+}%
}Q_{\boldsymbol{A}\leftarrow\boldsymbol{0}\leftarrow\bar{\boldsymbol{A}}%
}^{\left(  2n\right)  }.\label{Ec Def Accion M+ M-}%
\end{equation}

De esta forma, lo que tenemos son dos teor\'{\i}as de CS, cada una
\textquotedblleft viviendo\textquotedblright\ en cada lado de la variedad $M,
$ las cuales s\'{o}lo interact\'{u}an a trav\'{e}s del borde $\partial M.$

Debe notarse que ec.~(\ref{Ec Def Accion M+ M-}) es completamente
sim\'{e}trica en los roles de $\boldsymbol{A}$ y $\bar{\boldsymbol{A}}$ y las
orientaciones $M^{+}$ y $M^{-},$ debido a que $\partial M^{+}=-\partial M^{-}
$ y $Q_{\boldsymbol{A}\leftarrow\boldsymbol{0}\leftarrow\bar{\boldsymbol{A}}%
}^{\left(  2n\right)  }=-Q_{\bar{\boldsymbol{A}}\leftarrow\boldsymbol{0}%
\leftarrow\boldsymbol{A}}^{\left(  2n\right)  }.$ Que la teor\'{\i}a tenga la
misma forma en $\boldsymbol{A}$ y $\bar{\boldsymbol{A}}$ parece natural,
teniendo en cuenta que la orientaci\'{o}n de $M$ es una elecci\'{o}n. Una
forma diferente de decir lo mismo es que la acci\'{o}n completa ahora posee
una simetr\'{\i}a discreta extra, de invariancia bajo cambios de orientaci\'{o}n.

As\'{\i}, usando ec.~(\ref{Ec VarLagrangT}), llegamos a las ecuaciones del
movimiento%
\begin{align}
\left.  \left\langle \boldsymbol{T}_{A}\boldsymbol{F}^{n}\right\rangle
\right\vert _{M}  &  =0\label{Ec EcMov A General}\\
\left.  \left\langle \boldsymbol{T}_{A}\bar{\boldsymbol{F}}^{n}\right\rangle
\right\vert _{M}  &  =0\label{Ec EcMov A_ General}%
\end{align}
y a las condiciones de borde%
\begin{equation}
\left.  \int_{0}^{1}\mathrm{d}t~\left\langle \delta\boldsymbol{A}%
_{t}\boldsymbol{\Theta F}_{t}^{n-1}\right\rangle \right\vert _{\partial
M}=0.\label{Ec EcBorde General}%
\end{equation}

\subsection{\label{Sec Corrientes Noether}Corrientes de Noether}

La acci\'{o}n considerada presenta dos simetr\'{\i}as. La primera, es la
invariancia bajo difeomorfismos sobre $M$, debida a la utilizaci\'{o}n de
formas diferenciales para llevar a cabo la construcci\'{o}n. La segunda, es la
invariancia bajo transformaciones de gauge del grupo de simetr\'{\i}a,
garantizada por el hecho de que estamos usando una forma de transgresi\'{o}n
como Lagrangeano.

Dadas estas simetr\'{\i}as, el teorema de Noether (Ve\'{a}se
Ap\'{e}ndice~\ref{Apend TeoNoether}) nos garantiza la existencia de cargas
conservadas asociadas a ellas. Como veremos a continuaci\'{o}n, para el caso
de la forma del lagrangeano Transgresor las cargas de Noether asociadas a
difeomorfismos y a simetr\'{\i}as de gauge son no s\'{o}lo conservadas
\textit{on-shell} (\textit{i.e.}, sobre estados que satisfacen las ecuaciones
del moviemiento), sino que en general en forma \textit{off-shell}
(\textit{i.e.,} para cualquier configuraci\'{o}n). Esto esta de acuerdo con el
hecho de que, ya que el el Lagrangeano completo corresponde a una
anomal\'{\i}a, la teor\'{\i}a deber\'{\i}a ser en principio libre de anomal\'{\i}as.

Por otra parte, es interesante se\~{n}alar que estas cargas de Noether
conducen a resultados finitos y consistentes con el formalismo Hamiltoniano en
el caso de gravedad y en particular en el caso de Agujeros Negros (Ve\'{a}se
Refs.~\cite{CECS-Trans,CECS-VacuumOddDim,CECS-FiniteGrav}), haciendo as\'{\i}
de gravedad en dimensiones impares una teor\'{\i}a tan consistente como su
contraparte en dimensiones pares en este sentido (Ve\'{a}se Refs.~\cite{CECS-ConservCharg1,CECS-ConservCharg2}).

\subsubsection{Corriente de Difeomorfismos}

Comparando ec.~(\ref{Ec VarLagrangT}) y ec.~(\ref{Ec dL = E+dB}) del
Ap\'{e}ndice~\ref{Apend TeoNoether}, tenemos (en la notaci\'{o}n del
ap\'{e}ndice)%
\begin{align*}
E_{A}\varphi^{A}  &  =k\left(  n+1\right)  \left(  \left\langle \delta
\boldsymbol{AF}^{n}\right\rangle -\left\langle \delta\bar{\boldsymbol{A}}%
\bar{\boldsymbol{F}}^{n}\right\rangle \right)  ,\\
B_{A}\delta\varphi^{A}  &  =kn\left(  n+1\right)  \int_{0}^{1}\mathrm{d}%
t~\left\langle \delta\boldsymbol{A}_{t}\boldsymbol{\Theta F}_{t}%
^{n-1}\right\rangle .
\end{align*}

Utilizando ec~\ref{Ec J Dif On-Shell} para el presente caso, se tiene para la
corriente on-shell de difeomorfismos,%
\begin{equation}
\ast J^{\left(  \mathrm{dif-on}\right)  }=kn\left(  n+1\right)  \int_{0}%
^{1}\mathrm{d}t~\left\langle \pounds _{\xi}\boldsymbol{A}_{t}%
\boldsymbol{\Theta F}_{t}^{n-1}\right\rangle -k\left(  n+1\right)  \int
_{t=0}^{t=1}\mathrm{d}t~\mathrm{I}_{\xi}\left\langle \boldsymbol{\Theta F}%
_{t}^{n}\right\rangle .\label{Ec J dif on T ini}%
\end{equation}

Es posible escribir una expresi\'{o}n m\'{a}s sencilla para la corriente
escribiendo la derivada de Lie de la conexi\'{o}n como%

\begin{equation}
\pounds _{\xi}\boldsymbol{A}_{t}=\mathrm{I}_{\xi}\boldsymbol{F}_{t}%
+\mathrm{D}_{t}\mathrm{I}_{\xi}\boldsymbol{A}_{t}\label{Ec DerivLie At}%
\end{equation}
en donde%
\[
\mathrm{D}_{t}\mathrm{I}_{\xi}\boldsymbol{A}_{t}=\mathrm{dI}_{\xi
}\boldsymbol{A}_{t}+\left[  \boldsymbol{A}_{t},\mathrm{I}_{\xi}\boldsymbol{A}%
_{t}\right]  .
\]

Reemplazando ec.~(\ref{Ec DerivLie At}) en~(\ref{Ec J dif on T ini}) y usando
la regla de Leibniz para $\mathrm{I}_{\xi},$ tenemos que%
\[
\ast J^{\left(  \mathrm{dif-on}\right)  }=kn\left(  n+1\right)  \int_{0}%
^{1}\mathrm{d}t~\left\langle \mathrm{D}_{t}\mathrm{I}_{\xi}\boldsymbol{A}%
_{t}\boldsymbol{\Theta F}_{t}^{n-1}\right\rangle -k\left(  n+1\right)
\int_{t=0}^{t=1}\mathrm{d}t~\left\langle \mathrm{I}_{\xi}\boldsymbol{\Theta
F}_{t}^{n}\right\rangle
\]

Integrando por partes en $\mathrm{D}_{t}$ y $\frac{\mathrm{d}}{\mathrm{d}t},$
llegamos a%
\[
\ast J^{\left(  \mathrm{dif-on}\right)  }=kn\left(  n+1\right)  \mathrm{d}%
\int_{0}^{1}\mathrm{d}t~\left\langle \mathrm{I}_{\xi}\boldsymbol{A}%
_{t}\boldsymbol{\Theta F}_{t}^{n-1}\right\rangle -k\left(  n+1\right)  \left(
\left\langle \mathrm{I}_{\xi}\boldsymbol{AF}^{n}\right\rangle -\left\langle
\mathrm{I}_{\xi}\bar{\boldsymbol{\boldsymbol{A}}\bar{\boldsymbol{F}}}%
^{n}\right\rangle \right)  .
\]

Sobre una configuraci\'{o}n de $\boldsymbol{A}$ y $\bar
{\boldsymbol{\boldsymbol{A}}}$ que satisface las ecuaciones del movimiento, el
segundo t\'{e}rmino desaparece y la corriente conservada corresponder\'{\i}a
s\'{o}lo al primer t\'{e}rmino. Sin embargo, no haremos uso de las ecuaciones
del movimiento, si no que m\'{a}s bien veremos que para la transgresi\'{o}n se
satisface la condici\'{o}n ec.~(\ref{Ec CondOffShell Dif}) y por lo tanto, es
posible escribir una corriente conservada off-shell.

En efecto, tenemos que%
\begin{align*}
E_{A}\pounds _{\xi}\varphi^{A}  &  =k\left(  n+1\right)  \left(  \left\langle
\pounds _{\xi}\boldsymbol{AF}^{n}\right\rangle -\left\langle \pounds _{\xi
}\bar{\boldsymbol{A}}\bar{\boldsymbol{F}}^{n}\right\rangle \right)  ,\\
&  =k\left(  n+1\right)  \left(  \left\langle \mathrm{I}_{\xi}\boldsymbol{FF}%
^{n}\right\rangle -\left\langle \mathrm{I}_{\xi}\bar{\boldsymbol{F}}%
\bar{\boldsymbol{F}}^{n}\right\rangle \right)  +k\left(  n+1\right)  \left(
\left\langle \mathrm{D}_{\boldsymbol{A}}\mathrm{I}_{\xi}\boldsymbol{AF}%
^{n}\right\rangle -\left\langle \mathrm{D}_{\bar{\boldsymbol{A}}}%
\mathrm{I}_{\xi}\bar{\boldsymbol{A}}\bar{\boldsymbol{F}}^{n}\right\rangle
\right)  ,\\
&  =k\mathrm{I}_{\xi}\left(  \left\langle \boldsymbol{F}^{n+1}\right\rangle
-\left\langle \bar{\boldsymbol{F}}^{n+1}\right\rangle \right)  +k\left(
n+1\right)  \mathrm{d}\left(  \left\langle \mathrm{I}_{\xi}\boldsymbol{AF}%
^{n}\right\rangle -\left\langle \mathrm{I}_{\xi}\bar{\boldsymbol{A}}%
\bar{\boldsymbol{F}}^{n}\right\rangle \right)  ,\\
&  =\mathrm{I}_{\xi}\mathrm{d}L_{\mathrm{T}}^{\left(  2n+1\right)  }+k\left(
n+1\right)  \mathrm{d}\left(  \left\langle \mathrm{I}_{\xi}\boldsymbol{AF}%
^{n}\right\rangle -\left\langle \mathrm{I}_{\xi}\bar{\boldsymbol{A}}%
\bar{\boldsymbol{F}}^{n}\right\rangle \right)  ,\\
&  =k\left(  n+1\right)  \mathrm{d}\left(  \left\langle \mathrm{I}_{\xi
}\boldsymbol{AF}^{n}\right\rangle -\left\langle \mathrm{I}_{\xi}%
\bar{\boldsymbol{A}}\bar{\boldsymbol{F}}^{n}\right\rangle \right)  .
\end{align*}
y por lo tanto, tenemos $E_{A}\pounds _{\xi}\varphi^{A}=\mathrm{d}X$ con
$X=k\left(  n+1\right)  \mathrm{d}\left(  \left\langle \mathrm{I}_{\xi
}\boldsymbol{AF}^{n}\right\rangle -\left\langle \mathrm{I}_{\xi}%
\bar{\boldsymbol{A}}\bar{\boldsymbol{F}}^{n}\right\rangle \right)  .$

Por lo tanto, podemos usar la expresi\'{o}n para la corriente off-shell
ec.~(\ref{Ec Dif: Joff=Jon + X}),%
\begin{align*}
\ast J^{\left(  \mathrm{dif-off}\right)  }  &  =\ast J^{\left(
\mathrm{dif-on}\right)  }+X,\\
&  =kn\left(  n+1\right)  \mathrm{d}\int_{0}^{1}\mathrm{d}t~\left\langle
\mathrm{I}_{\xi}\boldsymbol{A}_{t}\boldsymbol{\Theta F}_{t}^{n-1}\right\rangle
-k\left(  n+1\right)  \left(  \left\langle \mathrm{I}_{\xi}\boldsymbol{AF}%
^{n}\right\rangle -\left\langle \mathrm{I}_{\xi}\bar
{\boldsymbol{\boldsymbol{A}}\bar{\boldsymbol{F}}}^{n}\right\rangle \right)
+\\
&  +k\left(  n+1\right)  \mathrm{d}\left(  \left\langle \mathrm{I}_{\xi
}\boldsymbol{AF}^{n}\right\rangle -\left\langle \mathrm{I}_{\xi}%
\bar{\boldsymbol{A}}\bar{\boldsymbol{F}}^{n}\right\rangle \right)  ,\\
&  =kn\left(  n+1\right)  \mathrm{d}\int_{0}^{1}\mathrm{d}t~\left\langle
\mathrm{I}_{\xi}\boldsymbol{A}_{t}\boldsymbol{\Theta F}_{t}^{n-1}\right\rangle
.
\end{align*}

\subsubsection{Corriente de Gauge}

En este caso, dada una transformaci\'{o}n infinitesimal de gauge generada por
el elemento $g=1+\boldsymbol{\lambda},$ las conexiones var\'{\i}an como%
\begin{align*}
\delta\boldsymbol{A}  &  =\mathrm{D}_{\boldsymbol{A}}\boldsymbol{\lambda},\\
\delta\bar{\boldsymbol{A}}  &  =\mathrm{D}_{\bar{\boldsymbol{A}}%
}\boldsymbol{\lambda}.
\end{align*}

As\'{\i}, en la notaci\'{o}n del Ap\'{e}ndice~\ref{Apend TeoNoether}, tenemos%
\begin{align}
E_{A}\epsilon^{A}  &  =k\left(  n+1\right)  \left(  \left\langle
\mathrm{D}_{\boldsymbol{A}}\boldsymbol{\lambda F}^{n}\right\rangle
-\left\langle \mathrm{D}_{\bar{\boldsymbol{A}}}\bar{\boldsymbol{F}}%
^{n}\right\rangle \right) \nonumber\\
&  =k\left(  n+1\right)  \mathrm{d}\left(  \left\langle \boldsymbol{\lambda
F}^{n}\right\rangle -\left\langle \boldsymbol{\lambda}\bar{\boldsymbol{F}}%
^{n}\right\rangle \right) \label{Ec Transg Eaea}\\
B_{A}\epsilon^{A}  &  =kn\left(  n+1\right)  \int_{0}^{1}\mathrm{d}%
t~\left\langle \mathrm{D}_{t}\boldsymbol{\lambda\Theta F}_{t}^{n-1}%
\right\rangle \label{Ec Transg Baea}%
\end{align}
y la corriente on-shell de gauge,%
\begin{align*}
\ast J^{\left(  \mathrm{gauge-on}\right)  }  &  =B_{A}\epsilon^{A},\\
&  =kn\left(  n+1\right)  \int_{0}^{1}\mathrm{d}t~\left\langle \mathrm{D}%
_{t}\boldsymbol{\lambda\Theta F}_{t}^{n-1}\right\rangle .
\end{align*}

Es posible escribir una expresi\'{o}n m\'{a}s compacta para la corriente de
gauge integrando por partes en $\mathrm{D}_{t}$ y $\frac{\mathrm{d}%
}{\mathrm{d}t},$%
\begin{align*}
\ast J^{\left(  \mathrm{gauge-on}\right)  }  &  =kn\left(  n+1\right)
\mathrm{d}\int_{0}^{1}\mathrm{d}t~\left\langle \boldsymbol{\lambda\Theta
F}_{t}^{n-1}\right\rangle -k\left(  n+1\right)  \int_{0}^{1}\mathrm{d}%
t~\frac{\mathrm{d}}{\mathrm{d}t}\left\langle \boldsymbol{\lambda F}_{t}%
^{n}\right\rangle ,\\
&  =kn\left(  n+1\right)  \mathrm{d}\int_{0}^{1}\mathrm{d}t~\left\langle
\boldsymbol{\lambda\Theta F}_{t}^{n-1}\right\rangle -k\left(  n+1\right)
\left(  \left\langle \boldsymbol{\lambda F}^{n}\right\rangle -\left\langle
\boldsymbol{\lambda\bar{\boldsymbol{F}}}^{n}\right\rangle \right)  .
\end{align*}

Ahora bien, de ec.~(\ref{Ec Transg Eaea}), tenemos que se satisface la
condici\'{o}n ec.~(\ref{Ec CondOffShell Gauge}), $E_{A}\epsilon^{A}%
=\mathrm{d}Y$ con%
\[
Y=k\left(  n+1\right)  \left(  \left\langle \boldsymbol{\lambda F}%
^{n}\right\rangle -\left\langle \boldsymbol{\lambda}\bar{\boldsymbol{F}}%
^{n}\right\rangle \right)  ,
\]

Por lo tanto, la corriente de gauge conservada off-shell viene dada por%
\begin{align*}
\ast J^{\left(  \mathrm{gauge-off}\right)  }  &  =\ast J^{\left(
\mathrm{gauge-on}\right)  }+Y,\\
&  =kn\left(  n+1\right)  \mathrm{d}\int_{0}^{1}\mathrm{d}t~\left\langle
\boldsymbol{\lambda\Theta F}_{t}^{n-1}\right\rangle -k\left(  n+1\right)
\left(  \left\langle \boldsymbol{\lambda F}^{n}\right\rangle -\left\langle
\boldsymbol{\lambda\bar{\boldsymbol{F}}}^{n}\right\rangle \right)  +\\
&  +k\left(  n+1\right)  \left(  \left\langle \boldsymbol{\lambda F}%
^{n}\right\rangle -\left\langle \boldsymbol{\lambda}\bar{\boldsymbol{F}}%
^{n}\right\rangle \right)  ,\\
&  =kn\left(  n+1\right)  \mathrm{d}\int_{0}^{1}\mathrm{d}t~\left\langle
\boldsymbol{\lambda\Theta F}_{t}^{n-1}\right\rangle .
\end{align*}

\subsubsection{Cargas Conservadas}

En la seccci\'{o}n anterior, demostramos la existencia de dos corrientes
conservadas off-shell, una para transformaciones de gauge y la otra para
difeomorfismos,%
\begin{align*}
\ast J^{\left(  \mathrm{gauge-off}\right)  }  &  =kn\left(  n+1\right)
\mathrm{d}\int_{0}^{1}\mathrm{d}t~\left\langle \boldsymbol{\lambda\Theta
F}_{t}^{n-1}\right\rangle ,\\
\ast J^{\left(  \mathrm{dif-off}\right)  }  &  =kn\left(  n+1\right)
\mathrm{d}\int_{0}^{1}\mathrm{d}t~\left\langle \mathrm{I}_{\xi}\boldsymbol{A}%
_{t}\boldsymbol{\Theta F}_{t}^{n-1}\right\rangle .
\end{align*}

Ambas corresponden a formas exactas, lo cual hace trivial verificar la
ecuaci\'{o}n de continuidad, $\mathrm{d}\ast J=0.$ A\'{u}n m\'{a}s, tal como
se discute en el Ap\'{e}ndice~\ref{Apend TeoNoether}, cuando $M$ tiene la
topolog\'{\i}a $M=\mathbb{R}\times\Sigma_{0},$ es posible escribir las cargas
como una integral sobre el borde de la secci\'{o}n espacial $\Sigma,$
\begin{align*}
Q^{\left(  \mathrm{gauge-off}\right)  }  &  =kn\left(  n+1\right)
\int_{\partial\Sigma_{0}}\int_{0}^{1}\mathrm{d}t~\left\langle
\boldsymbol{\lambda\Theta F}_{t}^{n-1}\right\rangle ,\\
Q^{\left(  \mathrm{dif-off}\right)  }  &  =kn\left(  n+1\right)
\int_{\partial\Sigma_{0}}\int_{0}^{1}\mathrm{d}t~\left\langle \mathrm{I}_{\xi
}\boldsymbol{A}_{t}\boldsymbol{\Theta F}_{t}^{n-1}\right\rangle ,
\end{align*}
en donde hemos omitido la imagen rec\'{\i}proca a $\partial\Sigma_{0}$ para no
recargar la notaci\'{o}n.

Ambas cargas son trivialmente invariantes de difeomorfismos, ya que han sido
escritas usando formas diferenciales. Por otra parte, bajo una
transformaci\'{o}n de gauge finita,%
\begin{align}
\boldsymbol{A}  &  \rightarrow\boldsymbol{A}^{\prime}=g^{-1}\boldsymbol{A}%
g+\boldsymbol{a}_{+},\label{Ec TransGaugeWea1}\\
\bar{\boldsymbol{A}}  &  \rightarrow\bar{\boldsymbol{A}}^{\prime}=g^{-1}%
\bar{\boldsymbol{A}}g+\boldsymbol{a}_{+}.\label{Ec TransGaugeWea2}%
\end{align}
la carga de difeomorfismos cambia de la forma%
\[
Q^{\left(  \mathrm{dif-off}\right)  \prime}=Q^{\left(  \mathrm{dif-off}%
\right)  }-kn\left(  n+1\right)  \int_{\partial\Sigma_{0}}\int_{0}%
^{1}\mathrm{d}t~\left\langle \mathrm{I}_{\xi}\boldsymbol{a}_{-}%
\boldsymbol{\Theta F}_{t}^{n-1}\right\rangle .
\]

As\'{\i}, la carga de difeomorfismos es invariante bajo transformaciones de
gauge que satisfagan la condici\'{o}n%
\[
\left.  \mathrm{I}_{\xi}\boldsymbol{a}_{-}\right\vert _{\partial\Sigma_{0}}=0
\]
en el borde de la secci\'{o}n espacial

Una forma alternativa de escribir esta condici\'{o}n es como%
\[
\left.  g\pounds _{\xi}g^{-1}\right\vert _{\partial\Sigma_{0}}=0,
\]
o para $g=1+\boldsymbol{\lambda}$ infinitesimalmente cercano a la identidad,%
\[
\left.  \pounds _{\xi}\boldsymbol{\lambda}\right\vert _{\partial\Sigma_{0}}=0.
\]

Bajo las transformaciones de gauge ec.~(\ref{Ec TransGaugeWea1}%
,~\ref{Ec TransGaugeWea2}), la carga de gauge $Q^{\left(  \mathrm{gauge-off}%
\right)  }$ transforma como%
\[
Q^{\left(  \mathrm{gauge-off}\right)  \prime}=kn\left(  n+1\right)
\int_{\partial\Sigma_{0}}\int_{0}^{1}\mathrm{d}t~\left\langle
g\boldsymbol{\lambda}g^{-1}\boldsymbol{\Theta F}_{t}^{n-1}\right\rangle .
\]

Para $g=1+\boldsymbol{\eta}$ infinitesimalmente cercano a la identidad,
tenemos%
\[
Q^{\left(  \mathrm{gauge-off}\right)  \prime}=Q^{\left(  \mathrm{gauge-off}%
\right)  }+kn\left(  n+1\right)  \int_{\partial\Sigma_{0}}\int_{0}%
^{1}\mathrm{d}t~\left\langle \left[  \boldsymbol{\eta},\boldsymbol{\lambda
}\right]  \boldsymbol{\Theta F}_{t}^{n-1}\right\rangle .
\]

Es posible reescribir esto como%
\[
\delta_{\boldsymbol{\eta}}Q^{\left(  \mathrm{gauge-off}\right)  }\left(
\boldsymbol{\lambda}\right)  =Q^{\left(  \mathrm{gauge-off}\right)  }\left(
\left[  \boldsymbol{\eta},\boldsymbol{\lambda}\right]  \right)
\]
o diciendo que las cargas reproducen el \'{a}lgebra de Lie,%
\[
\left\{  Q^{\left(  \mathrm{gauge-off}\right)  }\left(  \boldsymbol{\eta
}\right)  ,Q^{\left(  \mathrm{gauge-off}\right)  }\left(  \boldsymbol{\lambda
}\right)  \right\}  =Q^{\left(  \mathrm{gauge-off}\right)  }\left(  \left[
\boldsymbol{\eta},\boldsymbol{\lambda}\right]  \right)  .
\]

No hay cargas centrales extra, debido a que el lagrangeano transgresor es
completamente invariante.

\section{Caso de Variedades Cobordantes}

\begin{figure}[ptb]
\begin{center}
\includegraphics[width=.7\textwidth]{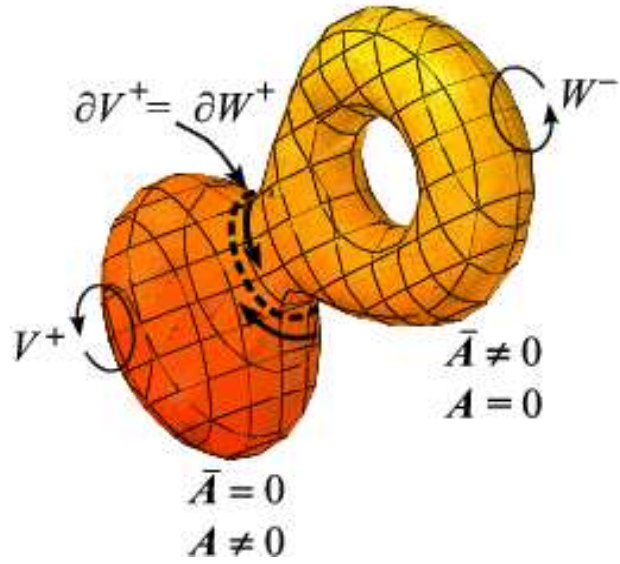}
\end{center}
\caption{Configuraci\'{o}n de variedades cobordantes.}%
\label{FigChupete}%
\end{figure}

Tal como fue mencionado en la introducci\'{o}n de este cap\'{\i}tulo, desde un
punto de vista f\'{\i}sico puede parecer perturbador la presencia de dos
conexiones. Una alternativa interesante es la configuraci\'{o}n de variedad
cobordante (Ve\'{a}se
Refs.~\cite{CECS-Trans,CECS-VacuumOddDim,CECS-FiniteGrav}). Sea el espacio
base $M$ una variedad orientable $\left(  2n+1\right)  $-dimensional y sin
bordes, y denotemos con $M^{+}$ al espacio base dotado de una cierta
orientaci\'{o}n y con $M^{-}$ para la orientaci\'{o}n contraria. Consideremos
$M^{+}$ como la uni\'{o}n de dos variedades $\left(  2n+1\right)
$-dimensionales, $V^{+}$ y $W^{-},$ $M^{+}=V^{+}\cup W^{-},$ tales que
$\partial V^{+}$ y $\partial W^{-} $ son cobordantes\footnote{Dos variedades
$\partial V^{+}$ y $\partial W^{-}$ se dicen \textit{cobordantes} cuando su
uni\'{o}n corresponde al borde de otra variedad, $\partial V^{+}\cup\partial
W^{-}=\partial U^{+}$. La variedad $U^{+}$ es llamada un \textit{cobordismo}
entre $\partial V^{+}$ y $\partial W^{-}.$} y que se cumple $\partial
V^{+}=\partial W^{+}$ (Ve\'{a}se Fig.~\ref{FigChupete}).

Por esta raz\'{o}n, en este contexto, se \textit{define} la acci\'{o}n como%
\begin{align*}
S_{\mathrm{T}}^{\left(  2n+1\right)  }\left[  \boldsymbol{A},\bar
{\boldsymbol{A}}\right]   &  =\int_{V^{+}}L_{\mathrm{CS}}^{\left(
2n+1\right)  }\left(  \boldsymbol{A}\right)  -\int_{W^{+}}L_{\mathrm{CS}%
}^{\left(  2n+1\right)  }\left(  \bar{\boldsymbol{A}}\right)  +k\int_{\partial
V^{+}}Q_{\boldsymbol{A}\leftarrow\boldsymbol{0}\leftarrow\bar{\boldsymbol{A}}%
}^{\left(  2n\right)  }\\
&  =\int_{V^{+}}L_{\mathrm{CS}}^{\left(  2n+1\right)  }\left(  \boldsymbol{A}%
\right)  +\int_{W^{-}}L_{\mathrm{CS}}^{\left(  2n+1\right)  }\left(
\bar{\boldsymbol{A}}\right)  +k\int_{\partial V^{+}}Q_{\boldsymbol{A}%
\leftarrow\boldsymbol{0}\leftarrow\bar{\boldsymbol{A}}}^{\left(  2n\right)  }%
\end{align*}

N\'{o}tese que como consecuencia la acci\'{o}n ya no se puede escribir como la
integral de una sola forma sobre una variedad determinada, a diferencia de en
ecs.~(\ref{Ec Def Accion M 1},~\ref{Ec Def Accion M+ M-}). Por otra parte, se
debe de tener en cuenta que en general, la imposici\'{o}n de una conexi\'{o}n
como cero sobre el espacio base s\'{o}lo puede hacerse en forma local.

Pese a ello, este tipo de configuraciones ha sido exit\'{o}samente utilizada
en el contexto de gravedad en dimensiones m\'{a}s altas
(Refs.~\cite{CECS-Trans,CECS-VacuumOddDim,CECS-FiniteGrav}).

Bajo esta configuraci\'{o}n, las ecuaciones del movimiento corresponden a%
\begin{align*}
\left.  \left\langle \boldsymbol{T}_{A}\boldsymbol{F}^{n}\right\rangle
\right\vert _{V}  &  =0\\
\left.  \left\langle \boldsymbol{T}_{A}\bar{\boldsymbol{F}}^{n}\right\rangle
\right\vert _{W}  &  =0
\end{align*}
y las condiciones de borde a%
\[
\left.  \int_{0}^{1}\mathrm{d}t~\left\langle \delta\boldsymbol{A}%
_{t}\boldsymbol{\Theta F}_{t}^{n-1}\right\rangle \right\vert _{\partial V}=0.
\]

\section{Caso de Chern--Simons}

De cierta manera, utilizar la forma de Chern--Simons como Lagrangeano parece
ser la forma m\'{a}s directa de atacar el problema de tener dos conexiones. Se
impone la condici\'{o}n $\bar{\boldsymbol{A}}=0$ sobre $M,$ y desembocamos a
una teor\'{\i}a con s\'{o}lo $\boldsymbol{A}$ como campo din\'{a}mico. Sin
embargo, pese a la sencillez de la idea, se presentan problemas no del todo
triviales. El primero es que en general no es posible imponer $\bar
{\boldsymbol{A}}=0$ globalmente, y por lo tanto, como fue explicado en
sec.~\ref{Sec Def Forma CS}, la forma de Chern--Simons s\'{o}lo est\'{a}
definida en forma local sobre el espacio base. En efecto, sea $\left\{
U_{\alpha}\right\}  $ un cubrimiento con abiertos del espacio base $\left(
2n+1\right)  $-dimensional $M.$ Entonces, sobre una intersecci\'{o}n
$U_{\alpha}\cap U_{\beta}$ en general se tiene
\[
Q_{\alpha}^{\left(  2n+1\right)  }\neq Q_{\beta}^{\left(  2n+1\right)  }%
\]
en donde $Q_{\alpha}^{\left(  2n+1\right)  }=\sigma_{\alpha}^{\ast}%
\mathbb{Q}^{\left(  2n+1\right)  }$ corresponde a la forma de Chern--Simons
sobre $U_{\alpha}$.

El primer problema que plantea la no localidad de la forma de Chern--Simons es
la ambiguedad al definir la acci\'{o}n. Sobre una intesecci\'{o}n $U_{\alpha
}\cap U_{\beta}$ no es lo mismo integrar $\left.  L_{\mathrm{CS}}^{\left(
2n+1\right)  }\right\vert _{U_{\alpha}}=kQ_{\alpha}^{\left(  2n+1\right)  }$
que $\left.  L_{\mathrm{CS}}^{\left(  2n+1\right)  }\right\vert _{U_{\beta}%
}=kQ_{\beta}^{\left(  2n+1\right)  }$ y por lo tanto en principio resulta
imposible definir la integral de acci\'{o}n en forma un\'{\i}voca.

Sin embargo, en el caso de las formas de Chern--Simons, es posible solucionar
este problema en forma parcial, debido a que las formas de Chern--Simons
var\'{\i}an en una forma cerrada bajo transformaciones de gauge, tal como
revisaremos brevemente en la pr\'{o}xima secci\'{o}n.

\subsection{Comportamiento bajo Transformaciones de Gauge}

Una forma de Chern--Simons $Q^{\left(  2n+1\right)  }\left(  \boldsymbol{A}%
\right)  =T_{\boldsymbol{A}\leftarrow\boldsymbol{0}}^{\left(  2n+1\right)  }$
no es invariante de gauge, si no que var\'{\i}a en un t\'{e}rmino que
localmente es cerrado. Llamaremos esta propiedad \textit{cuasi-invariancia}.

En efecto, usando ec.~(\ref{Ec <F^n+1>|Ua = d C-S}), y el hecho de que
$\left\langle \boldsymbol{F}^{n+1}\right\rangle $ es invariante bajo
transformaciones de gauge, tenemos que%
\[
\mathrm{d}\left[  Q^{\left(  2n+1\right)  }\left(  \boldsymbol{A}^{\prime
}\right)  -Q^{\left(  2n+1\right)  }\left(  \boldsymbol{A}\right)  \right]
=0,
\]
en donde $\boldsymbol{A}^{\prime}=g^{-1}\boldsymbol{A}g+\boldsymbol{a}_{+}.$
Por lo tanto, $Q^{\left(  2n+1\right)  }\left(  \boldsymbol{A}^{\prime
}\right)  $ y $Q^{\left(  2n+1\right)  }\left(  \boldsymbol{A}\right)  $
s\'{o}lo pueden diferir en una forma cerrada.

Para evaluar esta variaci\'{o}n en foma expl\'{\i}cita, basta considerar
$Q^{\left(  2n+1\right)  }\left(  \boldsymbol{A}^{\prime}\right)
=T_{\boldsymbol{A}^{\prime}\leftarrow\boldsymbol{0}}^{\left(  2n+1\right)  }$
con $\boldsymbol{A}^{\prime}=g^{-1}\boldsymbol{A}g+\boldsymbol{a}_{+}$ y
utilizar ec.~(\ref{EcTriangular}). As\'{\i}, se tiene%
\begin{equation}
T_{g^{-1}\boldsymbol{A}g+\boldsymbol{a}_{+}\leftarrow\boldsymbol{0}}^{\left(
2n+1\right)  }=T_{g^{-1}\boldsymbol{A}g+\boldsymbol{a}_{+}\leftarrow
\boldsymbol{a}_{+}}^{\left(  2n+1\right)  }+T_{\boldsymbol{a}_{+}%
\leftarrow\boldsymbol{0}}^{\left(  2n+1\right)  }+\mathrm{d}Q_{g^{-1}%
\boldsymbol{A}g+\boldsymbol{a}_{+}\leftarrow\boldsymbol{a}_{+}\leftarrow
\boldsymbol{0}}^{\left(  2n\right)  }.\label{Ec Variando CS}%
\end{equation}

Es posible evaluar cada t\'{e}rmino al lado derecho de
ec.~(\ref{Ec Variando CS}) en forma separada. El primer t\'{e}rmino
corresponde a una forma de Chern--Simons $T_{g^{-1}\boldsymbol{A}%
g+\boldsymbol{a}_{+}\leftarrow\boldsymbol{a}_{+}}^{\left(  2n+1\right)
}=T_{\boldsymbol{A}\leftarrow0}^{\left(  2n+1\right)  },$ pues
\[
T_{g^{-1}\boldsymbol{A}g+\boldsymbol{a}_{+}\leftarrow\boldsymbol{a}_{+}%
}^{\left(  2n+1\right)  }=\left(  n+1\right)  \int_{t=0}^{t=1}\mathrm{d}%
t\left\langle g^{-1}\boldsymbol{A}g\boldsymbol{F}_{t}^{n}\right\rangle
\]
en donde $\boldsymbol{F}_{t}=\mathrm{d}\boldsymbol{A}_{t}+\boldsymbol{A}%
_{t}^{2}$ con%
\[
\boldsymbol{A}_{t}=\boldsymbol{a}_{+}+tg^{-1}\boldsymbol{A}g.
\]

Utilizando ec.~(\ref{EcStructuraMC}), es directo demostrar que%
\[
\boldsymbol{F}_{t}=g^{-1}\left(  t\mathrm{d}\boldsymbol{A}+t^{2}%
\boldsymbol{A}^{2}\right)  g
\]
y as\'{\i}%
\[
T_{g^{-1}\boldsymbol{A}g+\boldsymbol{a}_{+}\leftarrow\boldsymbol{a}_{+}%
}^{\left(  2n+1\right)  }=T_{\boldsymbol{A}\leftarrow0}^{\left(  2n+1\right)
}.
\]

El segundo t\'{e}rmino del lado derecho de ec.~(\ref{Ec Variando CS}) es
extremadamente interesante. Usando ec.~(\ref{EcMC F=0}), se tiene%
\begin{align*}
T_{\boldsymbol{a}_{+}\leftarrow\boldsymbol{0}}^{\left(  2n+1\right)  }  &
=\left(  n+1\right)  \int_{t=0}^{t=1}\mathrm{d}t\left\langle \boldsymbol{a}%
_{+}\left(  t\mathrm{d}\boldsymbol{a}_{+}+t^{2}\boldsymbol{a}_{+}^{2}\right)
^{n}\right\rangle \\
&  =\left(  n+1\right)  \left(  -1\right)  ^{n}\int_{t=0}^{t=1}\mathrm{d}%
t~t^{n}\left(  1-t\right)  ^{n}\left\langle \boldsymbol{a}_{+}\left[
\boldsymbol{a}_{+}^{2}\right]  ^{n}\right\rangle
\end{align*}

Dado que%
\[
\int_{t=0}^{t=1}\mathrm{d}t~t^{n}\left(  1-t\right)  ^{n}=\frac{\left[
n!\right]  ^{2}}{\left(  2n+1\right)  !}%
\]
tenemos que%
\[
T_{\boldsymbol{a}_{+}\leftarrow\boldsymbol{0}}^{\left(  2n+1\right)  }=\left(
-1\right)  ^{n}\frac{\left(  n+1\right)  !n!}{\left(  2n+1\right)
!}\left\langle \boldsymbol{a}_{+}\left[  \boldsymbol{a}_{+}^{2}\right]
^{n}\right\rangle .
\]

Este t\'{e}rmino corresponde a la imagen rec\'{\i}proca sobre el espacio base
del $\left(  2n+1\right)  $-cociclo de Chevalley-Eilenberg $\Omega^{\left(
2n+1\right)  }\left(  \boldsymbol{a}_{+}\right)  $, por lo que
$T_{\boldsymbol{a}_{+}\leftarrow\boldsymbol{0}}^{\left(  2n+1\right)  }$ es
una forma cerrada $\mathrm{d}T_{\boldsymbol{a}_{+}\leftarrow\boldsymbol{0}%
}^{\left(  2n+1\right)  }=0$, pero no exacta. Por otra parte, es interesante
observar que este t\'{e}rmino aparecer\'{a} s\'{o}lo cuando el tensor
invariante posea una componente primitiva; en caso contrario, el cociclo se
anular\'{a} id\'{e}nticamente (Ve\'{a}se sec.~\ref{Sec TensInv Prim-NoPrim}).

El t\'{e}rmino restante en el lado derecho de ec.~(\ref{Ec Variando CS})
corresponde a,%
\[
Q_{g^{-1}\boldsymbol{A}g+\boldsymbol{a}\leftarrow\boldsymbol{a}\leftarrow
\boldsymbol{0}}^{\left(  2n\right)  }=n\left(  n+1\right)  \int_{0}%
^{1}\mathrm{d}t\int_{0}^{t}\mathrm{d}s\left\langle g^{-1}\boldsymbol{A}%
g\boldsymbol{aF}_{st}^{n-1}\right\rangle
\]
en donde $\boldsymbol{F}_{st}=\mathrm{d}\boldsymbol{A}_{st}+\boldsymbol{A}%
_{st}^{2},$ con
\[
\boldsymbol{A}_{st}=t\boldsymbol{a}+sg^{-1}\boldsymbol{A}g.
\]

As\'{\i}, se tiene que%
\[
\boldsymbol{F}_{st}=g^{-1}\left[  s\left(  t\mathrm{d}\boldsymbol{A}%
+s\boldsymbol{A}^{2}\right)  -t\left(  1-t\right)  \boldsymbol{a}_{-}%
^{2}\right]  g,
\]
y por lo tanto,%
\[
Q_{g^{-1}\boldsymbol{A}g+\boldsymbol{a}\leftarrow\boldsymbol{a}\leftarrow
\boldsymbol{0}}^{\left(  2n\right)  }=-n\left(  n+1\right)  \int_{0}%
^{1}\mathrm{d}t\int_{0}^{t}\mathrm{d}s\left\langle \boldsymbol{Aa}_{-}\left[
s\left(  t\mathrm{d}\boldsymbol{A}+s\boldsymbol{A}^{2}\right)  -t\left(
1-t\right)  \boldsymbol{a}_{-}^{2}\right]  ^{n-1}\right\rangle .
\]

As\'{\i}, tenemos que bajo la transformaci\'{o}n de gauge finita
$\boldsymbol{A}\rightarrow\boldsymbol{A}^{\prime}=g^{-1}\boldsymbol{A}%
g+\boldsymbol{a}_{+},$%
\[
Q^{\left(  2n+1\right)  }\left(  \boldsymbol{A}^{\prime}\right)  =Q^{\left(
2n+1\right)  }\left(  \boldsymbol{A}\right)  +\delta Q^{\left(  2n+1\right)
}\left(  \boldsymbol{A},\boldsymbol{a}_{-}\right)  ,
\]
con%
\begin{align*}
\delta Q^{\left(  2n+1\right)  }\left(  \boldsymbol{A},\boldsymbol{a}%
_{-}\right)   &  =-\frac{\left(  -1\right)  ^{n}\left(  n+1\right)
!n!}{\left(  2n+1\right)  !}\left\langle \boldsymbol{a}_{-}\left[
\boldsymbol{a}_{-}^{2}\right]  ^{n}\right\rangle +\\
&  -n\left(  n+1\right)  \mathrm{d}\int_{0}^{1}\mathrm{d}t\int_{0}%
^{t}\mathrm{d}s\left\langle \boldsymbol{Aa}_{-}\left[  s\left(  t\mathrm{d}%
\boldsymbol{A}+s\boldsymbol{A}^{2}\right)  -t\left(  1-t\right)
\boldsymbol{a}_{-}^{2}\right]  ^{n-1}\right\rangle .
\end{align*}

As\'{\i}, vemos que la forma de Chern--Simons var\'{\i}a s\'{o}lo en una forma
cerrada bajo transformaciones de gauge y por lo tanto corresponde localmente a
un t\'{e}rmino de borde. Por otra parte, es importante recalcar que esta
variaci\'{o}n est\'{a} \'{\i}ntimamente ligada con la estructura del tensor
invariante. En particular, cuando el tensor invariante es no primitivo, el
cociclo de Chevalley-Eilenberg es cero, y por lo tanto la variaci\'{o}n de la
forma de Chern--Simons es s\'{o}lo una derivada total.

\subsection{Acci\'{o}n de Chern--Simons}

El que bajo transformaciones de gauge una forma de Chern--Simons var\'{\i}e
localmente en un t\'{e}rmino de borde implica que una acci\'{o}n de
Chern--Simons puede definirse s\'{o}lo \textit{m\'{o}dulo t\'{e}rminos de
borde},%
\[
S_{\mathrm{CS}}^{\left(  2n+1\right)  }=\int_{M}L_{\mathrm{CS}}^{\left(
2n+1\right)  }\left(  \boldsymbol{A}\right)  +\int_{\partial M}X^{\left(
2n\right)  }.
\]

Esto significa que la acci\'{o}n de Chern--Simons tiene asociadas las
ecuaciones del movimiento para $\boldsymbol{A},$%
\[
\left.  \left\langle \boldsymbol{T}_{A}\boldsymbol{F}^{n}\right\rangle
\right\vert _{M}=0,
\]
pero no resulta posible definir condiciones de borde s\'{o}lo a partir del
principio de acci\'{o}n. De la misma forma, las corrientes de Noether y cargas
conservadas tampoco pueden ser definidas en forma inequ\'{\i}voca. Es
precisamente por estas razones por las que el estudio de la forma de
Transgresi\'{o}n como lagrangeano resulta interesante; a diferencia del caso
de Chern--Simons, tenemos una integral de acci\'{o}n inequ\'{\i}vocamente
definida, pero el precio a pagar es el de tener dos conexiones en la teor\'{\i}a.

\section[Forma Expl\'{\i}cita del Lagrangeano Transgresor y de Chern--Simons]%
{Forma Expl\'{\i}cita del Lagrangeano Transgresor y de Chern--Simons
\sectionmark{Forma Expl\'{\i}cita del Lagrangeano}}

\sectionmark{Forma Expl\'{\i}cita del Lagrangeano}

Al utilizar una forma de Transgresi\'{o}n o de Chern--Simons como Lagrangeano,
se debe de tener en cuenta que el \'{a}lgebra de Lie escogida tiene distintos
subespacios, cada uno de ellos con un significado f\'{\i}sico diferente. Por
ejemplo, un cierto subespacio puede corresponder a gravitaci\'{o}n, y otro, a
la presencia de campos fermi\'{o}nicos en la teor\'{\i}a. Adem\'{a}s,
dependiendo de la forma del tensor invariante y el \'{a}lgebra, un lagrangeano
de transgresi\'{o}n o Chern--Simons posee un trozo de volumen (\textit{bulk})
y un t\'{e}rmino de borde. Por lo tanto, pese a que en principio s\'{o}lo se
necesita de expresiones generales como ec.~(\ref{Ec LagrangT 2}), en la
pr\'{a}ctica se vuelve deseable disponer de un m\'{e}todo que permita separar
el Lagrangeano en forma expl\'{\i}cita en trozos que reflejen la
\textquotedblleft f\'{\i}sica\textquotedblright\ del problema (\textit{i.e.,}
la estructura de subespacios del \'{a}lgebra), as\'{\i} como separar la
acci\'{o}n en t\'{e}rminos de volumen y de borde.

Es posible construir un m\'{e}todo que realiza precisamente esta tarea
utilizando la f\'{o}rmula de homotop\'{\i}a ec.~(\ref{EcTriangular}) en forma
sistem\'{a}tica, como veremos a continuaci\'{o}n

\subsection{\label{Sec Metd Sep Sub Esp}M\'{e}todo de Separaci\'{o}n en
Subespacios}

Sea $\mathfrak{g}$ un \'{a}lgebra de Lie, $\boldsymbol{A}$ y $\bar
{\boldsymbol{A}}$ conexiones valuadas en $\mathfrak{g}$ y $L_{\mathrm{T}%
}^{\left(  2n+1\right)  }\left(  \boldsymbol{A},\bar{\boldsymbol{A}}\right)
=kT_{\boldsymbol{A}\mathbb{\leftarrow}\bar{\boldsymbol{A}}}^{\left(
2n+1\right)  }$ el correspondiente lagrangeano transgresor. Entonces, debemos

\begin{enumerate}
\item Separar el \'{a}lgebra $\mathfrak{g}$ en subespacios que reflejen la
f\'{\i}sica envuelta, $\mathfrak{g}=V_{0}\oplus\cdots\oplus V_{p}$.

\item Escribir las conexiones en trozos valuados en cada subespacio,
$\boldsymbol{A}=\boldsymbol{a}_{0}+\cdots+\boldsymbol{a}_{p},$ y
$\bar{\boldsymbol{A}}=\bar{\boldsymbol{a}}_{0}+\cdots+\bar{\boldsymbol{a}}%
_{p}.$

\item Utilizar ec.~(\ref{EcTriangular}) para escribir $L_{\mathrm{T}}^{\left(
2n+1\right)  }\left(  \boldsymbol{A},\bar{\boldsymbol{A}}\right)  $ (\'{o}
$L_{\mathrm{CS}}^{\left(  2n+1\right)  }\left(  \boldsymbol{A}\right)  ,$ con
$\bar{\boldsymbol{A}}=0$) con%
\begin{align*}
\boldsymbol{A}_{0}  &  =\bar{\boldsymbol{A}},\\
\boldsymbol{A}_{1}  &  =\boldsymbol{a}_{0}+\cdots+\boldsymbol{a}_{p-1},\\
\boldsymbol{A}_{2}  &  =\boldsymbol{A}%
\end{align*}

\item Iterar el paso 3 para la transgresi\'{o}n $T_{\boldsymbol{A}%
_{1}\mathbb{\leftarrow}\boldsymbol{A}_{0}}^{\left(  2n+1\right)  },$ y
as\'{\i} sucesivamente.
\end{enumerate}

Como un ejemplo del m\'{e}todo, mostraremos a continuaci\'{o}n como es posible
escribir expl\'{\i}citamente un lagrangeano transgresor o de Chern--Simons
para gravedad. En Sec.~\ref{Sec Lagrang M Alg} el mismo m\'{e}todo ser\'{a}
utilizado para escribir el Lagrangeano para el \'{A}lgebra~M.

\subsection{Ejemplo: Gravedad como Teor\'{\i}a de Transgresi\'{o}n y
Chern--Simons.}

En esta secci\'{o}n, consideraremos brevemente como puede reobtenerse gravedad
a trav\'{e}s de las herramientas matem\'{a}ticas desarrolladas en las
secciones anteriores. Para un an\'{a}lisis de la f\'{\i}sica asociada a
gravedad en dimensiones m\'{a}s altas, personalmente recomendar\'{\i}a las
Notas de Clase (\textit{Lecture Notes})
Refs.~\cite{CECS-Lect1,CECS-Lect2,CECS-Lect3} y las referencias dentro de ellas.

La gravedad de Transgresi\'{o}n/Chern--Simons contiene en forma natural altas
potencias de la curvatura, por lo que resulta interesante notar en este punto
una interesante relaci\'{o}n con teor\'{\i}a de cuerdas.

En el contexto de teor\'{\i}a de cuerdas, se genera una teor\'{\i}a efectiva
para gravedad en $D=10,$ la cual incluye altas potencias en la curvatura en
forma de una serie en $\alpha$. Sin embargo, en general este tipo de acciones
poseen fantasmas (\textit{ghosts}), lo cual est\'{a} en contradicci\'{o}n con
el hecho de que Teor\'{\i}a de Cuerdas est\'{a} libre de ellos. Este problema
fue estudiado en detalle por B.~Zwiebach y B.~Zumino en
Refs.~\cite{Zwiebach-R2inString} y~\cite{Zumino-GravMore4}, respectivamente,
concluy\'{e}ndose que la \'{u}nica soluci\'{o}n a este problema ser\'{\i}a que
este Lagrangeano coincidiese con el de la serie de Lanczos--Lovelock
(Ve\'{a}se Refs.~\cite{Lanczos,Lovelock}). Desde otro punto de vista, esto es
simplemente una consecuencia natural del hecho de que el lagrangeano de
Lanczos--Lovelock entrega ecuaciones de segundo orden para la m\'{e}trica.
Cuando se requiere que este lagrangeano posea una din\'{a}mica consistente, se
obtiene en dimensiones impares gravedad de Chern--Simons, y en dimensiones
pares, gravedad de Born--Infeld (Ve\'{a}se las
Refs.~\cite{CECS-Lect1,CECS-Lect2,CECS-Lect3}). Ambas gravedades pueden
relacionarse a trav\'{e}s de reducci\'{o}n dimensional (Ve\'{a}se
Ref.~\cite{Nosotros1-Lovelock}) y por lo tanto, es posible pensar en la
posibilidad de que una gravedad de Chern--Simons en $D=11$ genere la gravedad
efectiva de cuerdas en $D=10.$

Sea $\mathfrak{g}$ el \'{a}lgebra de Anti-de Sitter en $D=2n+1$ dimensiones,
$\mathfrak{g}=\mathfrak{so}\left(  2n,2\right)  $, y sea $M$ un espacio base
$D $-dimensional. Consideremos la separaci\'{o}n en subespacios $\mathfrak{g}%
=V_{0}\oplus V_{1},$ con $V_{0}$ la sub\'{a}lgebra de Lorentz, $V_{0}%
=\mathfrak{so}\left(  2n,1\right)  $ y el coseto sim\'{e}trico $V_{1}%
=\mathfrak{so}\left(  2n,2\right)  /\mathfrak{so}\left(  2n,1\right)  $
(\textit{Anti-de Sitter boosts}). Llamando $\boldsymbol{J}_{ab}$ a los
generadores de $\mathfrak{so}\left(  2n,1\right)  $, y $\boldsymbol{P}_{a}$ a
los generadores de $\mathfrak{so}\left(  2n,2\right)  /\mathfrak{so}\left(
2n,1\right)  $, el \'{a}lgebra de $\mathfrak{so}\left(  2n,2\right)  $ se
escribe como%
\begin{align*}
\left[  \boldsymbol{P}_{a},\boldsymbol{P}_{b}\right]   &  =\boldsymbol{J}%
_{ab},\\
\left[  \boldsymbol{J}_{ab},\boldsymbol{P}_{c}\right]   &  =\eta_{ce}%
\delta_{ab}^{de}\boldsymbol{P}_{d},\\
\left[  \boldsymbol{J}_{ab},\boldsymbol{J}_{cd}\right]   &  =\eta_{gh}%
\delta_{ab}^{eg}\delta_{cd}^{hf}\boldsymbol{J}_{ef}.
\end{align*}

As\'{\i}, podemos escribir las 1-formas conexi\'{o}n valuadas en el
\'{a}lgebra%
\begin{align*}
\boldsymbol{A}  &  =\boldsymbol{\omega}+\boldsymbol{e}\\
\bar{\boldsymbol{A}}  &  =\bar{\boldsymbol{\omega}}+\bar{\boldsymbol{e}}%
\end{align*}
con las componentes%
\begin{align*}
\boldsymbol{\omega}  &  =\frac{1}{2}\omega^{ab}\boldsymbol{J}_{ab}\\
\boldsymbol{e}  &  =e^{a}\boldsymbol{P}_{a}\\
\bar{\boldsymbol{\omega}}  &  =\frac{1}{2}\bar{\omega}^{ab}\boldsymbol{J}%
_{ab}\\
\bar{\boldsymbol{e}}  &  =\bar{e}^{a}\boldsymbol{P}_{a}%
\end{align*}

Bajo transformaciones de gauge valuadas en el subgrupo de Lorentz,
$\mathrm{so}\left(  2n,1\right)  $, $\omega^{ab}$, $\bar{\omega}^{ab}$
transforman como conexiones de Lorentz, y $e^{a}$, $\bar{e}^{a}$ como vectores
de Lorentz. Esto permite identificar $\omega^{ab}$, $\bar{\omega}^{ab}$ con
conexiones de esp\'{\i}n y $e^{a}$, $\bar{e}^{a}$ con \textit{vielbeine} sobre
$M.$

Consideremos ahora el tensor invariante de $\mathrm{so}\left(  2n,2\right)  $
con \'{u}nica componente no nula%
\begin{equation}
\left\langle \boldsymbol{J}_{a_{1}a_{2}}\cdots\boldsymbol{J}_{a_{2n-1}a_{2n}%
}\boldsymbol{P}_{a_{2n+1}}\right\rangle =2^{n}\varepsilon_{a_{1}\cdots
a_{2n+1}}.\label{Ec TensInv = Epsilon}%
\end{equation}

\subsubsection{Lagrangeano de Chern--Simons para Gravedad}

Una lagrangeano de Chern--Simons para $\mathrm{so}\left(  2n,2\right)  $
constru\'{\i}do con este tensor invariante tiene la forma expl\'{\i}cita%
\[
L_{\mathrm{CS}}^{\left(  2n+1\right)  }\left(  \boldsymbol{\omega
}+\boldsymbol{e}\right)  =k\left(  n+1\right)  \int_{t=0}^{t=1}\mathrm{d}%
t~\left\langle \left(  \boldsymbol{\omega}+\boldsymbol{e}\right)  \left(
t\mathrm{d}\left(  \boldsymbol{\omega}+\boldsymbol{e}\right)  +t^{2}\left(
\boldsymbol{\omega}+\boldsymbol{e}\right)  ^{2}\right)  ^{n}\right\rangle
\]

Es posible demostrar que este Lagrangeano corresponde a un lagrangeano de
Lovelock para gravedad, realizando integraciones por partes sucesivas. Este es
un proceso extremadamente laborioso en dimensiones mayores que 3, el cual
adem\'{a}s permite separar expl\'{\i}citamente el Lagrangeano en un
t\'{e}rmino de volumen y uno de borde. Este mismo proceso es realizado en
forma directa utilizando el m\'{e}todo de separaci\'{o}n en subespacios.
Escribiendo%
\begin{align*}
\boldsymbol{A}_{0}  &  =0\\
\boldsymbol{A}_{1}  &  =\boldsymbol{\omega}\\
\boldsymbol{A}_{2}  &  =\boldsymbol{\omega}+\boldsymbol{e}%
\end{align*}
tenemos que ec.~(\ref{EcTriangular}) toma la forma%
\[
L_{\mathrm{CS}}^{\left(  2n+1\right)  }\left(  \boldsymbol{\omega
}+\boldsymbol{e}\right)  =kT_{\left[  \boldsymbol{\omega}+\boldsymbol{e}%
\right]  \leftarrow\boldsymbol{\omega}}^{\left(  2n+1\right)  }%
+kT_{\boldsymbol{\omega}\leftarrow0}^{\left(  2n+1\right)  }+k\text{\textrm{d}%
}Q_{\left[  \boldsymbol{\omega}+\boldsymbol{e}\right]  \leftarrow
\boldsymbol{\omega}\leftarrow0}^{\left(  2n\right)  }.
\]

Dada la forma del tensor invariante ec.~(\ref{Ec TensInv = Epsilon}) tenemos
que $T_{\boldsymbol{\omega}\leftarrow0}^{\left(  2n+1\right)  }=0,$ y como un
Lagrangeano de Chern--Simons est\'{a} definido s\'{o}lo m\'{o}dulo
t\'{e}rminos de borde, podemos escribir%
\[
L_{\mathrm{CS}}^{\left(  2n+1\right)  }\left(  \boldsymbol{\omega
}+\boldsymbol{e}\right)  =kT_{\boldsymbol{\omega}+\boldsymbol{e}%
\leftarrow\boldsymbol{\omega}}^{\left(  2n+1\right)  }.
\]

El t\'{e}rmino $kT_{\left[  \boldsymbol{\omega}+\boldsymbol{e}\right]
\leftarrow\boldsymbol{\omega}}^{\left(  2n+1\right)  }$ corresponde
expl\'{\i}citamente a%
\[
kT_{\left[  \boldsymbol{\omega}+\boldsymbol{e}\right]  \leftarrow
\boldsymbol{\omega}}^{\left(  2n+1\right)  }=2^{n}\left(  n+1\right)
k\int_{t=0}^{t=1}\mathrm{d}t~\left\langle \boldsymbol{eF}_{t}^{n}%
\right\rangle
\]
con\footnote{En general, dadas dos conexiones $\boldsymbol{A}$ y
$\bar{\boldsymbol{A}},$ sus respectivas curvaturas est\'{a}n relacionadas a
trav\'{e}s de la identidad $\boldsymbol{F}=\bar{\boldsymbol{F}}+D_{\bar
{\boldsymbol{A}}}\boldsymbol{\Theta}+\boldsymbol{\Theta}^{2},$ con
$\boldsymbol{\Theta}=\boldsymbol{A}-\bar{\boldsymbol{A}}.$}%
\[
\boldsymbol{F}_{t}=\mathrm{d}\boldsymbol{\omega}+\frac{1}{2}\left[
\boldsymbol{\omega},\boldsymbol{\omega}\right]  +t\mathrm{D}_{\omega
}\boldsymbol{e}+t^{2}\frac{1}{2}\left[  \boldsymbol{e},\boldsymbol{e}\right]
\]
o en forma m\'{a}s compacta,%
\[
\boldsymbol{F}_{t}=\boldsymbol{R}+t\boldsymbol{T}+t^{2}\boldsymbol{e}^{2}%
\]
con%
\begin{align*}
\boldsymbol{R}  &  =\frac{1}{2}R^{ab}\boldsymbol{J}_{ab},\\
\boldsymbol{T}  &  =T^{a}\boldsymbol{P}_{a},\\
\boldsymbol{e}^{2}  &  =\frac{1}{2}e^{a}e^{b}\boldsymbol{J}_{ab},
\end{align*}
en donde $R^{ab}=\mathrm{d}\omega^{ab}+\omega_{\phantom{a}c}^{a}\omega^{cb}$ y
$T^{a}=\mathrm{d}e^{a}+\omega_{\phantom{a}b}^{a}e^{b}$ corresponden a la
curvatura de Riemmann\footnote{Debe de observarse sin embargo, que no hemos
impuesto (y no impondremos) la condici\'{o}n de nulidad de la torsi\'{o}n, por
lo que $\omega^{ab}$ y $e^{a} $ corresponden a campos independientes. En este
sentido, $R_{\phantom{\alpha}\beta\mu\nu}^{\alpha}=e_{a}^{\phantom{a}\alpha
}e_{\phantom{b}\beta}^{b}R_{\phantom{a}b\mu\nu}^{a}$ en general
corresponder\'{\i}a a un Riemann \textquotedblleft\'{a} la
Palatini\textquotedblright\ en t\'{e}rminos de una conexi\'{o}n independiente
de la m\'{e}trica.} y a la Torsi\'{o}n\footnote{El identificar $T^{a}%
=\mathrm{D}_{\omega}e^{a}$ con la Torsi\'{o}n es equivalente a imponer la
condici\'{o}n de que $\omega^{ab}$ y $\Gamma_{\mu\nu}^{\lambda}$ generen el
mismo transporte paralelo, o lo que es lo mismo, a imponer el lema de Weyl,
$\partial_{\mu}e_{\phantom{a}\nu}^{a}+\omega_{\_b\mu}^{a}e_{\phantom{b}\nu
}^{b}-\Gamma_{\mu\nu}^{\lambda}e_{\phantom{a}\lambda}^{a}=0.$}, respectivamente.

Tomando en cuenta la forma del tensor invariante
ec.~(\ref{Ec TensInv = Epsilon}), tenemos que%
\[
kT_{\left[  \boldsymbol{\omega}+\boldsymbol{e}\right]  \leftarrow
\boldsymbol{\omega}}^{\left(  2n+1\right)  }=2^{n}\left(  n+1\right)
k\int_{t=0}^{t=1}\mathrm{d}t~\left\langle \boldsymbol{e}\left(  \boldsymbol{R}%
+t^{2}\boldsymbol{e}^{2}\right)  ^{n}\right\rangle
\]
o m\'{a}s expl\'{\i}citamente,%
\[
L_{\mathrm{CS}}^{\left(  2n+1\right)  }\left(  \boldsymbol{\omega
}+\boldsymbol{e}\right)  =\left(  n+1\right)  k\int_{t=0}^{t=1}\mathrm{d}%
t~\varepsilon_{a_{1}\cdots a_{2n+1}}\left(  R^{a_{1}a_{2}}+t^{2}e^{a_{1}a_{2}%
}\right)  \cdots\left(  R^{a_{2n-1}a_{2n}}+t^{2}e^{a_{2n-1}a_{2n}}\right)
e^{a_{2n+1}}.
\]

Este lagrangeano corresponde a uno de la serie de Lovelock, en donde la
integraci\'{o}n en $t$ reproduce exactamente los coeficientes de
Refs.~\cite{Chamseddine-CS1,CECS-DimContBH} en la serie de Lovelock.

Es importante recalcar que pese a que el lagrangeano corresponde a una forma
de transgresi\'{o}n, $L_{\mathrm{CS}}^{\left(  2n+1\right)  }\left(
\boldsymbol{\omega}+\boldsymbol{e}\right)  =kT_{\left[  \boldsymbol{\omega
}+\boldsymbol{e}\right]  \leftarrow\boldsymbol{\omega}}^{\left(  2n+1\right)
},$ este es s\'{o}lo cuasi-invariante bajo transformaciones de gauge. Esto se
debe a que en la \textquotedblleft transgresi\'{o}n\textquotedblright%
\ $T_{\left[  \boldsymbol{\omega}+\boldsymbol{e}\right]  \leftarrow
\boldsymbol{\omega}}^{\left(  2n+1\right)  },$ $\boldsymbol{\omega
}+\boldsymbol{e}$ corresponde a una conexi\'{o}n, pero $\boldsymbol{\omega}$
no. En efecto, haciendo una transformaci\'{o}n de gauge infinitesimal,
$\delta\left(  \boldsymbol{\omega}+\boldsymbol{e}\right)  =\mathrm{D}%
_{\boldsymbol{\omega}+\boldsymbol{e}}\boldsymbol{\lambda},$ tenemos%
\begin{align*}
\delta e^{a}  &  =-\lambda_{\phantom{a}b}^{a}e^{b}+\mathrm{D}_{\omega}%
\lambda^{a},\\
\delta\omega^{ab}  &  =\mathrm{D}_{\omega}\lambda^{ab}+e^{a}\lambda^{b}%
-e^{b}\lambda^{a}.
\end{align*}

Estas ecuaciones corresponden a transformaciones de gauge infinitesimales para
$\boldsymbol{\omega}+\boldsymbol{e},$ pero $\delta\omega^{ab}=\mathrm{D}%
_{\omega}\lambda^{ab}+e^{a}\lambda^{b}-e^{b}\lambda^{a}$ \textit{no}
corresponde a una transformaci\'{o}n de gauge para $\boldsymbol{\omega},$ y
por lo tanto, estas transformaciones no dejan invariante a $T_{\left[
\boldsymbol{\omega}+\boldsymbol{e}\right]  \leftarrow\boldsymbol{\omega}%
}^{\left(  2n+1\right)  }.$

El hecho de que $T_{\left[  \boldsymbol{\omega}+\boldsymbol{e}\right]
\leftarrow\boldsymbol{\omega}}^{\left(  2n+1\right)  }$ sea cuasi-invariante
est\'{a} \'{\i}ntimamente ligado con la forma del tensor invariante. En
efecto, siguiendo el mismo razonamiento que en el teorema de Chern--Weil, es
posible escribir%
\[
\left\langle \left(  \boldsymbol{R}+\boldsymbol{T}+\boldsymbol{e}^{2}\right)
^{n+1}\right\rangle -\left\langle \boldsymbol{R}^{n+1}\right\rangle
=\mathrm{d}T_{\left[  \boldsymbol{\omega}+\boldsymbol{e}\right]
\leftarrow\boldsymbol{\omega}}^{\left(  2n+1\right)  }.
\]

Dado que $\left\langle \boldsymbol{R}^{n+1}\right\rangle $ \textit{no} es
invariante bajo la transformaci\'{o}n $\delta\omega^{ab}=\mathrm{D}_{\omega
}\lambda^{ab}+e^{a}\lambda^{b}-e^{b}\lambda^{a},$ en general la forma
$T_{\left[  \boldsymbol{\omega}+\boldsymbol{e}\right]  \leftarrow
\boldsymbol{\omega}}^{\left(  2n+1\right)  }$ var\'{\i}a en t\'{e}rminos no
cerrados. Sin embargo, con la elecci\'{o}n de tensor invariante
ec.~(\ref{Ec TensInv = Epsilon}), tenemos $\left\langle \boldsymbol{R}%
^{n+1}\right\rangle =0,$ valor inalterado incluso bajo la transformaci\'{o}n
$\delta\omega^{ab}=\mathrm{D}_{\omega}\lambda^{ab}+e^{a}\lambda^{b}%
-e^{b}\lambda^{a}.$ Por lo tanto, tenemos%
\[
\left\langle \left(  \boldsymbol{R}+\boldsymbol{T}+\boldsymbol{e}^{2}\right)
^{n+1}\right\rangle =\mathrm{d}T_{\left[  \boldsymbol{\omega}+\boldsymbol{e}%
\right]  \leftarrow\boldsymbol{\omega}}^{\left(  2n+1\right)  },
\]
y dado que $\left\langle \left(  \boldsymbol{R}+\boldsymbol{T}+\boldsymbol{e}%
^{2}\right)  ^{n+1}\right\rangle $ es invariante de gauge, $T_{\left[
\boldsymbol{\omega}+\boldsymbol{e}\right]  \leftarrow\boldsymbol{\omega}%
}^{\left(  2n+1\right)  }$ es cuasi-invariante.

\subsubsection{\label{Sec Gravedad Transg}Lagrangeano Transgresor para
Gravedad}

La elecci\'{o}n de Lagrangeano Transgresor m\'{a}s cercana al Lagrangeano de
Chern--Simons reci\'{e}n considerado es%
\[
L_{\mathrm{T}}^{\left(  2n+1\right)  }\left(  \boldsymbol{\omega
}+\boldsymbol{e},\bar{\boldsymbol{\omega}}\right)  =kT_{\left[
\boldsymbol{\omega}+\boldsymbol{e}\right]  \leftarrow\bar{\boldsymbol{\omega}%
}}^{\left(  2n+1\right)  },
\]
en donde es necesario recalcar que $\bar{\boldsymbol{\omega}}$ corresponde a
una conexi\'{o}n para $\mathfrak{so}\left(  2n,2\right)  .$

Nuevamente, es posible escribir una expresi\'{o}n cerrada para el Lagrangeano
utilizando el m\'{e}todo de separaci\'{o}n en subespacios; escogiendo%
\begin{align*}
\boldsymbol{A}_{0}  &  =\bar{\boldsymbol{\omega}}\\
\boldsymbol{A}_{1}  &  =\boldsymbol{\omega}\\
\boldsymbol{A}_{2}  &  =\boldsymbol{\omega}+\boldsymbol{e}%
\end{align*}
tenemos%
\begin{equation}
L_{\mathrm{T}}^{\left(  2n+1\right)  }\left(  \boldsymbol{\omega
}+\boldsymbol{e},\bar{\boldsymbol{\omega}}\right)  =kT_{\left[
\boldsymbol{\omega}+\boldsymbol{e}\right]  \leftarrow\boldsymbol{\omega}%
}^{\left(  2n+1\right)  }+kT_{\boldsymbol{\omega}\leftarrow\bar
{\boldsymbol{\omega}}}^{\left(  2n+1\right)  }+k\text{\textrm{d}}Q_{\left[
\boldsymbol{\omega}+\boldsymbol{e}\right]  \leftarrow\boldsymbol{\omega
}\leftarrow\bar{\boldsymbol{\omega}}}^{\left(  2n\right)  }%
.\label{Ec Lt Grav CON}%
\end{equation}

Considerando el tensor invariante ec.~(\ref{Ec TensInv = Epsilon}), tenemos%
\begin{equation}
L_{\mathrm{T}}^{\left(  2n+1\right)  }\left(  \boldsymbol{\omega
}+\boldsymbol{e},\bar{\boldsymbol{\omega}}\right)  =L_{\mathrm{CS}}^{\left(
2n+1\right)  }\left(  \boldsymbol{\omega}+\boldsymbol{e}\right)
+k\text{\textrm{d}}Q_{\left[  \boldsymbol{\omega}+\boldsymbol{e}\right]
\leftarrow\boldsymbol{\omega}\leftarrow\bar{\boldsymbol{\omega}}}^{\left(
2n\right)  }.\label{Ec Lt Grav SIN}%
\end{equation}

El Lagrangeano transgresor y el de Chern--Simons s\'{o}lo difieren en una
derivada total en este caso en particular. La forma expl\'{\i}cita del
t\'{e}rmino de borde viene dada por ec.~(\ref{EcQ(2n)=Intgrl_Doble}),%
\[
kQ_{\left[  \boldsymbol{\omega}+\boldsymbol{e}\right]  \leftarrow
\boldsymbol{\omega}\leftarrow\bar{\boldsymbol{\omega}}}^{\left(  2n\right)
}=n\left(  n+1\right)  k\int_{0}^{1}\mathrm{d}t\int_{0}^{t}\mathrm{d}%
s\left\langle \boldsymbol{e\theta F}_{st}^{n-1}\right\rangle
\]
con $\boldsymbol{\theta}=\boldsymbol{\omega}-\bar{\boldsymbol{\omega}}$ y
$\boldsymbol{F}_{st}$ la curvatura en la conexi\'{o}n $\boldsymbol{A}%
_{st}=\bar{\boldsymbol{\omega}}+s\boldsymbol{e}+t\boldsymbol{\theta},$%
\[
\boldsymbol{F}_{st}=\bar{\boldsymbol{R}}+\mathrm{D}_{\bar{\boldsymbol{\omega}%
}}\left(  s\boldsymbol{e}+t\boldsymbol{\theta}\right)  +s^{2}\boldsymbol{e}%
^{2}+st\left[  \boldsymbol{e},\boldsymbol{\theta}\right]  +t^{2}%
\boldsymbol{\theta}^{2}.
\]

Tomando en cuenta ec.~(\ref{EcQ(2n)=Intgrl_Doble}), tenemos%
\[
kQ_{\left[  \boldsymbol{\omega}+\boldsymbol{e}\right]  \leftarrow
\boldsymbol{\omega}\leftarrow\bar{\boldsymbol{\omega}}}^{\left(  2n\right)
}=n\left(  n+1\right)  k\int_{0}^{1}\mathrm{d}t\int_{0}^{t}\mathrm{d}%
s\left\langle \boldsymbol{e\theta}\left(  \bar{\boldsymbol{R}}+t\mathrm{D}%
_{\bar{\boldsymbol{\omega}}}\boldsymbol{\theta}+s^{2}\boldsymbol{e}^{2}%
+t^{2}\boldsymbol{\theta}^{2}\right)  ^{n-1}\right\rangle .
\]

El lagrangeano $L_{\mathrm{T}}^{\left(  2n+1\right)  }\left(
\boldsymbol{\omega}+\boldsymbol{e},\bar{\boldsymbol{\omega}}\right)  $
ec.~(\ref{Ec Lt Grav CON}) es completamente invariante bajo transformaciones
de gauge de $\mathfrak{so}\left(  2n,2\right)  $, dado que ambas,
$\boldsymbol{\omega}+\boldsymbol{e}$ y $\bar{\boldsymbol{\omega}},$
corresponden a conexiones de $\mathfrak{so}\left(  2n,2\right)  .$

Para fijar las condiciones de borde, se debe considerar
ec.~(\ref{Ec EcBorde General}). Para este caso, estas toman la forma%

\begin{equation}
\left.  \int_{0}^{1}dt\left\langle \left(  \delta\bar{\bm{\omega}}%
+t\delta\bm{\theta}+t\delta\bm{e}\right)  \left(  \bm{\theta}+\bm{e}\right)
\bm{F}_{t}^{n-1}\right\rangle \right\vert _{\partial M}=0,\label{bct}%
\end{equation}

en donde la conexi\'{o}n $\bm{A}_{t}$ y su curvatura $\bm{F}_{t}$ est\'{a}n
dadas por
\begin{align}
\bm{A}_{t}  &  =\bar{\bm{\omega}}+t\left(  \bm{e}+\bm{\theta}\right)  ,\\
\bm{F}_{t}  &  =\bar{\bm{R}}+t\text{D}_{\bar{\bm{\omega}}}\left(
\bm{e}+\bm{\theta}\right)  +t^{2}\left(  \bm{e}^{2}+\left[
\bm{e},\bm{\theta }\right]  +\bm{\theta}^{2}\right)  .
\end{align}

Hay muchas maneras de fijar las condiciones de contorno. Una de ellas
particularmente interesante (Ve\'{a}se
Ref.~\cite{CECS-FiniteGrav,Nosotros3-TransLargo,CECS-VacuumOddDim,CECS-Trans})
consiste en imponer que $\bar{\bm{\omega}}$ tenga un valor fijo sobre
$\partial M,$%
\begin{equation}
\left.  \delta\bar{\bm{\omega}}\right\vert _{\partial M}=0.
\end{equation}
y que $\bm{\omega}$ y $\bar{\bm{\omega}}$ induzcan el mismo concepto de
transporte paralelo a lo largo de $\partial M.$ As\'{\i}, ec.~(\ref{bct}) se
reduce a
\begin{equation}
\left.  \int_{0}^{1}dt\left\langle t\left(  \delta
\bm{\theta e}-\bm{\theta}\delta\bm{e}\right)  \left(  \bar{\bm {R}}%
+t^{2}\bm{e}^{2}+t^{2}\bm{\theta}^{2}\right)  ^{n-1}\right\rangle \right\vert
_{\partial M}=0,
\end{equation}
la cual, tomando en cuenta la forma del tensor invariante, puede ser
satisfecha requiriendo
\begin{equation}
\delta\theta^{\lbrack ab}e^{c]}=\theta^{\lbrack ab}\delta e^{c]}.
\end{equation}

El lagrangeano $L_{\mathrm{T}}^{\left(  2n+1\right)  }\left(
\boldsymbol{\omega}+\boldsymbol{e},\bar{\boldsymbol{\omega}}\right)  $ tienen
adem\'{a}s la propiedad extra de tener cargas conservadas de Noether bien
comportadas y completamente autoconsistentes (ve\'{a}se
Refs.~\cite{CECS-Trans,CECS-VacuumOddDim,CECS-FiniteGrav}). Esto tambi\'{e}n
es cierto para el caso m\'{a}s general, $L_{\mathrm{T}}^{\left(  2n+1\right)
}\left(  \boldsymbol{\omega}+\boldsymbol{e},\bar{\boldsymbol{\omega}}%
+\bar{\boldsymbol{e}}\right)  .$ Es interesante observar la relaci\'{o}n entre
invariancia y el buen comportamiento de las cargas. Una acci\'{o}n
cuasi-invariante, al estar definida s\'{o}lo m\'{o}dulo t\'{e}rminos de borde,
no permite definir cargas. En cambio, una acci\'{o}n invariante s\'{\i} lo permite.

Sin embargo, es necesario recalcar algunos aspectos sutiles en como act\'{u}an
las transformaciones de gauge sobre el lagrangeano. No es lo mismo calcular la
variaci\'{o}n bajo transformaciones de gauge de ec.~(\ref{Ec Lt Grav CON}) y
ec.~(\ref{Ec Lt Grav SIN}). En efecto, realizando una transformaci\'{o}n de
gauge correspondiente a $g=\exp\left(  \zeta^{a}\boldsymbol{P}_{a}\right)  ,$
mientras que el lagrangeano dado por ec.~(\ref{Ec Lt Grav CON}) permanece
invariante, el lagrangeano dado por ec.~(\ref{Ec Lt Grav SIN}) es s\'{o}lo
cuasi-invariante y cambia de la forma (Ve\'{a}se
Refs.~\cite{Nosotros3-TransLargo,Nosotros1-Lovelock,Nosotros2-GravAdS} para
manipulaciones finitas de AdS)%
\[
L_{\mathrm{T}}^{\left(  2n+1\right)  }\left(  \boldsymbol{\omega
}+\boldsymbol{e},\bar{\boldsymbol{\omega}}\right)  \rightarrow L_{\mathrm{T}%
}^{\left(  2n+1\right)  }\left(  \boldsymbol{\omega}^{\prime}+\boldsymbol{e}%
^{\prime},\bar{\boldsymbol{\omega}}^{\prime}\right)  =L_{\mathrm{T}}^{\left(
2n+1\right)  }\left(  \boldsymbol{\omega}+\boldsymbol{e},\bar
{\boldsymbol{\omega}}\right)  -kT_{\boldsymbol{\omega}+\Delta
\boldsymbol{\omega}\leftarrow\bar{\boldsymbol{\omega}}+\Delta\bar
{\boldsymbol{\omega}}}^{\left(  2n+1\right)  }%
\]
con%
\begin{align*}
\Delta\boldsymbol{\omega}  &  =\frac{1}{2}\left(  \frac{1-\cosh z}{z^{2}%
}\left(  \zeta^{a}\mathrm{D}_{\omega}\zeta^{b}-\zeta^{b}\mathrm{D}_{\omega
}\zeta^{a}\right)  -\frac{\operatorname*{senh}z}{z}\left(  \zeta^{a}%
e^{b}-\zeta^{b}e^{a}\right)  \right)  \boldsymbol{J}_{ab},\\
\Delta\bar{\boldsymbol{\omega}}  &  =\frac{1}{2}\frac{1-\cosh z}{z^{2}}\left(
\zeta^{a}\mathrm{D}_{\bar{\boldsymbol{\omega}}}\zeta^{b}-\zeta^{b}%
\mathrm{D}_{\bar{\boldsymbol{\omega}}}\zeta^{a}\right)  \boldsymbol{J}%
_{ab}+\Omega_{\phantom{a}b}^{a}\left(  \frac{\operatorname*{senh}z}{z}\right)
\mathrm{D}_{\bar{\boldsymbol{\omega}}}\zeta^{b}\boldsymbol{P}_{a}.
\end{align*}
en donde%
\[
z=\sqrt{\zeta^{a}\zeta_{a}}%
\]
y%
\[
\Omega_{\phantom{a}b}^{a}\left(  u\right)  =u\delta_{b}^{a}+\left(
1-u\right)  \frac{\zeta^{a}\zeta_{b}}{\zeta^{2}}.
\]

\textit{A priori}, esto puede parecer extra\~{n}o, ya que
ec.~(\ref{Ec Lt Grav CON}) y ec.~(\ref{Ec Lt Grav SIN}) corresponden al mismo
Lagrangeano. Sin embargo, se debe de tener en consideraci\'{o}n que la
expresi\'{o}n $T_{\boldsymbol{\omega}\leftarrow\bar{\boldsymbol{\omega}}%
}^{\left(  2n+1\right)  }=0$ no es invariante de gauge. En efecto, dado que
$\bar{\boldsymbol{\omega}}$ corresponde a una conexi\'{o}n de $\mathfrak{so}%
\left(  2n,2\right)  ,$ bajo una transformaci\'{o}n de $\mathrm{so}\left(
2n,2\right)  /\mathrm{so}\left(  2n,1\right)  $ aparece una componente valuada
en $\mathfrak{so}\left(  2n,2\right)  /\mathfrak{so}\left(  2n,1\right)  :$ un
vielbein de s\'{o}lo gauge: $T_{\boldsymbol{\omega}\leftarrow\bar
{\boldsymbol{\omega}}}^{\left(  2n+1\right)  }\rightarrow
T_{\boldsymbol{\omega}^{\prime}\leftarrow\bar{\boldsymbol{\omega}}^{\prime
}+\bar{\boldsymbol{e}}_{g}}^{\left(  2n+1\right)  }\neq0.$

Esta situaci\'{o}n puede ser visto como un abuso del s\'{\i}mbolo
\textquotedblleft$=$\textquotedblright. Denotando por $=$ igualdades
preservadas por transformaciones de gauge y con $\approx$ las que no,
escribir\'{\i}amos $T_{\boldsymbol{\omega}\leftarrow\bar{\boldsymbol{\omega}}%
}^{\left(  2n+1\right)  }\approx0$ y por lo tanto,%
\begin{align*}
L_{\mathrm{T}}^{\left(  2n+1\right)  }\left(  \boldsymbol{\omega
}+\boldsymbol{e},\bar{\boldsymbol{\omega}}\right)   &  =kT_{\left[
\boldsymbol{\omega}+\boldsymbol{e}\right]  \leftarrow\boldsymbol{\omega}%
}^{\left(  2n+1\right)  }+kT_{\boldsymbol{\omega}\leftarrow\bar
{\boldsymbol{\omega}}}^{\left(  2n+1\right)  }+k\text{\textrm{d}}Q_{\left[
\boldsymbol{\omega}+\boldsymbol{e}\right]  \leftarrow\boldsymbol{\omega
}\leftarrow\bar{\boldsymbol{\omega}}}^{\left(  2n\right)  },\\
&  \approx L_{\mathrm{CS}}^{\left(  2n+1\right)  }\left(  \boldsymbol{\omega
}+\boldsymbol{e}\right)  +k\text{\textrm{d}}Q_{\left[  \boldsymbol{\omega
}+\boldsymbol{e}\right]  \leftarrow\boldsymbol{\omega}\leftarrow
\bar{\boldsymbol{\omega}}}^{\left(  2n\right)  }.
\end{align*}

Desde un punto de vista m\'{a}s abstracto, esta es una problem\'{a}tica
general, originada por la no-trivialidad de la condici\'{o}n de tensor
invariante. En efecto, sea $\left\{  \boldsymbol{N}_{i}\right\}  $ una
colecci\'{o}n de $n_{i}$-formas valuadas en el \'{a}lgebra de Lie, las cuales
descompondremos como%
\[
\boldsymbol{N}_{i}=\boldsymbol{N}_{i}^{1}+\boldsymbol{N}_{i}^{0}%
\]
tal que se cumple que%
\[
\left\vert \boldsymbol{N}_{1}^{0}\cdots\boldsymbol{N}_{n}^{0}\right\vert =0,
\]
en general$.$ Esto significa que $\left\vert \boldsymbol{N}_{1}^{0}%
\cdots\boldsymbol{N}_{n}^{0}\right\vert $ est\'{a} valuado s\'{o}lo sobre
componentes nulas del tensor invariante. Escribiendo la condici\'{o}n de
invariancia a partir de ec.~(\ref{Ec ConmGenExplicito}) como%
\[
\left\vert \left[  \boldsymbol{M},\boldsymbol{N}_{1}\cdots\boldsymbol{N}%
_{n}\right]  \right\vert =\sum_{i=1}^{n}\left(  -1\right)  ^{m\left(
n_{1}+\cdots+n_{i-1}\right)  }\left\vert \boldsymbol{N}_{1}\cdots
\boldsymbol{N}_{i-1}\left[  \boldsymbol{M},\boldsymbol{N}_{i}\right]
\boldsymbol{N}_{i+1}\cdots\boldsymbol{N}_{n}\right\vert =0
\]
tenemos que pese a que $\left\vert \boldsymbol{N}_{1}^{0}\cdots\boldsymbol{N}%
_{n}^{0}\right\vert =0$, una componente del tipo%
\[
\left\vert \boldsymbol{N}_{1}^{0}\cdots\boldsymbol{N}_{i-1}^{0}\left[
\boldsymbol{M},\boldsymbol{N}_{i}^{0}\right]  \boldsymbol{N}_{i+1}^{0}%
\cdots\boldsymbol{N}_{n}^{0}\right\vert
\]
no tiene por qu\'{e} estar valuada en una componente nula del tensor
invariante. A\'{u}n m\'{a}s, la condici\'{o}n $\left\vert \left[
\boldsymbol{M},\boldsymbol{N}_{1}^{0}\cdots\boldsymbol{N}_{n}^{0}\right]
\right\vert =0$ \textit{no} \textit{sigue en forma trivial de }$\left\vert
\boldsymbol{N}_{1}^{0}\cdots\boldsymbol{N}_{n}^{0}\right\vert =0$. De hecho,
aunque el polinomio invariante puede escribirse como%
\[
\left\vert \boldsymbol{N}_{1}\cdots\boldsymbol{N}_{n}\right\vert
=\sum_{\substack{k_{1},\ldots k_{n}=0 \\k_{1}+\cdots+k_{n}\geq1}%
}^{1}\left\vert \boldsymbol{N}_{1}^{k_{1}}\cdots\boldsymbol{N}_{n}^{k_{n}%
}\right\vert ,
\]
tenemos que la condici\'{o}n de invariancia es de la forma%
\[
\left\vert \left[  \boldsymbol{M},\boldsymbol{N}_{1}\cdots\boldsymbol{N}%
_{n}\right]  \right\vert =\sum_{\substack{k_{1},\ldots k_{n}=0 \\k_{1}%
+\cdots+k_{n}\geq1}}^{1}\left\vert \left[  \boldsymbol{M},\boldsymbol{N}%
_{1}^{k_{1}}\cdots\boldsymbol{N}_{n}^{k_{n}}\right]  \right\vert +\left\vert
\left[  \boldsymbol{M},\boldsymbol{N}_{1}^{0}\cdots\boldsymbol{N}_{n}%
^{0}\right]  \right\vert .
\]
As\'{\i}, $\left\vert \boldsymbol{N}_{1}^{0}\cdots\boldsymbol{N}_{n}%
^{0}\right\vert $ juega un rol en la condici\'{o}n de polinomio invariante,
pese a que $\left\vert \boldsymbol{N}_{1}^{0}\cdots\boldsymbol{N}_{n}%
^{0}\right\vert =0$.

Esto hace que en general, resolver la condici\'{o}n de tensor invariante sea
una tarea altamente no trivial, y que encontrar todos los tensores invariantes
de un \'{a}lgebra dada permanezca como un problema abierto hasta ahora.

Un enfoque laborioso, aunque extremadamente interesante de este problema viene
dado por el m\'{e}todo de Noether. \'{E}ste consiste b\'{a}sicamente, en tomar
un Lagrangeano de Chern--Simons/Transgresi\'{o}n invariante bajo una cierta
\'{a}lgebra $\mathfrak{g}$ y luego considerar esta \'{a}lgebra como una
sub\'{a}lgebra de otra mayor, $\mathfrak{G}$. Realizando transformaciones de
gauge valuadas en $\mathfrak{G/g}$ es posible ir agregando t\'{e}rminos extra
al lagrangeano asegurando su invariancia bajo transformaciones de gauge
valuadas en $\mathfrak{G}$ (Para un interesante ejemplo relacionado con el
\'{A}lgebra~M, v\'{e}ase Refs.~\cite{CECS-MAlgNoether-1,CECS-MAlgNoether-2}%
)$.$

Ahora bien, la invariancia del lagrangeano bajo transformaciones de gauge y el
que \'{e}ste sea un polinomio invariante son afirmaciones equivalentes. Desde
el punto de vista del tensor invariante, esto equivale a tomar un polinomio
invariante de $\mathfrak{g}$ como \textquotedblleft semilla\textquotedblright,
y utilizar la condici\'{o}n de invariancia para ir completando el tensor
invariante con nuevos t\'{e}rminos. Es posible escribir un algoritmo general
para resolver este problema, el cual a cada paso da condiciones para el
polinomio. Sin embargo, no ha resultado posible probar la convergencia del
procedimiento y a\'{u}n m\'{a}s, las ecuaciones que van apareciendo a cada
paso se vuelven d\'{\i}ficiles de resolver a medida que crece $n.$ Por otra
parte, el que se deba teneren cuenta no s\'{o}lo que las componentes
\textquotedblleft visibles\textquotedblright\ del polinomio, sino que as\'{\i}
mismo las contribuciones de las componentes nulas del tipo $\left\vert
\boldsymbol{N}_{1}^{0}\cdots\boldsymbol{N}_{n}^{0}\right\vert ,$ vuelve este
procedimiento altamente no trivial.

\chapter{\label{SecS_Exp}\'{A}lgebras de Lie y Semigrupos Abelianos}

\begin{center}
\includegraphics[width=.5\textwidth]{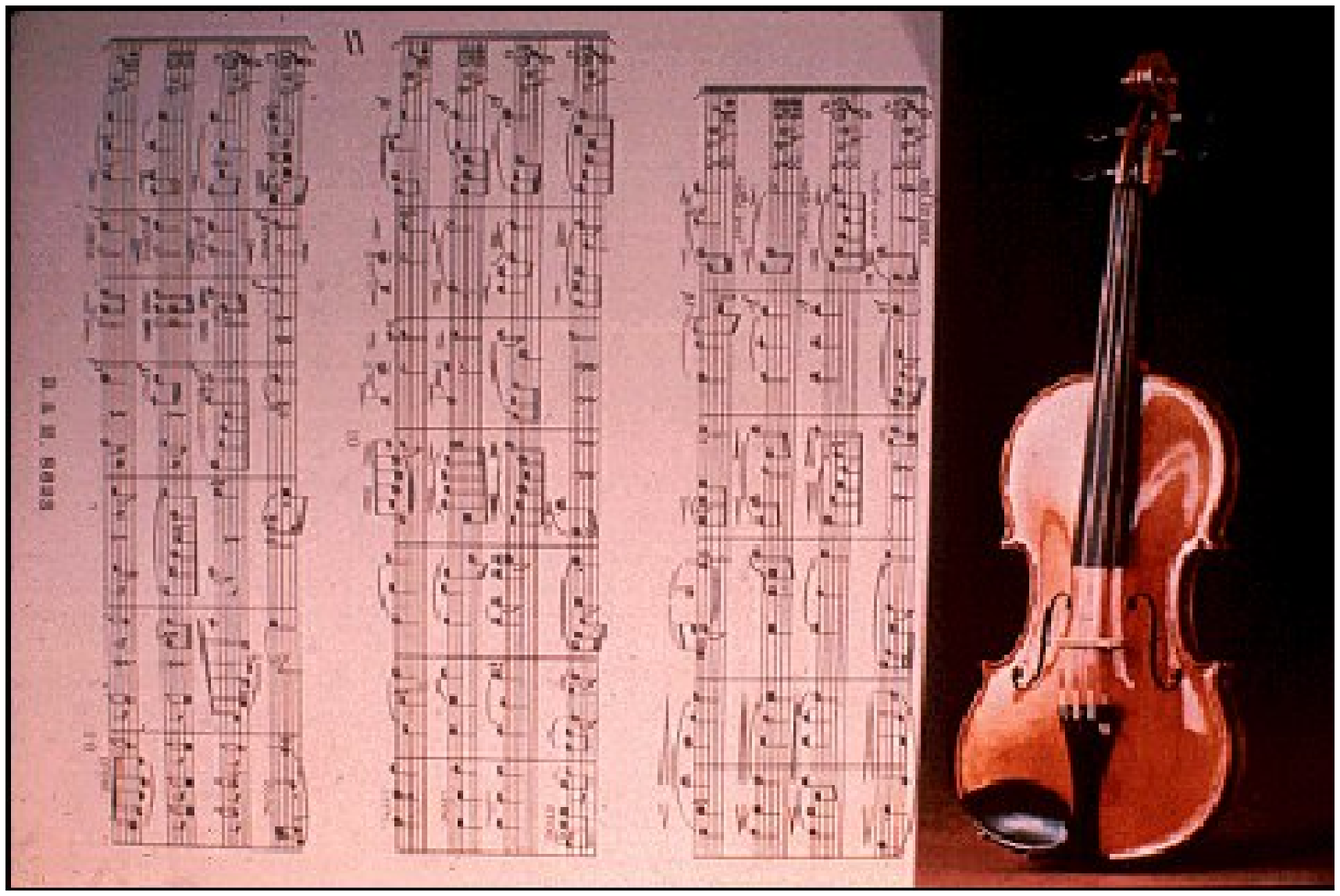}
\end{center}

\begin{quotation}
\textit{En la partitura musical de la imagen puede apreciarse un concepto
com\'{u}n a casi toda tradici\'{o}n musical: el uso simult\'{a}neo de
estructuras que est\'{a}n en armon\'{\i}a o resonancia mutua. }%
(\textit{As\'{\i} por ejemplo, puede observarse una l\'{\i}nea mel\'{o}dica en
clave de sol y otra en clave de fa, cada una de ellas arm\'{o}nica con la
otra})\textit{. El mismo concepto ser\'{a} de vital importancia en este
cap\'{\i}tulo para formular la idea de sub\'{a}lgebras resonante y reducci\'{o}n resonante. La partitura de la figura corresponde al Cuarteto de Cuerdas No.~13
en Si~}$\flat$\textit{\ mayor, Opus~130, por Ludwig van Beethoven,
\textquotedblleft Cavatina\textquotedblright. Esta imagen y su
interpretaci\'{o}n por el Cuarteto de Cuerdas de Budapest fueron enviadas en
discos adosados a las sondas Voyager I y II, como saludos de la Tierra hacia
otras inteligencias.}
\end{quotation}

En las secciones anteriores hemos revisado conceptos b\'{a}sicos de fibrados y
la construcci\'{o}n de teor\'{\i}as de gauge usando formas de transgresi\'{o}n
y de Chern--Simons. Para construir de esta manera una teor\'{\i}a de gauge de
Transgresi\'{o}n en 11 dimensiones para el \'{A}lgebra~M, uno de los
ingredientes principales es el tensor invariante. Sin embargo, justamente para
este caso, este se vuelve un problema dif\'{\i}cil de resolver. La supertraza
no es una buena opci\'{o}n para construir la teor\'{\i}a. \'{E}sta tiene
muchas componentes nulas, y un Lagrangeano Transgresor o de Chern--Simons
constru\'{\i}do a partir de ella s\'{o}lo depende de la conexi\'{o}n de
esp\'{\i}n y representa por lo tanto una especie de \textquotedblleft gravedad
ex\'{o}tica\textquotedblright. As\'{\i}, con la supertraza como tensor
invariante no se puede reproducir ni relatividad general ni incluir campos
fermi\'{o}nicos o campos asociados a las cargas centrales. El primer paso para
resolver este problema es entender cu\'{a}l es el origen del \'{A}lgebra~M, y
luego intentar construir un tensor invariante distinto de la supertraza para ella.

El \'{A}lgebra~M puede obtenerse a partir del \'{a}lgebra ortosimpl\'{e}ctica
$\mathfrak{osp}\left(  \mathfrak{32}|\mathfrak{1}\right)  $ a trav\'{e}s de
una \textit{expansi\'{o}n} en serie de potencias de las formas de
Maurer--Cartan (Ve\'{a}se
Refs.~\cite{Azcarraga-Expansion1,Azcarraga-Expansion2}). Sin embargo, dado que
este procedimiento est\'{a} basado en el uso de las formas de Maurer--Cartan,
resulta no trivial buscar tensores invariantes, los cuales est\'{a}n definido
sobre los generadores del \'{a}lgebra.

Debido a esto, resulta necesario buscar un m\'{e}todo alternativo, basado en
manipulaciones sobre los generadores del \'{a}lgebra. El resultado es lo que
hemos llamado $S$-Expansi\'{o}n, en donde la $S$ corresponde a un semigrupo
abeliano. La idea b\'{a}sica consiste en considerar el espacio del producto
directo entre un semigrupo abeliano y un \'{a}lgebra de Lie. Luego, es posible
obtener \'{a}lgebras de Lie sistem\'{a}ticamente desde dentro de este espacio
usando t\'{e}cnicas que describiremos en este cap\'{\i}tulo. Este m\'{e}todo
adem\'{a}s nos entrega tensores invariantes m\'{a}s generales que los
entregadas por la supertraza.

Para reobtener la expansi\'{o}n en formas de Maurer--Cartan en el contexto de
$S$-expansiones, se debe escoger un semigrupo en particular, al que llamaremos
$S_{\mathrm{E}}^{\left(  N\right)  }$. Por lo tanto, es posible reobtener
tambi\'{e}n el \'{A}lgebra~M dentro de este contexto, con la ventaja de que
este nuevo enfoque entrega un tensor invariante distinto de la supertraza.
Adem\'{a}s, para otras elecciones de semigrupos, es posible obtener nuevas
\'{a}lgebras, imposibles de obtener por otros m\'{e}todos. Esto permite
observar que el \'{A}lgebra~M pertenece a una amplia familia de \'{a}lgebras
que comparten ciertas caracter\'{\i}sticas.

Resulta interesante observar que el derivar nuevas \'{a}lgebras de Lie a
partir de otras previamente conocidas ha sido siempre un problema de gran
importancia en f\'{\i}sica y matem\'{a}ticas. Existen varios ejemplos
importantes de esto. As\'{\i} nos encontramos con el problema de las
\textit{extensiones} supersim\'{e}tricas, en la cual el objetivo es agrandar
el grupo de simetr\'{\i}a espaciotemporal de una forma no trivial. Otros
ejemplos interesantes lo constituyen las \textit{deformaciones}, las cuales
permiten por ejemplo obtener el \'{a}lgebra de Poincar\'{e} a partir del
\'{a}lgebra de Galileo\footnote{Es interesante observar que desde este punto
de vista, en principio hubiera sido posible obtener relatividad especial a
partir de la mec\'{a}nica cl\'{a}sica, s\'{o}lo a trav\'{e}s de argumentos
basados en teor\'{\i}a de grupos (Ve\'{a}se por ejemplo
Ref.~\cite{Azcarraga-Libro}).}, o las \textit{contracciones}, las cuales
permiten obtener el \'{a}lgebra dePoincar\'{e} a partir de la de AdS. Es
importante recalcar que estos tres procedimientos conservan la dimensi\'{o}n
del \'{a}lgebra, a diferencia de lo que ocurre con las expansiones en formas
de MC y $S$-expansiones.

Debido a la importancia que tienen en f\'{\i}sica los mecanismos que permiten
obtener nuevas \'{a}lgebra de Lie, desarrollaremos primero todo el formalismo
de $S$-expansiones en forma completamente general. Luego, veremos en detalle
como se reobtiene la expansi\'{o}n en formas de MC dentro de este contexto y
finalmente veremos algunos ejemplos expl\'{\i}citos para el caso de
supergravedad. Entre estos ejemplos reobtendremos el \'{A}lgebra~M a
trav\'{e}s de este mecanismo y escribiremos un tensor invariante para ella.

\section{Definiciones}

Antes de analizar el procedimiento de $S$-Expansiones, es necesario hacer
algunas definiciones, fijar notaci\'{o}n y demostrar algunos resultados
concernientes a semigrupos abelianos y \'{a}lgebras de Lie

\subsection{Semigrupos}

Un semigrupo es un conjunto dotado de un producto cerrado y
asociativo\footnote{Es necesario recalcar que \textit{no} exigiremos la
existencia de inverso o identidad, puesto que algunos autores utilizan
semigrupo como sin\'{o}nimo de \textit{monoide}, es decir, exigen la
existencia de identidad.}. Sea $S=\left\{  \lambda_{\alpha}\right\}  $ un
semigrupo \emph{finito} y \emph{discreto}. La informaci\'{o}n de la tabla de
multiplicaci\'{o}n del semigrupo puede codificarse a trav\'{e}s de la
siguiente definici\'{o}n

\begin{definition}
Sean $\lambda_{\alpha_{1}},\ldots\lambda_{\alpha_{n}}\in S,$ cuyo producto
viene dado por%
\[
\lambda_{\alpha_{1}}\cdots\lambda_{\alpha_{n}}=\lambda_{\gamma\left(
\alpha_{1}\cdots\alpha_{n}\right)  .}%
\]

Entonces, se define el $n$-Selector $K_{\alpha_{1}\cdots\alpha_{n}%
}^{\phantom{\alpha_{1} \cdots \alpha_{n}}\rho}$ como%
\begin{equation}
K_{\alpha_{1}\cdots\alpha_{n}}^{\phantom{\alpha_{1} \cdots \alpha_{n}}\rho
}=\left\{
\begin{array}
[c]{l}%
1\text{, cuando }\rho=\gamma\left(  \alpha_{1},\ldots,\alpha_{n}\right) \\
0\text{, en otro caso.}%
\end{array}
\right. \label{Ec n-Selector}%
\end{equation}

\end{definition}

De la definici\'{o}n vemos que el $2$-Selector codifica toda la
informaci\'{o}n contenida en la tabla de multiplicaci\'{o}n del semigrupo.
A\'{u}n m\'{a}s, siempre es posible escribir un $n$-selector en t\'{e}rminos
de $2$-selectores. Una manera quiz\'{a}s no del todo rigurosa\footnote{Esta
falta de \textquotedblleft rigurosidad\textquotedblright\ de
ec.~(\ref{Ec NoRigurosa}) consiste simplemente en que a\'{u}n no hemos
definido $1\times\lambda_{\alpha}$ \'{o} $0\times\lambda_{\alpha}$ (aunque
estas operaciones est\'{a}n definidas autom\'{a}ticamente cuando estamos
provistos de una representaci\'{o}n matricial del semigrupo, la cual siempre
existe para un semigrupo finito y discreto). Es posible demostrar todas las
propiedades de los $n$-selectores sin recurrir a ec.~(\ref{Ec NoRigurosa}),
pero resulta un poco m\'{a}s engorroso. Las propiedades de los $n$-selectores
descansan s\'{o}lo en la asociatividad y cierre del semigrupo.} pero s\'{\i}
bastante intuitiva de ver esto es escribiendo el producto de dos elementos del
semigrupo como%
\begin{equation}
\lambda_{\alpha}\lambda_{\beta}=K_{\alpha\beta}^{\phantom{\alpha\beta}\rho
}\lambda_{\rho}.\label{Ec NoRigurosa}%
\end{equation}

A partir de ec.~(\ref{Ec NoRigurosa}), es directo demostrar que la
asociatividad y clausura del producto,%
\[
\left(  \lambda_{\alpha}\lambda_{\beta}\right)  \lambda_{\gamma}%
=\lambda_{\alpha}\left(  \lambda_{\beta}\lambda_{\gamma}\right)
=\lambda_{\rho\left(  \alpha,\beta,\gamma\right)  }%
\]
es equivalente a la condici\'{o}n sobre los $2$-selectores%
\[
K_{\alpha\beta}^{\phantom{\alpha\beta}\rho}K_{\rho\gamma}%
^{\phantom{\rho\gamma}\sigma}=K_{\alpha\rho}^{\phantom{\alpha\beta}\sigma
}K_{\beta\gamma}^{\phantom{\beta\gamma}\rho}=K_{\alpha\beta\gamma
}^{\phantom{\alpha\beta\gamma}\sigma}.
\]

Esto, a su vez, significa que los $2$-selectores proveen de una
representaci\'{o}n matricial para el semigrupo. En efecto, definiendo%
\begin{equation}
\left[  \lambda_{\alpha}\right]  _{\mu}^{\ \nu}=K_{\mu\alpha}^{\quad\nu
},\label{Ec Rep Matricial = 2-selector}%
\end{equation}
se tiene%
\[
\left[  \lambda_{\alpha}\right]  _{\mu}^{\phantom{\mu}\sigma}\left[
\lambda_{\beta}\right]  _{\sigma}^{\phantom{\sigma}\nu}=K_{\alpha\beta
}^{\phantom{\alpha \beta}\sigma}\left[  \lambda_{\sigma}\right]  _{\mu
}^{\phantom{\mu}\nu}=\left[  \lambda_{\gamma\left(  \alpha,\beta\right)
}\right]  _{\mu}^{\phantom{\mu}\nu}.
\]

De la misma forma, es directo probar que los $n$-selectores satisfacen%
\begin{equation}
K_{\alpha_{1}\cdots\alpha_{n-1}}^{\phantom{\alpha_{1}\cdots\alpha_{n-1}}\sigma
}K_{\sigma\alpha_{n}}^{\phantom{\sigma\alpha_{n}}\rho}=K_{\alpha_{1}\sigma
}^{\phantom{\alpha_{1}\sigma}\rho}K_{\alpha_{2}\cdots\alpha_{n}}%
^{\phantom{\alpha_{2}\cdots\alpha_{n}}\sigma}=K_{\alpha_{1}\cdots\alpha_{n}%
}^{\phantom{\alpha_{1}\cdots\alpha_{n}}\rho}.\label{Ec Kn = K2 x Kn-1}%
\end{equation}

As\'{\i}, es posible expresar siempre un $n$-selector en t\'{e}rminos de
$2$-selectores, anidando ec.~(\ref{Ec Kn = K2 x Kn-1}).

De ahora en adelante, trabajaremos s\'{o}lo con semigrupos abelianos.
As\'{\i}, todo $n$-selector debe de ser completamente sim\'{e}trico en sus
\'{\i}ndices bajos.

Ahora bien, de la misma forma como desde un punto de vista f\'{\i}sico resulta
interesante dividir un \'{a}lgebra en subespacios, resultar\'{a} interesante
el considerar subconjuntos de $S.$ Esto induce la siguiente

\begin{definition}
\label{Def Prod Subsets}Sean $\left\{  S_{p}\right\}  _{p\in I}$ subconjuntos
del semigrupo $S,$ $S_{p}\subset S,$ $p\in I$, siendo $I$ un conjunto de
\'{\i}ndices. Cuando $S=\bigcup_{p\in I}S_{p},$ diremos que $\left\{
S_{p}\right\}  _{p\in I}$ es una \textit{descomposici\'{o}n} de $S.$ Dados dos
subconjuntos de $S,$ $S_{p}$ y $S_{q},$ definiremos su producto $S_{p}\times
S_{q}\subset S$ como%
\begin{equation}
S_{p}\times S_{q}=\left\{  \lambda_{\gamma}\in S\text{ tal que }%
\lambda_{\gamma}=\lambda_{\alpha_{p}}\lambda_{\alpha_{q}}\text{, con }%
\lambda_{\alpha_{p}}\in S_{p}\text{ y }\lambda_{\alpha_{q}}\in S_{q}\right\}
.\label{Ec Prod Subsets}%
\end{equation}

\end{definition}

En pocas palabras, $S_{p}\times S_{q}$ es el producto de todos los elementos
de $S_{p}$ con todos los elementos de $S_{q}.$

Es conveniente enfatizar que $S_{p}$ y $S_{q}$ no necesitan ser semigrupos
ellos mismos, y que dado que $S$ es abeliano, $S_{p}\times S_{q}=S_{q}\times
S_{p}$.

Algunos semigrupos est\'{a}n provistos de un elemento con un comportamiento especial,

\begin{definition}
Sea $S$ un semigrupo abeliano el cual posee un elemento $0_{S}\in S$ tal que
satisface%
\[
0_{S}\lambda_{\alpha}=\lambda_{\alpha}0_{S}=0_{S}%
\]
para todo $\lambda_{\alpha}\in S.$ Entonces, el elemento $0_{S}$ ser\'{a}
llamado el cero del semigrupo.
\end{definition}

Es trivial probar que cuando un semigrupo posee un elemento $0_{S},$ entonces
este es \'{u}nico. Por otra parte, un semigrupo provisto de un elemento
$0_{S}$ no es un grupo (este elemento no tiene inverso), con la excepci\'{o}n
trivial del grupo $S_{1},$ $e^{2}=e$ en donde $e=0_{S}.$

Dado que cuando existe, este elemento es \'{u}nico, resultar\'{a} conveniente
tener una notaci\'{o}n m\'{a}s sistem\'{a}tica para \'{e}l. Dado un semigrupo
provisto de un elemento $0_{S},$ $S=\left\{  \lambda_{\alpha}\right\}
_{\alpha=0}^{N+1},$ asignaremos el elemento $\lambda_{N+1}$ para el cero,
\[
\lambda_{N+1}=0_{S}.
\]

Los elementos restantes, $\lambda_{i},$ $i=0,\ldots,N$ se asignan para los que
son distintos de $0_{S},$ y ser\'{a}n denotados por un sub\'{\i}ndice latino
$i,j,k$.

\subsection{\'{A}lgebras de Lie Reducidas}

Es interesante observar que dada un \'{a}lgebra de Lie, en ciertos casos es
posible obtener \'{a}lgebras m\'{a}s peque\~{n}as a partir de ella, a
trav\'{e}s de lo que llamaremos \textquotedblleft
reducci\'{o}n\textquotedblright

\begin{definition}
\label{Def Alg Reducida}Sea $\mathfrak{g}$ un \'{a}lgebra de Lie de la forma
$\mathfrak{g}=V_{0}\oplus V_{1},$ siendo $\left\{  \bm{T}_{a_{0}}\right\}  $
los generadores de $V_{0}$ y $\left\{  \bm{T}_{a_{1}}\right\}  $ los de
$V_{1}. $ Cuando $\left[  V_{0},V_{1}\right]  \subset V_{1},$ \textit{i.e.,
cuando}%
\begin{align}
\left[  \bm{T}_{a_{0}},\bm{T}_{b_{0}}\right]   &  =C_{a_{0}b_{0}%
}^{\phantom{a_{0} b_{0}}c_{0}}\bm{T}_{c_{0}}+C_{a_{0}b_{0}}%
^{\phantom{a_{0} b_{0}}c_{1}}\bm{T}_{c_{1}},\label{ForcedAlgebra 00}\\
\left[  \bm{T}_{a_{0}},\bm{T}_{b_{1}}\right]   &  =C_{a_{0}b_{1}%
}^{\phantom{a_{0} b_{1}}c_{1}}\bm{T}_{c_{1}},\label{ForcedAlgebra 01}\\
\left[  \bm{T}_{a_{1}},\bm{T}_{b_{1}}\right]   &  =C_{a_{1}b_{1}%
}^{\phantom{a_{1} b_{1}}c_{0}}\bm{T}_{c_{0}}+C_{a_{1}b_{1}}%
^{\phantom{a_{1} b_{1}}c_{1}}\bm{T}_{c_{1}},\label{ForcedAlgebra 11}%
\end{align}

entonces las constantes de estructura $C_{a_{0}b_{0}}%
^{\phantom{a_{0} b_{0}}c_{0}}$ satisfacen la identidad de Jacobi por s\'{\i}
mismas, y por lo tanto,
\[
\left[  \bm{T}_{a_{0}},\bm{T}_{b_{0}}\right]  =C_{a_{0}b_{0}}%
^{\phantom{a_{0} b_{0}}c_{0}}\bm{T}_{c_{0}}%
\]
tambi\'{e}n corresponde a un \'{a}lgebra de Lie por s\'{\i} misma. Esta
\'{a}lgebra, de constantes de estructura $C_{a_{0}b_{0}}%
^{\phantom{a_{0} b_{0}}c_{0}}$ es llamada un \'{a}lgebra reducida de
$\mathfrak{g},$ y ser\'{a} simbolizada por $\left\vert V_{0}\right\vert .$
\end{definition}

Es directo probar que dado que las constantes de estructura $C_{AB}%
^{\phantom{AB}C}$ de $\mathfrak{g}$ satisfacen la identidad de Jacobi,%
\[
\left(  -1\right)  ^{\mathfrak{q}\left(  A\right)  \mathfrak{q}\left(
D\right)  }C_{AB}^{\phantom{AB}C}C_{CD}^{\phantom{CD}E}+\left(  -1\right)
^{\mathfrak{q}\left(  D\right)  \mathfrak{q}\left(  B\right)  }C_{DA}%
^{\phantom{DA}C}C_{CB}^{\phantom{CB}E}+\left(  -1\right)  ^{\mathfrak{q}%
\left(  B\right)  \mathfrak{q}\left(  A\right)  }C_{BD}^{\phantom{BD}C}%
C_{CA}^{\phantom{CA}E}=0,
\]
entonces $C_{a_{0}b_{0}}^{\phantom{a_{0} b_{0}}c_{0}}$ tambi\'{e}n lo
har\'{a}n por s\'{\i} mismas. En efecto, considerando la componente valuada en
$V_{0}$ de la identidad de Jacobi,%
\[
\left(  -1\right)  ^{\mathfrak{q}\left(  a_{0}\right)  \mathfrak{q}\left(
d_{0}\right)  }C_{a_{0}b_{0}}^{\phantom{a_{0}b_{0}}C}C_{Cd_{0}}%
^{\phantom{Cd_{0}}e_{0}}+\left(  -1\right)  ^{\mathfrak{q}\left(
d_{0}\right)  \mathfrak{q}\left(  b_{0}\right)  }C_{d_{0}a_{0}}%
^{\phantom{d_{0}a_{0}}C}C_{Cb_{0}}^{\phantom{Cb_{0}}e_{0}}+\left(  -1\right)
^{\mathfrak{q}\left(  b_{0}\right)  \mathfrak{q}\left(  a_{0}\right)
}C_{b_{0}d_{0}}^{\phantom{b_{0}d_{0}}C}C_{Ca_{0}}^{\phantom{Ca_{0}}e_{0}}=0,
\]
tenemos que%
\begin{multline*}
\left(  -1\right)  ^{\mathfrak{q}\left(  a_{0}\right)  \mathfrak{q}\left(
d_{0}\right)  }C_{a_{0}b_{0}}^{\phantom{a_{0}b_{0}}c_{0}}C_{c_{0}d_{0}%
}^{\phantom{Cd_{0}}e_{0}}+\left(  -1\right)  ^{\mathfrak{q}\left(
d_{0}\right)  \mathfrak{q}\left(  b_{0}\right)  }C_{d_{0}a_{0}}%
^{\phantom{d_{0}a_{0}}c_{0}}C_{c_{0}b_{0}}^{\phantom{Cb_{0}}e_{0}}+\left(
-1\right)  ^{\mathfrak{q}\left(  b_{0}\right)  \mathfrak{q}\left(
a_{0}\right)  }C_{b_{0}d_{0}}^{\phantom{b_{0}d_{0}}c_{0}}C_{c_{0}a_{0}%
}^{\phantom{Ca_{0}}e_{0}}+\\
+\left(  -1\right)  ^{\mathfrak{q}\left(  a_{0}\right)  \mathfrak{q}\left(
d_{0}\right)  }C_{a_{0}b_{0}}^{\phantom{a_{0}b_{0}}c_{1}}C_{c_{1}d_{0}%
}^{\phantom{Cd_{0}}e_{0}}+\left(  -1\right)  ^{\mathfrak{q}\left(
d_{0}\right)  \mathfrak{q}\left(  b_{0}\right)  }C_{d_{0}a_{0}}%
^{\phantom{d_{0}a_{0}}c_{1}}C_{c_{1}b_{0}}^{\phantom{Cb_{0}}e_{0}}+\left(
-1\right)  ^{\mathfrak{q}\left(  b_{0}\right)  \mathfrak{q}\left(
a_{0}\right)  }C_{b_{0}d_{0}}^{\phantom{b_{0}d_{0}}c_{1}}C_{c_{1}a_{0}%
}^{\phantom{Ca_{0}}e_{0}}=0.
\end{multline*}

As\'{\i}, vemos que $C_{a_{0}b_{0}}^{\phantom{a_{0}b_{0}}c_{0}}$ satisface la
identidad de Jacobi por s\'{\i} misma en dos situaciones: cuando
$C_{a_{0}b_{0}}^{\phantom{a_{0}b_{0}}c_{1}}=0$ (\textit{i.e.}, cuando $V_{0}$
es una sub\'{a}lgebra) o bien, cuando $C_{a_{0}b_{1}}%
^{\phantom{a_{0}b_{0}}c_{0}}=0,$ \textit{i.e.}, cuando $\left[  V_{0}%
,V_{1}\right]  \subset V_{1}$ (y por lo tanto, $\left\vert V_{0}\right\vert $
es un \'{a}lgebra reducida).

Es importante recalcar que en general $\left\vert V_{0}\right\vert $
\textit{no} es una sub\'{a}lgebra de $\mathfrak{g},$ sino que m\'{a}s bien
corresponder\'{\i}a a lo que se podr\'{\i}a llamar una \textquotedblleft
extensi\'{o}n inversa\textquotedblright\ de $\mathfrak{g}$, con la salvedad de
que $V_{1}$ no precisa ser un ideal.

\section{$S$-Expansi\'{o}n para un semigrupo arbitrario $S$.}

\subsection{Definici\'{o}n general}

\begin{theorem}
Sea $S=\left\{  \lambda_{\alpha}\right\}  $ un semigrupo abeliano, de
$2$-selector $K_{\alpha\beta}^{\phantom{\alpha \beta}\gamma}$ y sea
$\mathfrak{g}$ un \'{a}lgebra de Lie de base $\left\{  \bm{T}_{A}\right\}  $ y
constantes de estructura $C_{AB}^{\phantom{AB}C}$. Denotemos un elemento de la
base en el espacio del producto directo $S\otimes\mathfrak{g}$ por
$\bm{T}_{\left(  A,\alpha\right)  }=\lambda_{\alpha}\bm{T}_{A}$ y consideremos
el conmutador inducido $\left[  \bm{T}_{\left(  A,\alpha\right)
},\bm{T}_{\left(  B,\beta\right)  }\right]  \equiv\lambda_{\alpha}%
\lambda_{\beta}\left[  \bm{T}_{A},\bm{T}_{B}\right]  $. Entonces,
$S\otimes\mathfrak{g}$ es tambi\'{e}n un \'{a}lgebra de Lie, de constantes de
estructura
\begin{equation}
C_{\left(  A,\alpha\right)  \left(  B,\beta\right)  }%
^{\phantom{\left( A, \alpha \right) \left( B, \beta \right)}\left(
C,\gamma\right)  }=K_{\alpha\beta}^{\phantom{\alpha \beta}\gamma}%
C_{AB}^{\phantom{AB}C}.\label{Ec C=KC}%
\end{equation}

\end{theorem}

\begin{proof}
Utilizando la ley de multiplicaci\'{o}n del semigrupo, $\lambda_{\alpha
}\lambda_{\beta}=\lambda_{\rho\left(  \alpha,\beta\right)  },$ tenemos%
\begin{align*}
\left[  \bm{T}_{\left(  A,\alpha\right)  },\bm{T}_{\left(  B,\beta\right)
}\right]   &  \equiv\lambda_{\alpha}\lambda_{\beta}\left[  \bm{T}_{A}%
,\bm{T}_{B}\right]  ,\\
&  =\lambda_{\rho\left(  \alpha,\beta\right)  }C_{AB}^{\phantom{AB}C}%
\bm{T}_{C},\\
&  =C_{AB}^{\phantom{AB}C}\bm{T}_{\left(  C,\rho\left(  \alpha,\beta\right)
\right)  }.
\end{align*}

Utilizando la definici\'{o}n de $2$-selector [ec.~(\ref{Ec n-Selector})],%
\[
K_{\alpha\beta}^{\phantom{\alpha \beta}\rho}=\left\{
\begin{array}
[c]{l}%
1\text{, cuando }\rho=\gamma\left(  \alpha,\beta\right) \\
0\text{, en otro caso.}%
\end{array}
\right.
\]

tenemos que%
\[
\left[  \bm{T}_{\left(  A,\alpha\right)  },\bm{T}_{\left(  B,\beta\right)
}\right]  =K_{\alpha\beta}^{\phantom{\alpha \beta}\gamma}C_{AB}%
^{\phantom{AB}C}\bm{T}_{\left(  C,\gamma\right)  }.
\]

con lo que hemos probado que el \'{a}lgebra de $S\otimes\mathfrak{g}$ se cierra.

Se debe observar que ya que $S$ es abeliano, las constantes de estructura
$C_{\left(  A,\alpha\right)  \left(  B,\beta\right)  }%
^{\phantom{\left( A, \alpha
\right) \left( B, \beta \right)}\left(  C,\gamma\right)  }$ heredan su
simetr\'{\i}a de $C_{AB}^{\phantom{AB}C}$, de la forma%
\[
C_{\left(  A,\alpha\right)  \left(  B,\beta\right)  }%
^{\phantom{\left( A, \alpha \right) \left( B, \beta \right)}\left(
C,\gamma\right)  }=-\left(  -1\right)  ^{\mathfrak{q}\left(  A\right)
\mathfrak{q}\left(  B\right)  }C_{\left(  B,\beta\right)  \left(
A,\alpha\right)  }%
^{\phantom{\left( A, \alpha \right) \left( B, \beta \right)}\left(
C,\gamma\right)  },
\]
y por lo tanto, $\mathfrak{q}\left(  A,\alpha\right)  =\mathfrak{q}\left(
A\right)  .$

Utilizando el hecho de que $\mathfrak{q}\left(  A,\alpha\right)
=\mathfrak{q}\left(  A\right)  $ y que las constantes de estructura
$C_{AB}^{\phantom{AB}C}$ satisfacen por s\'{\i} mismas la identidad de Jacobi,
es directo demostrar que $C_{\left(  B,\beta\right)  \left(  A,\alpha\right)
}^{\phantom{\left( A,
\alpha \right) \left( B, \beta \right)}\left(  C,\gamma\right)  }$ tambi\'{e}n
lo hacen. En efecto,%
\begin{multline*}
\left(  -1\right)  ^{\mathfrak{q}\left(  A,\alpha\right)  \mathfrak{q}\left(
D,\delta\right)  }C_{\left(  A,\alpha\right)  \left(  B,\beta\right)
}^{\phantom{\left( A, \alpha \right) \left( B, \beta \right)}\left(
C,\gamma\right)  }C_{\left(  C,\gamma\right)  \left(  D,\delta\right)
}^{\phantom{\left( C, \gamma \right) \left( D, \delta \right)}\left(
E,\varepsilon\right)  }+\\
+\left(  -1\right)  ^{\mathfrak{q}\left(  D,\delta\right)  \mathfrak{q}\left(
B,\beta\right)  }C_{\left(  D,\delta\right)  \left(  A,\alpha\right)
}^{\phantom{\left( D, \delta \right) \left( A, \alpha \right)}C}C_{\left(
C,\gamma\right)  \left(  B,\beta\right)  }%
^{\phantom{\left( C, \gamma \right) \left( B, \beta \right)}\left(
E,\varepsilon\right)  }+\\
+\left(  -1\right)  ^{\mathfrak{q}\left(  B,\beta\right)  \mathfrak{q}\left(
A,\alpha\right)  }C_{\left(  B,\beta\right)  \left(  D,\delta\right)
}^{\phantom{BD}\left(  C,\gamma\right)  }C_{\left(  C,\gamma\right)  \left(
A,\alpha\right)  }%
^{\phantom{\left( C, \gamma \right) \left( A, \alpha \right)}\left(
E,\varepsilon\right)  }=\\
K_{\alpha\beta\delta}^{\phantom{\alpha \beta \delta}\varepsilon}\left(
\left(  -1\right)  ^{\mathfrak{q}\left(  A\right)  \mathfrak{q}\left(
D\right)  }C_{AB}^{\phantom{AB}C}C_{CD}^{\phantom{CD}E}+\right. \\
+\left(  -1\right)  ^{\mathfrak{q}\left(  D\right)  \mathfrak{q}\left(
B\right)  }C_{DA}^{\phantom{DA}C}C_{CB}^{\phantom{CB}E}+\\
\left.  +\left(  -1\right)  ^{\mathfrak{q}\left(  B\right)  \mathfrak{q}%
\left(  A\right)  }C_{BD}^{\phantom{BD}C}C_{CA}^{\phantom{CA}E}\right)  .
\end{multline*}
Dado que $C_{AB}^{\phantom{AB}C}$ satisface la identidad de Jacobi,
$C_{\left(  A,\alpha\right)  \left(  B,\beta\right)  }%
^{\phantom{\left( A, \alpha
\right) \left( B, \beta \right)}\left(  C,\gamma\right)  }$ tambi\'{e}n lo
hace,%
\begin{multline*}
\left(  -1\right)  ^{\mathfrak{q}\left(  A,\alpha\right)  \mathfrak{q}\left(
D,\delta\right)  }C_{\left(  A,\alpha\right)  \left(  B,\beta\right)
}^{\phantom{\left( A, \alpha \right) \left( B, \beta \right)}\left(
C,\gamma\right)  }C_{\left(  C,\gamma\right)  \left(  D,\delta\right)
}^{\phantom{\left( C, \gamma \right) \left( D, \delta \right)}\left(
E,\varepsilon\right)  }+\\
+\left(  -1\right)  ^{\mathfrak{q}\left(  D,\delta\right)  \mathfrak{q}\left(
B,\beta\right)  }C_{\left(  D,\delta\right)  \left(  A,\alpha\right)
}^{\phantom{\left( D, \delta \right) \left( A, \alpha \right)}C}C_{\left(
C,\gamma\right)  \left(  B,\beta\right)  }%
^{\phantom{\left( C, \gamma \right) \left( B, \beta \right)}\left(
E,\varepsilon\right)  }+\\
+\left(  -1\right)  ^{\mathfrak{q}\left(  B,\beta\right)  \mathfrak{q}\left(
A,\alpha\right)  }C_{\left(  B,\beta\right)  \left(  D,\delta\right)
}^{\phantom{BD}\left(  C,\gamma\right)  }C_{\left(  C,\gamma\right)  \left(
A,\alpha\right)  }%
^{\phantom{\left( C, \gamma \right) \left( A, \alpha \right)}\left(
E,\varepsilon\right)  }=0,
\end{multline*}
y por lo tanto, $S\otimes\mathfrak{g}$ es tambi\'{e}n un \'{a}lgebra de Lie.
\end{proof}

Este teorema induce naturalmente la siguiente

\begin{definition}
Sea $S$ un semigrupo abeliano y $\mathfrak{g}$ un \'{a}lgebra de Lie. El
\'{a}lgebra de Lie $\mathfrak{G}$ definida por $\mathfrak{G}=S\otimes
\mathfrak{g}$ es llamada \emph{\'{A}lgebra }$S$\emph{-Expandida de
}$\mathfrak{g}$.
\end{definition}

Es interesante observar que en general, para obtener un \'{a}lgebra a partir
de otra (por ejemplo, al realizar una contracci\'{o}n de
\.{I}n\"{o}n\"{u}--Wigner) es usual el multiplicar los generadores por un
cierto par\'{a}metro, y despu\'{e}s realizar alguna manipulaci\'{o}n, como
efectuar un cierto l\'{\i}mite. La $S$-Expansi\'{o}n puede ser vista como la
generalizaci\'{o}n natural de esta idea, en donde en lugar de multiplicar los
generadores por un par\'{a}metro n\'{u}merico, lo hacemos con elementos de un
semigrupo abeliano.

Un \'{a}lgebra $S$-expandida tiene una estructura bastante simple, en donde en
cierta forma tenemos una copia del \'{a}lgebra original $\mathfrak{g}$ para
cada elemento del semigrupo. Para poder encontrar estructuras m\'{a}s
complejas, debemos obtener \'{a}lgebras m\'{a}s peque\~{n}as a partir del
\'{a}lgebra $S$-expandida, \textit{i.e.}, sub\'{a}lgebras y \'{a}lgebras
reducidas. Como veremos a continuaci\'{o}n, es posible resolver este problema
en forma general y sistem\'{a}tica. Aplicado a casos particulares, estos
procedimientos permitir\'{a}n recuperar \'{a}lgebras de inter\'{e}s en
f\'{\i}sica, junto con sus correspondientes tensores invariantes en forma
directa, como veremos a continuaci\'{o}n.

\subsection{\'{A}lgebra $0_{S}$-Reducida}

Cuando el semigrupo $S$ est\'{a} provisto de un elemento $0_{S},$ sus
correspondientes $n$-selectores tienen una forma peculiar, y por ende,
tambi\'{e}n el \'{a}lgebra $S$-expandida asociada. En efecto, denotando con
$\lambda_{N+1}=0_{S}$ y con $\lambda_{i}$ los elementos distintos del cero,
tenemos que el $2$-selector satisface%
\begin{align}
K_{i,N+1}^{\phantom{i,N+1}j}  &  =K_{N+1,i}^{\phantom{N+1,i}j}=0,\\
K_{i,N+1}^{\phantom{i,N+1}N+1}  &  =K_{N+1,i}^{\phantom{N+1,i}N+1}=1,\\
K_{N+1,N+1}^{\phantom{N+1,N+1}j}  &  =0,\\
K_{N+1,N+1}^{\phantom{N+1,N+1}N+1}  &  =1.
\end{align}

Por lo tanto, el \'{a}lgebra $S$-expandida $\mathfrak{G}=S\otimes\mathfrak{g}
$ tiene la forma
\begin{align}
\left[  \bm{T}_{\left(  A,i\right)  },\bm{T}_{\left(  B,j\right)  }\right]
&  =K_{ij}^{\phantom{ij}k}C_{AB}^{\phantom{AB}C}\bm{T}_{\left(  C,k\right)
}+K_{ij}^{\phantom{ij}N+1}C_{AB}^{\phantom{AB}C}\bm{T}_{\left(  C,N+1\right)
},\label{Ec Ti Tj}\\
\left[  \bm{T}_{\left(  A,N+1\right)  },\bm{T}_{\left(  B,j\right)  }\right]
&  =C_{AB}^{\phantom{AB}C}\bm{T}_{\left(  C,N+1\right)  },\label{Ec T N+1 Tj}%
\\
\left[  \bm{T}_{\left(  A,N+1\right)  },\bm{T}_{\left(  B,N+1\right)
}\right]   &  =C_{AB}^{\phantom{AB}C}\bm{T}_{\left(  C,N+1\right)
}.\label{Ec T N+1 T N+1}%
\end{align}

Comparando el \'{a}lgebra ecs.~(\ref{Ec Ti Tj})-(\ref{Ec T N+1 T N+1}) con
ecs.~(\ref{ForcedAlgebra 00})-(\ref{ForcedAlgebra 11}), tenemos que las
componentes%
\[
\left[  \bm{T}_{\left(  A,i\right)  },\bm{T}_{\left(  B,j\right)  }\right]
=K_{ij}^{\phantom{ij}k}C_{AB}^{\phantom{AB}C}\bm{T}_{\left(  C,k\right)  }%
\]
forman un \'{a}lgebra reducida de $S\otimes\mathfrak{g}.$

De ecs.~(\ref{Ec Ti Tj})-(\ref{Ec T N+1 T N+1}), es claro que para este caso
en particular, el reducci\'{o}n es equivalente a imponer al condici\'{o}n
\begin{equation}
\bm{T}_{\left(  A,N+1\right)  }=0_{S}\bm{T}_{A}=\bm{0}.\label{0sxT=0}%
\end{equation}

Sin embargo, pese a que esta condici\'{o}n se ve muy natural, debido a que
hemos llamado \textquotedblleft cero\textquotedblright\ a este elemento en
particular del \'{a}lgebra, debe de recalcarse que esta condici\'{o}n est\'{a}
lejos de ser trivial, debido a que en general, $0_{S}$ \emph{no }es el cero
del campo sobre el cual est\'{a} definido el espacio vectorial $\mathfrak{g}.$
A\'{u}n m\'{a}s importante, reducir el \'{a}lgebra cambia la forma del
\'{a}lgebra, abelianizando muchos conmutadores. En general, para todo $i,j$
tal que $K_{ij}^{\phantom{ij}N+1}=1$ (i.e., $\lambda_{i}\lambda_{j}%
=\lambda_{N+1}$) tenemos que%
\[
\left[  \bm{T}_{\left(  A,i\right)  },\bm{T}_{\left(  B,j\right)  }\right]
=\bm{0}.
\]

As\'{\i}, se vuelve natural hacer la siguiente

\begin{definition}
Sea $S$ un semigrupo abeliano provisto de un elemento $0_{S}\in S,$ y sea
$\mathfrak{G}=S\otimes\mathfrak{g}$ un \'{a}lgebra $S$-expandida. El
\'{a}lgebra obtenida imponiendo la condici\'{o}n $0_{S}\bm{T}_{A}=\bm{0}$
ser\'{a} llamada \'{a}lgebra $0_{S}$-reducida.
\end{definition}

\subsection{Sub\'{a}lgebras Resonantes}

El problema de encontrar sub\'{a}lgebras de $\mathfrak{G}=S\otimes
\mathfrak{g}$ es uno no-trivial, pero soluble, cuando se enfoca desde la
apropiada perspectiva.

El primer paso para resolver este problema es codificar en forma
sistem\'{a}tica la estructura de subespacios del \'{a}lgebra original
$\mathfrak{g},$ de la siguiente manera.

Sea $\mathfrak{g}=\bigoplus_{p\in I}V_{p}$ una descomposici\'{o}n de
$\mathfrak{g}$ en subespacios $V_{p}$, donde $I$ corresponde a un conjunto de
\'{\i}ndices. Dado que el \'{a}lgebra es cerrada, para todo $p,q\in I$ es
posible definir los subconjuntos $i_{\left(  p,q\right)  }\subset I$ tales que%
\begin{equation}
\left[  V_{p},V_{q}\right]  \subset\bigoplus_{r\in i_{\left(  p,q\right)  }%
}V_{r}.\label{Ec [Vp , Vq] = Vr}%
\end{equation}

De esta forma, la colecci\'{o}n de subconjuntos $\left\{  i_{\left(
p,q\right)  }\right\}  _{p,q\in I}$ almacenan la informaci\'{o}n de la
estructura de subespacios de $\mathfrak{g}$.

En forma an\'{a}loga a como hemos descompuesto $\mathfrak{g}$ en subespacios,
es posible descomponer el semigrupo $S$ en subconjuntos $S_{p}\subset S$,
tales que%
\[
S=\bigcup_{p\in I}S_{p}.
\]

Esta descomposici\'{o}n en subconjuntos es en principio completamente
arbitraria; sin embargo, utilizando la definici\'{o}n de producto de
subconjuntos, ec.~(\ref{Ec Prod Subsets}), es posible escoger una
descomposici\'{o}n en particular, tal como veremos en la siguiente

\begin{definition}
Sea un \'{a}lgebra $\mathfrak{g},$ provista de una descomposici\'{o}n en
subespacios $\mathfrak{g}=\bigoplus_{p\in I}V_{p},$ con una estructura
provista por $\left\{  i_{\left(  p,q\right)  }\right\}  _{p,q\in I}$, tal
como en ec.~(\ref{Ec [Vp , Vq] = Vr}). Sea $S=\bigcup_{p\in I}S_{p}$ una
descomposici\'{o}n en subconjuntos de $S,$ tal que se satisface la
condici\'{o}n%
\begin{equation}
S_{p}\times S_{q}\subset\bigcap_{r\in i_{\left(  p,q\right)  }}S_{r}%
,\label{Ec [Sp , Sq] = Sr}%
\end{equation}
donde el producto de subconjuntos $\times$ es el de
Def.~\ref{Def Prod Subsets}. Cuando tal descomposici\'{o}n en subconjuntos de
$S$ existe, $S=\bigcup_{p\in I}S_{p}$, diremos que ella est\'{a} en
\emph{resonancia} con la descomposici\'{o}n en subespacios de $\mathfrak{g}$,
$\mathfrak{g}=\bigoplus_{p\in I}V_{p}.$
\end{definition}

\begin{theorem}
\label{Teo Subalg Resonante}Sea $\mathfrak{g}=\bigoplus_{p\in I}V_{p}$ una
descomposici\'{o}n en subespacios de $\mathfrak{g}$, con una estructura dada
por ec.~(\ref{Ec [Vp , Vq] = Vr}), y sea $S=\bigcup_{p\in I}S_{p}$ una
descomposici\'{o}n resonante del semigrupo abeliano $S$, \textit{i.e., }que
satisface ec.~(\ref{Ec [Sp , Sq] = Sr}). Sean los subespacios de
$\mathfrak{G}=S\otimes\mathfrak{g},$
\begin{equation}
W_{p}=S_{p}\otimes V_{p},\qquad p\in I.
\end{equation}
Entonces,
\begin{equation}
\mathfrak{G}_{\mathrm{R}}=\bigoplus_{p\in I}W_{p}%
\end{equation}
es una sub\'{a}lgebra de $\mathfrak{G}=S\otimes\mathfrak{g}$.
\end{theorem}

\begin{proof}
Utilizando ecs.~(\ref{Ec [Vp , Vq] = Vr}) y~(\ref{Ec [Sp , Sq] = Sr}), se
tiene que%
\begin{align*}
\left[  W_{p},W_{q}\right]   &  \subset\left(  S_{p}\times S_{q}\right)
\otimes\left[  V_{p},V_{q}\right]  ,\\
&  \subset\bigcap_{s\in i_{\left(  p,q\right)  }}S_{s}\otimes\bigoplus_{r\in
i_{\left(  p,q\right)  }}V_{r},\\
&  \subset\bigoplus_{r\in i_{\left(  p,q\right)  }}\left[  \bigcap_{s\in
i_{\left(  p,q\right)  }}S_{s}\right]  \otimes V_{r}.
\end{align*}

Ahora bien, dado que la intersecci\'{o}n de conjuntos es un subconjunto de
cada uno de ellos, tenemos que para todo $r\in i_{\left(  p,q\right)  },$%
\[
\bigcap_{s\in i_{\left(  p,q\right)  }}S_{s}\subset S_{r}.
\]

Por lo tanto, tenemos%
\[
\left[  W_{p},W_{q}\right]  \subset\bigoplus_{r\in i_{\left(  p,q\right)  }%
}S_{r}\otimes V_{r}%
\]
y as\'{\i}, finalmente,%
\[
\left[  W_{p},W_{q}\right]  \subset\bigoplus_{r\in i_{\left(  p,q\right)  }%
}W_{r}.
\]

As\'{\i}, el \'{a}lgebra se cierra y por lo tanto $\mathfrak{G}_{\mathrm{R}%
}=\bigoplus_{p\in I}W_{p}$ es una sub\'{a}lgebra de $\mathfrak{G}.$
\end{proof}

\begin{definition}
El \'{a}lgebra de Lie $\mathfrak{G}_{\mathrm{R}}=\bigoplus_{p\in I}W_{p}$
obtenida en Teorema~\ref{Teo Subalg Resonante} ser\'{a} llamada Sub\'{a}lgebra
Resonante de el \'{a}lgebra $S$-expandida $\mathfrak{G}=S\otimes\mathfrak{g}$.
\end{definition}

Es importante se\~{n}alar que todas las sub\'{a}lgebras triviales de
$\mathfrak{G}=S\otimes\mathfrak{g}$ corresponden a sub\'{a}lgebras resonantes.
As\'{\i} por ejemplo, la elecci\'{o}n $S_{p}=S$ para todo $p$ satisface en
forma trivial la condici\'{o}n de resonancia ec.~(\ref{Ec [Sp , Sq] = Sr}). Lo
mismo sucede cuando el semigrupo posee un elemento $0_{S};$ la elecci\'{o}n
$S_{p}=\left\{  0_{S}\right\}  $ tambi\'{e}n satisface autom\'{a}ticamente
ec.~(\ref{Ec [Sp , Sq] = Sr}). Este es tambi\'{e}n el caso cuando el semigrupo
est\'{a} provisto de un elemento identidad. Una curiosidad interesante es que
a lo largo de esta investigaci\'{o}n no hemos podido encontrar ning\'{u}n caso
en el que una sub\'{a}lgebra de $\mathfrak{G}=S\otimes\mathfrak{g}$ no haya
satisfecho la condici\'{o}n de resonancia; el problema de si toda
sub\'{a}lgebra de $\mathfrak{G}=S\otimes\mathfrak{g}$ corresponde a una
sub\'{a}lgebra resonante o no permanece como un problema abierto.

La elecci\'{o}n del nombre de \textquotedblleft resonancia\textquotedblright%
\ es debido a la similitud entre ecs.~(\ref{Ec [Vp , Vq] = Vr})
y~(\ref{Ec [Sp , Sq] = Sr}). La idea fundamental es que dada un \'{a}lgebra
$\mathfrak{g}$ con una cierta estructura de subespacios, debemos descomponer
el semigrupo $S$ de una forma armoniosa o \textquotedblleft
resonante\textquotedblright\ con la estructura original del \'{a}lgebra. De
ah\'{\i} tambi\'{e}n el paralelo con la (usual) armon\'{\i}a en estructuras
musicales a comienzos del cap\'{\i}tulo; el concepto es similar.

Sin embargo, es posible ir a\'{u}n m\'{a}s lejos; en la pr\'{o}xima
secci\'{o}n demostraremos como tambi\'{e}n puede efectuarse el reducci\'{o}n de
una forma \textquotedblleft resonante\textquotedblright\ con la estructura del
\'{a}lgebra de Lie original; el $0_{S}$-reducci\'{o}n ya considerado
corresponder\'{a} a un caso particular muy sencillo de este teorema. El mismo
procedimiento permitir\'{a} reproducir las contracciones de
\.{I}n\"{o}n\"{u}--Wigner generalizadas dentro del presente esquema.

Usando ec.~(\ref{Ec C=KC}), resulta sencillo escribir en foma expl\'{\i}cita
las constantes de estructura de la sub\'{a}lgebra resonante $\mathfrak{G}%
_{\mathrm{R}}$. Denotando la base de $V_{p}$ por $\left\{  \bm{T}_{a_{p}%
}\right\}  $, es posible escribir las constantes de estructura como
\begin{equation}
C_{\left(  a_{p},\alpha_{p}\right)  \left(  b_{q},\beta_{q}\right)
}%
^{\phantom{\left( a_{p}, \alpha_{p} \right) \left( b_{q}, \beta_{q} \right)}\left(
c_{r},\gamma_{r}\right)  }=K_{\alpha_{p}\beta_{q}}%
^{\phantom{\alpha_{p} \beta_{q}}\gamma_{r}}C_{a_{p}b_{q}}%
^{\phantom{a_{p} b_{q}}c_{r}}\text{ con }\alpha_{p},\beta_{q},\gamma_{r}\text{
tales que }\lambda_{\alpha_{p}}\in S_{p},\lambda_{\beta_{q}}\in S_{q}%
,\lambda_{\gamma_{r}}\in S_{r}.\label{Ec CtesStruct Resonantes}%
\end{equation}

El teorema de sub\'{a}lgebras resonantes resulta ser una herramienta
extremadamente poderosa para generar nuevas \'{a}lgebras a partir de
$\mathfrak{g}.$ El d\'{\i}ficil problema de encontrar sub\'{a}lgebras de
$\mathfrak{G}=S\otimes\mathfrak{g}$ se convierte en el sencillo problema de
encontrar una partici\'{o}n resonante de $S$ (tal como veremos con ejemplos
expl\'{\i}citos en Sec.~\ref{Sec S-Exp osp(32|1)}). En relaci\'{o}n con la
expansi\'{o}n en formas de Maurer--Cartan, resulta provechoso se\~{n}alar que
dentro del contexto de $S$-expansiones, ellas corresponden a sub\'{a}lgebras
resonantes para una elecci\'{o}n particular de semigrupo, con alg\'{u}n tipo
de reducci\'{o}n ($0_{S}$-reducci\'{o}n en general).

\subsection{Reducci\'{o}n Resonante}

Como ya hemos anunciado, no s\'{o}lo es posible encontrar sub\'{a}lgebras
utilizando la idea de una descomposici\'{o}n resonante, sino que tambi\'{e}n
es posible reducir \'{a}lgebras, tal como se muestra en el siguiente

\begin{theorem}
\label{Teo Forz Resonante}Sea $\mathfrak{G}_{\mathrm{R}}=\bigoplus_{p\in
I}S_{p}\otimes V_{p}$ una sub\'{a}lgebra resonante de $\mathfrak{G}%
=S\otimes\mathfrak{g}$, i.e. las ecs.~(\ref{Ec [Vp , Vq] = Vr})
y~(\ref{Ec [Sp , Sq] = Sr}) son satisfechas simult\'{a}neamente. Sea
$S_{p}=\hat{S}_{p}\cup\check{S}_{p}$ una partici\'{o}n de los subconjuntos
$S_{p}\subset S$ tal que se satisfacen las condiciones
\begin{align}
\hat{S}_{p}\cap\check{S}_{p}  &  =\varnothing,\label{Ec Cond Forz SpnSq=0}\\
\check{S}_{p}\times\hat{S}_{q}  &  \subset\bigcap_{r\in i_{\left(  p,q\right)
}}\hat{S}_{r}.\label{Ec Cond Forz SpSq=nSr}%
\end{align}

La descomposici\'{o}n de subconjuntos $S_{p}=\hat{S}_{p}\cup\check{S}_{p}$
induce a su vez la descomposici\'{o}n $\mathfrak{G}_{\mathrm{R}}%
=\mathfrak{\check{G}}_{\mathrm{R}}\oplus\hat{\mathfrak{G}}_{\mathrm{R}}$ sobre
la sub\'{a}lgebra resonante, donde%
\begin{align}
\mathfrak{\check{G}}_{\mathrm{R}}  &  =\bigoplus_{p\in I}\check{S}_{p}\otimes
V_{p},\\
\hat{\mathfrak{G}}_{\mathrm{R}}  &  =\bigoplus_{p\in I}\hat{S}_{p}\otimes
V_{p}.
\end{align}

Cuando se cumplen las condiciones ec.~(\ref{Ec Cond Forz SpnSq=0})
y~(\ref{Ec Cond Forz SpSq=nSr}), se tiene que
\begin{equation}
\left[  \mathfrak{\check{G}}_{\text{$\mathrm{R}$}},\hat{\mathfrak{G}%
}_{\text{$\mathrm{R}$}}\right]  \subset\hat{\mathfrak{G}}_{\text{$\mathrm{R}$%
}},
\end{equation}
y por lo tanto, a partir de Def.~\ref{Def Alg Reducida}, tenemos que
$\left\vert \mathfrak{\check{G}}_{\mathrm{R}}\right\vert $ corresponde a un
\'{a}lgebra reducida de la sub\'{a}lgebra $\mathfrak{G}_{\mathrm{R}}$.
\end{theorem}

\begin{proof}
Sean $\check{W}_{p}=\check{S}_{p}\otimes V_{p}$ y $\hat{W}_{p}=\hat{S}%
_{p}\otimes V_{p}$. Entonces, utilizando la condici\'{o}n
ec.~(\ref{Ec Cond Forz SpSq=nSr}), tenemos que%
\begin{align*}
\left[  \check{W}_{p},\hat{W}_{q}\right]   &  \subset\left(  \check{S}%
_{p}\times\hat{S}_{q}\right)  \otimes\left[  V_{p},V_{q}\right]  ,\\
&  \subset\bigcap_{s\in i_{\left(  p,q\right)  }}\hat{S}_{s}\otimes
\bigoplus_{r\in i_{\left(  p,q\right)  }}V_{r},\\
&  \subset\bigoplus_{r\in i_{\left(  p,q\right)  }}\left[  \bigcap_{s\in
i_{\left(  p,q\right)  }}\hat{S}_{s}\right]  \otimes V_{r}.
\end{align*}

Para todo $r\in i_{\left(  p,q\right)  }$ tenemos que $\bigcap_{s\in
i_{\left(  p,q\right)  }}\hat{S}_{s}\subset\hat{S}_{r}$, y por lo tanto,
\begin{align*}
\left[  \check{W}_{p},\hat{W}_{q}\right]   &  \subset\bigoplus_{r\in
i_{\left(  p,q\right)  }}\hat{S}_{r}\otimes V_{r}\\
&  \subset\bigoplus_{r\in i_{\left(  p,q\right)  }}\hat{W}_{r}.
\end{align*}

Ya que $\mathfrak{\check{G}}_{\mathrm{R}}=\bigoplus_{p\in I}\check{W}_{p}$ and
$\hat{\mathfrak{G}}_{\mathrm{R}}=\bigoplus\limits_{p\in I}\hat{W}_{p}$,
tenemos que
\[
\left[  \mathfrak{\check{G}}_{\text{$\mathrm{R}$}},\hat{\mathfrak{G}%
}_{\text{$\mathrm{R}$}}\right]  \subset\hat{\mathfrak{G}}_{\text{$\mathrm{R}$%
}},
\]
y por lo tanto, $\left\vert \mathfrak{\check{G}}_{\mathrm{R}}\right\vert $
corresponde a un \'{a}lgebra reducida de $\mathfrak{G}_{\mathrm{R}}$, a la que
llamaremos \textit{\'{a}lgebra reducida resonante de }$\mathfrak{G}%
_{\mathrm{R}}$.
\end{proof}

Usando las constantes de estructura ec.~(\ref{Ec CtesStruct Resonantes}) para
la sub\'{a}lgebra resonante, es posible escribir las constantes de estructura
para el \'{a}lgebra reducida resonante $\left\vert \mathfrak{\check{G}%
}_{\mathrm{R}}\right\vert $ como%
\begin{equation}
C_{\left(  a_{p},\alpha_{p}\right)  \left(  b_{q},\beta_{q}\right)
}^{\text{\qquad\qquad\qquad}\left(  c_{r},\gamma_{r}\right)  }=K_{\alpha
_{p}\beta_{q}}^{\quad\gamma_{r}}C_{a_{p}b_{q}}^{\text{\quad\ }c_{r}}\text{ con
}\alpha_{p},\beta_{q},\gamma_{r}\text{ tales que }\lambda_{\alpha_{p}}%
\in\check{S}_{p},\lambda_{\beta_{q}}\in\check{S}_{q},\lambda_{\gamma_{r}}%
\in\check{S}_{r}.\label{Ec Ctes Struc Forz}%
\end{equation}

Una consecuencia interesante del teorema~\ref{Teo Forz Resonante} es que el
procedimiento de $0_{S}$-reducci\'{o}n puede ser aplicado tambi\'{e}n sobre
sub\'{a}lgebras resonantes, tal como se demuestra en el siguiente

\begin{corollary}
\label{Cor O-Forz SubAlg Reson}Sea $S$ un semigrupo provisto de un elemento
$0_{S},$ y sea $\mathfrak{G}_{\mathrm{R}}=\bigoplus_{p\in I}S_{p}\otimes V_{p}
$ una sub\'{a}lgebra resonante de $\mathfrak{G}=S\otimes\mathfrak{g}$, tal que
para cada subconjunto $S_{p},$ $0_{S}\in S_{p}.$ Entonces, la
descomposici\'{o}n $S_{p}=\hat{S}_{p}\cup\check{S}_{p}$ con $\hat{S}%
_{p}=\left\{  0_{S}\right\}  $ y $\check{S}_{p}=S_{p}-\left\{  0_{S}\right\}
$ satisface las condiciones ecs.~(\ref{Ec Cond Forz SpnSq=0})
y~(\ref{Ec Cond Forz SpSq=nSr}) y por lo tanto, $\left\vert \mathfrak{\check
{G}}_{\mathrm{R}}\right\vert $ corresponde a un algebra reducida de
$\mathfrak{G}_{\mathrm{R}} $, la cual ser\'{a} llamada \emph{\'{a}lgebra
}$0_{S}$\emph{-reducida de }$\mathfrak{G}_{\mathrm{R}}.$
\end{corollary}

\begin{proof}
La descomposici\'{o}n $S_{p}=\hat{S}_{p}\cup\check{S}_{p}$ con $\hat{S}%
_{p}=\left\{  0_{S}\right\}  $ y $\check{S}_{p}=S_{p}-\left\{  0_{S}\right\}
$ satisface la condici\'{o}n ec.~(\ref{Ec Cond Forz SpnSq=0}) por
construcci\'{o}n, y dado que $\left\{  0_{S}\right\}  \times S_{p}=\left\{
0_{S}\right\}  ,$ tenemos que ec.~(\ref{Ec Cond Forz SpSq=nSr}) se satisface
en forma autom\'{a}tica. Por lo tanto, de acuerdo al
teorema~\ref{Teo Forz Resonante}, tenemos que se cumple%
\[
\left[  \mathfrak{\check{G}}_{\text{$\mathrm{R}$}},\hat{\mathfrak{G}%
}_{\text{$\mathrm{R}$}}\right]  \subset\hat{\mathfrak{G}}_{\text{$\mathrm{R}$%
}},
\]
con $\mathfrak{\check{G}}_{\text{$\mathrm{R}$}}=\bigoplus_{p\in I}\left(
S_{p}-\left\{  0_{S}\right\}  \right)  \otimes V_{p}$ y $\hat{\mathfrak{G}%
}_{\text{$\mathrm{R}$}}=\left\{  0_{S}\right\}  \otimes\mathfrak{g},$ y por lo
tanto, $\left\vert \mathfrak{\check{G}}_{\mathrm{R}}\right\vert $ corresponde
a un \'{a}lgebra de Lie reducida.
\end{proof}

\section[Expansi\'{o}n en formas de Maurer--Cartan y $S_{\mathrm{E}}^{\left(
N\right)  }$-Expansi\'{o}n]{Expansi\'{o}n en formas de Maurer--Cartan y
$S_{\mathrm{E}}^{\left(  N\right)  }$-Expansi\'{o}n
\sectionmark{Expansi\'{o}n en
formas de MC y $S_{\mathrm{E}}^{\left( N\right)  }$-Expansi\'{o}n}}

\sectionmark{Expansi\'{o}n en formas de MC y $S_{\mathrm{E}}^{\left(
N\right)  }$-Expansi\'{o}n}

\subsection{Expansi\'{o}n en formas de MC: Concepto general.}

La expansi\'{o}n en formas de Maurer--Cartan es un poderoso procedimiento, que
destaca en contraste con procedimientos como la contracci\'{o}n,
deformaci\'{o}n y extensi\'{o}n de \'{a}lgebras por aumentar el n\'{u}mero de
generadores del \'{a}lgebra de Lie. Muy brevemente, el procedimiento consiste
en describir el \'{a}lgebra de Lie $\mathfrak{g}$ a trav\'{e}s de las formas
de Maurer--Cartan $a_{+}^{A}$ definidas sobre la variedad del grupo $G$
[ec.~(\ref{EcStructuraMC})]. Sobre la variedad del grupo, se consideran las
coordenadas $\left\{  g^{A}\right\}  _{A=1}^{\dim\mathfrak{g}}.$ De entre
ellas, se debe de escojer un subconjunto $\left\{  g^{i}\right\}  $ y
reescalar \'{e}stas a trav\'{e}s de un factor arbitrario $\lambda,$%
\[
g^{i}\rightarrow g^{i\prime}=\lambda g^{i}%
\]

Ahora, se debe considerar la forma de Maurer--Cartan $\boldsymbol{a}%
_{+}=g^{-1}\mathrm{d}_{\text{{\tiny G}}}g$ con $g=\exp\left(  g^{A}%
\boldsymbol{T}_{A}\right)  ,$ y expandirla como una serie en el par\'{a}metro
$\lambda.$ Finalmente, esta serie debe de ser truncada de una forma que
asegure el cierre de las ecuaciones de estructura, ec.~(\ref{EcStructuraMC}).
El procedimiento esbozado es en general altamente no trivial, especialmente el
proceso de truncamiento, el cual toma distintas formas para distintas
estructuras del \'{a}lgebra de Lie original $\mathfrak{g}$. El tema ha sido
analizado en detalle por de~Azc\'{a}rraga and Izquierdo en
Ref.~\cite{Azcarraga-Libro} y por de~Azc\'{a}rraga, Izquierdo, Pic\'{o}n and
Varela en Ref.~\cite{Azcarraga-Expansion1}.

El caso m\'{a}s sencillo dentro de este procedimiento es cuando el \'{a}lgebra
de Lie $\mathfrak{g}$ no tiene ninguna estructura en particular. En este caso,
el Teorema~1 de Ref.~\cite{Azcarraga-Expansion1} muestra que las constantes de
estructura del \'{a}lgebra expandida corresponden a%
\begin{equation}
C_{\left(  A,i\right)  \left(  B,j\right)  }%
^{\phantom{\left( A, i \right) \left( B, j \right)}\left(  C,k\right)
}=\left\{
\begin{array}
[c]{rc}%
0\text{, } & \text{con }i+j\neq k\\
C_{AB}^{\text{\quad\ }C}\text{, } & \text{con }i+j=k
\end{array}
\right.  ,\label{Ec Ctes Azc}%
\end{equation}
donde los par\'{a}metros $i,j,k=0,\ldots,N$ corresponden a \'{o}rdenes en la
expansi\'{o}n en serie, y $N$ corresponde al orden de truncamiento.

Estas constantes de estructura pueden ser reobtenidas en el contexto de
$S$-expansiones para una elecci\'{o}n particular de semigrupo, al que
bautizaremos como $S_{\mathrm{E}}^{\left(  N\right)  },$ y aplicando $0_{S}$-reducci\'{o}n.

\subsection{$S_{\mathrm{E}}^{\left(  N\right)  }$-Expansi\'{o}n: Caso
general.}

\begin{definition}
Definamos $S_{\mathrm{E}}^{\left(  N\right)  }$ como el semigrupo de elementos%
\begin{equation}
S_{\mathrm{E}}^{\left(  N\right)  }=\left\{  \lambda_{\alpha},\alpha
=0,\ldots,N+1\right\}  ,\label{Ec SEN}%
\end{equation}
provisto con la regla de multiplicaci\'{o}n%
\begin{equation}
\lambda_{\alpha}\lambda_{\beta}=\lambda_{H_{N+1}\left(  \alpha+\beta\right)
},\label{Ec Multp SEN}%
\end{equation}
en donde $H_{N+1}$ corresponde a la funci\'{o}n%
\[
H_{n}\left(  x\right)  =\left\{
\begin{array}
[c]{l}%
x\text{, cuando }x<n,\\
n\text{, cuando }x\geq n.
\end{array}
\right.
\]

\begin{figure}[ptb]
\begin{center}
\includegraphics[width=.5\columnwidth]{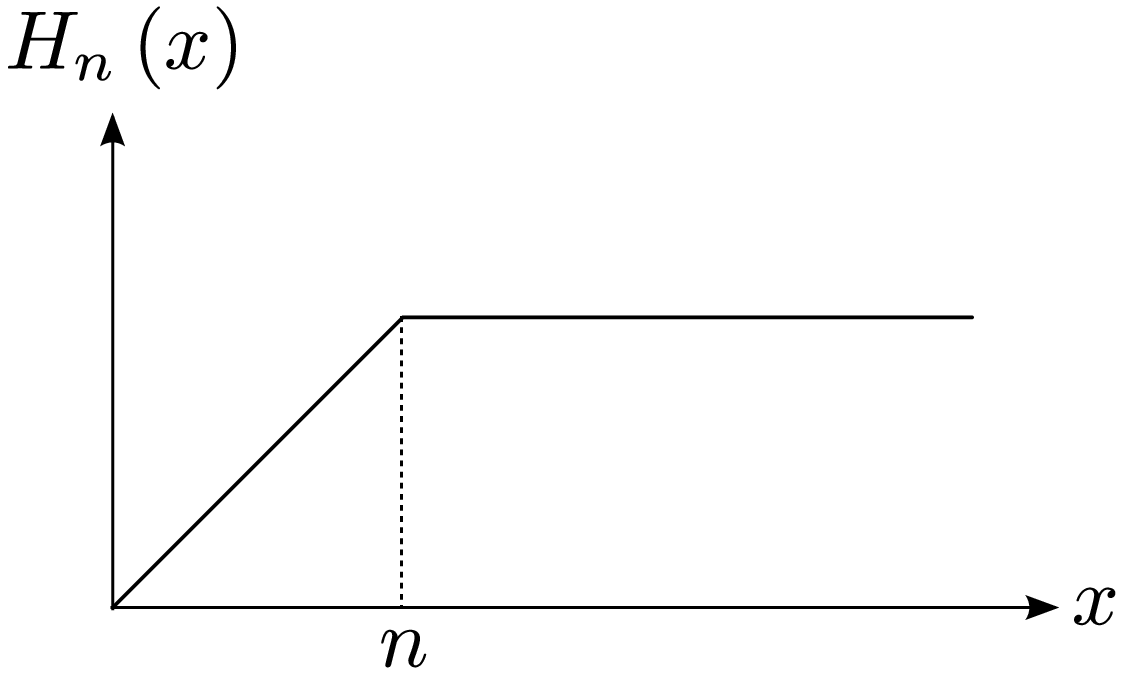}\caption{La funci\'{o}n $H_{n}(x)$}
\label{FigFuncionHn}
\end{center}
\end{figure}

Los $2$-selectores de $S_{\mathrm{E}}^{\left(  N\right)  }$ corresponden a%
\[
K_{\alpha\beta}^{\phantom{\alpha \beta}\gamma}=\delta_{H_{N+1}\left(
\alpha+\beta\right)  }^{\gamma},
\]
donde $\delta_{\sigma}^{\rho}$ es la delta de Kronecker. De
ec.~(\ref{Ec Multp SEN}), tenemos que $\lambda_{N+1}$ corresponde al elemento
cero de $S_{\mathrm{E}}^{\left(  N\right)  },$ $\lambda_{N+1}=0_{S}$.
\end{definition}

A partir de ec.~(\ref{Ec C=KC}), tenemos que las constantes de estructura para
el \'{a}lgebra $S_{\mathrm{E}}^{\left(  N\right)  }$-expandida corresponden a%
\[
C_{\left(  A,\alpha\right)  \left(  B,\beta\right)  }%
^{\phantom{\left( A, \alpha \right) \left( B, \beta \right)}\left(
C,\gamma\right)  }=\delta_{H_{N+1}\left(  \alpha+\beta\right)  }^{\gamma
}C_{AB}^{\phantom{AB}C},
\]
en donde $\alpha,\beta,\gamma=0,\ldots,N+1$. Cuando se impone la condici\'{o}n
de $0_{S}$-reducci\'{o}n, $\lambda_{N+1}\bm{T}_{A}=\bm{0},$ las constantes de
estructura toman la forma%
\[
C_{\left(  A,i\right)  \left(  B,j\right)  }%
^{\phantom{\left( A, i \right) \left( B, j \right)}\left(  C,k\right)
}=\delta_{i+j}^{k}C_{AB}^{\phantom{AB}C},
\]
lo que reproduce exactamente las constantes de estructura
ec.~(\ref{Ec Ctes Azc}).

As\'{\i}, la expansi\'{o}n general en formas de Maurer--Cartan para un
\'{a}lgebra $\mathfrak{g},$ con orden de truncamiento $N,$ coincide
precisamente con el $0_{S}$-reducci\'{o}n del \'{a}lgebra $S_{\mathrm{E}%
}^{\left(  N\right)  } $-expandida $S_{\mathrm{E}}^{(N)}\otimes\mathfrak{g}$.

Esta coincidencia se debe a que la ley de multiplicaci\'{o}n del semigrupo
$S_{\mathrm{E}}^{\left(  N\right)  }$ reproduce el \'{a}lgebra de las
potencias de una serie con orden de truncamiento $N.$ En efecto, en una serie
en el par\'{a}metro $\lambda$ se tiene que%
\[
\lambda^{\alpha}\lambda^{\beta}=\lambda^{\alpha+\beta},
\]
mientras que el truncamiento de la serie se impone como la condici\'{o}n%
\[
\lambda^{\alpha}=0\text{ cuando }\alpha>N.
\]

Puesto que una expansi\'{o}n en serie de la forma de Maurer--Cartan,%
\[
\boldsymbol{a}_{+}=\sum_{\alpha}\lambda^{\alpha}a_{+}^{\left(  A,\alpha
\right)  }\boldsymbol{T}_{A}%
\]
puede ser tambi\'{e}n re-escrita como%
\[
\boldsymbol{a}_{+}=\sum_{\alpha}a_{+}^{\left(  A,\alpha\right)  }%
\boldsymbol{T}_{\left(  A,\alpha\right)  }%
\]
con $\boldsymbol{T}_{\left(  A,\alpha\right)  }=\lambda^{\alpha}%
\boldsymbol{T}_{A},$ la coincidencia en las constantes de estructura
entregadas por ambos procedimientos parece natural.

Sin embargo, se puede ir mucho m\'{a}s lejos en esta analog\'{\i}a utilizando
los teoremas de sub\'{a}lgebra resonante y reducci\'{o}n resonante. En efecto, a
continuaci\'{o}n, algunos de los resultados presentados en el contexto de
expansiones de MC en Ref.~\cite{Azcarraga-Expansion1} para distintas
estructuras de $\mathfrak{g}$ son reobtenidos como sub\'{a}lgebras resonantes
de $S_{\mathrm{E}}^{\left(  N\right)  }\otimes\mathfrak{g}$ con $0_{S}%
$-reducci\'{o}n o reducci\'{o}n resonante (como es por ejemplo el caso de la
contracci\'{o}n de \.{I}n\"{o}n\"{u}--Wigner generalizada).

\subsection{Caso cuando $\mathfrak{g}=V_{0}\oplus V_{1}$, siendo $V_{0}$ una
Sub\'{a}lgebra y $V_{1}$ un Coseto Sim\'{e}trico.}

Sea $\mathfrak{g}=V_{0}\oplus V_{1}$ una descomposici\'{o}n en subespacios de
$\mathfrak{g},$ tal que se cumple%
\begin{align}
\left[  V_{0},V_{0}\right]   &  \subset V_{0},\label{Ec AlgCoseto 00}\\
\left[  V_{0},V_{1}\right]   &  \subset V_{1},\label{Ec AlgCoseto 01}\\
\left[  V_{1},V_{1}\right]   &  \subset V_{0}.\label{Ec AlgCoseto 11}%
\end{align}

Consideremos $S_{\mathrm{E}}^{\left(  N\right)  }\otimes\mathfrak{g}.$ Una
descomposici\'{o}n resonante de $S_{\mathrm{E}}^{\left(  N\right)  }=S_{0}\cup
S_{1}$, viene dada por\footnote{Aqu\'{\i} $\left[  x\right]  $ denota la parte
entera de $x$.}%
\begin{align}
S_{0}  &  =\left\{  \lambda_{2m},\text{ with }m=0,\ldots,\left[  \frac{N}%
{2}\right]  \right\}  \cup\left\{  \lambda_{N+1}\right\}  ,\\
S_{1}  &  =\left\{  \lambda_{2m+1},\text{ with }m=0,\ldots,\left[  \frac
{N-1}{2}\right]  \right\}  \cup\left\{  \lambda_{N+1}\right\}  .
\end{align}

En efecto, esta descomposici\'{o}n satisface la condici\'{o}n
ec.~(\ref{Ec [Sp , Sq] = Sr}), la cual para este caso corresponde
expl\'{\i}citamente a%
\begin{align}
S_{0}\times S_{0}  &  \subset S_{0},\label{Ec AlgCost S00}\\
S_{0}\times S_{1}  &  \subset S_{1},\label{Ec AlgCost S01}\\
S_{1}\times S_{1}  &  \subset S_{0}.\label{Ec AlgCost S11}%
\end{align}

[C\'{o}mparese ecs.~(\ref{Ec AlgCoseto 00})-(\ref{Ec AlgCoseto 11})
y~(\ref{Ec AlgCost S00})-(\ref{Ec AlgCost S11}) con
ecs.~(\ref{Ec [Vp , Vq] = Vr}) y~(\ref{Ec [Sp , Sq] = Sr})] As\'{\i}, de
acuerdo al teorema~\ref{Teo Subalg Resonante}, tenemos que%
\begin{equation}
\mathfrak{G}_{\text{$\mathrm{R}$}}=W_{0}\oplus W_{1},
\end{equation}
con
\begin{align}
W_{0}  &  =S_{0}\otimes V_{0},\\
W_{1}  &  =S_{1}\otimes V_{1},
\end{align}
es una sub\'{a}lgebra resonante de $\mathfrak{G}.$

Para escribir las constantes de estructura, basta con recurrir a
ec.~(\ref{Ec CtesStruct Resonantes}), la cual para este caso espec\'{\i}fico
toma la forma
\[
C_{\left(  a_{p},\alpha_{p}\right)  \left(  b_{q},\beta_{q}\right)
}%
^{\phantom{\left( a_{p}, \alpha_{p} \right) \left( b_{q}, \beta_{q} \right)}\left(
c_{r},\gamma_{r}\right)  }=\delta_{H_{N+1}\left(  \alpha_{p}+\beta_{q}\right)
}^{\gamma_{r}}C_{a_{p}b_{q}}^{\phantom{a_{p} b_{q}}c_{r}}\text{ con }\left\{
\begin{array}
[c]{l}%
p,q=0,1\\
\alpha_{p},\beta_{p},\gamma_{p}=2m+p,\\
m=0,\ldots,\left[  \frac{N-p}{2}\right]  ,\frac{N+1-p}{2}.
\end{array}
\right.
\]

Las constantes de estructura para el \'{a}lgebra $0_{S}$-reducida de
$\mathfrak{G}_{\text{$\mathrm{R}$}}$ [ve\'{a}se
corolario~\ref{Cor O-Forz SubAlg Reson} y ec.(\ref{Ec Ctes Struc Forz})]
corresponden a%
\[
C_{\left(  a_{p},\alpha_{p}\right)  \left(  b_{q},\beta_{q}\right)
}%
^{\phantom{\left( a_{p}, \alpha_{p} \right) \left( b_{q}, \beta_{q} \right)}\left(
c_{r},\gamma_{r}\right)  }=\delta_{H_{N+1}\left(  \alpha_{p}+\beta_{q}\right)
}^{\gamma_{r}}C_{a_{p}b_{q}}^{\phantom{a_{p} b_{q}}c_{r}}\text{ con }\left\{
\begin{array}
[c]{l}%
p,q=0,1\\
\alpha_{p},\beta_{p},\gamma_{p}=2m+p,\\
m=0,\ldots,\left[  \frac{N-p}{2}\right]  .
\end{array}
\right.
\]

En el contexto de expansiones de MC, estas constantes de estructura
corresponden en la notaci\'{o}n de Ref.~\cite{Azcarraga-Expansion1} a las del
\'{a}lgebra $\mathcal{G}\left(  N_{0},N_{1}\right)  $ (ec.~(3.31) en
Ref.~\cite{Azcarraga-Expansion1}) para el caso de coseto sim\'{e}trico, con%
\begin{align*}
N_{0}  &  =2\left[  \frac{N}{2}\right]  ,\\
N_{1}  &  =2\left[  \frac{N-1}{2}\right]  +1.
\end{align*}

\begin{figure}[ptb]
\includegraphics[width=\columnwidth]{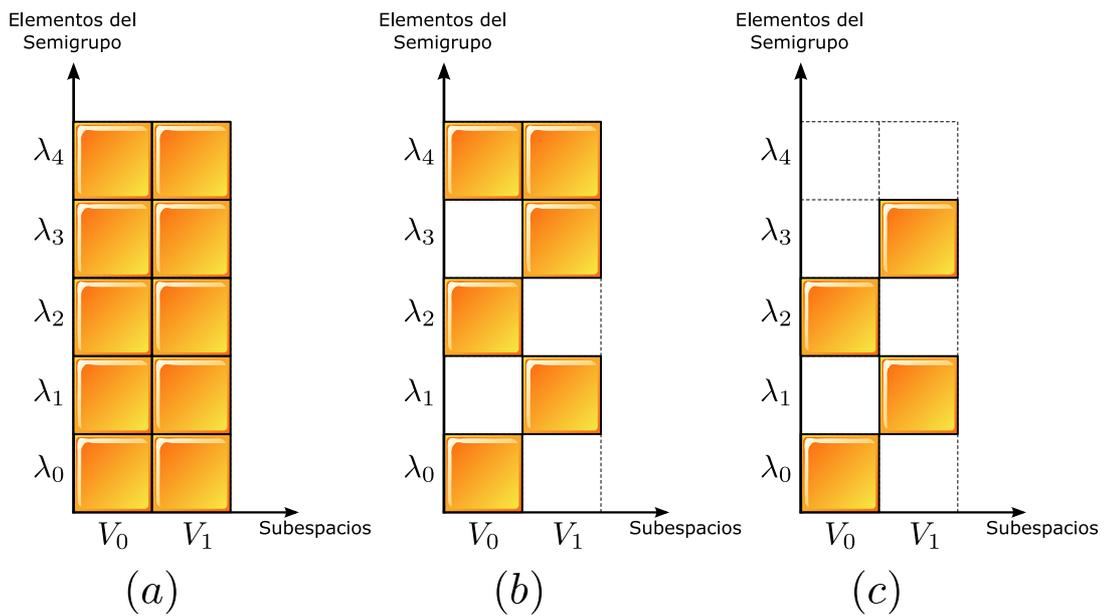}\caption{$S_{\mathrm{E}%
}^{(3)}$-expansi\'{o}n del \'{a}lgebra $\mathfrak{g}=V_{0}\oplus V_{1}$,
siendo $V_{0}$ una sub\'{a}lgebra y $V_{1}$ un coseto sim\'{e}trico.
(\textit{a}) La regi\'{o}n coloreada corresponde al \'{a}lgebra $S_{\mathrm{E}}^{(3)}$-expandida completa, $\mathfrak{G}=S_{\mathrm{E}}^{(3)}\otimes \mathfrak{g}$.
(\textit{b}) La regi\'{o}n destacada corresponde a la sub\'{a}lgebra resonante $\mathfrak{G}_{\mathrm{R}}$.
(\textit{c}) La regi\'{o}n naranja muestra ahora el $0_{S}$-reducci\'{o}n de la sub\'{a}lgebra resonante $\mathfrak{G}_{\mathrm{R}}$.}
\label{FigSymCosetRes}%
\end{figure}

Para tener una idea m\'{a}s intuitiva del procedimiento de $S$-Expansi\'{o}n,
sub\'{a}lgebra resonante y $0_{S}$-reducci\'{o}n, resulta de mucha ayuda un
diagrama como el de la Fig.~\ref{FigSymCosetRes}. Este diagrama corresponde
precisamente al caso que acabamos de analizar, para la elecci\'{o}n
$S=S_{\mathrm{E}}^{\left(  3\right)  }$.

Los subespacios de $\mathfrak{g}$ est\'{a}n representados en ele eje
horizontal, mientras que los elementos de $S_{\mathrm{E}}^{\left(  3\right)
}$ lo est\'{a}n en el eje vertical. As\'{\i}, el \'{a}lgebra $S_{\mathrm{E}%
}^{\left(  3\right)  }$-expandida completa, $S_{\mathrm{E}}^{\left(  3\right)
}\otimes\mathfrak{g}$ corresponde a la regi\'{o}n sombreada
Fig.~\ref{FigSymCosetRes}~(\textit{a}). En Fig.~\ref{FigSymCosetRes}%
~(\textit{b}), la regi\'{o}n gris representa la sub\'{a}lgebra resonante
$\mathfrak{G}_{\mathrm{R}}=W_{0}\oplus W_{1}$ con
\begin{align}
S_{0}  &  =\left\{  \lambda_{0},\lambda_{2},\lambda_{4}\right\}
,\label{partX1}\\
S_{1}  &  =\left\{  \lambda_{1},\lambda_{3},\lambda_{4}\right\}
.\label{partX2}%
\end{align}

Cada columna en el diagrama corresponde a cada uno de los subespacios $W_{p}$.
Fig.~\ref{FigSymCosetRes}~(\textit{c}) representa el \'{a}lgebra $0_{S}%
$-reducida, obtenida imponiendo $\hat{\mathfrak{G}}_{\text{$\mathrm{R}$}%
}=\left\{  \lambda_{4}\right\}  \otimes\mathfrak{g}=0$ . As\'{\i},
Fig.~\ref{FigSymCosetRes}~(\textit{c}) corresponde al \'{a}lgebra
$\mathcal{G}\left(  N_{0},N_{1}\right)  $.

Es evidente que el caso $N=1$, $\hat{\mathfrak{G}}_{\text{$\mathrm{R}$}%
}=\left\{  \lambda_{2}\right\}  \otimes\mathfrak{g}=0\ $reproduce la
contracci\'{o}n de \.{I}n\"{o}n\"{u}--Wigner para $\mathfrak{g}=V_{0}\oplus
V_{1}$. M\'{a}s sobre contracciones de \.{I}n\"{o}n\"{u}--Wigner y
contracciones generalizadas ser\'{a} considerado en
sec.~\ref{Sec Weimar-Woods}.

\subsection{\label{Sec Weimar-Woods}Caso cuando $\mathfrak{g}$ satisface las
condiciones de Weimar-Woods.}

Sea $\mathfrak{g}=\bigoplus_{p=0}^{n}V_{p}$ una descomposici\'{o}n en
subespacios de $\mathfrak{g}$. En t\'{e}rminos de ella, las condiciones de
Weimar-Woods (Refs.~\cite{WeimarWoods-Contract1, WeimarWoods-Contract1})
corresponden a
\begin{equation}
\left[  V_{p},V_{q}\right]  \subset\bigoplus_{r=0}^{H_{n}\left(  p+q\right)
}V_{r}.\label{Ec Weimar--Woods Condition}%
\end{equation}

La descomposici\'{o}n en subconjuntos
\begin{equation}
S_{\mathrm{E}}^{\left(  N\right)  }=\bigcup_{p=0}^{n}S_{p}%
\end{equation}
con%
\begin{equation}
S_{p}=\left\{  \lambda_{\alpha_{p}}\text{, tal que }\alpha_{p}=p,\ldots
,N+1\right\}  \qquad N+1\geq n
\end{equation}
est\'{a} en resonancia con ec.~(\ref{Ec Weimar--Woods Condition}), pues
satisface%
\begin{equation}
S_{p}\times S_{q}=S_{H_{n}\left(  p+q\right)  }\subset\bigcap_{r=0}%
^{H_{n}\left(  p+q\right)  }S_{r}.\label{Ec Sp x Sq = SHn(p+q)}%
\end{equation}
[comp\'{a}rese ecs.~(\ref{Ec Weimar--Woods Condition})
y~(\ref{Ec Sp x Sq = SHn(p+q)}) con ecs.~(\ref{Ec [Vp , Vq] = Vr})
y~(\ref{Ec [Sp , Sq] = Sr})]

As\'{\i}, del teorema~\ref{Teo Subalg Resonante}, tenemos que
\begin{equation}
\mathfrak{G}_{\mathrm{R}}=\bigoplus_{p=0}^{n}W_{p},
\end{equation}
con
\[
W_{p}=S_{p}\otimes V_{p},
\]
es una sub\'{a}lgebra resonante de $\mathfrak{G}$.

Usando ec.~(\ref{Ec CtesStruct Resonantes}), se tienen las siguientes
constantes de estructura para la sub\'{a}lgebra resonante%
\[
C_{\left(  a_{p},\alpha_{p}\right)  \left(  b_{q},\beta_{q}\right)
}%
^{\phantom{\left( a_{p}, \alpha_{p} \right) \left( b_{q}, \beta_{q} \right)}\left(
c_{r},\gamma_{r}\right)  }=\delta_{H_{N+1}\left(  \alpha_{p}+\beta_{q}\right)
}^{\gamma_{r}}C_{a_{p}b_{q}}^{\phantom{a_{p} b_{q}}c_{r}}\text{ con }\left\{
\begin{array}
[c]{l}%
p,q,r=0,\ldots,n\\
\alpha_{p},\beta_{p},\gamma_{p}=p,\ldots,N+1
\end{array}
\right.  .
\]
Al imponer $0_{S}$-reducci\'{o}n, se tiene
\begin{equation}
C_{\left(  a_{p},\alpha_{p}\right)  \left(  b_{q},\beta_{q}\right)
}%
^{\phantom{\left( a_{p}, \alpha_{p} \right) \left( b_{q}, \beta_{q} \right)}\left(
c_{r},\gamma_{r}\right)  }=\delta_{H_{N+1}\left(  \alpha_{p}+\beta_{q}\right)
}^{\gamma_{r}}C_{a_{p}b_{q}}^{\phantom{a_{p} b_{q}}c_{r}}\text{ con }\left\{
\begin{array}
[c]{l}%
p,q,r=0,\ldots,n\\
\alpha_{p},\beta_{p},\gamma_{p}=p,\ldots,N
\end{array}
\right.  .\label{xxxStrucConstZampForz}%
\end{equation}

Esta \'{a}lgebra $0_{S}$-reducida corresponde al caso $\mathcal{G}\left(
N_{0},\ldots,N_{n}\right)  $ del Teorema 3 de Ref.~\cite{Azcarraga-Expansion1}
con $N_{p}=N$ para todo $p=0,\ldots,n$. Las constantes de estructura
ec.~(\ref{xxxStrucConstZampForz}) corresponden a las de ec.~(4.8) en
Ref.~\cite{Azcarraga-Expansion1}. El caso m\'{a}s tambi\'{e}n puede ser
obtenido en el contexto de $S$-expansiones, pero a trav\'{e}s de reducci\'{o}n
resonante (Teorema~\ref{Teo Forz Resonante}). En efecto, particionemos cada
subconjunto $S_{p}$ de la forma
\begin{align}
\check{S}_{p}  &  =\left\{  \lambda_{\alpha_{p}}\text{, tal que }\alpha
_{p}=p,\ldots,N_{p}\right\}  ,\label{spdown}\\
\hat{S}_{p}  &  =\left\{  \lambda_{\alpha_{p}}\text{, tal que }\alpha
_{p}=N_{p}+1,\ldots,N+1\right\}  .\label{spup}%
\end{align}

Esta partici\'{o}n satisface por construcci\'{o}n la condici\'{o}n
ec.~(\ref{Ec Cond Forz SpnSq=0}); en el
ap\'{e}ndice~\ref{Apend ForzRes Zamponha} se demuestra que la condici\'{o}n
ec.~(\ref{Ec Cond Forz SpSq=nSr}) sobre ecs.~(\ref{spdown})--(\ref{spup}) es
equivalente a requerir sobre las constantes $N_{p}$'s:%
\begin{equation}
N_{p+1}=\left\{
\begin{array}
[c]{l}%
N_{p}\text{ \'{o}}\\
H_{N+1}\left(  N_{p}+1\right)
\end{array}
\right.  .\label{Ec ForzResZamp Np's}%
\end{equation}

Esta condici\'{o}n corresponde a la de Teorema~3 de
Ref.~\cite{Azcarraga-Expansion1}.

\begin{figure}[ptb]
\includegraphics[width=1.2\columnwidth]{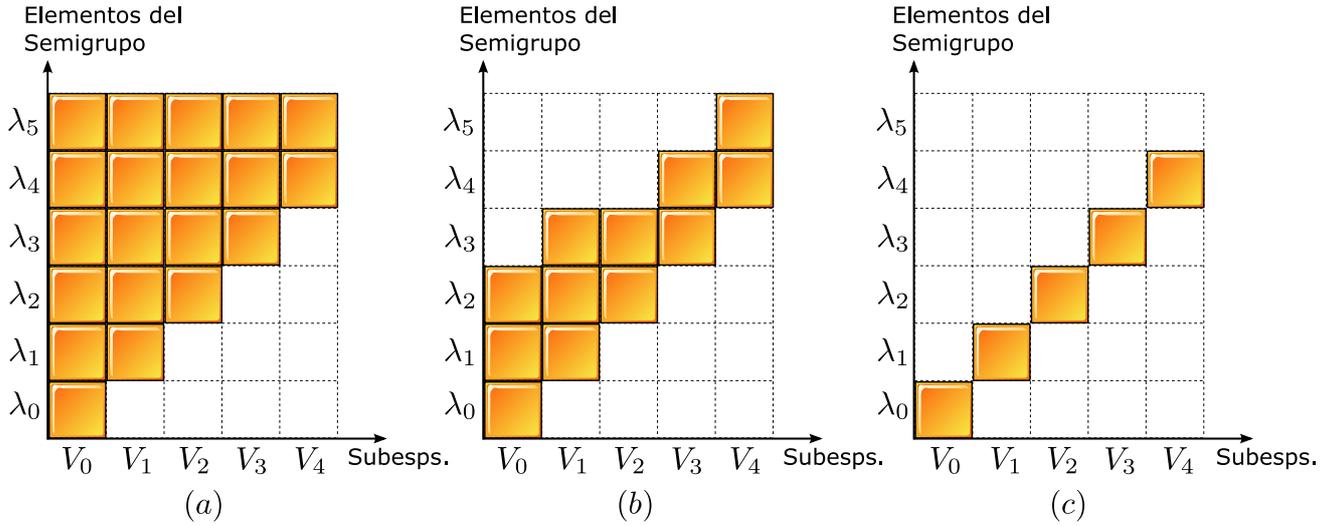}
\caption{(\textit{a}) $S_{\mathrm{E}}^{(4)}$-sub\'{a}lgebra resonante cuando $\mathfrak{g}=V_{0}\oplus V_{1}\oplus V_{2}\oplus V_{3}\oplus V_{4}$ satisface las condiciones de Weimar-Woods y
(\textit{b}) Un posible reducci\'{o}n de la sub\'{a}lgebra resonante, con $N_{0}=2$, $N_{1}=3$, $N_{2}=3$, $N_{3}=4$, $N_{4}=5$.
(\textit{c}) Contracci\'{o}n de In\"{o}n\"{u}--Wigner generalizada, correspondiente a un reducci\'{o}n resonante con $N_{p}=p$, $p=0,1,2,3,4$.}
\label{FigZamponhaForzIW}%
\end{figure}

En el contexto de $S$-expansiones el caso $N_{p+1}=N_{p}=N+1$ para todo $p$
corresponde a una sub\'{a}lgebra resonante, el caso $N_{p+1}=N_{p}=N$ al
$0_{S}$-reducci\'{o}n, y los restantes a reducci\'{o}n resonante. La
contracci\'{o}n de \.{I}n\"{o}n\"{u}--Wigner generalizada corresponde al caso
$N_{p}=p,$ y por lo tanto, corresponde a un reducci\'{o}n resonante (Compare con
Ref.~\cite{Azcarraga-Expansion1}). Fig.~\ref{FigZamponhaForzIW} muestra la
sub\'{a}lgebra resonante, un reducci\'{o}n resonante y la contracci\'{o}n de
\.{I}n\"{o}n\"{u}--Wigner con la elecci\'{o}n de semigrupo $S_{\mathrm{E}%
}^{\left(  4\right)  }.$

Hemos hecho fuerte incapi\'{e} en si un \'{a}lgebra corresponde a una
sub\'{a}lgebra resonante, un $0_{S}$-reducci\'{o}n o un reducci\'{o}n resonante,
pues esto ser\'{a} de vital importancia a la hora de escribir tensores
invariantes para el \'{a}lgebra correspondiente (V\'{e}ase
Sec.~\ref{Sec TensInv S-Exp})

\subsection{Caso cuando $\mathfrak{g}=V_{0}\oplus V_{1}\oplus V_{2}$ es una
super\'{a}lgebra}

Una super\'{a}lgebra $\mathfrak{g}$ est\'{a} dividida naturalmente en tres
subespacios, $\mathfrak{g}=V_{0}\oplus V_{1}\oplus V_{2}$ en donde
$V_{0}\oplus V_{2}$ corresponde al sector bos\'{o}nico y $V_{1}$ al
fermi\'{o}nico, siendo $V_{0}$ una sub\'{a}lgebra por s\'{\i} mismo.
Espec\'{\i}ficamente, l estructura de una super\'{a}lgebra puede ser escrita
como%
\begin{align}
\left[  V_{0},V_{0}\right]   &  \subset V_{0},\label{SuperV0V0=V0}\\
\left[  V_{0},V_{1}\right]   &  \subset V_{1},\label{SuperV0V1=V1}\\
\left[  V_{0},V_{2}\right]   &  \subset V_{2},\label{SuperV0V2=V2}\\
\left[  V_{1},V_{1}\right]   &  \subset V_{0}\oplus V_{2}%
,\label{SuperV1V1=V0+V2}\\
\left[  V_{1},V_{2}\right]   &  \subset V_{1},\label{SuperV1V2=V1}\\
\left[  V_{2},V_{2}\right]   &  \subset V_{0}\oplus V_{2}%
.\label{SuperV2V2=V0+V2}%
\end{align}

La descomposici\'{o}n en subconjuntos $S_{\mathrm{E}}^{(N)}=S_{0}\cup
S_{1}\cup S_{2}$ con%
\begin{equation}
S_{p}=\left\{  \lambda_{2m+p},\text{ con }m=0,\ldots,\left[  \frac{N-p}%
{2}\right]  \right\}  \cup\left\{  \lambda_{N+1}\right\}  ,\qquad
p=0,1,2.\label{ResSuperPartition Sp}%
\end{equation}

es una descomposici\'{o}n resonante, pues satisface
\begin{align}
S_{0}\times S_{0}  &  \subset S_{0},\label{SuperS0S0=S0}\\
S_{0}\times S_{1}  &  \subset S_{1},\label{SuperS0S1=S1}\\
S_{0}\times S_{2}  &  \subset S_{2},\label{SuperS0S2=S2}\\
S_{1}\times S_{1}  &  \subset S_{0}\cap S_{2},\label{SuperS1S1=S0nS2}\\
S_{1}\times S_{2}  &  \subset S_{1},\label{SuperS1S2=S1}\\
S_{2}\times S_{2}  &  \subset S_{0}\cap S_{2}.\label{SuperS2S2=S0nS2}%
\end{align}

[compare ecs.~(\ref{SuperS0S0=S0})--(\ref{SuperS2S2=S0nS2}) con
ecs.~(\ref{SuperV0V0=V0})--(\ref{SuperV2V2=V0+V2})]

As\'{\i}, de acuerdo al Teorema~\ref{Teo Subalg Resonante} tenemos que
$\mathfrak{G}_{\mathrm{R}}=W_{0}\oplus W_{1}\oplus W_{2}$, con $W_{p}%
=S_{p}\otimes V_{p}$, $p=0,1,2$, es una sub\'{a}lgebra resonante.

Las constantes de estructura ec.~(\ref{Ec CtesStruct Resonantes}),
corresponden en este caso a
\begin{equation}
C_{\left(  a_{p},\alpha_{p}\right)  \left(  b_{q},\beta_{q}\right)
}%
^{\phantom{\left( a_{p}, \alpha_{p} \right) \left( b_{q}, \beta_{q} \right)}\left(
c_{r},\gamma_{r}\right)  }=\delta_{H_{N+1}\left(  \alpha_{p}+\beta_{q}\right)
}^{\gamma_{r}}C_{a_{p}b_{q}}^{\phantom{a_{p} b_{q}}c_{r}}\text{ con }\left\{
\begin{array}
[c]{l}%
p,q,r=0,1,2,\\
\alpha_{p},\beta_{p},\gamma_{p}=2m+p,\\
m=0,\ldots,\left[  \frac{N-p}{2}\right]  ,\frac{N+1-p}{2}.
\end{array}
\right.
\end{equation}

Una vez aplicado el $0_{S}$-reducci\'{o}n, $\hat{\mathfrak{G}}%
_{\text{$\mathrm{R}$}}=\left\{  \lambda_{N+1}\right\}  \otimes\mathfrak{g}=0$,
las constantes de estructura corresponden a%
\begin{equation}
C_{\left(  a_{p},\alpha_{p}\right)  \left(  b_{q},\beta_{q}\right)
}%
^{\phantom{\left( a_{p}, \alpha_{p} \right) \left( b_{q}, \beta_{q} \right)}\left(
c_{r},\gamma_{r}\right)  }=\delta_{H_{N+1}\left(  \alpha_{p}+\beta_{q}\right)
}^{\gamma_{r}}C_{a_{p}b_{q}}^{\phantom{a_{p} b_{q}}c_{r}}\text{ con }\left\{
\begin{array}
[c]{l}%
p,q,r=0,1,2,\\
\alpha_{p},\beta_{p},\gamma_{p}=2m+p,\\
m=0,\ldots,\left[  \frac{N-p}{2}\right]  .
\end{array}
\right. \label{SuperResonantStructureConstants}%
\end{equation}

Esta \'{a}lgebra $0_{S}$-reducida corresponde al \'{a}lgebra $\mathcal{G}%
\left(  N_{0},N_{1},N_{2}\right)  $ del Teorema~$5$ de
Ref.~\cite{Azcarraga-Expansion1}, con%
\[
N_{p}=2\left[  \frac{N-p}{2}\right]  +p,\qquad p=0,1,2
\]
Las constantes de estructura ec.~(\ref{SuperResonantStructureConstants})
corresponden a las de ec.~(5.6) de Ref.~\cite{Azcarraga-Expansion1}.

\begin{figure}[ptb]
\includegraphics[width=\columnwidth]{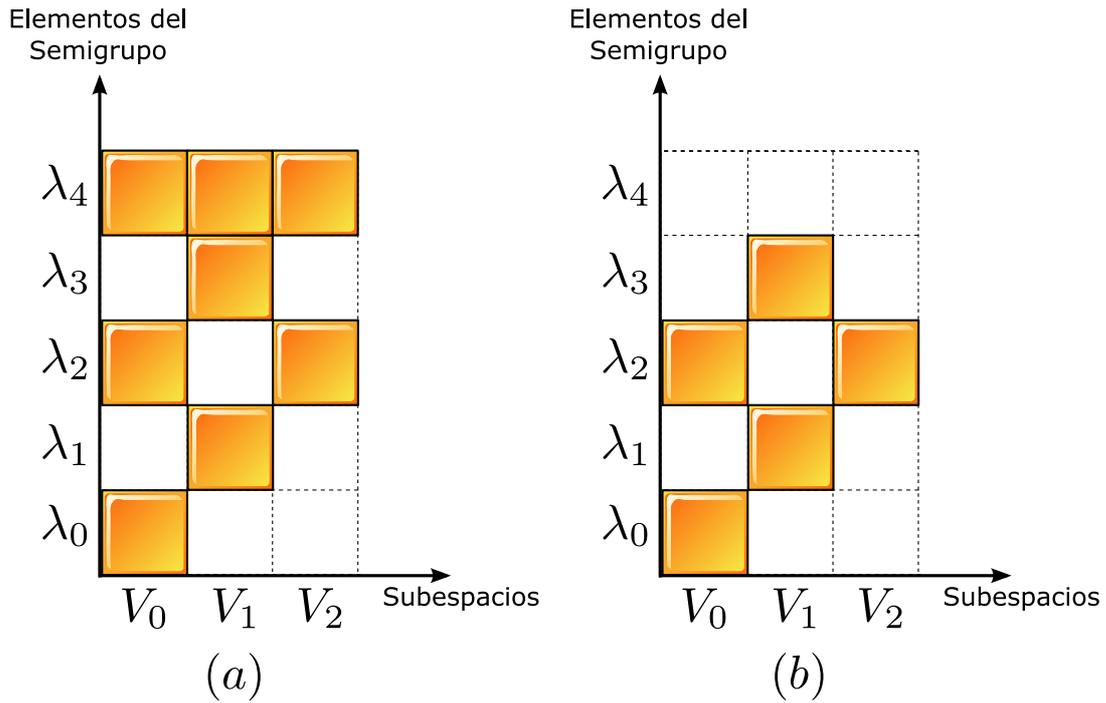}\caption{(\textit{a}%
) El \'{a}rea coloreada corresponde a una sub\'{a}lgebra resonante de
$\mathfrak{G}=S_{\mathrm{E}}^{(3)}\otimes\mathfrak{g}$ cuando $\mathfrak{g}$
es una superalgebra. (\textit{b}) La regi\'{o}n naranja muestra el $0_{S}%
$-reducci\'{o}n de la sub\'{a}lgebra resonante $\mathfrak{G}_{\mathrm{R}}$.
\'{E}ste corresponde a $\mathcal{G}(2,3,2)$ en el contexto de
Ref.~\cite{Azcarraga-Expansion1}.}%
\label{FigSupResSubAlg}%
\end{figure}

La Fig.~\ref{FigSupResSubAlg}~(\textit{a}) muestra la sub\'{a}lgebra resonante
de $S_{\mathrm{E}}^{(4)}\otimes\mathfrak{g}$, y Fig.~\ref{FigSupResSubAlg}%
~(\textit{b}), corresponde a su $0_{S}$-reducci\'{o}n.

\section[$S$-Expansiones de $\mathfrak{osp}\left(  \mathfrak{32}%
|\mathfrak{1}\right)  $ y Super\'{a}lgebras en $D=11$]%
{\label{Sec S-Exp osp(32|1)}$S$-Expansiones de $\mathfrak{osp}\left(
\mathfrak{32}|\mathfrak{1}\right)  $ y Super\'{a}lgebras en $D=11$
\sectionmark{$S$-Expansiones de
$\mathfrak{osp}\left( \mathfrak{32}|\mathfrak{1}\right)  $ y
Super\'{a}lgebras}}

\sectionmark{$S$-Expansiones de $\mathfrak{osp}\left(
\mathfrak{32}|\mathfrak{1}\right)  $ y Super\'{a}lgebras}

Hasta ahora hemos analizado el procedimiento de $S$-Expansi\'{o}n y su
relaci\'{o}n con la Expansi\'{o}n en formas de Maurer--Cartan en forma
completamente abstracta y general. Ahora, procederemos aplicar las
herramientas matem\'{a}ticas constru\'{\i}das para estudiar super\'{a}lgebras
en 11 dimensiones, y en particular, observaremos como es posible recuperar el
\'{A}lgebra~M dentro del esquema de las $S$-expansiones. Como veremos a
continuaci\'{o}n, dentro del esquema de las $S$-expansiones, el \'{A}lgebra~M
corresponde a un caso particular dentro de una familia de \'{a}lgebras, todas
las cuales se originan como distintas $S$-expansiones de $\mathfrak{osp}%
\left(  \mathfrak{32}|\mathfrak{1}\right)  ,$ y que comparten distintas
caracter\'{\i}sticas con el \'{A}lgebra M. Por otra parte, una de las grandes
ventajas del procedimiento de $S$-Expansiones es que \'{e}ste entrega en forma
autom\'{a}tica un tensor invariante para el \'{a}lgebra (Ve\'{a}se
Sec.~\ref{Sec TensInv S-Exp}), ingrediente clave en la construcci\'{o}n de un
lagrangeano de Chern--Simons o de Transgresi\'{o}n.

En las secciones subsiguientes reobtendremos el \'{A}lgebra~M (Ver
Refs.~\cite{Sezgin-MAlgebra,Azcarraga-Expansion2}) como un $0_{S}$-reducci\'{o}n (por lo que
se vuelve interesante considerar la sub\'{a}lgebra resonante a partir de la
cual esta proviene), y obtendremos un \'{a}lgebra similar a las de
D'Auria--Fr\'{e}~\cite{DAuria-Fre} utilizando $S=S_{\mathrm{E}}^{\left(
N\right)  }$ con $N=2$ y $N=3$, respectivamente. Como un ejemplo de una
$S$-expansi\'{o}n $S\neq S_{\mathrm{E}}^{\left(  N\right)  },$ consideraremos
$S=\mathbb{Z}_{4},$ con lo que se obtiene un \'{a}lgebra que es similar en
algunos aspectos al \'{A}lgebra~M, $\mathfrak{osp}\left(  32|1\right)
\otimes\mathfrak{osp}\left(  32|1\right)  $ y las super\'{a}lgebras de D'Auria--Fr\'{e}.

El punto de partida ser\'{a} siempre el \'{a}lgebra ortosimpl\'{e}ctica en
$D=11,$ $\mathfrak{osp}\left(  \mathfrak{32}|\mathfrak{1}\right)  .$
Consideraremos como base para el \'{a}lgebra $\left\{  \bm{P}_{a}%
,\bm{J}_{ab},\bm{Z}_{abcde},\bm{Q}\right\}  ,$ donde $\left\{  \bm{P}_{a}%
,\bm{J}_{ab}\right\}  $ corresponden a los generadores de AdS, $\bm{Z}_{abcde}%
$ corresponde a un tensor de Lorentz completamete antisim\'{e}trico de 5
\'{\i}ndices y $\bm{Q}$ es una carga espinorial de Majorana con 32
componentes. Las relaciones de (anti)conmutaci\'{o}n del \'{a}lgebra
$\mathfrak{osp}\left(  \mathfrak{32}|\mathfrak{1}\right)  $ son
expl\'{\i}citamente%
\begin{align*}
\left[  \boldsymbol{P}_{a},\boldsymbol{P}_{b}\right]   &  =\boldsymbol{J}%
_{ab},\\
\left[  \boldsymbol{J}_{ab},\boldsymbol{P}_{c}\right]   &  =\eta_{ce}%
\delta_{ab}^{de}\boldsymbol{P}_{d},\\
\left[  \boldsymbol{J}_{ab},\boldsymbol{J}_{cd}\right]   &  =\eta_{gh}%
\delta_{ab}^{eg}\delta_{cd}^{hf}\boldsymbol{J}_{ef}.
\end{align*}%
\begin{align}
\left[  \bm{P}_{a},\bm{Z}_{b_{1}\cdots b_{5}}\right]   &  =-\frac{1}%
{5!}\varepsilon_{ab_{1}\cdots b_{5}c_{1}\cdots c_{5}}\bm{Z}^{c_{1}\cdots
c_{5}},\\
\left[  \bm{J}^{ab},\bm{Z}_{c_{1}\cdots c_{5}}\right]   &  =\frac{1}{4!}%
\delta_{dc_{1}\cdots c_{5}}^{abe_{1}\cdots e_{4}}\bm{Z}_{\phantom{d}e_{1}%
\cdots e_{4}}^{d},
\end{align}%
\begin{align}
\left[  \bm{Z}^{a_{1}\cdots a_{5}},\bm{Z}_{b_{1}\cdots b_{5}}\right]   &
=\eta^{\left[  a_{1}\cdots a_{5}\right]  \left[  c_{1}\cdots c_{5}\right]
}\varepsilon_{c_{1}\cdots c_{5}b_{1}\cdots b_{5}e}\bm{P}^{e}+\delta
_{db_{1}\cdots b_{5}}^{a_{1}\cdots a_{5}e}\bm{J}_{\phantom{d}e}^{d}%
+\nonumber\\
&  -\frac{1}{3!3!5!}\varepsilon_{c_{1}\cdots c_{11}}\delta_{d_{1}d_{2}%
d_{3}b_{1}\cdots b_{5}}^{a_{1}\cdots a_{5}c_{4}c_{5}c_{6}}\eta^{\left[
c_{1}c_{2}c_{3}\right]  \left[  d_{1}d_{2}d_{3}\right]  }\bm{Z}^{c_{7}\cdots
c_{11}},
\end{align}%
\begin{align}
\left[  \bm{P}_{a},\bm{Q}\right]   &  =-\frac{1}{2}\Gamma_{a}\bm{Q},\\
\left[  \bm{J}_{ab},\bm{Q}\right]   &  =-\frac{1}{2}\Gamma_{ab}\bm{Q},\\
\left[  \bm{Z}_{abcde},\bm{Q}\right]   &  =-\frac{1}{2}\Gamma_{abcde}\bm{Q},
\end{align}%
\[
\left\{  \bm{Q},\bar{\bm{Q}}\right\}  =\frac{1}{8}\left(  \Gamma^{a}%
\bm{P}_{a}-\frac{1}{2}\Gamma^{ab}\bm{J}_{ab}+\frac{1}{5!}\Gamma^{abcde}%
\bm{Z}_{abcde}\right)
\]

en donde $\Gamma_{a}$ corresponde a las matrices de Dirac en $D=11.$

Antes de proceder con la $S$-Expansi\'{o}n, debemos dividir $\mathfrak{osp}%
\left(  \mathfrak{32}|\mathfrak{1}\right)  $ en subespacios,%
\[
\mathfrak{osp}\left(  32|1\right)  =V_{0}\oplus V_{1}\oplus V_{2}.
\]

En este caso, $V_{0}$ corresponder\'{a} a la sub\'{a}lgebra de Lorentz
(generada por $\bm{J}_{ab}$), $V_{1}$ corresponder\'{a} al subespacio
fermi\'{o}nico (generado por $\bm{Q}$) y $V_{2}$ a los \textquotedblleft%
\textit{boosts}\textquotedblright\ de AdS y a la M5-brana (generado por
$\bm{P}_{a}$ y $\bm{Z}_{abcde}$).

As\'{\i}, esta separaci\'{o}n en subespacios tiene precisamente la forma dada
en ecs.~(\ref{SuperV0V0=V0})--(\ref{SuperV2V2=V0+V2}), lo que se puede
chequear en forma directa.

\subsection{\label{Sec Algebra M}\'{A}lgebra~M}

Dado que el \'{A}lgebra~M puede ser obtenida a partir de $\mathfrak{osp}%
\left(  \mathfrak{32}|\mathfrak{1}\right)  $ a trav\'{e}s de una expansi\'{o}n
en formas de MC (Ve\'{a}se
Refs.~\cite{Azcarraga-Expansion1,Azcarraga-Expansion2}), entonces es posible
reobtenerla tambi\'{e}n dentro del contexto de $S$-expansiones. Para ello, se
debe escoger $S_{\mathrm{E}}^{\left(  2\right)  }=\left\{  \lambda_{0}%
,\lambda_{1},\lambda_{2},\lambda_{3}\right\}  ,$ utilizar la partici\'{o}n
resonante ec.~(\ref{ResSuperPartition Sp}), y aplicar $0_{S}$-reducci\'{o}n,
$\left\{  \lambda_{3}\right\}  \otimes\mathfrak{osp}\left(  \mathfrak{32}%
|\mathfrak{1}\right)  =0$, o lo que es equivalente, reemplazar las constantes
de estructura de $\mathfrak{osp}\left(  \mathfrak{32}|\mathfrak{1}\right)  $
en ec.~(\ref{SuperResonantStructureConstants}).

\begin{figure}[ptb]
\psfrag{Jab}{$\bm{J}_{ab}$}
\psfrag{Zab}{$\bm{Z}_{ab}$}
\psfrag{Q}{$\bm{Q}$}
\psfrag{Z5}{$\bm{Z}_{abcde}$}
\psfrag{Pa}{$\bm{P}_{a}$}
\includegraphics[width=\columnwidth]{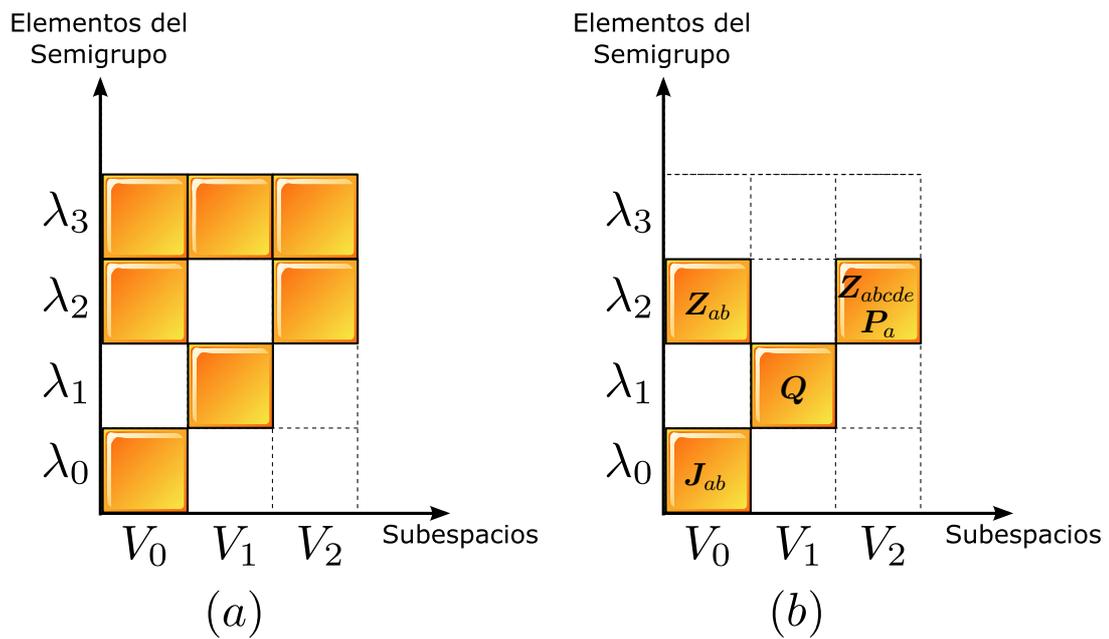}
\caption{El \'{A}lgebra~M en el contexto de $S$-Expansiones.
(\textit{a}) La regi\'{o}n coloreada corresponde a una sub\'{a}lgebra resonante de $\mathfrak{G}=S_{\mathrm{E}}^{\left(  2\right)  }\otimes\mathfrak{osp}\left(  32|1\right)  $.
(\textit{b}) El \'{A}lgebra M (\'{a}rea embaldosada) obtenida como un $0_{S}$-reducci\'{o}n de la sub\'{a}lgebra resonante. }
\label{Fig MAlgebra}
\end{figure}

Por simplicidad, re-etiquetemos los generadores del \'{a}lgebra como
$\bm{J}_{ab}=\bm{J}_{\left(  ab,0\right)  }$, $\bm{Q}_{\alpha}=\bm{Q}_{\left(
\alpha,1\right)  }$, $\bm{Z}_{ab}=\bm{J}_{\left(  ab,2\right)  }$,
$\bm{P}_{a}=\bm{P}_{\left(  a,2\right)  }$, $\bm{Z}_{abcde}=\bm{Z}_{\left(
abcde,2\right)  } $, tal como se muestra en la Fig.~\ref{Fig MAlgebra}. El
\'{a}lgebra resultante es de la forma%
\begin{align*}
\left[  \boldsymbol{P}_{a},\boldsymbol{P}_{b}\right]   &  =\boldsymbol{0},\\
\left[  \boldsymbol{J}_{ab},\boldsymbol{P}_{c}\right]   &  =\eta_{ce}%
\delta_{ab}^{de}\boldsymbol{P}_{d},\\
\left[  \boldsymbol{J}_{ab},\boldsymbol{J}_{cd}\right]   &  =\eta_{gh}%
\delta_{ab}^{eg}\delta_{cd}^{hf}\boldsymbol{J}_{ef}.
\end{align*}%
\begin{align}
\left[  \bm{J}^{ab},\bm{Z}_{cd}\right]   &  =\delta_{ecd}^{abf}%
\bm{Z}_{\phantom{e}f}^{e},\\
\left[  \bm{Z}^{ab},\bm{Z}_{cd}\right]   &  =\bm{0},\\
\left[  \bm{P}_{a},\bm{Z}_{b_{1}\cdots b_{5}}\right]   &  =\bm{0},\\
\left[  \bm{J}^{ab},\bm{Z}_{c_{1}\cdots c_{5}}\right]   &  =\frac{1}{4!}%
\delta_{dc_{1}\cdots c_{5}}^{abe_{1}\cdots e_{4}}\bm{Z}_{\phantom{d}e_{1}%
\cdots e_{4}}^{d},\\
\left[  \bm{Z}^{ab},\bm{Z}_{c_{1}\cdots c_{5}}\right]   &  =\bm{0},\\
\left[  \bm{Z}^{a_{1}\cdots a_{5}},\bm{Z}_{b_{1}\cdots b_{5}}\right]   &
=\bm{0},
\end{align}%
\begin{align}
\left[  \bm{P}_{a},\bm{Q}\right]   &  =\bm{0},\\
\left[  \bm{J}_{ab},\bm{Q}\right]   &  =-\frac{1}{2}\Gamma_{ab}\bm{Q},\\
\left[  \bm{Z}_{ab},\bm{Q}\right]   &  =\bm{0},\\
\left[  \bm{Z}_{abcde},\bm{Q}\right]   &  =\bm{0},
\end{align}%
\[
\left\{  \bm{Q},\bar{\bm{Q}}\right\}  =\frac{1}{8}\left(  \Gamma^{a}%
\bm{P}_{a}-\frac{1}{2}\Gamma^{ab}\bm{Z}_{ab}+\frac{1}{5!}\Gamma^{abcde}%
\bm{Z}_{abcde}\right)
\]

N\'{o}tese que el rol jgado por el $0_{S}$-reducci\'{o}n ha sido el de
abelianizar sectores de la sub\'{a}lgebra resonante.

\subsection{\'{A}lgebra estilo D'Auria--Fr\'{e}}

El ejemplo anterior hizo uso de la elecci\'{o}n de semigrupo $S_{\mathrm{E}%
}^{\left(  2\right)  }=\left\{  \lambda_{0},\lambda_{1},\lambda_{2}%
,\lambda_{3}\right\}  $. Para ver como cambia el resultado final escogiendo un
semigrupo ligeramente distinto, consideremos $S_{\mathrm{E}}^{\left(
3\right)  }=\left\{  \lambda_{0},\lambda_{1},\lambda_{2},\lambda_{3}%
,\lambda_{4}\right\}  ,$ y procedamos de la misma forma como lo hicimos con el
\'{A}lgebra~M (\textit{i.e.}, buscando la sub\'{a}lgebra resonante y luego
$0_{S}$-reduci\'{e}ndola).

\begin{figure}[ptb]
\psfrag{Jab}{$\bm{J}_{ab}$}
\psfrag{Zab}{$\bm{Z}_{ab}$}
\psfrag{Q}{$\bm{Q}$}
\psfrag{Q2}{$\bm{Q^{\prime}}$}
\psfrag{Z5}{$\bm{Z}_{abcde}$}
\psfrag{Pa}{$\bm{P}_{a}$}
\includegraphics[width=\columnwidth]{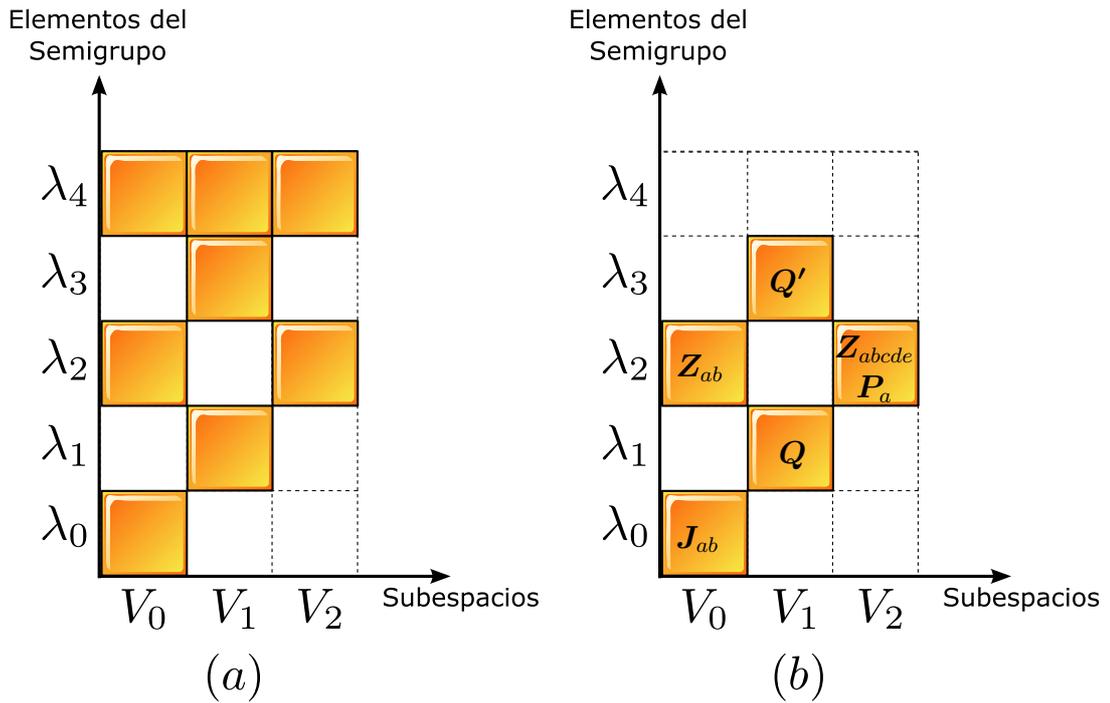}
\caption{Una super\'{a}lgebra similar a las D'Auria--Fr\'{e} en el contexto de $S$-Expansiones.
(\textit{a}) La regi\'{o}n embaldosada corresponde a una sub\'{a}lgebra resonante de $\mathfrak{G}=S_{\mathrm{E}}^{\left(  3\right)}\otimes\mathfrak{osp}\left(  32|1\right)  $.
(\textit{b}) Una super\'{a}lgebra similar a las de D'Auria and Fr\'{e} (Ref.~\cite{DAuria-Fre}) es obtenida a trav\'{e}s del $0_{S}$-reducci\'{o}n de la sub\'{a}lgebra resonante.}
\label{Fig Res DAuriaFre}
\end{figure}

El resultado es un \'{a}lgebra en el \textquotedblleft
estilo\textquotedblright\ de las de D'Auria--Fr\'{e}, la cual \textit{difiere}
de las presentadas en Ref.~\cite{DAuria-Fre} pero que posee el mismo
n\'{u}mero y tipo de conmutadores, valuados en los mismos subespacios.
As\'{\i} mismo, las relaciones de conmutaci\'{o}n nulas son las mismas.
Re-etiquetando los generadores de la forma $\bm{J}_{ab}=\bm{J}_{\left(
ab,0\right)  }$, $\bm{Q}_{\alpha}=\bm{Q}_{\left(  \alpha,1\right)  }$,
$\bm{Z}_{ab}=\bm{J}_{\left(  ab,2\right)  }$, $\bm{P}_{a}=\bm{P}_{\left(
a,2\right)  }$, $\bm{Z}_{abcde}=\bm{Z}_{\left(  abcde,2\right)  }$,
$\bm{Q}_{\alpha}^{\prime}=\bm{Q}_{\left(  \alpha,3\right)  },$ es posible
encontrar la estructura mostrada en Fig.~\ref{Fig Res DAuriaFre}.

Las relaciones de anticonmutaci\'{o}n, las cuales se pueden obtener en forma
directa de ec.~(\ref{SuperResonantStructureConstants}) son muy similares a
aquellas del \'{A}lgebra M. S\'{o}lo difieren las relaciones
\begin{align}
\left[  \bm{P}_{a},\bm{Q}\right]   &  =-\frac{1}{2}\Gamma_{a}\bm{Q}^{\prime
},\\
\left[  \bm{Z}_{ab},\bm{Q}\right]   &  =-\frac{1}{2}\Gamma_{ab}\bm{Q}^{\prime
},\\
\left[  \bm{Z}_{abcde},\bm{Q}\right]   &  =-\frac{1}{2}\Gamma_{abcde}%
\bm{Q}^{\prime}.
\end{align}
y se deben de agregar las relaciones que involucran al nuevo generador
fermi\'{o}nico $\bm{Q}^{\prime}$, las cuales son
\begin{align}
\left[  \bm{P}_{a},\bm{Q}^{\prime}\right]   &  =\bm{0},\label{pqp}\\
\left[  \bm{Z}_{ab},\bm{Q}^{\prime}\right]   &  =\bm{0},\label{z2qp}\\
\left[  \bm{Z}_{abcde},\bm{Q}^{\prime}\right]   &  =\bm{0},\label{z5qp}\\
\left\{  \bm{Q},\bm{Q}^{\prime}\right\}   &  =\bm{0},\\
\left\{  \bm{Q}^{\prime},\bm{Q}^{\prime}\right\}   &  =\bm{0},
\end{align}%
\begin{equation}
\left[  \bm{J}_{ab},\bm{Q}^{\prime}\right]  =-\frac{1}{2}\Gamma_{ab}%
\bm{Q}^{\prime}.
\end{equation}

El generador fermi\'{o}nico extra $\bm{Q}^{\prime}$ anticonmuta con todos los
generadores del \'{a}lgebra excepto con los de Lorentz, lo que resulta natural
considerando su car\'{a}cter espinorial.

\subsection{\'{A}lgebra de la 5~Brana}

Consideremos nuevamente \'{e}l \'{a}lgebra resonante que di\'{o} origen al
\'{A}lgebra M en Sec.~\ref{Sec Algebra M} y busquemos un reducci\'{o}n resonante
de ella. La partici\'{o}n resonante en ese caso correspondi\'{o} a
$S_{\mathrm{E}}^{\left(  2\right)  }=S_{0}\cup S_{1}\cup S_{2}$ con%
\begin{align}
S_{0}  &  =\left\{  \lambda_{0},\lambda_{2},\lambda_{3}\right\}  ,\\
S_{1}  &  =\left\{  \lambda_{1},\lambda_{3}\right\}  ,\\
S_{2}  &  =\left\{  \lambda_{2},\lambda_{3}\right\}  .
\end{align}

Observemos que la partici\'{o}n de los subconjuntos $S_{p},$ $S_{p}=\hat
{S}_{p}\cup\check{S}_{p}$,%
\begin{align}
\check{S}_{0}  &  =\left\{  \lambda_{0}\right\}  ,\qquad\hat{S}_{0}=\left\{
\lambda_{2},\lambda_{3}\right\}  ,\\
\check{S}_{1}  &  =\left\{  \lambda_{1}\right\}  ,\qquad\hat{S}_{1}=\left\{
\lambda_{3}\right\}  ,\\
\check{S}_{2}  &  =\left\{  \lambda_{2}\right\}  ,\qquad\hat{S}_{2}=\left\{
\lambda_{3}\right\}  .
\end{align}

\begin{figure}[ptb]
\psfrag{Jab}{$\bm{J}_{ab}$}
\psfrag{Q}{$\bm{Q}$}
\psfrag{Z5}{$\bm{Z}_{abcde}$}
\psfrag{Pa}{$\bm{P}_{a}$}
\includegraphics[width=\columnwidth]{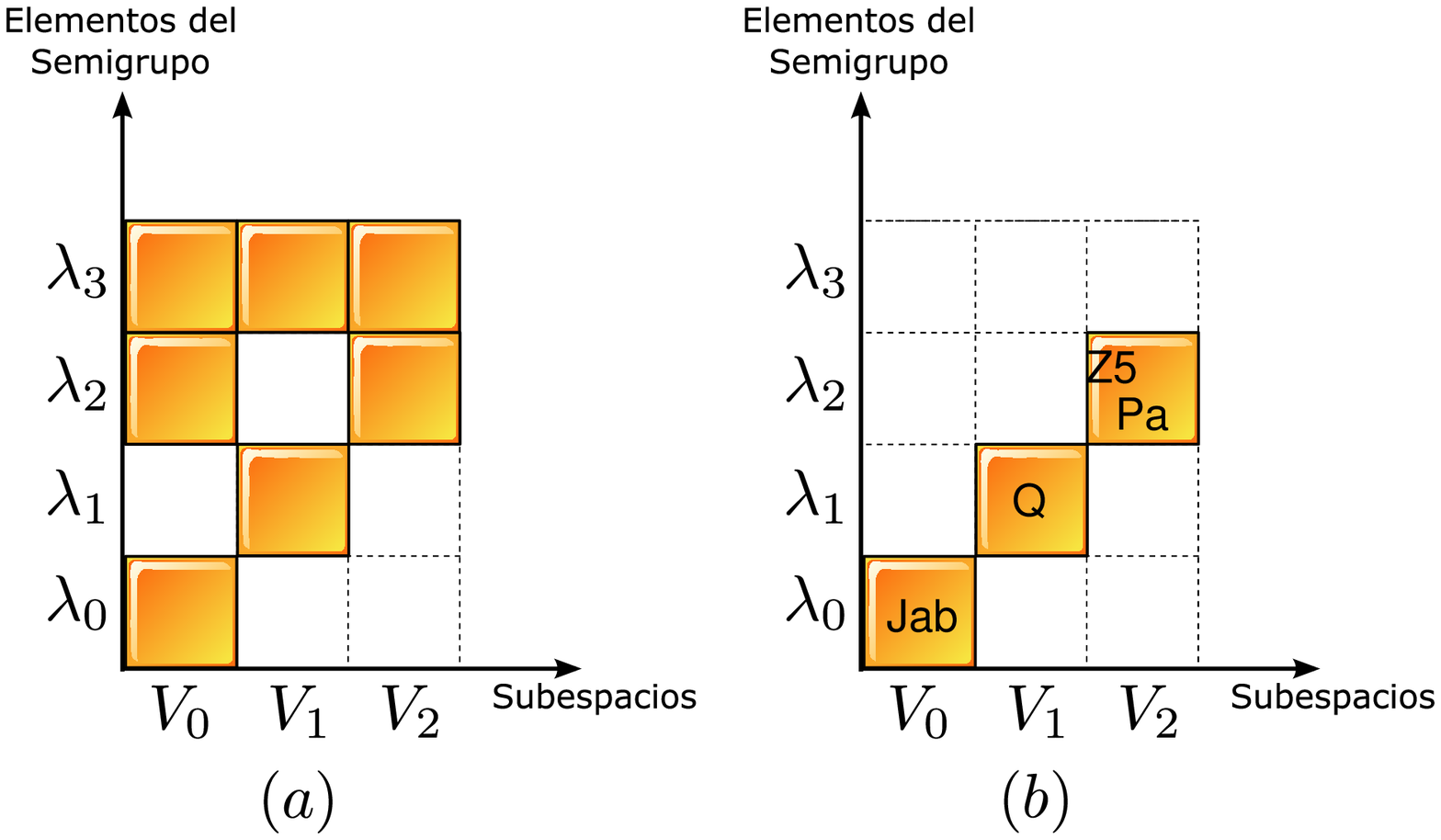}\caption{(\textit{a})
Sub\'{a}lgebra resonante de $S_{\text{E}}^{(2)}\otimes\mathfrak{osp}\left(
32|1\right)  $. (\textit{b}) La super\'{a}lgebra de la 5-brana como
reducci\'{o}n resonante.}
\label{Fig Forced5Brane}
\end{figure}

satisface las condiciones ecs.~(\ref{Ec Cond Forz SpnSq=0}%
)-(\ref{Ec Cond Forz SpSq=nSr}), pues en efecto, tenemos que%
\begin{align*}
\check{S}_{0}\times\hat{S}_{0}  &  \subset\hat{S}_{0},\\
\check{S}_{0}\times\hat{S}_{1}  &  \subset\hat{S}_{1},\\
\check{S}_{0}\times\hat{S}_{2}  &  \subset\hat{S}_{2},\\
\check{S}_{1}\times\hat{S}_{1}  &  \subset\hat{S}_{0}\cap\hat{S}_{2},\\
\check{S}_{1}\times\hat{S}_{2}  &  \subset\hat{S}_{1},\\
\check{S}_{2}\times\hat{S}_{2}  &  \subset\hat{S}_{0}\cap\hat{S}_{2},
\end{align*}
[Compare con ecs.~(\ref{SuperS0S0=S0})--(\ref{SuperS2S2=S0nS2})
y~(\ref{SuperV0V0=V0})--(\ref{SuperV2V2=V0+V2})].

El reducci\'{o}n resonante del \'{a}lgebra es representado en forma
expl\'{\i}cita en Fig.~\ref{Fig Forced5Brane}, y de
ec.~(\ref{Ec Ctes Struc Forz}) podemos obtener las constantes de estructura
para este caso, con lo que obtenemos%
\begin{align*}
\left[  \boldsymbol{P}_{a},\boldsymbol{P}_{b}\right]   &  =\boldsymbol{0},\\
\left[  \boldsymbol{J}_{ab},\boldsymbol{P}_{c}\right]   &  =\eta_{ce}%
\delta_{ab}^{de}\boldsymbol{P}_{d},\\
\left[  \boldsymbol{J}_{ab},\boldsymbol{J}_{cd}\right]   &  =\eta_{gh}%
\delta_{ab}^{eg}\delta_{cd}^{hf}\boldsymbol{J}_{ef}.
\end{align*}%
\begin{align}
\left[  \bm{P}_{a},\bm{Z}_{b_{1}\cdots b_{5}}\right]   &  =\bm{0},\\
\left[  \bm{J}^{ab},\bm{Z}_{c_{1}\cdots c_{5}}\right]   &  =\frac{1}{4!}%
\delta_{dc_{1}\cdots c_{5}}^{abe_{1}\cdots e_{4}}\bm{Z}_{\phantom{d}e_{1}%
\cdots e_{4}}^{d},\\
\left[  \bm{Z}^{a_{1}\cdots a_{5}},\bm{Z}_{b_{1}\cdots b_{5}}\right]   &
=\bm{0},
\end{align}%
\begin{align}
\left[  \bm{P}_{a},\bm{Q}\right]   &  =\bm{0},\\
\left[  \bm{J}_{ab},\bm{Q}\right]   &  =-\frac{1}{2}\Gamma_{ab}\bm{Q},\\
\left[  \bm{Z}_{abcde},\bm{Q}\right]   &  =\bm{0},
\end{align}%
\[
\left\{  \bm{Q},\bar{\bm{Q}}\right\}  =\frac{1}{8}\left(  \Gamma^{a}%
\bm{P}_{a}+\frac{1}{5!}\Gamma^{abcde}\bm{Z}_{abcde}\right)
\]

Esta es la Superalgebra de la 5 Brana (Ver Refs.~\cite{vanHo82,Azcarraga89})

\subsection{\label{Sec Z4 x osp SubAlg Reson}Sub\'{a}lgebra Resonante de
$\mathbb{Z}_{4}\otimes\mathfrak{osp}\left(  \mathfrak{32}|\mathfrak{1}\right)
$}

Hasta ahora, todas los ejemplos expl\'{\i}citos constru\'{\i}dos han usado
como semigrupo a $S_{\mathrm{E}}^{\left(  N\right)  }.$ Ahora, por mera
curiosidad, vamos a experimentar con otra alternativa. Una elecci\'{o}n
interesante es la de grupos abelianos discretos. \'{E}sta es especialmente
sencilla, pues un grupo no-trivial no tiene elemento $0_{S};$ por lo tanto, no
es posible aplicar $0_{S}$-reducci\'{o}n. Entre los grupos abelianos, la
alternativa m\'{a}s cercana a $S_{\mathrm{E}}^{\left(  N\right)  }$ consiste
en los grupos c\'{\i}clicos $\mathbb{Z}_{N},$ tal como se puede ver al
considerar sus respectivas leyes de multiplicaci\'{o}n, las cuales son muy
similares\footnote{\'{O}bservese as\'{\i} mismo la similitud entre las
funciones $\operatorname{mod}_{N}$ y $H_{N+1}.$}.%
\begin{align}
S_{\mathrm{E}}^{\left(  N\right)  }  &  :\lambda_{\alpha}\lambda_{\beta
}=\lambda_{H_{N+1}\left(  \alpha+\beta\right)  }\\
\mathbb{Z}_{N}  &  :\lambda_{\alpha}\lambda_{\beta}=\lambda
_{\operatorname{mod}_{N}\left(  \alpha+\beta\right)  }.
\end{align}

\begin{figure}[ptb]
\begin{center}
\includegraphics[width=.5\columnwidth]{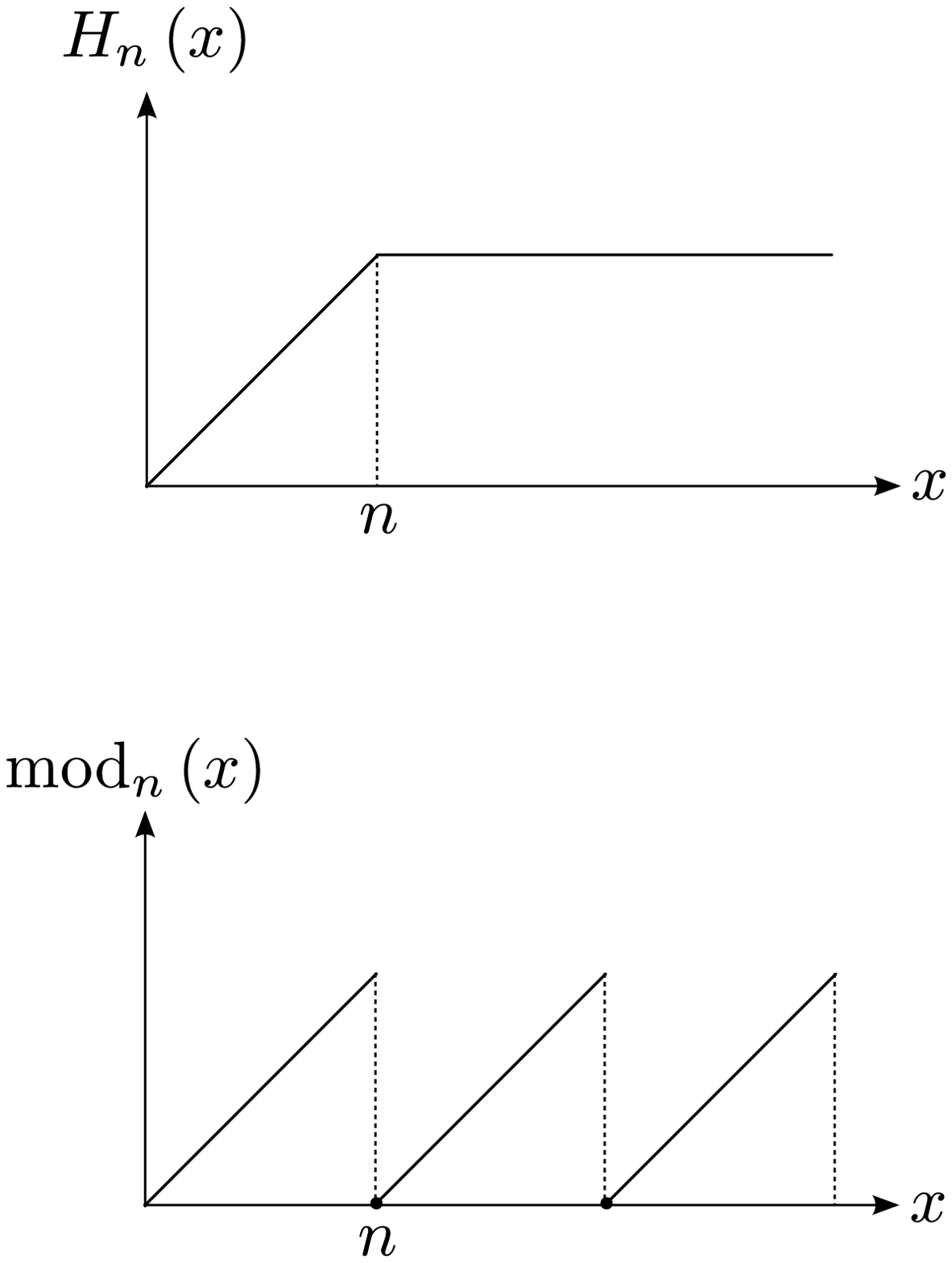}\caption{Las funciones $H_{n}(x)$ y $\operatorname{mod}_{N}(x)$}
\label{FigHn_vs_mod}
\end{center}
\end{figure}

Para este ejemplo en particular, se escogi\'{o} el grupo c\'{\i}clico
$\mathbb{Z}_{4},$ porque para el caso de super\'{a}lgebras, este es el ejemplo
no-trivial m\'{a}s simple. El caso $\mathbb{Z}_{2}$ es trivial [la
sub\'{a}lgebra resonante es $\mathfrak{osp}\left(  \mathfrak{32}%
|\mathfrak{1}\right)  $ misma], y $\mathbb{Z}_{3}$ parece no tener una
partici\'{o}n resonante. Es importante recordar que dado que ya que estamos
usando un semigrupo distinto de $S_{\mathrm{E}}^{\left(  N\right)  },$ el
\'{a}lgebra que obtendremos \textit{no} corresponde a una expansi\'{o}n en
formas de Maurer--Cartan.

Dada una super\'{a}lgebra $\mathfrak{g}=V_{0}\oplus V_{1}\oplus V_{2}$ con la
estructura ec.~(\ref{SuperV0V0=V0})--(\ref{SuperV2V2=V0+V2}), una
partici\'{o}n resonante [\textit{i.e.}, que satisface ecs.(\ref{SuperS0S0=S0}%
)-(\ref{SuperS2S2=S0nS2})] de $\mathbb{Z}_{4}=\left\{  \lambda_{0},\lambda
_{1},\lambda_{2},\lambda_{3}\right\}  $ es dada por%
\begin{align}
S_{0}  &  =\left\{  \lambda_{0},\lambda_{2}\right\}  ,\\
S_{1}  &  =\left\{  \lambda_{1},\lambda_{3}\right\}  ,\\
S_{2}  &  =\left\{  \lambda_{0},\lambda_{2}\right\}  .
\end{align}

\begin{figure}[ptb]
\begin{center}
\psfrag{Jab}{$\bm{J}_{ab}$}
\psfrag{Zab}{$\bm{Z}_{ab}$}
\psfrag{Q}{$\bm{Q}$}
\psfrag{Q2}{$\bm{Q^{\prime}}$}
\psfrag{Z5}{$\bm{Z}_{abcde}$}
\psfrag{Pa}{$\bm{P}_{a}$}
\psfrag{Zc5}{$\bm{Z^{\prime}}_{abcde}$}
\psfrag{Pca}{$\bm{P^{\prime}}_{a}$}
\includegraphics[width=.5\columnwidth]{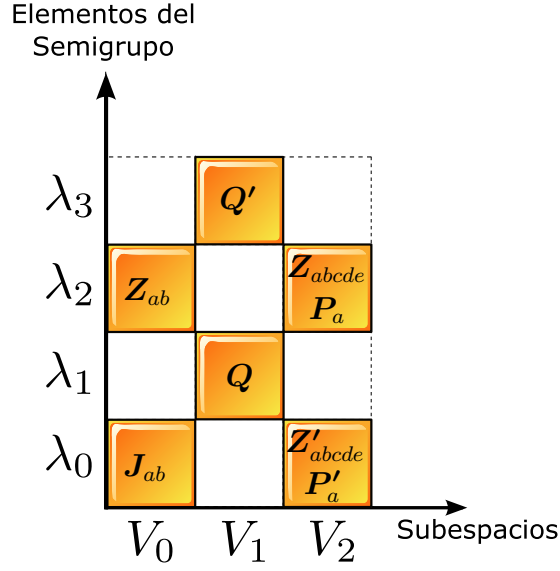}\caption{Nueva
super\'{a}lgebra en $d=11,$ la cual comparte algunas caracter\'{\i}sticas con
el \'{A}lgebra~M y las super\'{a}lgebras de D'Auria--Fr\'{e}, obtenida
directamente como un reducci\'{o}n resonante de $\mathbb{Z}_{4}\otimes
\mathfrak{osp}\left(  32|1\right)  $.}%
\label{Fig ResonantZ4}
\end{center}
\end{figure}

Para no recargar la notaci\'{o}n, re-etiquetemos los generadores como
$\bm{J}_{ab}=\bm{J}_{\left(  ab,0\right)  }$, $\bm{Z}_{a_{1}\cdots a_{5}%
}^{\prime}=\bm{Z}_{\left(  a_{1}\cdots a_{5},0\right)  }$, $\bm{P}_{a}%
^{\prime}=\bm{P}_{\left(  a,0\right)  }$, $\bm{Q}_{\alpha}=\bm{Q}_{\left(
\alpha,1\right)  }$, $\bm{Z}_{ab}=\bm{J}_{\left(  ab,2\right)  }$,
$\bm{Z}_{a_{1}\cdots a_{5}}=\bm{Z}_{\left(  a_{1}\cdots a_{5},2\right)  }$,
$\bm{P}_{a}=\bm{P}_{\left(  a,2\right)  }$ and $\bm{Q}_{\alpha}^{\prime
}=\bm{Q}_{\left(  \alpha,3\right)  } $, tal como se muestra en
Fig.~\ref{Fig ResonantZ4}.

La correspondiente sub\'{a}lgebra resonante puede ser escrita f\'{a}cilmente
reemplazando las constantes de estructura de $\mathfrak{osp}\left(
\mathfrak{32}|\mathfrak{1}\right)  $ en ec.~(\ref{Ec CtesStruct Resonantes}%
); aqu\'{\i} s\'{o}lo mostraremos algunos de los sectores que parecen m\'{a}s interesantes.

Una de las caracter\'{\i}sticas m\'{a}s interesantes de esta \'{a}lgebra es
que \emph{ambos} generadores fermi\'{o}nicos, $\bm{Q}$ y $\bm{Q}^{\prime},$
tienen anticonmutadores que coinciden con aquellos del \'{A}lgebra~M,%
\[
\left\{  \bm{Q}^{\prime},\bar{\bm{Q}}^{\prime}\right\}  =\left\{
\bm{Q},\bar{\bm{Q}}\right\}  =\frac{1}{8}\left(  \Gamma^{a}\bm{P}_{a}-\frac
{1}{2}\Gamma^{ab}\bm{Z}_{ab}+\frac{1}{5!}\Gamma^{abcde}\bm{Z}_{abcde}\right)
\]

Una caracter\'{\i}stica \textquotedblleft exrta\~{n}a\textquotedblright\ de
esta \'{a}lgebra es que posee \emph{dos} generadores de \textquotedblleft%
\textit{AdS-boosts}\textquotedblright\ para la misma \'{a}lgebra de Lorentz,%
\begin{equation}%
\begin{tabular}
[c]{l}%
\begin{tabular}
[c]{ll}%
$%
\begin{array}
[c]{l}%
\left[  \bm{P}_{a},\bm{P}_{b}\right]  =\bm{J}_{ab},\\
\left[  \bm{J}_{ab},\bm{P}_{c}\right]  =\eta_{ce}\delta_{ab}^{de}\bm{P}_{d},
\end{array}
$ & $%
\begin{array}
[c]{l}%
\left[  \bm{P}_{a}^{\prime},\bm{P}_{b}^{\prime}\right]  =\bm{J}_{ab},\\
\left[  \bm{J}_{ab},\bm{P}_{c}^{\prime}\right]  =\eta_{ce}\delta_{ab}%
^{de}\bm{P}_{d}^{\prime},
\end{array}
$%
\end{tabular}
\\
\multicolumn{1}{c}{$\left[  \bm{J}_{ab},\bm{J}_{cd}\right]  =\eta_{gh}%
\delta_{ab}^{eg}\delta_{cd}^{hf}\bm{J}_{ef}.$}\\
\multicolumn{1}{c}{$\left[  \bm{P}_{a},\bm{P}_{b}^{\prime}\right]
=\bm{Z}_{ab}.$}%
\end{tabular}
\end{equation}

Las `cargas' $\bm{Z}_{ab}$, $\bm{Z}_{a_{1}\cdots a_{5}}$, $\bm{Z}_{ab}%
^{\prime}$ y $\bm{Z}_{a_{1}\cdots a_{5}}^{\prime}$ son tensores de Lorentz,
pero ya no son m\'{a}s \textquotedblleft centrales\textquotedblright%
\ (\textit{no} son abelianas)%
\begin{equation}
\left[  \bm{Z}^{ab},\bm{Z}_{cd}\right]  =\delta_{ecd}^{abf}%
\bm{J}_{\phantom{e}f}^{e},
\end{equation}%
\begin{align}
\left[  \bm{Z}^{a_{1}\cdots a_{5}},\bm{Z}_{b_{1}\cdots b_{5}}\right]   &
=\left[  \bm{Z}^{\prime a_{1}\cdots a_{5}},\bm{Z}_{b_{1}\cdots b_{5}}^{\prime
}\right] \nonumber\\
&  =\eta^{\left[  a_{1}\cdots a_{5}\right]  \left[  c_{1}\cdots c_{5}\right]
}\varepsilon_{c_{1}\cdots c_{5}b_{1}\cdots b_{5}e}\bm{P}^{\prime e}%
+\delta_{db_{1}\cdots b_{5}}^{a_{1}\cdots a_{5}e}\bm{J}_{\phantom{d}e}%
^{d}+\nonumber\\
&  -\frac{1}{3!3!5!}\varepsilon_{c_{1}\cdots c_{11}}\delta_{d_{1}d_{2}%
d_{3}b_{1}\cdots b_{5}}^{a_{1}\cdots a_{5}c_{4}c_{5}c_{6}}\eta^{\left[
c_{1}c_{2}c_{3}\right]  \left[  d_{1}d_{2}d_{3}\right]  }\bm{Z}^{\prime
c_{7}\cdots c_{11}}.
\end{align}

En conmutadores involucrando un generador fermi\'{o}nico, esta
super\'{a}lgebra tiene alguna similitud con la D'Auria--Fr\'{e}; as\'{\i}, los
conmutadores entre los generadores $\bm{P}_{a}$, $\bm{Z}_{ab}$, $\bm{Z}_{a_{1}%
\cdots a_{5}}$ y un generador fermi\'{o}nico $\bm{Q}$ est\'{a}n valuados en
$\bm{Q}^{\prime}$, pero sin embargo, su conmutador con $\bm{Q}^{\prime}$
est\'{a} valuado en $\bm{Q}$ en vez de anularse.

\section{\label{Sec TensInv S-Exp}Tensores Invariantes para \'{A}lgebras \\
$S$-Expandidas}

Encontrar todos los tensores invariantes para un \'{a}lgebra no semisimple ha permanecido
como un problema abierto hasta ahora. Esto no s\'{o}lo desde un
punto de vista matem\'{a}tico, sino tambi\'{e}n desde el punto de vista de la
f\'{\i}sica, pues el tensor invariante es el ingrediente clave en la
construcci\'{o}n de un lagrangeano de Chern--Simons o Transgresor. En general,
distintas elecciones de tensor invariante dan origen a teor\'{\i}as con
din\'{a}micas completamente diferentes.

Sin embargo, la supertraza siempre provee de un tensor invariante. Esta
\textquotedblleft disponibilidad inmediata\textquotedblright\ hace de la
supertraza la elecci\'{o}n m\'{a}s popular en la literatura. Sin embargo, este
procedimiento presenta importantes limitaciones en el caso de las
$S$-expansiones, y por lo tanto, se vuelve fundamental buscar tensores
invariantes distintos de la traza para este tipo \'{a}lgebras.

Estos tensores invariantes distintos de la supertraza, ser\'{a}n entregados a
trav\'{e}s de la siguiente serie de teoremas:

\begin{theorem}
\label{Teo TensInv S-Exp}Sea $S$ un semigrupo abeliano, $\mathfrak{g}$ un
\'{a}lgebra de Lie de base $\left\{  \bm{T}_{A}\right\}  $,y sea $\left\vert
\bm{T}_{A_{1}}\cdots\bm{T}_{A_{n}}\right\vert $ un tensor invariante para
$\mathfrak{g}$. Entonces, la expresi\'{o}n%
\begin{equation}
\left\vert \bm{T}_{\left(  A_{1},\alpha_{1}\right)  }\cdots\bm{T}_{\left(
A_{n},\alpha_{n}\right)  }\right\vert =\alpha_{\gamma}K_{\alpha_{1}%
\cdots\alpha_{n}}^{\phantom{\alpha_{1}\cdots\alpha_{n}}\gamma}\left\vert
\bm{T}_{A_{1}}\cdots\bm{T}_{A_{n}}\right\vert \label{Ec InvTensorS-Exp}%
\end{equation}
siendo $\alpha_{\gamma}$ constantes arbitrarias y $K_{\alpha_{1}\cdots
\alpha_{n}}^{\phantom{\alpha_{1}\cdots\alpha_{n}}\gamma}$ el $n$-selector de
$S$, corresponde a un tensor invariante para el \'{A}lgebra $S$-Expandida
$\mathfrak{G}=S\otimes\mathfrak{g}$.
\end{theorem}

\begin{proof}
Sea%
\begin{align*}
X_{\left(  A_{0},\alpha_{0}\right)  \cdots\left(  A_{n},\alpha_{n}\right)
}^{\left(  p\right)  }  &  =\left(  -1\right)  ^{\mathfrak{q}\left(
A_{0},\alpha_{0}\right)  \left(  \mathfrak{q}\left(  A_{1},\alpha_{1}\right)
+\cdots+\mathfrak{q}\left(  A_{p-1},\alpha_{p-1}\right)  \right)  }\times\\
&  \times C_{\left(  A_{0},\alpha_{0}\right)  \left(  A_{p},\alpha_{p}\right)
}^{\text{\qquad\qquad\qquad}\left(  B,\beta\right)  }\left\vert
\bm{T}_{\left(  A_{1},\alpha_{1}\right)  }\cdots\bm{T}_{\left(  A_{p-1}%
,\alpha_{p-1}\right)  }\bm{T}_{\left(  B,\beta\right)  }\bm{T}_{\left(
A_{p+1},\alpha_{p+1}\right)  }\cdots\bm{T}_{\left(  A_{n},\alpha_{n}\right)
}\right\vert .
\end{align*}

Utilizando la expresi\'{o}n para las constantes de estructura
ec.~(\ref{Ec C=KC}), el hecho de que $\mathfrak{q}\left(  A,\alpha\right)
=\mathfrak{q}\left(  A\right)  $ y ec.~(\ref{Ec InvTensorS-Exp}), tenemos que%
\[
X_{\left(  A_{0},\alpha_{0}\right)  \cdots\left(  A_{n},\alpha_{n}\right)
}^{\left(  p\right)  }=\alpha_{\gamma}K_{\alpha_{0}\cdots\alpha_{n}}%
^{\qquad\;\gamma}X_{A_{0}\cdots A_{n}}^{\left(  p\right)  }.
\]
en donde $X_{A_{0}\cdots A_{n}}^{\left(  p\right)  }$ viene dado por
ec.~(\ref{Ec X(p) = C x TensInv}). Dado que $\left\vert \bm{T}_{A_{1}}%
\cdots\bm{T}_{A_{n}}\right\vert $ es un tensor invariante para $\mathfrak{g},$
satisface ec.~(\ref{EcCondInvCtesStruc}), y as\'{\i},%
\begin{equation}
\sum_{p=1}^{n}X_{\left(  A_{0},\alpha_{0}\right)  \cdots\left(  A_{n}%
,\alpha_{n}\right)  }^{\left(  p\right)  }=\alpha_{\gamma}K_{\alpha_{0}%
\cdots\alpha_{n}}^{\qquad\;\gamma}\sum_{p=1}^{n}X_{A_{0}\cdots A_{n}}^{\left(
p\right)  }=0.\label{Ec X(p) = 0 S-Exp}%
\end{equation}

Por lo tanto, $\left\vert \bm{T}_{\left(  A_{1},\alpha_{1}\right)  }%
\cdots\bm{T}_{\left(  A_{n},\alpha_{n}\right)  }\right\vert =\alpha_{\gamma
}K_{\alpha_{1}\cdots\alpha_{n}}^{\phantom{\alpha_{1}\cdots\alpha_{n}}\gamma
}\left\vert \bm{T}_{A_{1}}\cdots\bm{T}_{A_{n}}\right\vert $ tambi\'{e}n
satisface ec.~(\ref{EcCondInvCtesStruc}), y por lo tanto, es un tensor
invariante para $\mathfrak{G}=S\otimes\mathfrak{g}$.
\end{proof}

\begin{corollary}
\label{Cor TensInv Subalg Reson}Sea $\mathfrak{G}_{\mathrm{R}}$ una
sub\'{a}lgebra resonante de $\mathfrak{G}=S\otimes\mathfrak{g}$, siendo
$\mathfrak{g}=\bigoplus_{p\in I}V_{p}$ la subdivisi\'{o}n en subespacios de
$\mathfrak{g}$ y $S=\bigcup_{p\in I}S_{p}$ la correspondiente
descomposici\'{o}n resonante de $S.$ Entonces,%
\begin{equation}
\left\vert \bm{T}_{\left(  A_{p_{1}},\alpha_{p_{1}}\right)  }\cdots
\bm{T}_{\left(  A_{p_{n}},\alpha_{p_{n}}\right)  }\right\vert =\alpha_{\gamma
}K_{\alpha_{p_{1}}\cdots\alpha_{p_{n}}}%
^{\phantom{\alpha_{1}\cdots\alpha_{n}}\gamma}\left\vert \bm{T}_{A_{p_{1}}%
}\cdots\bm{T}_{A_{p_{n}}}\right\vert \label{Ec InvTensor Subalg Reson}%
\end{equation}
es un tensor invariante para $\mathfrak{G}_{\mathrm{R}},$ con $\alpha_{p}$ tal
que $\lambda_{\alpha_{p}}\in S_{p}$ y siendo $\left\{  \boldsymbol{T}_{A_{p}%
}\right\}  $ los generadores de $V_{p}.$
\end{corollary}

\begin{proof}
Dado un tensor invariante para un \'{a}lgebra, sus componentes valuadas en una
sub\'{a}lgebra siempre forman un tensor invariante para \'{e}sta [como es
sencillo verificar por simple inspecci\'{o}n de ec.~(\ref{EcCondInvCtesStruc}%
)]. Sin embargo, en general estas componentes del tensor invariante pudieran
anularse id\'{e}nticamente. En el caso de la sub\'{a}lgebra resonante, esto
nunca sucede (provistos que todos los $\alpha_{\gamma} $ son no nulos). En
efecto, dado que $S$ es cerrado bajo el producto, para cada elecci\'{o}n de
$\alpha_{p_{1}},\ldots,\alpha_{p_{n}}$ siempre existe un valor de $\gamma$ tal
que $K_{\alpha_{p_{1}}\cdots\alpha_{p_{n}}}%
^{\phantom{\alpha_{p_{1}} \cdots \alpha_{p_{n}}}\gamma}=1,$ y por lo tanto, el
tensor invariante ec.~(\ref{Ec InvTensor Subalg Reson}) es no nulo.
\end{proof}

Resulta interesante comparar estos tensores con la supertraza. Para
constru\'{\i}rla, consideremos los generadores $\left\{  \left[
\bm{T}_{A}\right]  _{a}^{\phantom{a}b}\right\}  $ en alguna representaci\'{o}n
matricial de $\mathfrak{g},$ y los elementos del semigrupo en la
representaci\'{o}n dada por los $2$-selectores
[ec.(\ref{Ec Rep Matricial = 2-selector})],%
\[
\left[  \lambda_{\alpha}\right]  _{\mu}^{\ \nu}=K_{\mu\alpha}^{\quad\nu}.
\]

Entonces, esto induce en forma natural una representaci\'{o}n matricial para
los generadores del \'{a}lgebra $S$-expandida, de la forma%
\begin{equation}
\left[  \bm{T}_{\left(  A,\alpha\right)  }\right]  _{\left(  a,\mu\right)
}^{\phantom{\left( a,\mu \right)}\left(  b,\nu\right)  }=\left[
\lambda_{\alpha}\right]  _{\mu}^{\ \nu}\left[  \bm{T}_{A}\right]
_{a}^{\phantom{a}b}.\label{Ec RepMatricial S-Exp}%
\end{equation}

Cuando los generadores de $\mathfrak{g}$ se encuentran en la
representaci\'{o}n adjunta, entonces la representaci\'{o}n matricial dada por
ec.~(\ref{Ec RepMatricial S-Exp}) corresponde tambi\'{e}n a la
representaci\'{o}n adjunta del \'{a}lgebra $S$-expandida. Calculando la
supertraza en esta representaci\'{o}n, tenemos
\begin{equation}
\operatorname*{STr}\left(  \bm{T}_{\left(  A_{1},\alpha_{1}\right)  }%
\cdots\bm{T}_{\left(  A_{n},\alpha_{n}\right)  }\right)  =K_{\gamma\alpha
_{1}\cdots\alpha_{n}}^{\phantom{\gamma \alpha_{1} \cdots \alpha_{n}}\gamma
}\operatorname*{Str}\left(  \bm{T}_{A_{1}}\cdots\bm{T}_{A_{n}}\right)
,\label{Ec STr (Con Cero)}%
\end{equation}
en donde hemos usado $\operatorname*{STr}$ para la supertraza en los
generadores $\bm{T}_{\left(  A,\alpha\right)  }$ y $\operatorname*{Str}$ para
la supertraza en los generadores $\bm{T}_{A}$.

La supertraza corresponde a un caso particular de la expresi\'{o}n%
\begin{equation}
\left\vert \bm{T}_{\left(  A_{1},\alpha_{1}\right)  }\cdots\bm{T}_{\left(
A_{n},\alpha_{n}\right)  }\right\vert =\sum_{m=0}^{M}\alpha_{\gamma}%
^{\beta_{1}\cdots\beta_{m}}K_{\beta_{1}\cdots\beta_{m}\alpha_{1}\cdots
\alpha_{n}}%
^{\phantom{\beta_{1} \cdots \beta_{m} \alpha_{1} \cdots \alpha_{n}}\gamma
}\left\vert \bm{T}_{A_{1}}\cdots\bm{T}_{A_{n}}\right\vert
,\label{AlfaGammaBeta1...m}%
\end{equation}
(siendo $M$ el n\'{u}mero de elementos de $S$) la cual tambi\'{e}n corresponde
a un tensor invariante de $\mathfrak{G}=S\otimes\mathfrak{g}$, en donde hemos
escogido como \'{u}nica constante $\alpha_{\gamma}^{\beta_{1}\cdots\beta_{m}}$
no nula una delta de Kronecker,%
\[
\left\vert \bm{T}_{\left(  A_{1},\alpha_{1}\right)  }\cdots\bm{T}_{\left(
A_{n},\alpha_{n}\right)  }\right\vert =\delta_{\gamma}^{\rho}K_{\rho\alpha
_{1}\cdots\alpha_{n}}^{\phantom{\rho \alpha_{1} \cdots \alpha_{n}}\gamma
}\left\vert \bm{T}_{A_{1}}\cdots\bm{T}_{A_{n}}\right\vert .
\]

Pese a que el tensor invariante ec.~(\ref{AlfaGammaBeta1...m}) parece m\'{a}s
general que el de ec.~(\ref{Ec InvTensorS-Exp}), en realidad este no es el
caso. Usando s\'{o}lo la asocitividad y cierre del producto del semigrupo, es
directo probar que es siempre posible reducir ec.~(\ref{AlfaGammaBeta1...m})
en ec.~(\ref{Ec InvTensorS-Exp}), el cual es de esta forma el tensor
invariante \textquotedblleft fundamental\textquotedblright. Por consiguiente,
as\'{\i} mismo la supertraza corresponde a un caso particular de
ec.~(\ref{Ec InvTensorS-Exp}), para una elecci\'{o}n particular de las
constantes $\alpha_{\gamma}.$

Para \'{a}lgebras expandidas y sus sub\'{a}lgebras resonantes, pese a que la
supertraza corresponde a un caso particular de ec.~(\ref{Ec InvTensorS-Exp})
\'{o} ec.~(\ref{Ec InvTensor Subalg Reson}), igualmente permite constru\'{\i}r
lagrangeanos de Chern--Simons o Transgresi\'{o}n bien comportados. Este no
ser\'{a} el caso cuando hay $0_{S}$-reducci\'{o}n, tal como veremos despu\'{e}s.
Debemos recordar que un \'{a}lgebra reducida \textit{no} es una sub\'{a}lgebra
de $S\otimes\mathfrak{g},$ y por lo tanto, en general las respectivas
componentes de ec.~(\ref{Ec InvTensorS-Exp}) \textit{no} corresponden a un
tensor invariante para ella. Para el caso de \'{a}lgebras $0_{S}$-reducidas, el
siguiente teorema y su corolario proveen con una soluci\'{o}n:

\begin{theorem}
\label{Teo TensInv 0-Forz}Sea $S$ un semigrupo abeliano de elementos no nulos
$\lambda_{i},$ $i=0,\ldots,N,$ y $\lambda_{N+1}=0_{S}$. Sea $\mathfrak{g}$ un
\'{a}lgebra de Lie de base $\left\{  \bm{T}_{A}\right\}  $, y sea $\left\vert
\bm{T}_{A_{1}}\cdots\bm{T}_{A_{n}}\right\vert $ un tensor invariante para
$\mathfrak{g}$. Entonces, la expresi\'{o}n%
\begin{equation}
\left\vert \bm{T}_{\left(  A_{1},i_{1}\right)  }\cdots\bm{T}_{\left(
A_{n},i_{n}\right)  }\right\vert =\alpha_{j}K_{i_{1}\cdots i_{n}}^{\quad
\ \ j}\left\vert \bm{T}_{A_{1}}\cdots\bm{T}_{A_{n}}\right\vert
,\label{Ec InvTensor 0s-Forz}%
\end{equation}
con $\alpha_{j}$ constantes arbitrarias, corresponde a un tensor invariante
para el \'{a}lgebra $0_{S}$-reducida.
\end{theorem}

\begin{proof}
Este teorema puede ser visto como un corolario de
Teorema~\ref{Teo TensInv S-Exp}. En efecto, considerando la componente
$i_{0}\cdots i_{n}$, de ec.~(\ref{Ec X(p) = 0 S-Exp}), y escribi\'{e}ndola en
forma expl\'{\i}cita, tenemos que para el \'{A}lgebra $S$-Expandida se cumple
que%
\begin{multline*}
\sum_{p=0}^{n}\left(  -1\right)  ^{\mathfrak{q}\left(  A_{0},i_{0}\right)
\left(  \mathfrak{q}\left(  A_{1},i_{1}\right)  +\cdots+\mathfrak{q}\left(
A_{p-1},i_{p-1}\right)  \right)  }\times\\
\times\left(  K_{i_{0}i_{p}}^{\quad\ k}C_{A_{0}A_{p}}^{\quad\ B}\alpha
_{j}K_{i_{1}\cdots i_{p-1}ki_{p+1}\cdots i_{n}}^{\quad\qquad\qquad
\;\;j}\left\vert \bm{T}_{A_{1}}\cdots\bm{T}_{p-1}\bm{T}_{B}\bm{T}_{p+1}%
\cdots\bm{T}_{A_{n}}\right\vert +\right. \\
+K_{i_{0}i_{p}}^{\quad\ N+1}C_{A_{0}A_{p}}^{\quad\ B}\alpha_{j}K_{i_{1}\cdots
i_{p-1}\left(  N+1\right)  i_{p+1}\cdots i_{n}}^{\quad\qquad\qquad\qquad
\;j}\left\vert \bm{T}_{A_{1}}\cdots\bm{T}_{p-1}\bm{T}_{B}\bm{T}_{p+1}%
\cdots\bm{T}_{A_{n}}\right\vert +\\
+K_{i_{0}i_{p}}^{\quad\ k}C_{A_{0}A_{p}}^{\quad\ B}\alpha_{N+1}K_{i_{1}\cdots
i_{p-1}ki_{p+1}\cdots i_{n}}^{\quad\qquad\qquad\qquad\;N+1}\left\vert
\bm{T}_{A_{1}}\cdots\bm{T}_{p-1}\bm{T}_{B}\bm{T}_{p+1}\cdots\bm{T}_{A_{n}%
}\right\vert +\\
\left.  +K_{i_{0}i_{p}}^{\quad\ N+1}C_{A_{0}A_{p}}^{\quad\ B}\alpha
_{N+1}K_{i_{1}\cdots i_{p-1}\left(  N+1\right)  i_{p+1}\cdots i_{n}}%
^{\quad\qquad\qquad\qquad\;N+1}\left\vert \bm{T}_{A_{1}}\cdots\bm{T}_{p-1}%
\bm{T}_{B}\bm{T}_{p+1}\cdots\bm{T}_{A_{n}}\right\vert \right)  =0,
\end{multline*}
en donde hemos separado las componentes valuadas en $\lambda_{N+1}$ y
$\lambda_{k}$ y $\alpha_{N+1}$ y $\alpha_{j}.$

Ahora bien, dado que%
\[
\lambda_{i_{1}}\cdots\lambda_{i_{p-1}}\lambda_{N+1}\lambda_{i_{p+1}}%
\cdots\lambda_{i_{n}}=\lambda_{N+1},
\]
entonces se tiene que
\[
K_{i_{1}\cdots i_{p-1}\left(  N+1\right)  i_{p+1}\cdots i_{n}}^{\quad
\qquad\qquad\qquad\;\;j}=0.
\]
As\'{\i}, cuando imponemos $\alpha_{N+1}=0,$ tenemos que
\begin{multline*}
\sum_{p=0}^{n}\left(  -1\right)  ^{\mathfrak{q}\left(  A_{0},i_{0}\right)
\left(  \mathfrak{q}\left(  A_{1},i_{1}\right)  +\cdots+\mathfrak{q}\left(
A_{p-1},i_{p-1}\right)  \right)  }\times\\
\times K_{i_{0}i_{p}}^{\quad\ k}C_{A_{0}A_{p}}^{\quad\ B}\alpha_{j}%
K_{i_{1}\cdots i_{p-1}ki_{p+1}\cdots i_{n}}^{\quad\qquad\qquad\;\;j}\left\vert
\bm{T}_{A_{1}}\cdots\bm{T}_{p-1}\bm{T}_{B}\bm{T}_{p+1}\cdots\bm{T}_{A_{n}%
}\right\vert =0
\end{multline*}
la cual corresponde justamente a la condici\'{o}n de invariancia para el
\'{a}lgebra $0_{S}$-reducida, de constantes de estructura $K_{ij}^{\quad
\ k}C_{AB}^{\quad\ C}.$ Por lo tanto, ec.~(\ref{Ec InvTensor 0s-Forz})
corresponde a un tensor invariante para \'{e}l \'{a}lgebra $0_{S}$-reducida.
\end{proof}

\begin{corollary}
Sea $S$ un semigrupo provisto de un elemento $0_{S},$ y sea $\left\vert
\mathfrak{\check{G}}_{\mathrm{R}}\right\vert $ el \'{a}lgebra $0_{S}$-reducida
de la sub\'{a}lgebra resonante $\mathfrak{G}_{\mathrm{R}}=\bigoplus_{p\in
I}S_{p}\otimes V_{p}.$ Sea $\left\{  \boldsymbol{T}_{A_{p}}\right\}  $ un
generador de $V_{p}$ y denotemos con un sub\'{\i}ndice $i_{p}$ a un elemento
arbitrario de $\check{S}_{p}=S_{p}-\left\{  0_{S}\right\}  ,$ $\lambda_{i_{p}%
}\in\check{S}_{p}.$ Entonces, la expresi\'{o}n%
\begin{equation}
\left\vert \bm{T}_{\left(  A_{p_{1}},i_{p_{1}}\right)  }\cdots\bm{T}_{\left(
A_{p_{n}},i_{p_{n}}\right)  }\right\vert =\alpha_{j}K_{i_{p_{1}}\cdots
i_{p_{n}}}^{\quad\ \ j}\left\vert \bm{T}_{A_{p_{1}}}\cdots\bm{T}_{A_{p_{n}}%
}\right\vert \label{Ec InvTens 0-Forz Subalg Reson}%
\end{equation}
corresponde a un tensor invariante para el $0_{S}$-reducci\'{o}n de la
sub\'{a}lgebra resonante.
\end{corollary}

\begin{proof}
La prueba sigue exactamente la misma l\'{\i}nea que la demostraci\'{o}n del
Teorema~\ref{Teo TensInv 0-Forz}; la \'{u}nica diferencia estriba en que se
debe utilizar como punto d partida el tensor invariante de la sub\'{a}lgebra
resonante, ec.~(\ref{Ec InvTensor Subalg Reson}).
\end{proof}

Para comparar estos tensores invariantes con la supertraza del \'{a}lgebra
$0_{S}$-reducida, consideremos su respectiva representaci\'{o}n matricial,%
\[
\left[  \bm{T}_{\left(  A,i\right)  }\right]  _{\left(  a,j\right)
}^{\phantom{\left( a,\mu \right)}\left(  b,k\right)  }=\left[  \lambda
_{i}\right]  _{j}^{\ k}\left[  \bm{T}_{A}\right]  _{a}^{\phantom{a}b}%
\]
con%
\[
\left[  \lambda_{i}\right]  _{j}^{\ k}=K_{ji}^{\quad k}.
\]

Entonces, tenemos que%
\[
\operatorname*{STr}\left(  \bm{T}_{\left(  A_{1},i_{1}\right)  }%
\cdots\bm{T}_{\left(  A_{n},i_{n}\right)  }\right)  =K_{j_{1}i_{1}%
}^{\phantom{j_{1}i_{1}}j_{2}}K_{j_{2}i_{2}}^{\phantom{j_{2}i_{2}}j_{3}}\cdots
K_{j_{n-1}i_{n-1}}^{\phantom{j_{n-1}i_{n-1}}j_{n}}K_{j_{n}i_{n}}%
^{\phantom{j_{n}i_{n}}j_{1}}\operatorname*{Str}\left(  \bm{T}_{A_{1}}%
\cdots\bm{T}_{A_{n}}\right)  ,
\]
y por lo tanto, dado que para todo $\lambda_{i},\lambda_{j}$ tales que
$\lambda_{i}\lambda_{j}=\lambda_{k\left(  i,j\right)  },$ se tiene que
$\lambda_{i},\lambda_{j}\neq\lambda_{N+1},$ tenemos que%
\[
\operatorname*{STr}\left(  \bm{T}_{\left(  A_{1},i_{1}\right)  }%
\cdots\bm{T}_{\left(  A_{n},i_{n}\right)  }\right)  =K_{j_{1}i_{1}\cdots
i_{n}}^{\phantom{j_{1} i_{1} \cdots i_{n}}j_{1}}\operatorname*{Str}\left(
\bm{T}_{A_{1}}\cdots\bm{T}_{A_{n}}\right)  .
\]

En general, esta expresi\'{o}n tiene muy pocas componentes no nulas, cuando se
compara con ecs.~(\ref{Ec STr (Con Cero)}) \'{o}~(\ref{Ec InvTensorS-Exp}). En
efecto, por simplicidad consideremos un semigrupo $S=\left\{  \lambda_{\alpha
}\right\}  _{\alpha=0}^{N+1}$ en donde se tiene que $\lambda_{N+1}=0_{S},$
$\lambda_{i}\neq0,$ $i=0,\ldots,N,$ y en donde adem\'{a}s impodremos que
existe un elemento identidad, $\lambda_{0}=e$, y que para todo $\lambda
_{i},\lambda_{j}\neq e,.\lambda_{i}\lambda_{j}\neq\lambda_{j}$ (este es por
ejemplo justamente el caso de $S_{\mathrm{E}}^{\left(  N\right)  }$).
Entonces, tenemos que $K_{j_{1}i_{1}\cdots i_{n}}%
^{\phantom{j_{1} i_{1} \cdots i_{n}}j_{1}}=K_{i_{1}\cdots i_{n}}%
^{\phantom{i_{1}\cdots i_{n}}0},$ y por lo tanto, se tiene como \'{u}nica
componente no nula
\[
\operatorname*{STr}\left(  \bm{T}_{\left(  A_{1},i_{1}\right)  }%
\cdots\bm{T}_{\left(  A_{n},i_{n}\right)  }\right)  =K_{i_{1}\cdots i_{n}%
}^{\phantom{i_{1}\cdots i_{n}}0}\operatorname*{Str}\left(  \bm{T}_{A_{1}%
}\cdots\bm{T}_{A_{n}}\right)  .
\]

Para el caso en particular de $S_{\mathrm{E}}^{\left(  N\right)  },$ esto
corresponde a s\'{o}lo%
\[
\operatorname*{STr}\left(  \bm{T}_{\left(  A_{1},0\right)  }\cdots
\bm{T}_{\left(  A_{n},0\right)  }\right)  =\operatorname*{Str}\left(
\bm{T}_{A_{1}}\cdots\bm{T}_{A_{n}}\right)  .
\]

Una teor\'{\i}a de Chern--Simons constru\'{\i}da a partir de este tensor
invariante para el \'{A}lgebra M tendr\'{\i}a como \'{u}nico campo
din\'{a}mico la conexi\'{o}n de esp\'{\i}n; no ser\'{\i}a posible incluir el
vielbein, campos bos\'{o}nicos extra ni campos fermi\'{o}nicos en la
acci\'{o}n. Por ende, la existencia de un tensor invariante distinto de la
supertraza, como el de ec.~(\ref{Ec InvTens 0-Forz Subalg Reson}) es de vital
importancia en la construcci\'{o}n de un Lagrangeano, tal como veremos en la
pr\'{o}xima secci\'{o}n. Sin embargo, debe de observarse que pese a que
ec.~(\ref{Ec InvTens 0-Forz Subalg Reson}) tiene muchas m\'{a}s componentes
que la supertraza, a\'{u}n as\'{\i} tiene muchas menos componentes que las de
ec.~(\ref{Ec InvTensor Subalg Reson}). Esto parece ser inherente al proceso de
$0_{S}$-reducci\'{o}n en s\'{\i} mismo, y limitar\'{a} la din\'{a}mica de la
correspondiente teor\'{\i}a, como veremos en el pr\'{o}ximo cap\'{\i}tulo.
Esto parece sugerir fuertemente el uso de \'{a}lgebras que no correspondan a
\'{a}lgebras $0_{S}$-reducidas, sino m\'{a}s bien a sub\'{a}lgebras resonantes
cuando se trate de construir principios de acci\'{o}n.

\chapter{\label{SecAcc_M_Alg}Principio de Acci\'{o}n para el \'{A}lgebra~M.}

\bigskip

\begin{center}
\textit{\textquotedblleft A possible explanation of the physicist's use of
mathematics to formulate his laws of nature is that he is a somewhat
irresponsible person. As a result, when he finds a connection between two
quantities which resembles a connection well-known from mathematics, he will
jump at the conclusion that the connection}\textbf{is} \textit{that
discussed in mathematics simply because he does not know of any other similar
connection. It is not the intention of the present discussion to refute the
charge that the physicist is a somewhat irresponsible person. Perhaps he is.
However, it is important to point out that the mathematical formulation of the
physicist's often crude experience leads in an uncanny number of cases to an
amazingly accurate description of a large class of phenomena. This shows that
the mathematical language has more to commend it than being the only language
which we can speak; it show that it is, in a very real sense, the correct
language.\textquotedblright}

(Eugene P. Wigner, en The Unreasonable Effectiveness of Mathematics in the
Natural Sciences\footnote{\textquotedblleft \textit{Una posible explicaci\'{o}n de el por qu\'{e} el f\'{\i}sico usa las matem\'{a}ticas para formular sus leyes de la naturaleza es que \'{e}l es una persona un tanto irresponsable. Como resultado, cuando encuentra una conexi\'{o}n entre dos cantidades que se parece a una conexi\'{o}n bien conocida de las matem\'{a}ticas, el saltar\'{a} a la conclusi\'{o}n de que la conexi\'{o}n }\textbf{es} \textit{aquella discutida en matem\'{a}ticas simplemente porque no conoce alguna otra conexi\'{o}n similar. No es la intenci\'{o}n de la presente discusi\'{o}n refutar la acusaci\'{o}n de que el f\'{\i}sico es algo irresponsable. Quiz\'{a}s lo es. Sin embargo, es importante se\~{n}alar que la formulaci\'{o}n matem\'{a}tica de la a menudo cruda experiencia dirige en un n\'{u}mero extraordinario de casos a una descripci\'{o}n asombrosamente exacta de un gran conjunto de fen\'{o}menos. Esto muestra que el lenguaje matem\'{a}tico tiene m\'{a}s a su favor que ser tan s\'{o}lo el lenguaje que sabemos hablar; al contrario, esto demuestra en un sentido muy real que \'{e}ste es el lenguaje correcto}\textquotedblright})
\end{center}

\begin{quotation}
\medskip
\end{quotation}

En este cap\'{\i}tulo, utilzaremos las herramientas matem\'{a}ticas
desarrolladas a lo largo de la tesis para escribir una teor\'{\i}a de gauge
para el \'{A}lgebra~M en $D=11,$ utilizando un lagrangeano transgresor. En
pocas palabras, utilizaremos el tensor invariante
ec.~(\ref{Ec InvTens 0-Forz Subalg Reson}) para construir un Lagrangeano
Transgresor con dos conexiones de gauge, tal como fue descrito en
Sec.~\ref{Sec Trans y CS como Lagrang}, en donde el principio de acci\'{o}n
estar\'{a} definido como la integral de la forma de transgresi\'{o}n sobre una
variedad $11$ dimensional, [ve\'{a}se ec.~(\ref{Ec Def Accion M+ M-})]. La
din\'{a}mica de una teor\'{\i}a de transgresi\'{o}n es un problema altamente
no trivial, debido a la no-linealidad del
lagrangeano. En particular, la obtenci\'{o}n de la din\'{a}mica en cuatro
dimensiones a partir de la acci\'{o}n en 11 dimensiones resulta en este
contexto un problema no-trivial particularmente interesante. Aqu\'{\i}
utilizaremos el enfoque mostrado en
Refs.~\cite{CECS-MAlgNoether-1,CECS-MAlgNoether-2}, en donde el criterio
fundamental para buscar un \textquotedblleft vac\'{\i}o\textquotedblright%
\ para la teor\'{\i}a es la existencia de perturbaciones propagantes en torno
a \'{e}ste. La forma de esta din\'{a}mica cuadridimensional estar\'{a}
fuertemente restringida como consecuencia directa del $0_{S}$-reducci\'{o}n
necesario para constru\'{\i}r el \'{A}lgebra~M, por lo que se vuelve
interesante considerar otras \'{a}lgebras de la misma \textquotedblleft
familia\textquotedblright\ las cuales no sean $0_{S}$-reducidas.

\section{El \'{A}lgebra~M}

\subsection{\label{Sec Sub Alg Reson Original}Sub\'{a}lgebra Resonante
original de $S_{\mathrm{E}}^{\left(  2\right)  }\otimes\mathfrak{osp}\left(
\mathfrak{32}|\mathfrak{1}\right)  $ y $0_{S}$-reducci\'{o}n}

En Sec.~\ref{Sec Algebra M}, reobtuvimos el \'{A}lgebra~M como el $0_{S}%
$-reducci\'{o}n de una sub\'{a}lgebra resonante de $S_{\mathrm{E}}^{\left(
2\right)  }\otimes\mathfrak{osp}\left(  \mathfrak{32}|\mathfrak{1}\right)  .$
Recordemos que la descomposici\'{o}n resonante de $S_{\mathrm{E}}^{\left(
2\right)  }=\left\{  \lambda_{0},\lambda_{1},\lambda_{2},\lambda_{3}\right\}
,$ $S_{\mathrm{E}}^{\left(  2\right)  }=S_{0}\cup S_{1}\cup S_{2}$ corresponde
a%
\begin{align*}
S_{0}  &  =\left\{  \lambda_{0},\lambda_{2},\lambda_{3}\right\}  ,\\
S_{1}  &  =\left\{  \lambda_{1},\lambda_{3}\right\}  ,\\
S_{2}  &  =\left\{  \lambda_{2},\lambda_{3}\right\}  .
\end{align*}

Para evitar una proliferaci\'{o}n de sub\'{\i}ndices, re-etiquetemos los
generadores tal como se muestra en Tabla~\ref{Tabla AlgReson Proto M},

\begin{table}[ptb]
\begin{center}%
\begin{tabular}
[c]{cr@{$\; = \;$}l}\hline\hline
Subespacios de $\mathfrak{G}_{\text{R}}$ & \multicolumn{2}{c}{Generadores}%
\\\hline
\multirow{3}*{$S_{0} \otimes V_{0}$} & $\bm{J}_{ab}$ & $\lambda_{0}
\bm{J}_{ab}^{\left(  \mathfrak{osp} \right)  }$\\
& $\bm{Z}_{ab}$ & $\lambda_{2} \bm{J}_{ab}^{\left(  \mathfrak{osp} \right)  }%
$\\
& $\bm{Z}_{ab}^{\prime}$ & $\lambda_{3} \bm{J}_{ab}^{\left(  \mathfrak{osp}
\right)  }$\\\hline
\multirow{2}*{$S_{1} \otimes V_{1}$} & $\bm{Q}$ & $\lambda_{1} \bm{Q}^{\left(
\mathfrak{osp} \right)  }$\\
& $\bm{Q}^{\prime}$ & $\lambda_{3} \bm{Q}^{\left(  \mathfrak{osp} \right)  }%
$\\\hline
\multirow{4}*{$S_{2} \otimes V_{2}$} & $\bm{P}_{a}$ & $\lambda_{2}
\bm{P}_{a}^{\left(  \mathfrak{osp} \right)  }$\\
& $\bm{Z}_{abcde}$ & $\lambda_{2} \bm{Z}_{abcde}^{\left(  \mathfrak{osp}
\right)  }$\\
& $\bm{P}_{a}^{\prime}$ & $\lambda_{3} \bm{P}_{a}^{\left(  \mathfrak{osp}
\right)  }$\\
& $\bm{Z}_{abcde}^{\prime}$ & $\lambda_{3} \bm{Z}_{abcde}^{\left(
\mathfrak{osp} \right)  }$\\\hline\hline
\end{tabular}
\end{center}
\caption{El \'{A}lgebra M proviene del $0_{S}$-reducci\'{o}n de una
sub\'{a}lgebra resonante de $S_{\mathrm{E}}^{(2)}\otimes\mathfrak{osp}\left(
\mathfrak{32}|\mathfrak{1}\right)  .$ La tabla muestra la relaci\'{o}n entre
los generadores de esta sub\'{a}lgebra resonante y los generadores de
$\mathfrak{osp}\left(  \mathfrak{32}|\mathfrak{1}\right)  .$ Los tres niveles
corresponden a las tres columnas de Fig. \ref{Fig ProtoAlgM}, o lo que es lo mismo, a cada uno
de los subconjuntos en los cuales $S_{\mathrm{E}}^{(2)}$ fue subdividido. }%
\label{Tabla AlgReson Proto M}%
\end{table}

Los conmutadores de esta sub\'{a}lgebra resonante tienen la forma expl\'{\i}cita%
\begin{align}
\left[  \boldsymbol{P}_{a},\boldsymbol{P}_{b}\right]   &  =\boldsymbol{Z}%
_{ab}^{\prime},\label{Ec ProtoAlgM EcPP=Z}\\
\left[  \boldsymbol{J}_{ab},\boldsymbol{P}_{c}\right]   &  =\eta_{ce}%
\delta_{ab}^{de}\boldsymbol{P}_{d},\\
\left[  \boldsymbol{J}_{ab},\boldsymbol{J}_{cd}\right]   &  =\eta_{gh}%
\delta_{ab}^{eg}\delta_{cd}^{hf}\boldsymbol{J}_{ef}.
\end{align}%
\begin{align}
\left[  \boldsymbol{J}_{ab},\boldsymbol{Z}_{cd}\right]   &  =\eta_{gh}%
\delta_{ab}^{eg}\delta_{cd}^{hf}\boldsymbol{Z}_{ef},\\
\left[  \boldsymbol{Z}_{ab},\boldsymbol{Z}_{cd}\right]   &  =\eta_{gh}%
\delta_{ab}^{eg}\delta_{cd}^{hf}\boldsymbol{Z}_{ef}^{\prime},\\
\left[  \boldsymbol{P}_{a},\boldsymbol{Z}_{b_{1}\cdots b_{5}}\right]   &
=-\frac{1}{5!}\varepsilon_{ab_{1}\cdots b_{5}c_{1}\cdots c_{5}}\boldsymbol{Z}%
^{\prime c_{1}\cdots c_{5}},\\
\left[  \boldsymbol{J}^{ab},\boldsymbol{Z}_{c_{1}\cdots c_{5}}\right]   &
=\frac{1}{4!}\delta_{dc_{1}\cdots c_{5}}^{abe_{1}\cdots e_{4}}\boldsymbol{Z}%
_{\phantom{d}e_{1}\cdots e_{4}}^{d},\\
\left[  \boldsymbol{Z}^{ab},\boldsymbol{Z}_{c_{1}\cdots c_{5}}\right]   &
=\frac{1}{4!}\delta_{dc_{1}\cdots c_{5}}^{abe_{1}\cdots e_{4}}\boldsymbol{Z}%
_{\phantom{d}e_{1}\cdots e_{4}}^{\prime d}.
\end{align}%
\begin{align}
\left[  \boldsymbol{Z}^{a_{1}\cdots a_{5}},\boldsymbol{Z}_{b_{1}\cdots b_{5}%
}\right]   &  =\eta^{\left[  a_{1}\cdots a_{5}\right]  \left[  c_{1}\cdots
c_{5}\right]  }\varepsilon_{c_{1}\cdots c_{5}b_{1}\cdots b_{5}e}%
\boldsymbol{P}^{\prime e}+\delta_{db_{1}\cdots b_{5}}^{a_{1}\cdots a_{5}%
e}\boldsymbol{Z}_{\phantom{d}e}^{\prime d}+\nonumber\\
&  -\frac{1}{3!3!5!}\varepsilon_{c_{1}\cdots c_{11}}\delta_{d_{1}d_{2}%
d_{3}b_{1}\cdots b_{5}}^{a_{1}\cdots a_{5}c_{4}c_{5}c_{6}}\eta^{\left[
c_{1}c_{2}c_{3}\right]  \left[  d_{1}d_{2}d_{3}\right]  }\boldsymbol{Z}%
^{\prime c_{7}\cdots c_{11}},
\end{align}%
\begin{align}
\left[  \boldsymbol{P}_{a},\boldsymbol{Q}\right]   &  =-\frac{1}{2}\Gamma
_{a}\boldsymbol{Q}^{\prime},\\
\left[  \boldsymbol{J}_{ab},\boldsymbol{Q}\right]   &  =-\frac{1}{2}%
\Gamma_{ab}\boldsymbol{Q},\\
\left[  \boldsymbol{Z}_{ab},\boldsymbol{Q}\right]   &  =-\frac{1}{2}%
\Gamma_{ab}\boldsymbol{Q}^{\prime},\\
\left[  \boldsymbol{Z}_{abcde},\boldsymbol{Q}\right]   &  =-\frac{1}{2}%
\Gamma_{abcde}\boldsymbol{Q}^{\prime},
\end{align}%
\begin{equation}
\left\{  \boldsymbol{Q},\bar{\boldsymbol{Q}}\right\}  =\frac{1}{8}\left(
\Gamma^{a}\boldsymbol{P}_{a}-\frac{1}{2}\Gamma^{ab}\boldsymbol{Z}_{ab}%
+\frac{1}{5!}\Gamma^{abcde}\boldsymbol{Z}_{abcde}\right)
\label{Ec ProtoAlgM EcQQ=de todo}%
\end{equation}
y los conmutadores involucrando un generador con $\prime$ est\'{a}n valuados
en los correspondientes generadores con $\prime;$ en particular,
$\boldsymbol{Z}_{ab}^{\prime},$ $\boldsymbol{P}_{a}^{\prime},$ $\boldsymbol{Z}%
_{a_{1}\cdots a_{5}}^{\prime}$ y $\boldsymbol{Q}^{\prime}$ forman una
sub\'{a}lgebra $\mathfrak{osp}\left(  \mathfrak{32}|\mathfrak{1}\right)  .$

\begin{figure}[ptb]
\begin{center}
\psfrag{Jab}{$\bm{J}_{ab}$}
\psfrag{Zab}{$\bm{Z}_{ab}$}
\psfrag{Zcab}{$\bm{Z^{\prime}}_{ab}$}
\psfrag{Q}{$\bm{Q}$}
\psfrag{Q2}{$\bm{Q^{\prime}}$}
\psfrag{Z5}{$\bm{Z}_{abcde}$}
\psfrag{Pa}{$\bm{P}_{a}$}
\psfrag{Zc5}{$\bm{Z^{\prime}}_{abcde}$}
\psfrag{Pca}{$\bm{P^{\prime}}_{a}$}
\includegraphics[width=.5\textwidth]{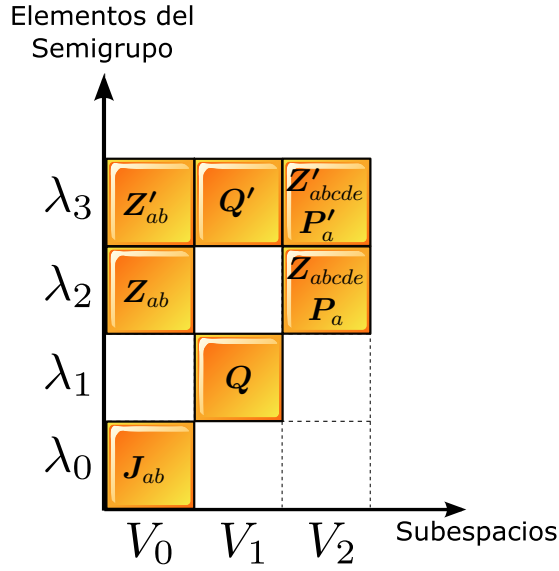}
\caption{Sub\'{a}lgebra resonante de la cual proviene el \'{A}lgebra M a trav\'{e}s de $0_{S}$-Reducci\'{o}n; en particular, obs\'{e}rvese que la fila superior es isom\'{o}rfica a $\mathfrak{osp}\left(  \mathfrak{32}|\mathfrak{1}\right).$}
\label{Fig ProtoAlgM}
\end{center}
\end{figure}

Imponiendo la condici\'{o}n de $0_{S}$-reducci\'{o}n sobre la sub\'{a}lgebra
resonante, $\boldsymbol{Z}_{ab}^{\prime}=\boldsymbol{P}_{a}^{\prime
}=\boldsymbol{Z}_{a_{1}\cdots a_{5}}^{\prime}=\boldsymbol{Q}^{\prime}=0,$
recuperamos el \'{A}lgebra~M [Sec.~\ref{Sec Algebra M}].

\subsection{Conexiones y Curvaturas}

En la construcci\'{o}n se utilizar\'{a}n dos $1$-formas conexi\'{o}n, $\bm{A}
$ y $\bar{\bm{A}}$ valuadas en el \'{A}lgebra M. Descompondremos estos campos
de la forma
\begin{align}
\bm{A}  &  =\bm{\omega}+\bm{e}+\bm{b}_{2}+\bm{b}_{5}+\bar{\bm{\psi}},\\
\bar{\bm{A}}  &  =\bar{\bm{\omega}}+\bar{\bm{e}}+\bar{\bm{b}}_{2}+\bar
{\bm{b}}_{5}+\bar{\bm{\chi}},
\end{align}
donde cada t\'{e}rmino est\'{a} valuado en un subespacio distinto,%
\begin{align}
\bm{\omega}  &  =\frac{1}{2}\omega^{ab}\bm{J}_{ab},\\
\bm{e}  &  =e^{a}\bm{P}_{a},\\
\bm{b}_{2}  &  =\frac{1}{2}b_{2}^{ab}\bm{Z}_{ab},\\
\bm{b}_{5}  &  =\frac{1}{5!}b_{5}^{abcde}\bm{Z}_{abcde},\\
\bar{\bm{\psi}}  &  =\bar{\psi}_{\alpha}\bm{Q}^{\alpha},
\end{align}
y as\'{\i} mismo para $\bar{\bm{A}}$. Para recuperar gravedad en este
contexto, debemos identificar $\omega^{ab}$ con una conexi\'{o}n de
esp\'{\i}n, y $e^{a}$ con un \textit{elfbein}. Lo mismo ocurre para
$\bar{\omega}^{ab} $ y $\bar{e}^{a},$ los cuales estar\'{a}n asociados a la
orientaci\'{o}n contraria de la variedad. La curvatura para $\bm{A}$ es de la
forma
\begin{equation}
\bm{F}=\bm{R}+\bm{F}_{P}+\bm{F}_{2}+\bm{F}_{5}+\mathrm{D}_{\omega}%
\bar{\bm{\psi}},
\end{equation}
donde cada t\'{e}rmino corresponde a
\begin{align}
\bm{R}  &  =\frac{1}{2}R^{ab}\bm{J}_{ab},\\
\bm{F}_{P}  &  =\left(  T^{a}+\frac{1}{16}\bar{\psi}\Gamma^{a}\psi\right)
\bm{P}_{a},\\
\bm{F}_{2}  &  =\frac{1}{2}\left(  \mathrm{D}_{\omega}b^{ab}-\frac{1}{16}%
\bar{\psi}\Gamma^{ab}\psi\right)  \bm{Z}_{ab},\\
\bm{F}_{5}  &  =\frac{1}{5!}\left(  \mathrm{D}_{\omega}b^{a_{1}\cdots a_{5}%
}+\frac{1}{16}\bar{\psi}\Gamma^{a_{1}\cdots a_{5}}\psi\right)  \bm{Z}_{a_{1}%
\cdots a_{5}},\\
\mathrm{D}_{\omega}\bar{\bm{\psi}}  &  =\mathrm{D}_{\omega}\bar{\psi}\bm{Q}.
\end{align}

Aqu\'{\i}, hemos escrito la derivada covarienate de lorentz para los espinores
de la forma usual,
\begin{align}
\mathrm{D}_{\omega}\psi &  =\mathrm{d}\psi+\frac{1}{4}\omega^{ab}\Gamma
_{ab}\psi,\\
\mathrm{D}_{\omega}\bar{\psi}  &  =\mathrm{d}\bar{\psi}-\frac{1}{4}\omega
^{ab}\bar{\psi}\Gamma_{ab}.
\end{align}

\subsection{Tensor Invariante}

Un lagrangeano transgresor en 11 dimensiones corresponde a la elecci\'{o}n
$n=5$ en ec.(\ref{Ec LagrangT 2}),%
\[
L_{\mathrm{T}}^{\left(  11\right)  }\left(  \boldsymbol{A},\bar{\boldsymbol{A}%
}\right)  =6k\int_{t=0}^{t=1}\mathrm{d}t~\left\langle \boldsymbol{\Theta
F}_{t}^{5}\right\rangle ,
\]
y por lo tanto, se requiere un tensor invariante con $6$ componentes. Dado que
el \'{A}lgebra M corresponde al $0_{S}$-reducci\'{o}n de una sub\'{a}lgebra
resonante, debemos de utilizar el corolario~\ref{Cor O-Forz SubAlg Reson} para
constru\'{\i}r un tensor invariante para \'{e}sta, a partir de un tensor
invariante de $\mathfrak{osp}\left(  \mathfrak{32}|\mathfrak{1}\right)  $.
Como tensor invariante de $\mathfrak{osp}\left(  \mathfrak{32}|\mathfrak{1}%
\right)  $ usaremos la supertraza en la representaci\'{o}n en supermatrices
est\'{a}ndar en $11$ dimensiones, y en donde usaremos como base para el sector
bos\'{o}nico del \'{a}lgebra, $\mathfrak{sp}\left(  \mathfrak{32}\right)  ,$
matrices de Dirac de $32\times32$ componentes\footnote{Hay otras alternativas
para realizar el \'{a}lgebra $\mathfrak{osp}\left(  \mathfrak{32}%
|\mathfrak{1}\right)  $ en forma matricial adem\'{a}s de \'{e}sta; una
alternativa interesante consiste en considerar como base matrices de Dirac en 12
dimensiones ($64\times64$ componentes) y usar la representaci\'{o}n asociada de supermatrices. (Ve\'{a}se por ejemplo el
enfoque seguido en Ref. \cite{TesisTroncoso})} en la representaci\'{o}n $\Gamma_{1}%
\cdots\Gamma_{11}=+\mathbbm{1}$ (Ve\'{a}se Ap\'{e}ndice~\ref{Apend Dirac}).
Para $S_{\mathrm{E}}^{\left(  2\right)  },$ las componentes de los
$6$-selectores son de la forma%
\[
K_{\alpha_{1}\cdots\alpha_{6}}^{\quad\ \ \gamma}=\delta_{H_{3}\left(
\alpha_{1}+\cdots+\alpha_{6}\right)  }^{\gamma},
\]
y por lo tanto las componentes $K_{i_{1}\cdots i_{6}}^{\quad\ \ j}$
corresponden a $K_{i_{1}\cdots i_{6}}^{\quad\ \ j}=\delta_{i_{1}+\cdots+i_{6}%
}^{j}$. As\'{\i}, el tensor sim\'{e}trico invariante tiene la forma
[ec.~(\ref{Ec InvTens 0-Forz Subalg Reson})],%
\[
\left\langle \bm{T}_{\left(  A_{p_{1}},i_{p_{1}}\right)  }\cdots
\bm{T}_{\left(  A_{p_{6}},i_{p_{6}}\right)  }\right\rangle =\alpha_{j}%
\delta_{i_{p_{1}}+\cdots+i_{p_{6}}}^{j}\left\langle \bm{T}_{A_{p_{1}}}%
\cdots\bm{T}_{A_{p_{6}}}\right\rangle ,\qquad p=0,1,2,
\]
o m\'{a}s expl\'{\i}citamente,
\begin{align}
\left\langle \bm{J}_{a_{1}b_{1}}\cdots\bm{J}_{a_{6}b_{6}}\right\rangle
_{\mathrm{M}}  &  =\alpha_{0}\left\langle \bm{J}_{a_{1}b_{1}}\cdots
\bm{J}_{a_{6}b_{6}}\right\rangle _{\mathfrak{osp}},\label{itmalg1}\\
\left\langle \bm{J}_{a_{1}b_{1}}\cdots\bm{J}_{a_{5}b_{5}}\bm{P}_{c}%
\right\rangle _{\mathrm{M}}  &  =\alpha_{2}\left\langle \bm{J}_{a_{1}b_{1}%
}\cdots\bm{J}_{a_{5}b_{5}}\bm{P}_{c}\right\rangle _{\mathfrak{osp}},\\
\left\langle \bm{J}_{a_{1}b_{1}}\cdots\bm{J}_{a_{5}b_{5}}\bm{Z}_{a_{6}b_{6}%
}\right\rangle _{\mathrm{M}}  &  =\alpha_{2}\left\langle \bm{J}_{a_{1}b_{1}%
}\cdots\bm{J}_{a_{6}b_{6}}\right\rangle _{\mathfrak{osp}},\\
\left\langle \bm{J}_{a_{1}b_{1}}\cdots\bm{J}_{a_{5}b_{5}}\bm{Z}_{c_{1}\cdots
c_{5}}\right\rangle _{\mathrm{M}}  &  =\alpha_{2}\left\langle \bm{J}_{a_{1}%
b_{1}}\cdots\bm{J}_{a_{5}b_{5}}\bm{Z}_{c_{1}\cdots c_{5}}\right\rangle
_{\mathfrak{osp}},\\
\left\langle \bm{QJ}_{a_{1}b_{1}}\cdots\bm{J}_{a_{4}b_{4}}\bar{\bm{Q}}%
\right\rangle _{\mathrm{M}}  &  =\alpha_{2}\left\langle \bm{QJ}_{a_{1}b_{1}%
}\cdots\bm{J}_{a_{4}b_{4}}\bar{\bm{Q}}\right\rangle _{\mathfrak{osp}%
},\label{itmalg5}%
\end{align}
en donde $\alpha_{0}$ y $\alpha_{2}$ son constantes arbitrarias. No existe un
trozo proporcional a una constante $\alpha_{1}$ pues este hubiese involucrado
un tensor invariante de $\mathfrak{osp}\left(  \mathfrak{32}|\mathfrak{1}%
\right)  $ valuado en un s\'{o}lo generador fermi\'{o}nico y 5 generadores de
Lorentz. De la representaci\'{o}n en supermatrices de los generadores de
$\mathfrak{osp}\left(  \mathfrak{32}|\mathfrak{1}\right)  $ es directo ver que
cualquier supertraza valuada en un n\'{u}mero impar de generadores
fermi\'{o}nicos es cero, y por lo tanto, tampoco puede haber una componente
tal en el tensor invariante para el \'{A}lgebra~M. A\'{u}n m\'{a}s, a\'{u}n si
existiese un tensor invariante no nulo valuado en un n\'{u}mero impar de
generadores fermi\'{o}nicos, este dar\'{\i}a origen a un Lagrangeano de
Chern--Simons o Transgresi\'{o}n que corresponder\'{\i}a a un n\'{u}mero de
Grassmann.

Es importante notar que el tensor invariante ecs.~(\ref{itmalg1}%
)-(\ref{itmalg5}) tiene muchas menos componentes que el de la sub\'{a}lgebra
resonante de $S_{\mathrm{E}}^{\left(  2\right)  }\otimes\mathfrak{osp}\left(
\mathfrak{32}|\mathfrak{1}\right)  $ antes del $0_{S}$-reducci\'{o}n (faltan
todas las componentes proporcionales a $\alpha_{3}$). Tal como veremos en
sec.~\ref{Sec Cuadridinamics}, esto tendr\'{a} una gran influencia en la forma
de la din\'{a}mica inducida en cuatro dimensiones.

Puesto que siempre estaremos interesados en \textit{polinomios} invariantes
(El lagrangeano por ejemplo), m\'{a}s
\'{u}til que la forma expl\'{\i}cita del \textit{tensor} invariante de
$\mathfrak{osp}\left(  \mathfrak{32}|\mathfrak{1}\right)  $ resulta el
considerar algunas de las componentes de un polinomio invariante arbitrario
valuado en la supertraza simetrizada. Un c\'{a}lculo expl\'{\i}cito entrega
(El c\'{a}lculo de la supertraza y su simetrizaci\'{o}n est\'{a} lejos de ser
trivial, ve\'{a}se Ap\'{e}ndice~\ref{Apend Dirac})%
\begin{equation}
L_{1}^{a_{1}b_{1}}\cdots L_{5}^{a_{5}b_{5}}B_{1}^{c}\left\langle
\bm{J}_{a_{1}b_{1}}\cdots\bm{J}_{a_{5}b_{5}}\bm{P}_{c}\right\rangle
_{\mathfrak{osp}}=\frac{1}{2}\varepsilon_{a_{1}\cdots a_{11}}L_{1}^{a_{1}%
a_{2}}\cdots L_{5}^{a_{9}a_{10}}B_{1}^{a_{11}},\label{L5B1}%
\end{equation}%
\begin{align}
&  L_{1}^{a_{1}b_{1}}\cdots L_{6}^{a_{6}b_{6}}\left\langle \bm{J}_{a_{1}b_{1}%
}\cdots\bm{J}_{a_{6}b_{6}}\right\rangle _{\mathfrak{osp}}\nonumber\\
&  =\frac{1}{3}\sum_{\sigma\in S_{6}}\left[  \frac{1}{4}\operatorname*{Tr}%
\left(  L_{\sigma\left(  1\right)  }L_{\sigma\left(  2\right)  }\right)
\operatorname*{Tr}\left(  L_{\sigma\left(  3\right)  }L_{\sigma\left(
4\right)  }\right)  \operatorname*{Tr}\left(  L_{\sigma\left(  5\right)
}L_{\sigma\left(  6\right)  }\right)  +\right. \nonumber\\
&  -\operatorname*{Tr}\left(  L_{\sigma\left(  1\right)  }L_{\sigma\left(
2\right)  }L_{\sigma\left(  3\right)  }L_{\sigma\left(  4\right)  }\right)
\operatorname*{Tr}\left(  L_{\sigma\left(  5\right)  }L_{\sigma\left(
6\right)  }\right)  +\nonumber\\
&  \left.  +\frac{16}{15}\operatorname*{Tr}\left(  L_{\sigma\left(  1\right)
}L_{\sigma\left(  2\right)  }L_{\sigma\left(  3\right)  }L_{\sigma\left(
4\right)  }L_{\sigma\left(  5\right)  }L_{\sigma\left(  6\right)  }\right)
\right]  ,\label{L6}%
\end{align}%
\begin{align}
&  L_{1}^{a_{1}b_{1}}\cdots L_{5}^{a_{5}b_{5}}B_{5}^{c_{1}\cdots c_{5}%
}\left\langle \bm{J}_{a_{1}b_{1}}\cdots\bm{J}_{a_{5}b_{5}}\bm{Z}_{c_{1}\cdots
c_{5}}\right\rangle _{\mathfrak{osp}}\nonumber\\
&  =\frac{1}{3}\varepsilon_{a_{1}\cdots a_{11}}\sum_{\sigma\in S_{5}}\left[
-\frac{5}{4}L_{\sigma\left(  1\right)  }^{a_{1}a_{2}}\cdots L_{\sigma\left(
4\right)  }^{a_{7}a_{8}}\left[  L_{\sigma\left(  5\right)  }\right]
_{bc}B_{5}^{bca_{9}a_{10}a_{11}}+\right. \nonumber\\
&  +10L_{\sigma\left(  1\right)  }^{a_{1}a_{2}}L_{\sigma\left(  2\right)
}^{a_{3}a_{4}}L_{\sigma\left(  3\right)  }^{a_{5}a_{6}}\left[  L_{\sigma
\left(  4\right)  }\right]  _{\;\;b}^{a_{7}}\left[  L_{\sigma\left(  5\right)
}\right]  _{\;\;c}^{a_{8}}B_{5}^{bca_{9}a_{10}a_{11}}+\nonumber\\
&  +\frac{1}{4}L_{\sigma\left(  1\right)  }^{a_{1}a_{2}}L_{\sigma\left(
2\right)  }^{a_{3}a_{4}}L_{\sigma\left(  3\right)  }^{a_{5}a_{6}}B_{5}%
^{a_{7}\cdots a_{11}}\operatorname*{Tr}\left(  L_{\sigma\left(  4\right)
}L_{\sigma\left(  5\right)  }\right)  +\nonumber\\
&  \left.  -L_{\sigma\left(  1\right)  }^{a_{1}a_{2}}L_{\sigma\left(
2\right)  }^{a_{3}a_{4}}\left[  L_{\sigma\left(  3\right)  }L_{\sigma\left(
4\right)  }L_{\sigma\left(  5\right)  }\right]  ^{a_{5}a_{6}}B_{5}%
^{a_{7}\cdots a_{11}}\right]  ,\label{L5B5}%
\end{align}%
\begin{align}
&  L_{1}^{a_{1}b_{1}}\cdots L_{4}^{a_{4}b_{4}}\bar{\chi}_{\alpha}\zeta^{\beta
}\left\langle \bm{Q}^{\alpha}\bm{J}_{a_{1}b_{1}}\cdots\bm{J}_{a_{4}b_{4}}%
\bar{\bm{Q}}_{\beta}\right\rangle _{\mathfrak{osp}}\nonumber\\
&  =-\frac{1}{240}\varepsilon_{a_{1}\cdots a_{8}abc}L_{1}^{a_{1}a_{2}}\cdots
L_{4}^{a_{7}a_{8}}\bar{\chi}\Gamma^{abc}\zeta+\nonumber\\
&  +\frac{1}{60}\sum_{\sigma\in S_{4}}\left[  \frac{3}{4}\operatorname*{Tr}%
\left(  L_{\sigma\left(  1\right)  }L_{\sigma\left(  2\right)  }\right)
L_{\sigma\left(  3\right)  }^{a_{1}a_{2}}L_{\sigma\left(  4\right)  }%
^{a_{3}a_{4}}\bar{\chi}\Gamma_{a_{1}\cdots a_{4}}\zeta+\right. \nonumber\\
&  -2L_{\sigma\left(  1\right)  }^{a_{1}a_{2}}\left[  L_{\sigma\left(
2\right)  }L_{\sigma\left(  3\right)  }L_{\sigma\left(  4\right)  }\right]
^{a_{3}a_{4}}\bar{\chi}\Gamma_{a_{1}\cdots a_{4}}\zeta+\nonumber\\
&  +\frac{3}{4}\operatorname*{Tr}\left(  L_{\sigma\left(  1\right)  }%
L_{\sigma\left(  2\right)  }\right)  \operatorname*{Tr}\left(  L_{\sigma
\left(  3\right)  }L_{\sigma\left(  4\right)  }\right)  \bar{\chi}%
\zeta+\nonumber\\
&  \left.  -\operatorname*{Tr}\left(  L_{\sigma\left(  1\right)  }%
L_{\sigma\left(  2\right)  }L_{\sigma\left(  3\right)  }L_{\sigma\left(
4\right)  }\right)  \bar{\chi}\zeta\right]  ,\label{L4FF}%
\end{align}
donde $\operatorname*{Tr}$ quiere decir traza en los \'{\i}ndices de Lorentz,
\textit{i.e.} $\operatorname*{Tr}\left(  L_{i}L_{j}\right)  =\left(
L_{i}\right)  _{\;b}^{a}\left(  L_{j}\right)  _{\;a}^{b}$ y en donde $\sigma$
corresponde a una permutaci\'{o}n de $S_{4},$ $S_{5}$ \'{o} $S_{6}$ seg\'{u}n corresponda.

\section{Lagrangeano Transgresor para el \'{A}lgebra~M}

\subsection{\label{Sec Lagrang M Alg}Forma Expl\'{\i}cita del Lagrangeano y el
M\'{e}todo de Separaci\'{o}n en Subespacios}

El c\'{a}lculo de una expresi\'{o}n expl\'{\i}cita cerrada para el lagrangeano
transgresor es una tarea engorrosa, incluso provistos de una expresi\'{o}n
para el tensor invariante. Sin embargo, esto se vuelve una tarea mucho m\'{a}s
sencilla provistos de la herramienta apropiada, la cual es el M\'{e}todo de
Separaci\'{o}n en Subespacios desarrollado en Sec.~\ref{Sec Metd Sep Sub Esp}.

Primero que nada, observemos que dado que el tensor invariante en s\'{\i}
mismo lleva constantes multiplicativas $\alpha_{0}$ y $\alpha_{2},$ resulta
superfluo agregar delante la constante $k,$ la cual puede ser reabsorbida en
ellas. Por otra parte, utilizando ec.~(\ref{EcTriangular}), el Lagrangeano
transgresor puede separarse de la forma%
\[
L_{\mathrm{T}}^{\left(  11\right)  }\left(  \boldsymbol{A},\bar{\boldsymbol{A}%
}\right)  =T_{\boldsymbol{A}\leftarrow\bar{\boldsymbol{\omega}}}^{\left(
11\right)  }+T_{\bar{\boldsymbol{\omega}}\leftarrow\bar{\boldsymbol{A}}%
}^{\left(  11\right)  }+\text{\textrm{d}}Q_{\boldsymbol{A}\leftarrow
\bar{\boldsymbol{\omega}}\leftarrow\bar{\boldsymbol{A}}}^{\left(  10\right)  }%
\]
y a su vez,%
\[
T_{\boldsymbol{A}\leftarrow\bar{\boldsymbol{\omega}}}^{\left(  11\right)
}=T_{\boldsymbol{A}\leftarrow\omega}^{\left(  11\right)  }+T_{\omega
\leftarrow\bar{\boldsymbol{\omega}}}^{\left(  11\right)  }+\text{\textrm{d}%
}Q_{\boldsymbol{A}\leftarrow\boldsymbol{\omega}\leftarrow\bar
{\boldsymbol{\omega}}}^{\left(  10\right)  }%
\]
con lo que tenemos%
\[
L_{\mathrm{T}}^{\left(  11\right)  }\left(  \boldsymbol{A},\bar{\boldsymbol{A}%
}\right)  =T_{\boldsymbol{A}\leftarrow\omega}^{\left(  11\right)  }%
-T_{\bar{\boldsymbol{A}\leftarrow}\bar{\boldsymbol{\omega}}}^{\left(
11\right)  }+T_{\omega\leftarrow\bar{\boldsymbol{\omega}}}^{\left(  11\right)
}+\text{\textrm{d}}Q_{\boldsymbol{A}\leftarrow\bar{\boldsymbol{\omega}%
}\leftarrow\bar{\boldsymbol{A}}}^{\left(  10\right)  }+\text{\textrm{d}%
}Q_{\boldsymbol{A}\leftarrow\boldsymbol{\omega}\leftarrow\bar
{\boldsymbol{\omega}}}^{\left(  10\right)  }%
\]

As\'{\i}, vemos que antes de calcular el Lagrangeano completo $L_{\mathrm{T}%
}^{\left(  11\right)  }\left(  \boldsymbol{A},\bar{\boldsymbol{A}}\right)  $,
resulta importante calcular una expresi\'{o}n cerrada para el trozo
$T_{\bm{A}\leftarrow\bm{\omega}}^{\left(  11\right)  }$. Para ello,
introducimos las conexiones intermedias
\begin{align}
\bm{A}_{0}  &  =\bm{\omega},\\
\bm{A}_{1}  &  =\bm{\omega}+\bm{e},\\
\bm{A}_{2}  &  =\bm{\omega}+\bm{e}+\bm{b}_{2},\\
\bm{A}_{3}  &  =\bm{\omega}+\bm{e}+\bm{b}_{2}+\bm{b}_{5},\\
\bm{A}_{4}  &  =\bm{\omega}+\bm{e}+\bm{b}_{2}+\bm{b}_{5}+\bar{\bm{\psi}}.
\end{align}
La ecuaci\'{o}n triangular, ec.~(\ref{EcTriangular}) nos permite separar
$T_{\bm{A}_{4}\leftarrow\bm{A}_{0}}^{\left(  11\right)  }$ de la forma
\begin{align}
T_{\bm{A}_{4}\leftarrow\bm{A}_{0}}^{\left(  11\right)  }  &  =T_{\bm{A}_{4}%
\leftarrow\bm{A}_{3}}^{\left(  11\right)  }+T_{\bm{A}_{3}\leftarrow\bm{A}_{0}%
}^{\left(  11\right)  }+\mathrm{d}Q_{\bm{A}_{4}\leftarrow\bm{A}_{3}%
\leftarrow\bm{A}_{0}}^{\left(  10\right)  },\\
T_{\bm{A}_{3}\leftarrow\bm{A}_{0}}^{\left(  11\right)  }  &  =T_{\bm{A}_{3}%
\leftarrow\bm{A}_{2}}^{\left(  11\right)  }+T_{\bm{A}_{2}\leftarrow\bm{A}_{0}%
}^{\left(  11\right)  }+\mathrm{d}Q_{\bm{A}_{3}\leftarrow\bm{A}_{2}%
\leftarrow\bm{A}_{0}}^{\left(  10\right)  },\\
T_{\bm{A}_{2}\leftarrow\bm{A}_{0}}^{\left(  11\right)  }  &  =T_{\bm{A}_{2}%
\leftarrow\bm{A}_{1}}^{\left(  11\right)  }+T_{\bm{A}_{1}\leftarrow\bm{A}_{0}%
}^{\left(  11\right)  }+\mathrm{d}Q_{\bm{A}_{2}\leftarrow\bm{A}_{1}%
\leftarrow\bm{A}_{0}}^{\left(  10\right)  }.
\end{align}

Es importante observar que debido a la estructura particular del \'{a}lgebra y
del tensor invariante, los t\'{e}rminos de borde $Q_{\bm{A}_{4}\leftarrow
\bm{A}_{3}\leftarrow\bm{A}_{0}}^{\left(  10\right)  },$ $Q_{\bm{A}_{3}%
\leftarrow\bm{A}_{2}\leftarrow\bm{A}_{0}}^{\left(  10\right)  }$ y
$Q_{\bm{A}_{2}\leftarrow\bm{A}_{1}\leftarrow\bm{A}_{0}}^{\left(  10\right)  }$
se cancelan id\'{e}nticamente. As\'{\i}, se llega en forma directa a la
expresi\'{o}n
\begin{equation}
T_{\bm{A}_{4}\leftarrow\bm{A}_{0}}^{\left(  11\right)  }=6\left[  H_{a}%
e^{a}+\frac{1}{2}H_{ab}b_{2}^{ab}+\frac{1}{5!}H_{abcde}b_{5}^{abcde}-\frac
{5}{2}\bar{\psi}\mathcal{R}\mathrm{D}_{\omega}\psi\right]  .\label{q40}%
\end{equation}
en donde los tensores $H_{a}$, $H_{ab}$, $H_{abcde}$ y $\mathcal{R}$ est\'{a}n
definidos como
\begin{align}
H_{a}  &  \equiv\left\langle \bm{R}^{5}\bm{P}_{a}\right\rangle _{\mathrm{M}%
},\label{H1}\\
H_{ab}  &  \equiv\left\langle \bm{R}^{5}\bm{Z}_{ab}\right\rangle _{\mathrm{M}%
},\label{H2}\\
H_{abcde}  &  \equiv\left\langle \bm{R}^{5}\bm{Z}_{abcde}\right\rangle
_{\mathrm{M}},\label{H5}\\
\mathcal{R}_{\;\beta}^{\alpha}  &  \equiv\left\langle \bm{Q}^{\alpha
}\bm{R}^{4}\bar{\bm{Q}}_{\beta}\right\rangle _{\mathrm{M}},\label{Rcal}%
\end{align}

Utilizando las componentes del polinomio invariante ec.~(\ref{L5B1}%
)--(\ref{L4FF}) y la definici\'{o}n ecs.~(\ref{itmalg1})-(\ref{itmalg5}),
resulta sencillo encontrar expresiones expl\'{\i}citas para los tensores
$H_{a}$, $H_{ab}$, $H_{abcde}$ y $\mathcal{R},$%
\begin{align}
H_{a}  &  =\frac{\alpha_{2}}{64}R_{a}^{\left(  5\right)  },\label{H1ex}\\
H_{ab}  &  =\alpha_{2}\left[  \frac{5}{2}\left(  R^{4}-\frac{3}{4}R^{2}%
R^{2}\right)  R_{ab}+5R^{2}R_{ab}^{3}-8R_{ab}^{5}\right]  ,\label{H2ex}\\
H_{abcde}  &  =-\frac{5}{16}\alpha_{2}\left[  5R_{[ab}R_{cde]}^{\left(
4\right)  }-40R_{\;[a}^{f}R_{\;b}^{g}R_{cde]fg}^{\left(  3\right)  }%
-R^{2}R_{abcde}^{\left(  3\right)  }+4R_{abcdefg}^{\left(  2\right)  }\left(
R^{3}\right)  ^{fg}\right]  ,\label{H5ex}\\
\mathcal{R}  &  =-\frac{\alpha_{2}}{40}\left\{  \left(  R^{4}-\frac{3}{4}%
R^{2}R^{2}\right)  \mathbbm{1}+\frac{1}{96}R_{abc}^{\left(  4\right)  }%
\Gamma^{abc}-\frac{3}{4}\left[  R^{2}R^{ab}-\frac{8}{3}\left(  R^{3}\right)
^{ab}\right]  R^{cd}\Gamma_{abcd}\right\}  ,\label{Rcalex}%
\end{align}

en donde hemos usado las abreviaciones
\begin{align}
R^{2n}  &  =R_{\;\;a_{2}}^{a_{1}}\cdots R_{\quad a_{1}}^{a_{2n}},\\
R_{ab}^{2n+1}  &  =R_{ac_{1}}R_{\;\;c_{2}}^{c_{1}}\cdots R_{\quad b}^{c_{2n}%
},\\
R_{a_{1}\cdots a_{11-2n}}^{\left(  n\right)  }  &  =\varepsilon_{a_{1}\cdots
a_{11-2n}b_{1}\cdots b_{2n}}R^{b_{1}b_{2}}\cdots R^{b_{2n-1}b_{2n}}.
\end{align}

El t\'{e}rmino $T_{\bar{\boldsymbol{A}\leftarrow}\bar{\boldsymbol{\omega}}%
}^{\left(  2n+1\right)  }$ tiene por supuesto, misma forma que
$T_{\bm{A}\leftarrow\bm{\omega}}^{\left(  11\right)  };$ as\'{\i}, s\'{o}lo
nos va quedando considerar el t\'{e}rmino de volumen $T_{\omega\leftarrow
\bar{\boldsymbol{\omega}}}^{\left(  2n+1\right)  }.$ Tomando en cuenta la
forma del tensor invariante ecs.~(\ref{itmalg1})-(\ref{itmalg5}), es directo
escribir%
\begin{equation}
T_{\bm{\omega }\leftarrow\bar{\bm{\omega}}}^{\left(  11\right)  }=3\int
_{0}^{1}\mathrm{d}t~\theta^{ab}L_{ab}\left(  t\right)  ,
\end{equation}
donde
\begin{equation}
L_{ab}\left(  t\right)  =\left\langle \bm{R}_{t}^{5}\bm{J}_{ab}\right\rangle
_{\mathrm{M}}%
\end{equation}
con
\begin{align}
\bm{R}_{t}  &  =\frac{1}{2}\left[  R_{t}\right]  ^{ab}\bm{J}_{ab},\\
\left[  R_{t}\right]  ^{ab}  &  =\bar{R}^{ab}+t\mathrm{D}_{\bar{\bm{\omega}}%
}\theta^{ab}+t^{2}\theta_{\;\;c}^{a}\theta^{cb},\\
\theta^{ab}  &  =\omega^{ab}-\bar{\omega}^{ab}.
\end{align}

Utilizando ec.~(\ref{L6}), se tiene en forma expl\'{\i}cita
\begin{equation}
L_{ab}\left(  t\right)  =\alpha_{0}\left[  \frac{5}{2}\left(  R_{t}^{4}%
-\frac{3}{4}R_{t}^{2}R_{t}^{2}\right)  \left[  R_{t}\right]  _{ab}+5R_{t}%
^{2}\left[  R_{t}\right]  _{ab}^{3}-8\left[  R_{t}\right]  _{ab}^{5}\right]
.\label{Lab(t)}%
\end{equation}

Es importante observar que $T_{\bm{\omega }\leftarrow\bar{\bm{\omega}}%
}^{\left(  11\right)  }$ es proporcional a $\alpha_{0}$ [ec.~(\ref{Lab(t)})],
a diferencia de todos los otros t\'{e}rminos, que son proporcionales a
$\alpha_{2}.$ Esto es una consecuencia directa de la forma del tensor
invariante. El hecho de que $\alpha_{0}$ y $\alpha_{2}$ sean constantes
arbitrarias independientes significa que el trozo proporcional a $\alpha_{0}$
y el proporcional a $\alpha_{2}$ son invariantes por s\'{\i} mismos bajo el
\'{A}lgebra~M. Por otra parte esto est\'{a} en armon\'{\i}a con el hecho de
que el trozo proporcional a $\alpha_{0}$ corresponde a la elecci\'{o}n de la
supertraza como tensor invariante, y por lo tanto, tiene que ser invariante
por s\'{\i} mismo.

Debido a su forma, pareciera que el t\'{e}rmino $T_{\bm{\omega }\leftarrow
\bar{\bm{\omega}}}^{\left(  11\right)  }$ es un t\'{e}rmino de interacci\'{o}n
en el volumen de las conexiones de esp\'{\i}n $\bm{\omega }$ y $\bar
{\bm{\omega}}.$ Esto es s\'{o}lo aparente; basta con utilizar
ec.~(\ref{EcTriangular}) para escribir%
\[
T_{\bm{\omega }\leftarrow\bar{\bm{\omega}}}^{\left(  11\right)  }%
=T_{\bm{\omega }\leftarrow0}^{\left(  11\right)  }-T_{\bar{\bm{\omega}}%
\leftarrow0}^{\left(  11\right)  }+\mathrm{d}Q_{\bm{\omega }\leftarrow
0\leftarrow\bar{\bm{\omega}}}^{\left(  10\right)  }%
\]
y observar as\'{\i} que estamos en presencia de dos \textquotedblleft
gravedades ex\'{o}ticas\textquotedblright\ para $\bm{\omega }$ y
$\bar{\bm{\omega}},$ las cuales interact\'{u}an s\'{o}lo a trav\'{e}s de un
t\'{e}rmino de borde.

\subsection{\label{Sec Relajando Ctes Acopl}Relajando las Constantes de
Acoplamiento}

Para escribir el lagrangeano, hemos usado como tensor invariante de
$\mathfrak{osp}\left(  \mathfrak{32}|\mathfrak{1}\right)  $ la supertraza
simetrizada de seis generadores. Pero \'{e}sta no es la \'{u}nica alternativa;
recordemos que el producto de tensores invariantes es tambi\'{e}n un tensor
invariante, ec.(\ref{Ec Prod InvTensor = InvTensor}). As\'{\i}, dado que el
tensor invariante que estamos considerando involucra seis generadores, vamos a
considerar las componentes de tensor invariante de $\mathfrak{osp}\left(
\mathfrak{32}|\mathfrak{1}\right)  $ dadas por las particiones de $6$,%
\begin{align*}
6  &  =6,\\
6  &  =4+2,\\
6  &  =3+3,\\
6  &  =2+2+2,\\
6  &  =1+1+1+1+1+1,
\end{align*}
o m\'{a}s expl\'{\i}citamente,%
\begin{align*}
\left\langle \boldsymbol{T}_{A_{1}}\cdots\boldsymbol{T}_{A_{6}}\right\rangle
_{\mathfrak{osp}}  &  =\operatorname*{Str}\left(  \boldsymbol{T}_{A_{1}}%
\cdots\boldsymbol{T}_{A_{6}}\right)  +\beta_{4+2}\operatorname*{Str}\left(
\boldsymbol{T}_{A_{1}}\cdots\boldsymbol{T}_{A_{4}}\right)  \operatorname*{Str}%
\left(  \boldsymbol{T}_{A_{5}}\boldsymbol{T}_{A_{6}}\right)  +\\
&  +\beta_{3+3}\operatorname*{Str}\left(  \boldsymbol{T}_{A_{1}}%
\cdots\boldsymbol{T}_{A_{3}}\right)  \operatorname*{Str}\left(  \boldsymbol{T}%
_{A_{4}}\cdots\boldsymbol{T}_{A_{6}}\right)  +\\
&  +\beta_{2+2+2}\operatorname*{Str}\left(  \boldsymbol{T}_{A_{1}%
}\boldsymbol{T}_{A_{2}}\right)  \operatorname*{Str}\left(  \boldsymbol{T}%
_{A_{3}}\boldsymbol{T}_{A_{4}}\right)  \operatorname*{Str}\left(
\boldsymbol{T}_{A_{5}}\boldsymbol{T}_{A_{6}}\right)  +\\
&  +\beta_{1+1+1+1+1+1}\operatorname*{Str}\left(  \boldsymbol{T}_{A_{1}%
}\right)  \operatorname*{Str}\left(  \boldsymbol{T}_{A_{2}}\right)
\operatorname*{Str}\left(  \boldsymbol{T}_{A_{3}}\right)  \operatorname*{Str}%
\left(  \boldsymbol{T}_{A_{4}}\right)  \operatorname*{Str}\left(
\boldsymbol{T}_{A_{5}}\right)  \operatorname*{Str}\left(  \boldsymbol{T}%
_{A_{6}}\right)  .
\end{align*}

Ahora bien, los generadores de $\mathfrak{osp}\left(  \mathfrak{32}%
|\mathfrak{1}\right)  $ son sin traza; adem\'{a}s, es directo ver que para las
componentes que nos interesan para constru\'{\i}r el tensor invariante para el
\'{A}lgebra M [ecs.~(\ref{itmalg1})-(\ref{itmalg5})] son todas nulas en la
partici\'{o}n $6=3+3$ [Ve\'{a}se Ap\'{e}ndice~\ref{Apend Dirac}]. As\'{\i},
s\'{o}lo debemos considerar
\begin{align*}
\left\langle \boldsymbol{T}_{A_{1}}\cdots\boldsymbol{T}_{A_{6}}\right\rangle
_{\mathfrak{osp}}  &  =\operatorname*{Str}\left(  \boldsymbol{T}_{A_{1}}%
\cdots\boldsymbol{T}_{A_{6}}\right)  +\beta_{4+2}\operatorname*{Str}\left(
\boldsymbol{T}_{A_{1}}\cdots\boldsymbol{T}_{A_{4}}\right)  \operatorname*{Str}%
\left(  \boldsymbol{T}_{A_{5}}\boldsymbol{T}_{A_{6}}\right)  +\\
&  +\beta_{2+2+2}\operatorname*{Str}\left(  \boldsymbol{T}_{A_{1}%
}\boldsymbol{T}_{A_{2}}\right)  \operatorname*{Str}\left(  \boldsymbol{T}%
_{A_{3}}\boldsymbol{T}_{A_{4}}\right)  \operatorname*{Str}\left(
\boldsymbol{T}_{A_{5}}\boldsymbol{T}_{A_{6}}\right)  .
\end{align*}
(es posible imponer $\beta_{6}=1$ sin perder generalidad, ya que el tensor
invariante para el \'{A}lgebra~M incluye constantes arbitrarias globales.)

Es interesante observar que no aparecer\'{a}n nuevos trozos en el lagrangeano
de volumen, sino que m\'{a}s bien algunos t\'{e}rminos en el lagrangeano
considerado en la secci\'{o}n anterior adquirir\'{a}n nuevas constantes de
acoplamiento. En efecto, el resultado neto es simplemente una redifinici\'{o}n
en $H_{a}$, $H_{ab}$, $H_{abcde}$ y $\mathcal{R},$ de la forma
\begin{align}
H_{a}  &  =\frac{\alpha_{2}}{64}R_{a}^{\left(  5\right)  },\label{H1re}\\
H_{ab}  &  =\alpha_{2}\left[  \frac{5}{2}\left(  \kappa_{15}R^{4}-\frac{3}%
{4}\gamma_{5}R^{2}R^{2}\right)  R_{ab}+5\kappa_{15}R^{2}R_{ab}^{3}-8R_{ab}%
^{5}\right]  ,\label{H2re}\\
H_{abcde}  &  =-\frac{5}{16}\alpha_{2}\left[  5R_{[ab}R_{cde]}^{\left(
4\right)  }+40R_{\;[a}^{f}R_{\;b}^{g}R_{cde]fg}^{\left(  3\right)  }%
-\kappa_{15}R^{2}R_{abcde}^{\left(  3\right)  }+4R_{abcdefg}^{\left(
2\right)  }\left(  R^{3}\right)  ^{fg}\right]  ,\label{H5re}%
\end{align}%
\begin{align}
\mathcal{R}  &  =-\frac{\alpha_{2}}{40}\left\{  \left[  \kappa_{3}R^{4}%
-\frac{3}{4}\left(  5\gamma_{9}-4\right)  R^{2}R^{2}\right]  \mathbbm{1}+\frac
{1}{96}R_{abc}^{\left(  4\right)  }\Gamma^{abc}+\right. \nonumber\\
&  \left.  -\frac{3}{4}\left[  \kappa_{9}R^{2}R^{ab}-\frac{8}{3}\left(
R^{3}\right)  ^{ab}\right]  R^{cd}\Gamma_{abcd}\right\}  .\label{Rcalre}%
\end{align}
en donde hemos inclu\'{\i}do nuevas constantes de acoplamiento $\kappa_{n}$ y
$\gamma_{n}.$ Estas nuevas constantes de acoplamiento est\'{a}n ligadas entre
s\'{\i} a trav\'{e}s de las relaciones%
\begin{align}
\kappa_{m}  &  =1+\frac{n}{m}\left(  \kappa_{n}-1\right)  ,\label{km}\\
\gamma_{m}  &  =\gamma_{n}+\left(  \frac{n}{m}-1\right)  \left(  \kappa
_{n}-1\right)  .\label{gm}%
\end{align}

Estos dos conjuntos de constantes reemplazan al par de constantes $\beta
_{4+2}$ y $\beta_{2+2+2}.$ Observemos que una vez que se fija un $\kappa_{n},
$ todos los otros $\kappa$'s quedan fijados, y luego, fijando un $\gamma_{n},
$ lo propio sucede con todos los $\gamma$'s. La relaci\'{o}n entre $\kappa
_{n}$ y $\gamma_{n}$ con las constantes $\beta_{4+2}$ y $\beta_{2+2+2}$ viene
dada por
\begin{align}
\beta_{4+2}  &  =\frac{1}{\mathrm{Tr}\left(  \mathbbm{1}\right)  }n\left(
\kappa_{n}-1\right)  ,\\
\beta_{2+2+2}  &  =\frac{15}{\left[  \mathrm{Tr}\left(  \mathbbm{1}\right)
\right]  ^{2}}\left(  \gamma_{n}-\kappa_{n}\right)  .
\end{align}
en donde debemos recordar que $\mathbbm{1}$ es una matriz de $32\times32,$ y
as\'{\i} $\mathrm{Tr}\left(  \mathbbm{1}\right)  =32.$

Anular las constantes de acoplamiento $\beta_{4+2}$ y $\beta_{2+2+2}$ equivale
a%
\begin{align}
\beta_{4+2}  &  =0\qquad\Leftrightarrow\qquad\kappa_{n}=1,\\
\beta_{2+2+2}  &  =0\qquad\Leftrightarrow\qquad\gamma_{n}=\kappa_{n}.
\end{align}
con lo que recuperamos ecs.~(\ref{H1ex})-(\ref{Rcalex}). En lo que sigue,
seguiremos utilizando los polinomios invariantes ecs.~(\ref{H1ex}%
)-(\ref{Rcalex}) en lugar de ecs.~(\ref{H1re})-(\ref{Rcalre}), simplemente
para no sobrecargar la notaci\'{o}n, pero es interesante observar que en
principio toda la construcci\'{o}n puede ser llevada a cabo con
ecs.~(\ref{H1re})-(\ref{Rcalre}), desembocando en un resultado m\'{a}s general.

\subsection{Principio de Acci\'{o}n y Ecuaciones de Campo}

Para construir el principio de acci\'{o}n, usaremos el procedimiento explicado
en Sec.~\ref{Sec Accion T General}, en particular
ec.~(\ref{Ec Def Accion M+ M-}). Sea $M$ una variedad orientable $11$
dimensional, y asociemos la conexi\'{o}n $\boldsymbol{A}$ con la
orientaci\'{o}n $M^{+}$ y $\bar{\boldsymbol{A}}$ con la orientaci\'{o}n
contraria, $M^{-}.$ Entonces, la acci\'{o}n transgresora completa para el
\'{A}lgebra~M viene dada por%
\begin{align*}
S_{\mathrm{M}}^{\left(  11\right)  }\left[  \boldsymbol{A},\bar{\boldsymbol{A}%
}\right]   &  =\int_{M^{+}}\left(  T_{\boldsymbol{A}\leftarrow\omega}^{\left(
11\right)  }+T_{\bm{\omega }\leftarrow0}^{\left(  11\right)  }\right)
+\int_{M^{-}}\left(  T_{\bar{\boldsymbol{A}\leftarrow}\bar{\boldsymbol{\omega
}}}^{\left(  11\right)  }+T_{\bar{\bm{\omega}}\leftarrow0}^{\left(  11\right)
}\right)  +\\
&  +\int_{\partial M^{+}}\left(  Q_{\boldsymbol{A}\leftarrow\bar
{\boldsymbol{\omega}}\leftarrow\bar{\boldsymbol{A}}}^{\left(  10\right)
}+Q_{\boldsymbol{A}\leftarrow\boldsymbol{\omega}\leftarrow\bar
{\boldsymbol{\omega}}}^{\left(  10\right)  }+Q_{\bm{\omega }\leftarrow
0\leftarrow\bar{\bm{\omega}}}^{\left(  10\right)  }\right)  .
\end{align*}

Ambas conexiones tienen el mismo estatus y juegan roles completamente
sim\'{e}tricos en la teor\'{\i}a, e interact\'{u}an s\'{o}lo a trav\'{e}s del
borde la variedad, $\partial M.$ La imagen mental asociada es la de
Fig.~\ref{FigOrientacion}, en donde cada conexi\'{o}n \textquotedblleft vive
en un lado\textquotedblright\ de la variedad, y s\'{o}lo se \textquotedblleft
tocan\textquotedblright\ en el borde de $M.$

Para calcular las ecuaciones de campo, basta con utilizar
ecs.~(\ref{Ec EcMov A General}) y~(\ref{Ec EcMov A_ General}) para el presente
caso. Dada la simetr\'{\i}a entre $\boldsymbol{A}$ y $\bar{\boldsymbol{A}},$
consideraremos expl\'{\i}citamente s\'{o}lo ec.~(\ref{Ec EcMov A General}).

Primero que nada, debemos observar que en la acci\'{o}n no aparecen derivadas
de $e^{a}$, $b_{2}^{ab}$ y $b_{5}^{a_{1}\cdots a_{5}}$, debido a la forma del
tensor invariante para el \'{A}lgebra~M inducida por el $0_{S}$-reducci\'{o}n.
Por lo tanto, no es una sorpresa que sus variaciones induzcan las ecuaciones
del movimiento%
\begin{align}
H_{a}  &  =0,\label{HaDeltaVielbein=0}\\
H_{ab}  &  =0,\label{DeltaB2}\\
H_{abcde}  &  =0,\label{DeltaB5}%
\end{align}

Similarmente, a la variaci\'{o}n de $\psi$ corresponde la ecuaci\'{o}n del
movimiento

\begin{align}
\mathcal{R}\mathrm{D}_{\omega}\psi=0.\label{DeltaPsi}
\end{align}

Por otra parte, la variaci\'{o}n de la conexi\'{o}n de $\omega^{ab}$
corresoponde a la ecuaci\'{o}n de campo%
\begin{align}
L_{ab}-10\left(  \mathrm{D}_{\omega}\bar{\psi}\right)  \mathcal{Z}_{ab}\left(
\mathrm{D}_{\omega}\psi\right)  +5H_{abc}\left(  T^{c}+\frac{1}{16}\bar{\psi
}\Gamma^{c}\psi\right)  +  & \nonumber\\
+\frac{5}{2}H_{abcd}\left(  \mathrm{D}_{\omega}b^{cd}-\frac{1}{16}\bar{\psi
}\Gamma^{cd}\psi\right)  +\frac{1}{24}H_{abc_{1}\cdots c_{5}}\left(
\mathrm{D}_{\omega}b^{c_{1}\cdots c_{5}}+\frac{1}{16}\bar{\psi}\Gamma
^{c_{1}\cdots c_{5}}\psi\right)   &  =0,\label{DeltaOmega}%
\end{align}
donde
\begin{align}
L_{ab}  &  \equiv\left\langle \bm{R}^{5}\bm{J}_{ab}\right\rangle _{\mathrm{M}%
},\\
\left(  \mathcal{Z}_{ab}\right)  _{\;\beta}^{\alpha}  &  \equiv\left\langle
\bm{Q}^{\alpha}\bm{R}^{3}\bm{J}_{ab}\bar{\bm{Q}}_{\beta}\right\rangle
_{\mathrm{M}},\\
H_{abc}  &  \equiv\left\langle \bm{R}^{4}\bm{J}_{ab}\bm{P}_{c}\right\rangle
_{\mathrm{M}},\\
H_{abcd}  &  \equiv\left\langle \bm{R}^{4}\bm{J}_{ab}\bm{Z}_{cd}\right\rangle
_{\mathrm{M}},\\
H_{abcdefg}  &  \equiv\left\langle \bm{R}^{4}\bm{J}_{ab}\bm{Z}_{cdefg}%
\right\rangle _{\mathrm{M}}.
\end{align}

En forma expl\'{\i}cita, ellas corresponden a [v\'{e}ase el polinomio
invariante ecs.~(\ref{L5B1})-(\ref{L4FF})],%
\begin{equation}
L_{ab}=\alpha_{0}\left[  \frac{5}{2}\left(  R^{4}-\frac{3}{4}R^{2}%
R^{2}\right)  R_{ab}+5R^{2}R_{ab}^{3}-8R_{ab}^{5}\right]  ,
\end{equation}%
\begin{align}
\mathcal{Z}_{ab}  &  =\frac{\alpha_{2}}{40}\left(  2\left[  R_{ab}^{3}%
-\frac{3}{4}R^{2}R_{ab}\right]  \mathbbm{1}-\frac{1}{48}R_{abcde}^{\left(
3\right)  }\Gamma^{cde}+\right. \nonumber\\
&  -\frac{3}{4}\left(  R_{ab}R^{cd}-\frac{1}{2}R^{2}\delta_{ab}^{cd}\right)
R^{ef}\Gamma_{cdef}+\nonumber\\
&  \left.  -\left[  \delta_{ab}^{cg}R_{gh}R^{hd}R^{ef}-R_{\;a}^{c}R_{\;b}%
^{d}R^{ef}+\frac{1}{2}\delta_{ab}^{ef}\left(  R^{3}\right)  ^{cd}\right]
\Gamma_{cdef}\right)  ,
\end{align}
\begin{equation}
H_{abc} = \frac{\alpha_{2}}{32} R_{abc}^{\left( 4 \right)},
\end{equation}
\begin{align}
H_{abcd}  &  =\alpha_{2}\delta_{ab}^{ef}\delta_{cd}^{gh}\left[  \frac{3}%
{4}R^{2}R_{ef}R_{gh}-R_{ef}^{3}R_{gh}-R_{ef}R_{gh}^{3}+\right. \nonumber\\
&  -\frac{4}{5}\left(  R_{eh}R_{fg}^{3}+R_{eh}^{3}R_{fg}-R_{eh}^{2}R_{fg}%
^{2}\right)  +\frac{1}{2}R^{2}R_{eh}R_{fg}+\nonumber\\
&  \left.  +\frac{1}{8}\eta_{\left[  ef\right]  \left[  gh\right]  }\left(
R^{4}-\frac{3}{4}R^{2}R^{2}\right)  -\eta_{fg}\left(  R^{2}R_{eh}^{2}-\frac
{8}{5}R_{eh}^{4}\right)  \right]  ,
\end{align}%
\begin{align}
H_{abc_{1}\cdots c_{5}}  &  =\frac{\alpha_{2}}{80}\delta_{c_{1}\cdots c_{5}%
}^{cdefg}\left(  -\frac{5}{3}R_{abcde}^{\left(  3\right)  }R_{fg}%
+10R_{abcdepq}^{\left(  2\right)  }R_{\;f}^{p}R_{\;g}^{q}+\right. \nonumber\\
&  -\frac{1}{6}R_{ab}R_{cdefg}^{\left(  3\right)  }+\frac{1}{4}R^{2}%
R_{abcdefg}^{\left(  2\right)  }-\frac{2}{3}R_{abcdefgpq}^{\left(  1\right)
}\left(  R^{3}\right)  ^{pq}+\nonumber\\
&  +\frac{1}{3}R_{\;a}^{p}R_{\;b}^{q}R_{cdefgpq}^{\left(  2\right)  }-\frac
{1}{3}R_{\;a}^{q}R_{bcdefgp}^{\left(  2\right)  }R_{\;q}^{p}+\frac{1}%
{3}R_{\;b}^{q}R_{acdefgp}^{\left(  2\right)  }R_{\;q}^{p}+\nonumber\\
&  \left.  -\frac{10}{3}\eta_{ga}R_{bcdep}^{\left(  3\right)  }R_{\;f}%
^{p}+\frac{10}{3}\eta_{gb}R_{acdep}^{\left(  3\right)  }R_{\;f}^{p}-\frac
{5}{24}\eta_{\left[  ab\right]  \left[  cd\right]  }R_{efg}^{\left(  4\right)
}\right)  .
\end{align}

Estas cantidades est\'{a}n relacionadas con las de ecs.~(\ref{H1ex}%
)-(\ref{Rcalex}) a trav\'{e}s de
\begin{align}
H_{c}  &  =\frac{1}{2}R^{ab}H_{abc},\\
H_{cd}  &  =\frac{1}{2}R^{ab}H_{abcd},\\
H_{cdefg}  &  =\frac{1}{2}R^{ab}H_{abcdefg},\\
\mathcal{R}_{\;\beta}^{\alpha}  &  =\frac{1}{2}R^{ab}\left(  \mathcal{Z}%
_{ab}\right)  _{\;\beta}^{\alpha}.
\end{align}

\subsection{Condiciones de Borde}

En general, existen muchas maneras de satisfacer las condiciones de contorno
ec.~(\ref{Ec EcBorde General}). Lo primero que debemos observar es que las
condiciones de borde ec.~(\ref{Ec EcBorde General}) aparentemente son
\textquotedblleft\'{a}simetricas\textquotedblright\ en los roles de
$\boldsymbol{A}$ y $\bar{\boldsymbol{A}},$%
\[
\left.  \int_{0}^{1}\mathrm{d}t\left\langle \left(  \delta\bar{\bm{A}}%
+t\delta\bm{\Theta}\right)  \bm{\Theta }\bm{F}_{t}^{n-1}\right\rangle
\right\vert _{\partial M}=0.
\]
Por lo tanto, parece natural resolver la condiciones de borde de la forma
\begin{align}
\left\langle \delta\bar{\bm{A}}\bm{\Theta}\bm{T}_{A_{1}}\cdots\bm{T}_{A_{n-1}%
}\right\rangle  &  =0,\\
\left\langle \delta\bm{\Theta}\bm{\Theta}\bm{T}_{A_{1}}\cdots\bm{T}_{A_{n-1}%
}\right\rangle  &  =0,
\end{align}
e imponer
\begin{align}
\delta\bar{\bm{A}}  &  =0,\\
\delta\bm{\Theta}  &  =\delta\bm{A},\\
\left\langle \delta\bm{\Theta}\bm{\Theta}\bm{T}_{A_{1}}\cdots\bm{T}_{A_{n-1}%
}\right\rangle  &  =0.
\end{align}

Por ejemplo, este tipo de solci\'{o}n es la que se utiliz\'{o} para el caso
de gravedad en Sec.~\ref{Sec Gravedad Transg} (Ve\'{a}se tambi\'{e}n
Refs.~\cite{CECS-FiniteGrav,Nosotros3-TransLargo,CECS-Trans}.)

Sin embargo, dado que en el enfoque que estamos utilizando ambas conexiones
cumplen roles completamente sim\'{e}tricos, parece natural buscar condiciones
de borde sim\'{e}tricas en $\boldsymbol{A}$ y $\bar{\boldsymbol{A}}.$ A\'{u}n
m\'{a}s, estamos asociando a cada conexi\'{o}n con una orientaci\'{o}n de la
variedad. Una de las componentes de la conexiones corresponde a un elfbein, y
por lo tanto, cuando consideramos un gauge en donde \'{e}ste es invertible, el
elemento de volumen de la variedad puede expresarse en t\'{e}rminos de
\'{e}ste. As\'{\i}, debemos probar la existencia de una manera de fijar las
condiciones de borde que sea compatible con la asociaci\'{o}n de cada una de
las conexiones con una orientaci\'{o}n de la variedad.

La asimetr\'{\i}a en ec.~(\ref{Ec EcBorde General}) es s\'{o}lo aparente;
basta con realizar el cambio de variables $t=1-\tau$ para obtener condiciones
de borde equivalentes, pero en las cuales los roles de $\boldsymbol{A}$ y
$\bar{\boldsymbol{A}}$ aparecen intercambiados.

As\'{\i}, una forma muy sencilla (pero no la \'{u}nica) de resolver las
condiciones de borde en forma sim\'{e}trica es considerar ambas versiones de
ec.~(\ref{Ec EcBorde General}) y resolverlas requiriendo simult\'{a}neamente%
\begin{align}
\left\langle \left(  \delta\bar{\bm{A}}+t\delta\bm{\Theta }\right)
\bm{\Theta}\bm{T}_{A_{1}}\cdots\bm{T}_{A_{n-1}}\right\rangle  &  =0,\\
\left\langle \left(  \delta\bm{A}-t\delta\bm{\Theta}\right)
\bm{\Theta}\bm{T}_{A_{1}}\cdots\bm{T}_{A_{n-1}}\right\rangle  &  =0,
\end{align}
con $t$ arbitrario. Sumando y restando estas ecuaciones, llegamos a las
condiciones expl\'{\i}citamente sim\'{e}tricas%
\begin{align}
\left\langle \left(  \delta\bar{\bm{A}}+\delta\bm{A}\right)
\bm{\Theta}\bm{T}_{A_{1}}\cdots\bm{T}_{A_{n-1}}\right\rangle  &  =0,\\
\left\langle \delta\bm{\Theta}\bm{\Theta}\bm{T}_{A_{1}}\cdots\bm{T}_{A_{n-1}%
}\right\rangle  &  =0.
\end{align}

Estas condiciones pueden ser resueltas requiriendo
\begin{gather}
\delta\bm{A}=-\delta\bar{\bm{A}}=\frac{1}{2}\delta
\bm{\Theta},\label{DeltaA=-DeltaA_}\\
\left\langle \delta\bm{\Theta\Theta T}_{A_{1}}\cdots\bm{T}_{A_{n-1}%
}\right\rangle =0.\label{DeltaThetaTheta=0}%
\end{gather}

Debido a la forma del tensor invariante [ecs.~(\ref{itmalg1})--(\ref{itmalg5}%
)], tenemos que ec.~(\ref{DeltaThetaTheta=0}) es satisfecha al requerir que se
cumpla%
\begin{align}
\left\langle \left(  \delta\bm{\theta e}+\delta\bm{e\theta }\right)
\bm{J}_{a_{1}b_{1}}\cdots\bm{J}_{a_{4}b_{4}}\right\rangle  &
=0,\label{Bound e}\\
\left\langle \left(  \delta\bm{\theta}\bar{\bm{e}}+\delta\bar{\bm{e}}%
\bm{\theta}\right)  \bm{J}_{a_{1}b_{1}}\cdots\bm{J}_{a_{4}b_{4}}\right\rangle
&  =0,\label{Bound e_}\\
\left\langle \left(  \delta\bm{\theta b}_{2}+\delta\bm{b}_{2}%
\bm{\theta}\right)  \bm{J}_{a_{1}b_{1}}\cdots\bm{J}_{a_{4}b_{4}}\right\rangle
&  =0,\label{Bound b2}\\
\left\langle \left(  \delta\bm{\theta}\bar{\bm{b}}_{2}+\delta\bar{\bm{b}}%
_{2}\bm{\theta}\right)  \bm{J}_{a_{1}b_{1}}\cdots\bm{J}_{a_{4}b_{4}%
}\right\rangle  &  =0,\label{Bound b2_}\\
\left\langle \left(  \delta\bm{\theta b}_{5}+\delta\bm{b}_{5}%
\bm{\theta}\right)  \bm{J}_{a_{1}b_{1}}\cdots\bm{J}_{a_{4}b_{4}}\right\rangle
&  =0,\label{Bound b5}\\
\left\langle \left(  \delta\bm{\theta}\bar{\bm{b}}_{5}+\delta\bar{\bm{b}}%
_{5}\bm{\theta}\right)  \bm{J}_{a_{1}b_{1}}\cdots\bm{J}_{a_{4}b_{4}%
}\right\rangle  &  =0,\label{Bound b5_}\\
\left\langle \left(  \delta\bm{\theta}\bar{\bm{\psi}}+\delta\bar
{\bm{\psi}}\bm{\theta}\right)  \bm{J}_{a_{1}b_{1}}\bm{J}_{a_{2}b_{2}%
}\bm{J}_{a_{3}b_{3}}\bar{\bm{Q}}\right\rangle  &  =0,\label{Bound Psi}\\
\left\langle \left(  \delta\bm{\theta}\bar{\bm{\chi}}+\delta\bar
{\bm{\chi}}\bm{\theta}\right)  \bm{J}_{a_{1}b_{1}}\bm{J}_{a_{2}b_{2}%
}\bm{J}_{a_{3}b_{3}}\bar{\bm{Q}}\right\rangle  &  =0,\label{Bound Chai}%
\end{align}
y%
\begin{align}
\left\langle \delta\bm{\theta\theta J}_{a_{1}b_{1}}\cdots\bm{J}_{a_{4}b_{4}%
}\right\rangle  &  =0,\label{Bound Theta J4}\\
\left\langle \delta\bm{\theta\theta J}_{a_{1}b_{1}}\bm{J}_{a_{2}b_{2}%
}\bm{Q}\bar{\bm{Q}}\right\rangle  &  =0,\label{Bound Theta J2Q2}\\
\left\langle \left(  \delta\bar{\bm{\psi}}+\delta\bar{\bm{\chi }}\right)
\left(  \bm{\psi}+\bm{\chi}\right)  \bm{J}_{a_{1}b_{1}}\cdots\bm{J}_{a_{4}%
b_{4}}\right\rangle  &  =0.\label{Bound Theta+Chai J4}%
\end{align}

Dado que $\delta\bm{A}=-\delta\bar{\bm{A}}$, es natural resolver el sistema
ecs.~(\ref{Bound e})--(\ref{Bound Chai}) requiriendo%
\begin{align}
\left.  \bar{\bm{e}}\right\vert _{\partial M}  &  =-\left.  \bm{e}\right\vert
_{\partial M},\label{e_=-e}\\
\left.  \bar{\bm{b}}_{2}\right\vert _{\partial M}  &  =-\left.  \bm{b}_{2}%
\right\vert _{\partial M},\label{b2_=-b2}\\
\left.  \bar{\bm{b}}_{5}\right\vert _{\partial M}  &  =-\left.  \bm{b}_{5}%
\right\vert _{\partial M},\label{b5_=-b5}\\
\left.  \bm{\chi}\right\vert _{\partial M}  &  =-\left.  \bm{\psi}\right\vert
_{\partial M}.\label{Chai=-Psi}%
\end{align}

Es importante notar que la condici\'{o}n sobre el elfbein $\left.
\bar{\bm{e}}\right\vert _{\partial M}=-\left.  \bm{e}\right\vert _{\partial
M}$ resulta perfectamente consistente con la idea de asociar cada conexi\'{o}n
con una orientaci\'{o}n de $M$ (en dimensiones impares, la transformaci\'{o}n
$e^{a}\rightarrow-e^{a}$ implica un cambio de orientaci\'{o}n de la variedad).

Usando la condici\'{o}n $\delta\bm{A}=-\delta\bar{\bm{A}}$, la
ec.~(\ref{Bound Theta+Chai J4}) se satisface en forma autom\'{a}tica, y
as\'{\i}, s\'{o}lo nos quedan las condiciones de borde
\begin{align*}
\left\langle \left(  \delta\bm{\theta e}+\delta\bm{e\theta }\right)
\bm{J}_{a_{1}b_{1}}\cdots\bm{J}_{a_{4}b_{4}}\right\rangle  &  =0,\\
\left\langle \left(  \delta\bm{\theta b}_{2}+\delta\bm{b}_{2}%
\bm{\theta}\right)  \bm{J}_{a_{1}b_{1}}\cdots\bm{J}_{a_{4}b_{4}}\right\rangle
&  =0,\\
\left\langle \left(  \delta\bm{\theta b}_{5}+\delta\bm{b}_{5}%
\bm{\theta}\right)  \bm{J}_{a_{1}b_{1}}\cdots\bm{J}_{a_{4}b_{4}}\right\rangle
&  =0,\\
\left\langle \left(  \delta\bm{\theta}\bar{\bm{\psi}}+\delta\bar
{\bm{\psi}}\bm{\theta}\right)  \bm{J}_{a_{1}b_{1}}\bm{J}_{a_{2}b_{2}%
}\bm{J}_{a_{3}b_{3}}\bar{\bm{Q}}\right\rangle  &  =0,\\
\left\langle \delta\bm{\theta\theta J}_{a_{1}b_{1}}\cdots\bm{J}_{a_{4}b_{4}%
}\right\rangle  &  =0,\\
\left\langle \delta\bm{\theta\theta J}_{a_{1}b_{1}}\bm{J}_{a_{2}b_{2}%
}\bm{Q}\bar{\bm{Q}}\right\rangle  &  =0,
\end{align*}
las cuales son suplementadas con $\delta\bm{A}=-\delta\bar{\bm{A}}$ y
ecs.~(\ref{e_=-e})--(\ref{Chai=-Psi}).

Por otra parte, resulta interesante observar que ec.~(\ref{Bound e}) coincide
con las condiciones de borde utilizadas en el caso de gravedad (v\'{e}ase
Refs.~\cite{CECS-FiniteGrav,Nosotros3-TransLargo,CECS-Trans} y Sec.~\ref{Sec Gravedad Transg});
las condiciones restantes parecen ser una extensi\'{o}n natural de las mismas.

Una forma completamente gen\'{e}rica de resolver estas condiciones es
requiriendo $\delta\bm{A}=\delta\tau\bm{A}$, siendo $\delta\tau$ un
par\'{a}metro infinitesimal arbitrario. Sin embargo, esto se parece ser
innecesariamente restrictivo; soluciones m\'{a}s generales pueden ser
obtenidas utilizando en forma expl\'{\i}cita la forma del tensor invariante.

Por otra parte, resulta interesante observar que es posible resolver las
condiciones de borde en forma completamente sim\'{e}trica en los roles de
$\boldsymbol{A}$ y $\bar{\boldsymbol{A}},$ y a\'{u}n m\'{a}s, de una forma
compatible con la idea de asociar $\boldsymbol{A}$ y $\bar{\boldsymbol{A}}$ a
orientaciones opuestas de la variedad base.

\section{\label{Sec Cuadridinamics}Din\'{a}mica Cuadridimensional}

Encontrar el \textquotedblleft vac\'{\i}o\textquotedblright\ para una
teor\'{\i}a constru\'{\i}da utilizando una forma de Transgresi\'{o}n o de
Chern--Simons es un problema no trivial. En general, dado que las ecuaciones
del movimiento son de la forma%
\begin{align}
\left.  \left\langle \boldsymbol{T}_{A}\boldsymbol{F}^{n}\right\rangle
\right\vert _{M}  &  =0,\\
\left.  \left\langle \boldsymbol{T}_{A}\bar{\boldsymbol{F}}^{n}\right\rangle
\right\vert _{M}  &  =0,
\end{align}
la elecci\'{o}n de $\boldsymbol{F}=0,$ $\bar{\boldsymbol{F}}=0$ como
vac\'{\i}o parece natural, ya que esta configuraci\'{o}n satisface las
ecuaciones de campo, es estable, con cargas nulas y adem\'{a}s, es invariante
de gauge. Sin embargo, este vac\'{\i}o se vuelve problem\'{a}tico al
considerar la propagaci\'{o}n de perturbaciones. Cuando\footnote{Por otra
parte, cuando $n=1$ ($D=3$) y el \'{a}lgebra posee una m\'{e}trica de Killing
invertible, la \'{u}nica soluci\'{o}n de las ecuaciones del movimiento es
$\boldsymbol{F}=0,$ y por lo tanto, en tres dimensiones no hay grados de
libertad propagantes.} $n\geq2$ (o sea, $D\geq5$) la ecuaci\'{o}n para las
perturbaciones se reduce a simplemente $0=0$ sobre el fondo
(\textit{background}) dado por $\boldsymbol{F}=0,$ $\bar{\boldsymbol{F}}=0$
(Ve\'{a}se Refs.~\cite{CECS NewSUGRA}%
,\cite{CECS-MAlgNoether-1,CECS-MAlgNoether-2}\cite{Horava,Nastase}). Se han
ofrecido diferentes soluciones para esta no-propagaci\'{o}n de grados de
libertad locales. Por ejemplo, en Ref.~\cite{Horava}, se agregan al
Lagrangeano de Chern--Simons t\'{e}rminos extra de interacci\'{o}n con
corrientes de \textquotedblleft materia\textquotedblright\ dada en
t\'{e}rminos de l\'{\i}neas de Wilson, y as\'{\i} se pueden propagar
perturbaciones en torno a $\boldsymbol{F}=0$.

Si bien esto resuleve el problema, por otra parte parece un poco artificioso
agregar de esta forma nuevos t\'{e}rminos al Lagrangeano transgresor. Una
soluci\'{o}n alternativa particularmente elegante ha sido propuesta en
Refs.~\cite{CECS-MAlgNoether-1,CECS-MAlgNoether-2}. En ellas no se agrega
ning\'{u}n t\'{e}rmino nuevo al lagrangeano, sino que m\'{a}s bien se busca
una topolog\'{\i}a de $M$ tal que admita un vac\'{\i}o con $\boldsymbol{F}%
\neq0$ en el cual las ecuaciones del movimiento se satisfagan como un cero
simple, y de esta forma, permitiendo propagaci\'{o}n de las pertubaciones.

En el contexto de gravedad de Chern--Simons de Poincar\'{e} en $D=11,$ al
buscar una topolog\'{\i}a tal que permita a las perturbaciones propagarse en
el contexto de gravedad de Poincar\'{e} en $D=11,$ el resultado es que la
teor\'{\i}a \textit{escoge} por s\'{\i} misma como soluci\'{o}n una
topolog\'{\i}a que incluye un universo cuadridimensional con constante
cosmol\'{o}gica posititiva (Ve\'{a}se
Refs.~\cite{CECS-MAlgNoether-1,CECS-MAlgNoether-2}). Esto es un muestra de lo
altamente no-trivial que puede ser la reducci\'{o}n dimensional en el contexto
de teor\'{\i}as con lagrangeanos altamente no-lineales. Por otra parte, es
interesante hacer el contraste con el procedimiento usual en supergravedad o
teor\'{\i}a de cuerdas, en donde se asume \textit{a priori} la existencia del
\'{u}niverso cuadridimensional para luego compactificar las dimensiones
restantes en alguna forma conveniente. En contraste, en el contexto de
Refs.~\cite{CECS-MAlgNoether-1,CECS-MAlgNoether-2}, la existencia de un
universo cuadridimensional es una predicci\'{o}n de la teor\'{\i}a.

Aqu\'{\i}, seguiremos de cerca el enfoque de
Refs.~\cite{CECS-MAlgNoether-1,CECS-MAlgNoether-2}, aplicado al lagrangeano
para el \'{A}lgebra~M que ha sido constru\'{\i}do en
Sec.~\ref{Sec Lagrang M Alg} (En
Refs.~\cite{CECS-MAlgNoether-1,CECS-MAlgNoether-2} el lagrangeano analizado
utiliza otro tensor invariante, por lo que se debe ser cuidadoso) y explorando
algunas de las consecuencias de permitir valores no nulos para la torsi\'{o}n.

Consideremos el ansatz geom\'{e}trico $M=X_{d+1}\times S^{10-d},$ donde
$X_{d}$ corresponde al producto combado (\textit{warped product}) entre una
variedad $d$-dimensional $M_{d}$ y $\mathbb{R}$, y en donde $S^{10-d}$
corresponde a una variedad $\left(  10-d\right)  $-dimensional de curvatura
constante, no-plana y con torsi\'{o}n cero.

Para este caso, la m\'{e}trica toma la forma de ec.~(4.3) from
Ref.~\cite{CECS-MAlgNoether-2},
\[
\mathrm{d}s^{2}=e^{2f\left(  \left\vert z\right\vert \right)  }\left(
\mathrm{d}z^{2}+\tilde{g}_{\mu\nu}^{\left(  d\right)  }\mathrm{d}x^{\mu
}\mathrm{d}x^{\nu}\right)  +\gamma_{mn}^{\left(  10-d\right)  }\mathrm{d}%
y^{m}\mathrm{d}y^{n}%
\]

\begin{figure}
\begin{center}
\includegraphics[width=\textwidth]{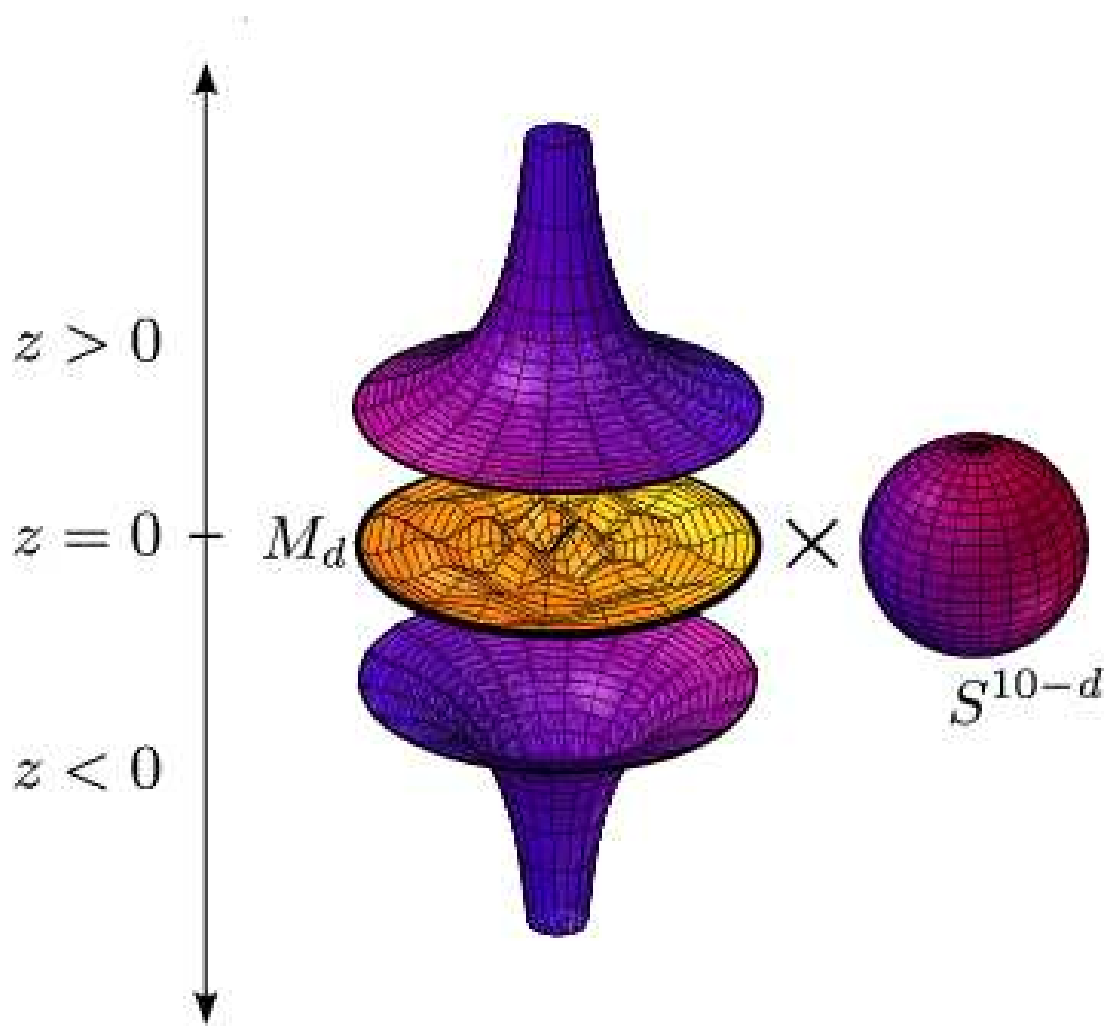}
\caption{Ansatz geom\'{e}trico $M=X_{d+1}\times S^{10-d}$}
\end{center}
\end{figure}

De ahora en adelante, usaremos en esta secci\'{o}n, el \'{\i}ndice $Z$ para el
espacio tangente de $\mathbb{R},$ $a,b,c,\ldots$ para el espacio tangente de
$M_{d}$ y $i,j,k,\ldots$ para el espacio tangente de $S^{10-d}$. Usaremos
$A,B,C$ como \'{\i}ndices para el espacio tangente total en 11 dimensiones.

As\'{\i}, tenemos que las componentes del vielbein vienen dadas por%
\begin{align}
E^{Z}  &  =e^{f\left(  \left\vert z\right\vert \right)  }\mathrm{d}%
z,\label{Ec Ez=wea}\\
E^{a}  &  =e^{f\left(  \left\vert z\right\vert \right)  }\tilde{e}%
^{a},\label{Ec Ea=wea}%
\end{align}
en donde $\tilde{e}^{a}$ corresponde al vielbein sobre $M_{d}.$ y las
componentes $E^{i}$ satisfacen $R^{ij}\propto E^{i}E^{j}.$

Es posible descomponer la conexi\'{o}n de esp\'{\i}n como $W^{AB}=W_{0}%
^{AB}+K^{AB}$ siendo $W_{0}^{AB}$ la conexi\'{o}n sin torsi\'{o}n ,
$\mathrm{d}E^{A}+\left[  W_{0}\right]  _{\phantom{A}B}^{A}E^{B}=0.$ Utilizando
las componentes del elfbein, ecs.~(\ref{Ec Ez=wea}) y~(\ref{Ec Ea=wea}),
tenemos que%
\begin{align*}
\left[  W_{0}\right]  ^{ab}  &  =\left[  \tilde{\omega}_{0}\right]  ^{ab},\\
W^{aZ}  &  =2f^{\prime}\left(  \left\vert z\right\vert \right)  \theta\left(
z\right)  \tilde{e}^{a},
\end{align*}
en donde $\theta\left(  z\right)  $ corresponde a la funci\'{o}n de Heaviside
y $\left[  \tilde{\omega}_{0}\right]  ^{ab}$ es la conexi\'{o}n de esp\'{\i}n
sin torsi\'{o}n de $M_{d}$. Ahora bien, un c\'{a}lculo directo entrega que la
torsi\'{o}n $11$-dimensional viene dada por%
\[
T^{a}=e^{f\left(  \left\vert z\right\vert \right)  }K^{aZ}\mathrm{d}%
z+e^{f\left(  \left\vert z\right\vert \right)  }\left(  \mathrm{d}\tilde
{e}^{a}+\left[  W_{\mathrm{T}}\right]  _{\_b}^{a}\tilde{e}^{b}\right)
\]

As\'{\i}, identificando las componentes de la contorsi\'{o}n, $K^{ab}%
=\tilde{\kappa}^{ab},$ tenemos que%
\[
T^{a}=e^{f\left(  \left\vert z\right\vert \right)  }\kappa^{a}\mathrm{d}%
z+e^{f\left(  \left\vert z\right\vert \right)  }\tilde{T}^{a}%
\]
en donde $\tilde{T}^{a}$ corresponde a la torsi\'{o}n de $M_{d}\ $y en donde
hemos definido el vector $\kappa^{a}=K^{aZ}.$ De la misma forma, es posible
calcular todas las componentes de la curvatura y torsi\'{o}n, llegando a%
\begin{align*}
R^{ab}  &  =\tilde{R}^{ab}-\left[  f^{\prime}\left(  \left\vert z\right\vert
\right)  \right]  ^{2}\tilde{e}^{a}\tilde{e}^{b}-2f^{\prime}\left(  \left\vert
z\right\vert \right)  \theta\left(  z\right)  \left(  \tilde{e}^{a}\kappa
^{b}-\tilde{e}^{b}\kappa^{a}\right)  -\kappa^{a}\kappa^{b},\\
R^{aZ}  &  =e^{-f\left(  \left\vert z\right\vert \right)  }\partial_{z}%
^{2}f\left(  \left\vert z\right\vert \right)  E^{Z}\tilde{e}^{a}+2f^{\prime
}\left(  \left\vert z\right\vert \right)  \theta\left(  z\right)  \tilde
{T}^{a}+\mathrm{D}_{\tilde{\omega}_{\mathrm{T}}}\kappa^{a},\\
R^{ij}  &  \propto E^{i}E^{j},\\
T^{a}  &  =\kappa^{a}E^{Z}+e^{f\left(  \left\vert z\right\vert \right)
}\tilde{T}^{a},\\
T^{Z}  &  =-e^{f\left(  \left\vert z\right\vert \right)  }\kappa^{a}\tilde
{e}_{a},\\
T^{i}  &  =0.
\end{align*}

Un detalle de sumo interer\'{e}s es que anular la torsi\'{o}n sobre $M$ y
sobre $M_{d}$ son dos cosas completamente distintas; en particular, es posible
anular la torsi\'{o}n sobre $M_{d}$ sin anular la torsi\'{o}n $11$-dimensional
cuando $\kappa^{a}\neq0$. Para recuperar gravedad, bastar\'{a} con imponer que
$f^{\prime}\left(  \left\vert z\right\vert \right)  $ sea constante
$f^{\prime}\left(  \left\vert z\right\vert \right)  =-\xi;$ as\'{\i}, tenemos
$f\left(  \left\vert z\right\vert \right)  =-\xi\left\vert z\right\vert $ y
por lo tanto la m\'{e}trica toma la forma%

\[
\mathrm{d}s^{2}=e^{-2\xi  \left\vert z\right\vert   }\left(
\mathrm{d}z^{2}+\tilde{g}_{\mu\nu}^{\left(  d\right)  }\mathrm{d}x^{\mu
}\mathrm{d}x^{\nu}\right)  +\gamma_{mn}^{\left(  10-d\right)  }\mathrm{d}%
y^{m}\mathrm{d}y^{n},
\]
y las componentes de la curvatura y torsi\'{o}n,%
\begin{align*}
R^{ab}  &  =\tilde{R}^{ab}-\xi^{2}\tilde{e}^{a}\tilde{e}^{b}+2\xi\theta\left(
z\right)  \left(  \tilde{e}^{a}\kappa^{b}-\tilde{e}^{b}\kappa^{a}\right)
-\kappa^{a}\kappa^{b},\\
R^{aZ}  &  =-2e^{\xi\left\vert z\right\vert }\xi\delta\left(  z\right)
E^{Z}\tilde{e}^{a}-2\xi\theta\left(  z\right)  \tilde{T}^{a}+\mathrm{D}%
_{\tilde{\omega}_{\mathrm{T}}}\kappa^{a},\\
R^{ij}  &  \propto E^{i}E^{j},\\
T^{a}  &  =\kappa^{a}E^{Z}+e^{-\xi\left\vert z\right\vert }\tilde{T}^{a},\\
T^{Z}  &  =-e^{-\xi\left\vert z\right\vert }\kappa^{a}\tilde{e}_{a},\\
T^{i}  &  =0.
\end{align*}

Concentr\'{e}mosnos por ahora en una componente particular de las ecuaciones
del moviemiento, por ejemplo, en la obtenida variando el elfbein,%
\[
\varepsilon_{A_{1}\cdots A_{11}}R^{A_{1}A_{2}}\cdots R^{A_{9}A_{10}}=0
\]
y debemos buscar bajo que valores de $d$ esta ecuaci\'{o}n tiene ceros simples.

Utilizando el hecho de que $R^{ij}\propto E^{i}E^{j},$ es posible descomponer
esta ecuaci\'{o}n en las componentes%
\begin{align}
\varepsilon_{Zb_{1}\cdots b_{d}j_{1}\cdots j_{10-d}}R^{b_{1}b_{2}}\cdots
R^{b_{d-1}b_{d}}E^{j_{1}}\cdots E^{j_{10-d}}  &  =0,\label{Ec EcMov Modo 1}\\
\varepsilon_{a_{1}b_{1}\cdots b_{d-2}cZj_{1}\cdots j_{10-d}}R^{b_{1}b_{2}%
}\cdots R^{b_{1}b_{d-2}}R^{cZ}E^{j_{1}}\cdots E^{j_{10-d}}  &
=0,\label{Ec EcMov Modo 2}\\
\varepsilon_{i_{1}b_{1}\cdots b_{d-1}cZj_{1}\cdots j_{9-d}}R^{b_{1}b_{2}%
}\cdots R^{b_{1}b_{d-1}}R^{cZ}E^{j_{1}}\cdots E^{j_{9-d}}  &
=0.\label{Ec EcMov Modo 3}%
\end{align}

En ec.~(\ref{Ec EcMov Modo 1}) tenemos ceros simples s\'{o}lo para $d=2;$ para
ec.~(\ref{Ec EcMov Modo 2}) cuando $d=4$ y para ec.~(\ref{Ec EcMov Modo 3}),
cuando $d=3.$ As\'{\i}, la \'{u}nica alternativa que posee grados de libertad
propagantes es $d=4.$ En este caso, la ecuaci\'{o}n de campo
ec.~(\ref{HaDeltaVielbein=0}) toma la forma expl\'{\i}cita de%
\begin{align}
\xi e^{\xi\left\vert z\right\vert }\delta\left(  z\right)  E^{Z}%
\varepsilon_{abcd}\left(  \tilde{R}^{ab}-\xi^{2}\tilde{e}^{a}\tilde{e}%
^{b}\right)  \tilde{e}^{c}  &  =\mathcal{T}_{d},\label{Ec D=4 Einstein}\\
\varepsilon_{abcd}\left(  \tilde{R}^{ab}-\xi^{2}\tilde{e}^{a}\tilde{e}%
^{b}\right)  \left(  \tilde{R}^{cd}-\xi^{2}\tilde{e}^{c}\tilde{e}^{d}\right)
&  =\mathcal{T},\label{Ec D=4 (R-e2)(R-e2)=0}%
\end{align}
con
\begin{align}
\mathcal{T}_{d}  &  =2E^{Z}e^{\xi\left\vert z\right\vert }\xi\delta\left(
z\right)  \varepsilon_{abcd}\left(  \frac{1}{2}\kappa^{a}\kappa^{b}-\xi
\theta\left(  z\right)  \left(  \tilde{e}^{a}\kappa^{b}-\tilde{e}^{b}%
\kappa^{a}\right)  \right)  \tilde{e}^{c}+\nonumber\\
&  +\varepsilon_{abcd}\left[  \tilde{R}^{ab}-\xi^{2}\tilde{e}^{a}\tilde{e}%
^{b}+2\xi\theta\left(  z\right)  \left(  \tilde{e}^{a}\kappa^{b}-\tilde{e}%
^{b}\kappa^{a}\right)  -\kappa^{a}\kappa^{b}\right]  \left(  \frac{1}%
{2}\mathrm{D}_{\tilde{\omega}}\kappa^{c}-\xi\theta\left(  z\right)  \tilde
{T}^{c}\right)  ,
\end{align}%
\begin{align}
\mathcal{T}  &  =-4\varepsilon_{abcd}\left(  \tilde{R}^{ab}-\xi^{2}\tilde
{e}^{a}\tilde{e}^{b}+\xi\theta\left(  z\right)  \left(  \tilde{e}^{a}%
\kappa^{b}-\tilde{e}^{b}\kappa^{a}\right)  -\frac{1}{2}\kappa^{a}\kappa
^{b}\right)  \times\nonumber\\
&  \times\left(  \xi\theta\left(  z\right)  \left(  \tilde{e}^{c}\kappa
^{d}-\tilde{e}^{d}\kappa^{c}\right)  -\frac{1}{2}\kappa^{c}\kappa^{d}\right)
.
\end{align}

La ec.(\ref{Ec D=4 Einstein}) corresponde a las ecuaciones de campo de
Einstein, con soporte s\'{o}lo sobre $M_{4}$. El lado derecho incluye
acoplamiento con la torsi\'{o}n. A\'{u}n cuando se impone la torsi\'{o}n
cuadridimensional como nula, las componentes $\kappa^{a}$ se comportan como un
campo de materia desde un punto de vista cuadridimensional.

Por otra parte, ec.(\ref{Ec D=4 (R-e2)(R-e2)=0}) impone relaciones extra entre
la curvatura y $\kappa^{a}$. A\'{u}n m\'{a}s, las ecuaciones de campo
restantes, ecs.~(\ref{DeltaB2}) y~(\ref{DeltaB5}) imponen a\'{u}n m\'{a}s
restricciones sobre la geometr\'{\i}a cuadridimensional.
Por otra parte, las ecuaciones del movimiento ec.~(\ref{DeltaPsi})
y~(\ref{DeltaOmega}) relacionan la geometr\'{\i}a cuadridimensional con
$\kappa^{a}$, los campos fermi\'{o}nicos y los correspondientes a las cargas centrales.

As\'{\i}, pese a que no anular la torsi\'{o}n en $11$ dimensiones parece
restringir menos la geometr\'{\i}a, aparentemente hay demasiados v\'{\i}nculos
sobre la geometr\'{\i}a como para reproducir Relatividad General pura en
cuatro dimensiones (Ve\'{a}se un caso similar en 5 dimensiones en
Refs.~\cite{CECS-Exp1,CECS-Exp2}).

Es interesante observar que esta \textquotedblleft rigidez\textquotedblright%
\ en la din\'{a}mica cuadridimensional tiene su origen en el proceso de
$0_{S}$-reducci\'{o}n del \'{a}lgebra, el cual anul\'{o} una cantidad
substancial de componentes del tensor invariante. \'{O}bservese por ejemplo
que las ecuaciones de campo ecs.~(\ref{DeltaB2}) y~(\ref{DeltaB5}) siguen
siendo v\'{a}lidas a\'{u}n cuando $b_{2}^{ab}=0$ y $b_{5}^{a_{1}\cdots a_{5}%
}=0,$ y s\'{o}lo involucran a la curvatura, con lo que restringen la
geometr\'{\i}a. Si estuviesen presentes las componentes correspondientes a
$\alpha_{3}$ en el tensor invariante, estas hubiesen acoplado lo que ahora son
restricciones sobre la geometr\'{\i}a con los campos restantes, y por lo
tanto, la din\'{a}mica cuadridimensional hubise perdido su \textquotedblleft
rigidez\textquotedblright.

As\'{\i}, se vuelve interesante considerar otras \'{a}lgebras, las cuales
comparten algunas de las caracter\'{\i}sticas del \'{A}lgebra~M pero que son
originadas sin recurrir al $0_{S}$-reducci\'{o}n. Este es el caso por ejemplo de
la sub\'{a}lgebra resonante de la cual proviene el \'{A}lgebra~M,
ecs.~(\ref{Ec ProtoAlgM EcPP=Z})-(\ref{Ec ProtoAlgM EcQQ=de todo}) o la
sub\'{a}lgebra resonante de $\mathbb{Z}_{4}\otimes\mathfrak{osp}\left(
\mathfrak{32}|\mathfrak{1}\right)  $ considerada en
Sec.~\ref{Sec Z4 x osp SubAlg Reson}, las cuales en principio podr\'{\i}an
originar din\'{a}micas menos r\'{\i}gidas que las del \'{A}lgebra~M.

\chapter{\label{SecConcl}Conclusiones}

\bigskip

\begin{center}
\textquotedblleft\textit{This is not the end. It is not even the beginning of
the end.}

\textit{But it is, perhaps, the end of the
beginning\footnote{\textquotedblleft Este no es el final. Ni siquiera es el
comienzo del final. Pero, quiz\'{a}s, es el final del
comienzo\textquotedblright}}\textquotedblright

(Winston Churchill, 1942, despu\'{e}s de que las tropas aliadas rodearon el
puerto de Casablanca, Marruecos)

\bigskip
\end{center}

En la presente tesis, hemos tenido la oportunidad de analizar en el
Cap\'{\i}tulo~\ref{SecTrans} la construcci\'{o}n de una teor\'{\i}a de gauge a
trav\'{e}s de formas de transgresi\'{o}n, en el Cap\'{\i}tulo~\ref{SecS_Exp}
se ha desarrollado el m\'{e}todo de las $S$-Expansiones y finalmente en el
Cap\'{\i}tulo~\ref{SecAcc_M_Alg} se han utilizado ambas herramientas en la
construcci\'{o}n de una teor\'{\i}a de gauge para el \'{A}lgebra~M. En cada
uno de estos puntos se han contestado interrogantes, pero tambi\'{e}n han
aparecido nuevas e interesantes direcciones en las cuales esta
investigaci\'{o}n puede ser extendida.

\begin{itemize}
\item \textbf{Forma de Transgresi\'{o}n como Lagrangeano:} El lagrangeano
transgresor posee en general muy buenas cualidades, como ser
\textit{completamente} invariante de gauge (puesto que la transgresi\'{o}n
sobre el fibrado es una forma proyectable), y permite generar en en forma
natural una teor\'{\i}a de gauge supersim\'{e}trica \textit{sin m\'{e}trica de
fondo}, la cual sin embargo en general incluye gravitaci\'{o}n cuando existe
una sub\'{a}lgebra del tipo $\mathfrak{so}\left(  D+1\right)  $ \'{o}
$\mathfrak{iso}\left(  D\right)  $. Sin embargo, no deja de ser misterioso el
rol de la estructura de dos conexiones necesaria para llevar a cabo la
construcci\'{o}n, especialmente la no interacci\'{o}n de volumen
(\textit{bulk}) entre ellas. Pese a ello, la existencia de una segunda
conexi\'{o}n est\'{a} lejos de ser una hip\'{o}tesis superflua, ya que tiene
un gran efecto en lo relacionado a la finitud de cargas conservadas y a la
termodin\'{a}mica asociada a ellas (Ve\'{a}se
Refs.~\cite{CECS-FiniteGrav,CECS-VacuumOddDim,CECS-Trans}).

Por otra parte, parece interesante considerar este problema desde el punto de
vista desarrollado en Sec.~\ref{Sec Accion T General}, en donde hemos asociado
a cada orientaci\'{o}n del espaciotiempo una conexi\'{o}n. La resultante
simetr\'{\i}a discreta bajo cambios de orientaci\'{o}n parece natural, pero
ser\'{\i}a provechoso comparar en profundidad esta soluci\'{o}n con la de la
configuraci\'{o}n de variedades cobordantes.

Por otra parte, resulta interesante comprobar que la forma de Transgresi\'{o}n
genera cargas de Noether conservadas \textit{off-shell}, lo cual est\'{a} en
armon\'{\i}a con indicios que se\~{n}alan que una teor\'{\i}a de
Chern--Simons/Transgresi\'{o}n deber\'{\i}a ser libre de anomal\'{\i}as.

Es importante se\~{n}alar sin embargo, que pese a haber fuertes indicios de
que las teor\'{\i}as de Chern--Simons/Transgresi\'{o}n son renormalizables,
muy poco se conoce sobre su cuantizaci\'{o}n en dimensiones m\'{a}s altas que
tres. Esto se debe a lo complejo del t\'{e}rmino cin\'{e}tico y del potencial
en el lagrangiano; las autointeracciones resultan ser altamente no-lineales.
En efecto, incluso la din\'{a}mica cl\'{a}sica de este tipo de teor\'{\i}as
presente un sistema de v\'{\i}nculos inusualmente complejo para una
teor\'{\i}a de campos, el cual es tema de investigaci\'{o}n (Ve\'{a}se por
ejemplo Ref.~\cite{CECS Degeneracion1,CECS Deg Olivera}). Por lo tanto, la
mec\'{a}nica cu\'{a}ntica de esta clase teor\'{\i}as constituye un problema abierto,
el cual pese a verse soluble, no parece saludable enfrentar a trav\'{e}s del
m\'{e}todo de fuerza bruta; probablemente, se requerir\'{a} la creaci\'{o}n de
nuevos m\'{e}todos de cuantizaci\'{o}n, sino una revisi\'{o}n del concepto de
cuantizaci\'{o}n en s\'{\i} mismo, para ser resuelto. Pese a ello, todo parece
indicar que la teor\'{\i}a es cuantizable y renormalizable (Ve\'{a}se
Refs.~\cite{Chamseddine-CS1,Chamseddine-CS2,Witten-(2+1)Grav,Witten-QFT Jones
Poly,Witten-TopQuantFieldTheory,Mora-Tesis,CECS Quant Const}). En efecto,
debemos observar que no hay constantes de acoplamiento arbitrarias con
dimensiones, y a\'{u}n m\'{a}s, la constante $k$ en frente del lagrangeano
est\'{a} cuantizada en forma natural cuando el espacio base $M$ constituye el
borde de otra variedad (Ve\'{a}se por ejemplo Ref.~\cite{CECS Quant Const})

Otro problema interesante relacionado con la no-linealidad de la
transgresi\'{o}n es el del vac\'{\i}o de la teor\'{\i}a. La soluci\'{o}n de
reducci\'{o}n dimensional din\'{a}mica presentada en
Refs.~\cite{CECS-MAlgNoether-1,CECS-MAlgNoether-2} y extendida aqu\'{\i}
permitiendo torsi\'{o}n distinta de cero parece ser prometedora, especialmente
por el hecho de \textquotedblleft predecir\textquotedblright\ un universo
cuadridimensional. Ser\'{\i}a interesante plantear este m\'{e}todo en una
forma m\'{a}s general, para un \'{a}lgebra de Lie arbitraria, y encontrar que
topolog\'{\i}as induce/admite cada \'{a}lgebra de Lie bajo el criterio de
permitir propagaci\'{o}n de las perturbaciones. Este tipo de reducci\'{o}n
dimensional podr\'{\i}a quiz\'{a}s tener importantes aplicaciones en
cosmolog\'{\i}a inflacionaria (trabajo en progreso), pues parece probable que
se puedan generar modelos del tipo de $k$-esencia a trav\'{e}s de este
procedimiento cuando se consideren simetr\'{\i}as con $\left[  \boldsymbol{P}%
_{a},\boldsymbol{P}_{b}\right]  \neq0,$ \textit{i.e., (}A)dS \'{o} alguna de
las simetr\'{\i}as correspondientes a una $S$-Expansi\'{o}n (por ejemplo, la
sub\'{a}lgebra resonante de la cual proviene el \'{A}lgebra~M a trav\'{e}s de
$0_{S}$-reducci\'{o}n)

Desde un punto de vista pr\'{a}ctico, es interesante observar que para
escribir el Lagrangeno Transgresor en forma expl\'{\i}cita ha sido necesario
desarrollar el M\'{e}todo de Separaci\'{o}n en Subespacios
[Sec.~\ref{Sec Metd Sep Sub Esp}]. Por supuesto, este procedimiento no afecta
la din\'{a}mica de la teor\'{\i}a en forma alguna, pero resulta importante para:

\begin{enumerate}
\item separar el lagrangeano en un t\'{e}rmino de volumen y un t\'{e}rmino de
borde, y

\item para poder separar el lagrangeano en trozos correspondientes a las
distintas interacciones presentes en la teor\'{\i}a, y as\'{\i} tener una idea
clara de la f\'{\i}sica asociada con \'{e}ste.
\end{enumerate}

Cuando no se usa este procedimiento, el lagrangeano es una complicada,
no-intuitiva y altamente-no-lineal madeja de interacciones. La alternativa a
este m\'{e}todo es realizar esta separaci\'{o}n en t\'{e}rminos
correspondientes a cada interacci\'{o}n a trav\'{e}s de sucesivas
integraciones por partes, lo cual en la pr\'{a}ctica se vuelve una tarea larga
y engorrosa para dimensiones m\'{a}s altas que tres.

\item $S$-\textbf{Expansi\'{o}n}: El problema original a partir del cual
surgi\'{o} el procedimiento de $S$-Expansiones era el de encontrar una
expresi\'{o}n expl\'{\i}cita para el tensor invariante del \'{A}lgebra~M, lo
cual es necesario para poder aplicar el M\'{e}todo de Separaci\'{o}n de
Subespacios y as\'{\i} poder escribir el correspondiente lagrangeano de
Chern--Simons/Transgresi\'{o}n\footnote{Debe de se\~{n}alarse que en el
contexto de expansi\'{o}n en formas de Maurer--Cartan \textbf{s\'{\i}} es
posible escribir una forma de Chern--Simons/Transgresi\'{o}n, realizando una
expansi\'{o}n en \'{a}lgebras diferenciales libres, sin requerir una
expresi\'{o}n expl\'{\i}cita para el tensor invariante del \'{a}lgebra
expandida (Ve\'{a}se Ref.~\cite{Azcarraga-Expansion1}). Ahora bien, como
consecuencia de ello, ya no resulta posible utilizar el M\'{e}todo de
Separaci\'{o}n en Subespacios para el \'{a}lgebra expandida.
\par
En torno a estos puntos en especial agradezco grandemente las claras y
detalladas explicaciones del Prof.~J.~A~de~Azc\'{a}rraga durante la
estad\'{\i}a en la Universidad de Valencia, Espa\~{n}a, las cuales resultaron
cruciales para poder realizar esta investigaci\'{o}n.}. En este sentido, es un
interesante ejemplo de interacci\'{o}n entre F\'{\i}sica y Matem\'{a}tica
pura; para resolver un problema f\'{\i}sico (escribir un lagrangeano), ha sido
necesario crear una herramienta matem\'{a}tica abstracta ($S$-Expansi\'{o}n),
con la cual luego ha resultado sencillo resolver el problema original como un
caso particular.

Resulta interesante contrastar el procedimiento de $S$-Expansiones con los
procedimientos utilizados a principios de la dec\'{a}da de los 80's en
supergravedad para generar nuevas \'{a}lgebras de Lie con ciertas
caracter\'{\i}sticas (Ve\'{a}se por ejemplo, el trabajo de D'Auria y Fr\'{e}
en Ref.~\cite{DAuria-Fre}). En general, constru\'{\i}r un \'{a}lgebra de Lie
con una cierta estructura dada \textit{a priori} es un problema altamente no
trivial, pues se debe de resolver la identidad de Jacobi manteniendo el
\'{a}lgebra cerrada. En efecto, debemos de tener en cuenta que cuando nuestra
\'{a}lgebra original esta separada en $n$ subespacios, $\mathfrak{g}%
=\bigoplus_{p=1}^{n}V_{n},$ la identidad de Jacobi corresponder\'{a} a un
\textquotedblleft sistema\textquotedblright\ de $n^{4}$ ecuaciones, lo cual
r\'{a}pidamente se torna un problema muy d\'{\i}ficil de manejar.

En este sentido, el procedimiento de $S$-Expansiones se muestra una
herramienta particularmente \'{u}til. Para generar un \'{a}lgebra de Lie con
una cierta estructura debemos buscar un semigrupo con un producto apropiado.
Luego, en lugar de resolver la identidad de Jacobi, basta con resolver las
condiciones de resonancia [Ve\'{a}se ecs.~(\ref{Ec [Sp , Sq] = Sr})
y~(\ref{Ec Cond Forz SpSq=nSr})], lo cual en la pr\'{a}ctica muestra ser
sencillo de resolver.

Este nuevo procedimiento no s\'{o}lo permite obtener tensores invariantes para
\'{a}lgebras expandidas, sino que tambi\'{e}n permite obtener nuevas
\'{a}lgebras,junto con sus correspondientes tensores invariantes (m\'{a}s
generales que la traza) para elecciones de semigrupo $S\neq S_{\mathrm{E}%
}^{\left(  N\right)  }.$ Como un peque\~{n}o ejemplo \textquotedblleft de
juguete\textquotedblright, hemos obtenido una sub\'{a}lgebra resonante de
$\mathbb{Z}_{4}\otimes\mathfrak{osp}\left(  \mathfrak{32}|\mathfrak{1}\right)
, $ para mostrar las posibilidades del m\'{e}todo.

En este sentido, los teoremas del Cap\'{\i}tulo~\ref{SecS_Exp} forman una
\'{u}til \textquotedblleft caja de herramientas para
f\'{\i}sicos\textquotedblright\ la cual permite f\'{a}cilmente generar nuevas
\'{a}lgebras y sus correspondientes lagrangeanos de
Chern--Simons/Transgresi\'{o}n cuando se utilizan en conjunto con el
M\'{e}todo de Separaci\'{o}n de Subespacios. Por otra parte, resulta
interesante observar c\'{o}mo el origen de un \'{a}lgebra (si es que
corresponde a una sub\'{a}lgebra resonante, a un $0_{S}$-reducci\'{o}n, etc.)
tiene consecuencias en la forma del correspondiente tensor invariante, y
as\'{\i}, en la din\'{a}mica de la teor\'{\i}a constru\'{\i}da a partir de
\'{e}ste. El procedimiento de reducci\'{o}n en s\'{\i} mismo
[\ref{Def Alg Reducida}] resulta interesante; en cierta forma, permite
f\'{a}cilmente \textquotedblleft podar\textquotedblright\ un \'{a}lgebra y
obtener as\'{\i} una m\'{a}s peque\~{n}a. As\'{\i} por ejemplo, ha resultado
posible entender la contracci\'{o}n de \.{I}n\"{o}n\"{u}--Wigner no como un
proceso de l\'{\i}mite, sino m\'{a}s bien como una cuidadosa \textquotedblleft
poda\textquotedblright\ (reducci\'{o}n resonante) de una sub\'{a}lgebra
resonante. As\'{\i}, resulta claro qu\'{e} informaci\'{o}n es la que se pierde
al realizar la contracci\'{o}n, ya que es posible ver precisamente qu\'{e}
sectores del \'{a}lgebra original est\'{a}n siendo eliminados, lo cual a
trav\'{e}s del procedimiento de l\'{\i}mite original resulta d\'{\i}ficil de observar.

Sin embargo, a partir de los teoremas del Cap\'{\i}tulo~\ref{SecS_Exp}, uno
queda con la impresi\'{o}n de que de alguna forma, cuando un \'{a}lgebra de
Lie corresponde a alg\'{u}n tipo de $S$-Expansi\'{o}n de otra, entonces el
\'{a}lgebra original corresponder\'{\i}a en cierta forma a una simetr\'{\i}a
\textquotedblleft m\'{a}s fundamental\textquotedblright. As\'{\i}, se vuelve
importante resolver la pregunta inversa: \textquotedblleft dada un
\'{a}lgebra, \textquestiondown es \'{e}sta una $S$-Expansi\'{o}n de otra? Y si
lo es, \textquestiondown de c\'{u}al?\textquotedblright. Un ejemplo sencillo
de esto es por ejemplo recordar que $\mathfrak{so}\left(  \mathfrak{4}\right)
=\mathbb{Z}_{2}\otimes\mathfrak{so}\left(  \mathfrak{3}\right)  .$ Resolver
esta pregunta permitir\'{\i}a encontrar tensores invariantes distintos de la
traza, m\'{a}s generales. Resulta evidente que para resolver este problema, un
primer paso es realizar una clasificaci\'{o}n de \'{a}lgebras $S$-Expandidas,
basada a su vez en una clasificaci\'{o}n de semigrupos abelianos. As\'{\i}
mismo, resultar\'{\i}a extremadamente interesante considerar las consecuencias
sobre la din\'{a}mica asociada a cada elecci\'{o}n de semigrupo y reducci\'{o}n
posible, y en particular sobre el \textquotedblleft problema del
vac\'{\i}o\textquotedblright\ considerado en el punto anterior. En general,
disponer de una clasificaci\'{o}n de la din\'{a}mica/topolog\'{\i}a del
vac\'{\i}o para distintas simetr\'{\i}as podr\'{\i}a ser de gran ayuda en la
b\'{u}squeda de una simetr\'{\i}a fundamental.

Por otra parte, resulta interesante considerar c\'{o}mo es posible generalizar
el procedimiento de $S$-Expansi\'{o}n en s\'{\i} mismo. Hemos escogido un
semigrupo discreto, finito y abeliano. Relajando la hip\'{o}stesis de
discretitud, el resultado ser\'{\i}a un \'{a}lgebra continua con la forma de
un \'{a}lgebra de corrientes,%
\[
\left[  \boldsymbol{T}_{A}\left(  x\right)  ,\boldsymbol{T}_{B}\left(
y\right)  \right]  =C_{AB}^{\phantom{AB}C}\int\mathrm{d}zK\left(
x,y,z\right)  \boldsymbol{T}_{A}\left(  z\right)  .
\]

Escogiendo un semigrupo discreto pero de infinitos elementos,
obtendr\'{\i}amos un \'{a}lgebra de dimensi\'{o}n infinita. Por ejemplo, las
constantes de estructura de $S_{\mathrm{E}}^{\left(  \infty\right)  }$
tendr\'{\i}an la forma%
\[
\left[  \boldsymbol{T}_{\left(  A,\alpha\right)  },\boldsymbol{T}_{\left(
B,\beta\right)  }\right]  =C_{AB}^{\phantom{AB}C}\boldsymbol{T}_{\left(
C,\alpha+\beta\right)  },
\]
con lo que obtendr\'{\i}amos un \'{a}lgebra del tipo de las \'{a}lgebras de
Kac--Moody\footnote{En torno a estos puntos, debo de agradecer especialmente
algunas iluminadoras discusiones con el Prof.~S.~Bruce, Universidad de
Concepci\'{o}n, Chile.}. Por otra parte, generalizar el requerimiento de
abelianidad parece tener consecuencias bastante m\'{a}s dram\'{a}ticas. Este
requisito es bastante m\'{a}s fundamental que los anteriores, pues es el que
nos permite escribir el conmutador%
\[
\left[  \boldsymbol{T}_{\left(  A,\alpha\right)  },\boldsymbol{T}_{\left(
B,\beta\right)  }\right]  =\lambda_{\alpha}\lambda_{\beta}\left[
\boldsymbol{T}_{A},\boldsymbol{T}_{B}\right]  .
\]

Sin embargo, es posible relajar este requerimiento utilizando en lugar de un
semigrupo $S$ un anillo $R=B\cup F,$ que cumpla con los siguientes
requerimientos extra:

\begin{itemize}
\item $B\cap F=\left\{  0\right\}  ,$

\item Para todo elemento $x,y\in B$ y $\xi,\theta\in F$, sus productos
satisfagan
\begin{align*}
xy  &  =yx,\\
x\theta &  =\theta x,\\
\xi\theta &  =-\theta\xi,
\end{align*}

\item y que utilizando el producto de subconjuntos Def.~\ref{Def Prod Subsets}
se cumpla que%
\begin{align*}
B\times B  &  \subset B,\\
B\times F  &  \subset F,\\
F\times F  &  \subset B.
\end{align*}
\textit{i.e.}, que todo elemento en $R$ sea o bien abeliano o bien un
n\'{u}mero de Grassmann. Utilizando el anillo $R$ en forma an\'{a}loga a como
fue utilizado el semigrupo abeliano $S,$ es posible extender el procedimiento
de expansiones en formas extremadamente interesantes; por ejemplo, permite
obtener super\'{a}lgebras a partir de \'{a}lgebras y probablemente tambi\'{e}n
a la inversa. Sin embargo, el procedimiento parece no trivial cuando no se
posee el \'{a}lgebra envolvente universal en forma expl\'{\i}cita (trabajo en progreso).
\end{itemize}

\item \textbf{Lagrangeano de Transgresi\'{o}n para el \'{A}lgebra~M:}

El ingrediente crucial en la construcci\'{o}n de un lagrangeano de
Chern--Simons/Transgresi\'{o}n es el tensor invariante. Tal como hemos visto
\'{e}ste es justamente el ingrediente \textquotedblleft dif\'{\i}cil de
conseguir\textquotedblright\ en el caso del \'{A}lgebra~M. Con anterioridad a
nuestro trabajo, se hab\'{\i}a investigado en
Refs.~\cite{CECS-MAlgNoether-1,CECS-MAlgNoether-2} la construcci\'{o}n de un
tensor invariante a trav\'{e}s del m\'{e}todo de Noether para el
\'{A}lgebra~M. Este interesante resultado nos motiv\'{o} a buscar un
m\'{e}todo distinto para escribir tensores invariantes distintos de la traza,
lo que desemboc\'{o} finalmente en el procedimiento de $S$-Expansiones. Como
resultado final, utilizando la \textquotedblleft caja de herramientas para
f\'{\i}sicos\textquotedblright\ del Cap\'{\i}tulo~\ref{SecS_Exp}, result\'{o}
finalmente sencillo escribir el lagrangeano. Al proceder de la misma forma que
en Refs.~\cite{CECS-MAlgNoether-1,CECS-MAlgNoether-2}, es posible encontrar la
emergencia de un universo cuadridimensional como \textit{consecuencia} del
requerimiento de propagaci\'{o}n de las perturbaciones, incluso cuando no se
impone la nulidad de la torsi\'{o}n. Es interesante observar que en este
contexto, es posible anular la torsi\'{o}n cuadridimensional sin anular la
torsi\'{o}n en $D=11;$ las componentes de la torsi\'{o}n juegan el rol de un
campo de materia en cuatro dimensiones. Tambi\'{e}n se debe de observar que
aunque el \'{A}lgebra~M es una extensi\'{o}n supersim\'{e}trica del
\'{a}lgebra de Poincar\'{e}, se obtiene un universo cuadridimensional con
constante cosmol\'{o}gica positiva. Esto es una muestra de lo no-trivial que
se vuelve el predecir la f\'{\i}sica cuadridimensional inducida por un
lagrangeano altamente no-lineal como el de Chern--Simons/Transgresi\'{o}n. Por
esta misma raz\'{o}n, se debe ser extremadamente cuidadoso al tomar prestados
argumentos de Supergravedad Est\'{a}ndar o Teor\'{\i}a de Cuerdas basados en
condiciones de consistencia con la f\'{\i}sica cuadridimensional conocida. Los
lagrangeanos de estas teor\'{\i}as suelen no presentar el grado de
no-linealidad presente en el de Chern--Simons/Transgresi\'{o}n, y por lo
tanto, la relaci\'{o}n [f\'{\i}sica en dimensiones m\'{a}s altas] -
[f\'{\i}sica cuadridimensional] en estos casos es m\'{a}s directa (o m\'{a}s
ingenua, seg\'{u}n el punto de vista que se prefiera). En efecto, en
Refs.~\cite{CECS-Lect2,CECS-Lect3}, se listan algunas de las
caracter\'{\i}sticas m\'{a}s \textquotedblleft inusuales\textquotedblright\ de
las teor\'{\i}as de Chern--Simons/Transgresi\'{o}n, de las cuales citaremos un par:

\begin{enumerate}
\item \emph{El n\'{u}mero de grados de libertad bos\'{o}nicos y
fermi\'{o}nicos no es el mismo}: En las teor\'{\i}as supersim\'{e}tricas
usuales, el n\'{u}mero de grados de libertad es el mismo, bajo las
\emph{hip\'{o}tesis} (lo cual es rara vez mencionado) de que el grupo de
simetr\'{\i}a del espaciotiempo es Poincar\'{e} y de que los campos pertenecen
a un supermultiplete. En el caso de las teor\'{\i}as de
Chern--Simons/Transgresi\'{o}n, los campos pertenecen a una conexi\'{o}n en
lugar de a un supermultiplete, y la simetr\'{\i}a del espacio tiempo puede ser
distinta de Poincar\'{e} (por ejemplo, este es el caso cuando se utiliza
$\mathfrak{osp}\left(  \mathfrak{32}|\mathfrak{1}\right)  ,$ en donde se tiene
AdS). Por lo tanto, en general en las teor\'{\i}as de
Chern--Simons/Transgresi\'{o}n el n\'{u}mero de grados de libertad
bos\'{o}nicos y fermi\'{o}nicos no precisa ser el mismo.\footnote{Es
interesante observar que en la literatura muchas veces ni siquiera se
mencionan las hip\'{o}tesis en las que descansa la afirmaci\'{o}n de que ambos
n\'{u}meros de grados de libertad deben ser iguales.
\par
A\'{u}n m\'{a}s, en los hasta ahora infructuosos experimentos en busca de
supersimetr\'{\i}a, esto ha sido siempre dado por supuesto; se espera que
sobre la escala de quiebre de supersimetr\'{\i}a, aparezcan estados apareados
que difieran s\'{o}lo en el n\'{u}mero fermi\'{o}nico. Si este tipo de
experimentos (por ejemplo en el LHC) continuara arrojando resultados negativos,
esto no deber\'{\i}a interpretarse necesariamente como la no existencia de
supersimetr\'{\i}a, sino m\'{a}s bien, como la violaci\'{o}n de las susodichas
hip\'{o}tesis.}

\item \emph{Todos los campos fundamentales son de esp\'{\i}n menor que 2
autom\'{a}ticamente}: Dado que todos los campos fundamentales son parte de una
$1$-forma conexi\'{o}n, y todos son tensores antisim\'{e}tricos de Lorentz
(como las cargas centrales del \'{A}lgebra~M, por ejemplo), se tiene en forma
directa que todos los campos poseen un esp\'{\i}n menor o igual dos, sin
importar en que dimensi\'{o}n se est\'{e} trabajando. Esto est\'{a} de acuerdo
con la f\'{\i}sica cuadridimensional observada, pero debemos recordar que el
argumento usado para escoger $D=11$ como el n\'{u}mero de dimensiones de una
teor\'{\i}a fundamental est\'{a} basado en parte en impedir la aparici\'{o}n de espines
mayores que dos en las teor\'{\i}as supersim\'{e}tricas usuales. Esto
significa que en el contexto de Chern--Simons/Transgresi\'{o}n, ser\'{\i}a conveniente revisar cuidadosamente los argumentos que impiden la aparici\'{o}n de dimensiones mayores que 11.
\end{enumerate}

En el contexto del lagrangeano obtenido a trav\'{e}s del m\'{e}todo de
$S$-Expansiones para el \'{A}lgebra~M, debemos observar que las variaciones de
$e^{a}$, $b_{2}^{ab}$ y $b_{5}^{a_{1}\cdots a_{5}}$ en general generan
restricciones en la din\'{a}mica cuadridimensional inducida. Todo parece
indicar que el permitir torsi\'{o}n no nula en dimensiones m\'{a}s altas ayuda
a introducir m\'{a}s libertad en la f\'{\i}sica cuadridimensional, pero no
parece ser suficiente como para tener Relatividad General est\'{a}ndar en
cuatro dimensiones. En lugar de esto, lo que se se tiene es una din\'{a}mica
cuadridimensional de Relatividad General con restricciones extra; una
din\'{a}mica \textquotedblleft congelada\textquotedblright.

Esto es una consecuencia directa del proceso de $0_{S}$-reducci\'{o}n necesario
para obtener el \'{A}lgebra~M, pues el $0_{S}$-reducci\'{o}n elimina las
componentes del tensor invariante que justamente brindar\'{\i}an m\'{a}s
\textquotedblleft m\'{a}s elasticidad\textquotedblright\ a la teor\'{\i}a. El
procedimiento de $S$-Expansiones entrega una soluci\'{o}n natural a este
problema, la cual es simplemente considerar otras super\'{a}lgebras (v\'{e}ase
por ejemplo la sub\'{a}lgebra resonante de donde proviene el \'{A}lgebra~M,
Sec.~\ref{Sec Sub Alg Reson Original}, o la sub\'{a}lgebra resonante de
$\mathbb{Z}_{4}\otimes\mathfrak{osp}\left(  \mathfrak{32}|\mathfrak{1}\right)
,$ Sec.~\ref{Sec Z4 x osp SubAlg Reson}) las cuales reproducen diversas
caracter\'{\i}sticas del \'{A}lgebraM sin necesitar del $0_{S}$-reducci\'{o}n.

\item \textbf{Relaci\'{o}n con Supergravedad CJS y Teor\'{\i}a de Cuerdas}: La
relaci\'{o}n entre una teor\'{\i}a de Chern--Simons/Transgresi\'{o}n para el
\'{A}lgebra~M u otra \'{a}lgebra relacionada y Supergravedad est\'{a}ndar y
Teor\'{\i}a de Cuerdas es un problema altamente no trivial, el cual se
complica a\'{u}n m\'{a}s debido a nuestra ignorancia en el comportamiento de
teor\'{\i}as de Chern--Simons/Transgresi\'{o}n a nivel cu\'{a}ntico. Sin
embargo, es claro que para hacer contacto con Supergravedad est\'{a}ndar,
algunas identificaciones de los campos involucrados pueden hacerse; por
ejemplo resulta natural constru\'{\i}r a partir de la contorsi\'{o}n $k^{ab}$
una tres forma potencial, $A_{3}=e^{a}e^{b}k_{ab}$; as\'{\i} mismo, es posible
constru\'{\i}r una $6$-forma a partir de $b_{5}^{a_{1}\cdots a_{5}},$
$A_{6}=e_{a_{1}}\cdots e_{a_{5}}b_{5}^{a_{1}\cdots a_{5}};$ en $D=11,$ sus
respectivas \textquotedblleft curvaturas\textquotedblright\ $F_{4}%
=\mathrm{d}A_{3}$ y $F_{7}=\mathrm{d}A_{6}$ pueden ser Hodge-duales (Ve\'{a}se
por ejemplo, Refs.~\cite{CECS-Lect2,CECS-Lect3} y~\cite{Banh01}). Dado que una
teor\'{\i}a de Chern--Simons/Transgresi\'{o}n es una teor\'{\i}a de gauge
invariante off-shell, y supergravedad de CJS no lo es, resulta claro que en el
proceso de recuperar supergravedad de CJS (si es que esto es posible) debe de
haber un quiebre de simetr\'{\i}a, adem\'{a}s de que resultar\'{a} necesario
imponer restricciones extra (como la nulidad de la torsi\'{o}n por ejemplo).
Podemos imaginar que el resultado final ser\'{\i}a probablemente CJS m\'{a}s
otros tipos de t\'{e}rminos; en cierta forma, cualquier relaci\'{o}n entre
ambas teor\'{\i}as debe ser bastante indirecta. Sin embargo, resulta
interesante observar que en el paper original de CJS (Ve\'{a}se
Ref.~\cite{Cre78}), se afirma que probablemente una supergravedad m\'{a}s
fundamental en $D=11$ deber\'{\i}a ser invariante bajo $\mathfrak{osp}\left(
\mathfrak{32}|\mathfrak{1}\right)  $. Por otra parte, pese a no constituir
teor\'{\i}as de supergravedad est\'{a}ndar, estamos frente a un tipo de
teor\'{\i}a que incluye gravedad (de Lanczos--Lovelock) y que es invariante
\textit{off-shell} bajo transformaciones de supersimetr\'{\i}a\footnote{Esto
est\'{a} en fuerte contraste con lo que ocurre en las teor\'{\i}as de
supergravedad usuales, las cuales son sim\'{e}tricas s\'{o}lo
\textit{on-shell}, y por lo tanto, pueden poseer anomal\'{\i}as a nivel
cu\'{a}ntico.}, y por lo tanto, constituyen supergravedades perfectamente bien
definidas y autoconsistentes. As\'{\i} aparte de razones hist\'{o}ricas, no
parece haber argumentos \textit{a priori} para preferir supergravedad de CJS
por sobre una supergravedad de Chern--Simons/Transgresi\'{o}n.

Por otra parte, las relaciones con Teor\'{\i}a de Cuerdas tambi\'{e}n
constituyen un problema abierto. Sin embargo, debemos observar que los campos
$b_{2}^{ab}$ y $b_{5}^{a_{1}\cdots a_{5}}$ en el lagrangeano constru\'{\i}do
en la presente tesis se asocian en forma natural con $2$ y $5$-branas; por
otra parte, es posible engrosar la superficie de mundo (\textit{world sheet})
de cuerdas para asociarles acciones de Chern--Simons/Transgresi\'{o}n en tres
dimensiones (Ve\'{a}se Ref.~\cite{SeibergCuerdas}). Un trabajo extremadamente
interesante con respecto a acciones de branas y formas de Transgresi\'{o}n fue
llevado a cabo por P.~Mora en Refs.~\cite{Mora-BraneCS}%
,\cite{MoraNishino-BranaCS1}; la idea principal es tomar branas de un
n\'{u}mero impar de dimensiones y asociarles a cada una de ellas una
acci\'{o}n transgresora, sobre un fondo dado tambi\'{e}n por una transgresi\'{o}n.

Un enfoque un poco m\'{a}s tradicional al respecto ser\'{\i}a la b\'{u}squeda
de soluciones de branas en teor\'{\i}as de transgresi\'{o}n en $D=11$ o
a\'{u}n mejor, el obtener \textit{acciones} de brana a partir de una
acci\'{o}n de transgresi\'{o}n utilizando deltas de Dirac generalizadas como
forma diferencial\footnote{En este aspecto en particular, deseo agradecer
iluminadoras conversaciones con B. Jurco (Arnold Sommerfeld Center de la
Ludwig-Maximilians Universit\"{a}t, M\'{u}nich), el cual me gui\'{o} hacia
importantes trabajos de De~Rham en este tema.} (trabajo en progreso). Este
punto de vista quiz\'{a}s pudiera entregarnos cierta informaci\'{o}n extra en
relaci\'{o}n con Teor\'{\i}a de Cuerdas, pese a no disponer a\'{u}n de una
teor\'{\i}a cu\'{a}ntica para una acci\'{o}n transgresora en altas dimensiones.

Por otra parte, no est\'{a} del todo claro cuan exitosa ser\'{a} Teor\'{\i}a
de Cuerdas como marco final para describir en forma unificada todas las
interacciones de la naturaleza. En efecto, en el \'{u}ltimo tiempo, se han
alzado fuertes cr\'{\i}ticas se\~{n}alando algunos serios defectos de
Teor\'{\i}a de Cuerdas (Ve\'{a}se por ejemplo, Ref.~\cite{P. Woit}). A\'{u}n
cuando Teor\'{\i}a de Cuerdas resolviera estos inconvenientes y resultara ser
este promisorio marco unificado, el estudio de gravedad de
Chern--Simons/Transgresi\'{o}n en dimensiones m\'{a}s altas continuar\'{\i}a
siendo de extrema utilidad, pues en dimensiones impares estas constituyen la
\'{u}nica alternativa din\'{a}micamente consistente (Ve\'{a}se
Ref.~\cite{CECS-DimContBH}) y libre de fantasmas (Ve\'{a}se
Ref.~\cite{Zumino-GravMore4}) para una teor\'{\i}a efectiva para gravedad que
incluya potencias m\'{a}s altas de la curvatura. Por otra parte, una
teor\'{\i}a de gauge de Chern--Simons/Transgresi\'{o}n para $\mathfrak{osp}\left(\mathfrak{32}|\mathfrak{1}\right)$,
$\mathfrak{osp}\left(\mathfrak{32}|\mathfrak{1}\right)\times\mathfrak{osp}\left(\mathfrak{32}|\mathfrak{1}\right)$ \'{o} el \'{A}lgebra~M parece ser un candidato interesante para la
postulada Teor\'{\i}a~M de Witten (Ve\'{a}se Refs.~\cite{CECS
NewSUGRA,Horava,Nastase}).

En el caso de que estuvi\'{e}ramos equivocados, y Teor\'{\i}a de Cuerdas no
fuese este marco final, este tipo de teor\'{\i}as siguen siendo una
interesante alternativa, pues por s\'{\i} mismas constituyen supergravedades
invariantes off-shell independientemente de s\'{\i} Teor\'{\i}a de Cuerdas
funciona o no. Sin embargo, parece interesante en este caso abrirse a la
posibilidad de trabajar en dimensiones distintas de once y a investigar nuevas
simetr\'{\i}as, como las que han sido investigadas en la presente tesis.
\end{itemize}

\medskip

Mi opini\'{o}n personal con respecto a la relaci\'{o}n entre Teor\'{\i}a de
Cuerdas y supergravedades de Transgresi\'{o}n/Chern--Simons es en cierta forma
una opci\'{o}n intermedia entre los dos puntos de vista extremos ya
discutidos. Por una parte, me parece que la idea de utilizar cuerdas y
membranas para construir una teor\'{\i}a fundamental es bell\'{\i}sima; la
sencillez matem\'{a}tica de describir todas las interacciones de la naturaleza
como distintos modos de vibraci\'{o}n de un solo objeto fundamental es
irresistible. Pero as\'{\i} mismo, debemos recordar que para constru\'{\i}r
una teor\'{\i}a cu\'{a}ntica, debemos recurrir a la igualmente elegante
formulaci\'{o}n de Integrales de Camino de Feynman, para lo cual debemos de
constru\'{\i}r el apropiado funcional de acci\'{o}n. Es en este punto en
particular en donde me parece que en la literatura existe una alta dosis de
ingenuidad; por ejemplo, es claro que la acci\'{o}n de Polyakov utilizada como
acci\'{o}n para la cuerda constituye en el mejor de los casos simplemente una
acci\'{o}n efectiva sobre una m\'{e}trica de fondo. La situaci\'{o}n se repite
de una forma u otra cuando se escriben acciones para branas. El punto es que
los principios de acci\'{o}n se construyen en cierta forma \textquotedblleft a
mano\textquotedblright, simplemente para intentar de alguna forma generar algo
que se parezca a la f\'{\i}sica cuadridimensional conocida (\textit{i.e.}, el
Modelo Est\'{a}ndar). Este camino es en cierta forma el camino tradicional de
la F\'{\i}sica (constru\'{\i}r una teor\'{\i}a que permita reproducir los
datos experimentales); basta con considerar por ejemplo la historia de la
Mec\'{a}nica de Newton o de la Electrodin\'{a}mica de Maxwell. Sin embargo, me
parece que este camino a cierto nivel de profundidad se vuelve impracticable,
tal como explica P.~A.~M.~Dirac en la introducci\'{o}n de su famoso paper del
Monopolo Magn\'{e}tico (Ve\'{a}se Ref.~\cite{Dirac}):

\begin{quotation}
\textquotedblleft\textit{The steady progress of physics requires for its
theoretical formulations a mathematics that gets continually more advanced.
This is only natural and to be expected. What, however, was not expected by
the scientific workers of the last century was the particular form that the
line of advancement of the mathematics would take, namely, it was expected
that the mathematics would get more and more complicated, but would rest on a
permanent basis of axioms and definitions, while actually the modern physical
developments have required a mathematics that continually shifts its
foundations and gets more abstract. Non.euclidean geometry and non-commutative
algebra, which were at one time considered to be purely fictions of the mind
and pastimes for logical thinkers, have now been found to be necessary for the
description of general facts of the physical world. It seems likely that this
process of increasing abstraction will continue in the future and that advance
in physics is to be associated with a continual modification and
generalisation of the axioms at the base of the mathematics rather than with a
logical development of any one mathematical scheme on a fixed foundation.}

\textit{There are at present fundamental problems in theoretical physics
awaiting solution, }\textbf{e.g.}\textit{, the relativistic formulation of
quantum mechanics and the nature of atomic nuclei (to be followed by more
difficult ones such as the problem of life), the solution of which problems
will presumably require a more drastic revision of our fundamental concepts
than any that have gone before. Quite likely these changes will be so great
that it will be beyond the power of human intelligence to get the necessary
new ideas by direct attempts to formulate the experimental data in
mathematical terms. The theoretical worker in the future will therefore have
to proceed in a more indirect way. The most powerful method of advance that
can be suggested at present is to employ all the resources of pure mathematics
in attempts to perfect and generalise the mathematical formalism that forms
the existing basis of theoretical physics, and} \textbf{after}\textit{\ each
succes in this direction, to try to interpret the new mathematical features in
terms of physical entities.\footnote{\textquotedblleft\textit{El constante
progreso en f\'{\i}sica requiere para su formulaci\'{o}n te\'{o}rica una
matem\'{a}tica que continuamente se vuelve m\'{a}s avanzada. Esto es natural y
esperable. Lo que, sin embargo, no fue esperado por los trabajadores
cient\'{\i}ficos del \'{u}ltimo siglo fue forma particular que la l\'{\i}nea
de avances matem\'{a}ticos tomar\'{\i}a, a saber, se esperaba que las
matem\'{a}ticas se volver\'{\i}an m\'{a}s y m\'{a}s complicadas, pero
descansando en una base permanente de axiomas y definiciones, mientras que en
realidad los desarrollos de la f\'{\i}sica moderna han requerido una
matem\'{a}tica que continuamente mueve sus fundamentos y se vuelve m\'{a}s
abstracta. Geometr\'{\i}a no-eucl\'{\i}dea y \'{a}lgebra no-conmutativa, las
cuales en su momento fueron consideradas puramente imaginaciones y pasatiempos
para pensadores l\'{o}gicos, ahora han pasado ha ser necesarias para la
descripci\'{o}n general del mundo f\'{\i}sico. Parece probable que este
proceso de incremento de la abstracci\'{o}n contin\'{u}e en el futuro y que el
avance en f\'{\i}sica estar\'{a} asociado con una continua modificaci\'{o}n y
generalizaci\'{o}n de los axiomas en las bases matem\'{a}ticas en vez de con
desarrollos l\'{o}gicos de cualquier esquema matem\'{a}tico con un fundamento
fijo.} \textit{En el presente, hay problemas fundamentales en f\'{\i}sica
te\'{o}rica esperando por soluci\'{o}n, }\textbf{e.g.},\textit{\ la
formulaci\'{o}n relativista de la mec\'{a}nica cu\'{a}ntica y la naturaleza
del n\'{u}cleo at\'{o}mico (seguidos de problemas m\'{a}s d\'{\i}ficiles tales
como el problema de la vida), cuya soluci\'{o}n presumiblemente requerir\'{a}
una revisi\'{o}n de nuestros conceptos fundamentales m\'{a}s dr\'{a}stica que
cualquiera ocurrida antes. Muy probablemente estos cambios ser\'{a}n tan
grandes que estar\'{a} m\'{a}s all\'{a} del alcance de la inteligencia humana
obtener las nuevas ideas necesarias a trav\'{e}s de intentos directos de
formular los datos experimentales en t\'{e}rminos matem\'{a}ticos. Por lo
tanto, el trabajador te\'{o}rico en el futuro tendr\'{a} que proceder de una
forma m\'{a}s indirecta. El m\'{e}todo de avance m\'{a}s poderoso que puede
ser sugerido en el presente es emplear todos los recursos de la matem\'{a}tica
pura en intentos por perfeccionar y generalizar el formalismo matem\'{a}tico
que constituye la base existente de la f\'{\i}sica te\'{o}rica, y}
\textbf{despu\'{e}s} \textit{de cada \'{e}xito en esta direcci\'{o}n, intentar
interpretar las nuevas propiedades matem\'{a}ticas en t\'{e}rminos de
entidades f\'{\i}sicas.}\textquotedblright}}\textquotedblright
\end{quotation}

Siguiendo esa l\'{\i}nea de pensamiento, debemos recordar que Einstein al
formular Relatividad General, o Dirac al predecir la existencia de la
antipart\'{\i}cula, no estaban intentando simplemente reproducir f\'{\i}sica
conocida, sino que estaban siendo guiados a trav\'{e}s de principios de
belleza matem\'{a}tica, los cuales les permitieron constru\'{\i}r una
teor\'{\i}a consistente y \textit{posteriormente}, efectuar predicciones
experimentales. De la misma forma, me parece probable que la mejor manera de
buscar un principio de acci\'{o}n fundamental, que unifique todas las
interacciones conocidas, no sea buscar \textquotedblleft
algo\textquotedblright\ que reproduzca la f\'{\i}sica conocida despu\'{e}s de
un complicado proceso de compactificaci\'{o}n, sino m\'{a}s bien, buscar una
entidad que tenga un significado matem\'{a}tico especial por s\'{\i} mismo, y
que se preste en forma natural para la construcci\'{o}n de la teor\'{\i}a. En
ese sentido, la profunda relaci\'{o}n con topolog\'{\i}a de las formas de
transgresi\'{o}n y su relaci\'{o}n con el teorema del \'{\i}ndice parece
especialmente seductora: este tipo de acciones tiene un claro significado
matem\'{a}tico \textit{per se}, y se presta en forma natural para
constru\'{\i}r teor\'{\i}as de gauge, y a\'{u}n m\'{a}s, supergravedad.

Por ello, me parece probable que una teor\'{\i}a de campos constru\'{\i}da a
partir de una supergravedad de transgresi\'{o}n contenga de alguna manera
alg\'{u}n tipo de teor\'{\i}a de cuerdas (probablemente con una acci\'{o}n
m\'{a}s compleja que la de Poliakov) y branas en forma similar a como la
acci\'{o}n de Dirac contiene la acci\'{o}n para la part\'{\i}cula libre
cl\'{a}sica. En otras palabras, lo que se necesita es una Teor\'{\i}a de
Campos Cu\'{a}ntica para cuerdas consistente, y personalmente, me parece que
una acci\'{o}n del tipo de Transgresi\'{o}n podr\'{\i}a hacer este trabajo,
dada su profunda belleza matem\'{a}tica.

\chapter{Ap\'{e}ndices}

\bigskip

\bigskip

\begin{center}
\textquotedblleft\textit{Long is the way}

\textit{And hard, that out of Hell leads up to Light\footnote{\textit{Largo y
d\'{\i}ficil es el camino, que desde el Infierno lleva hacia la Luz}.}%
}\textquotedblright

(John Milton, Paradise Lost)

\newpage
\end{center}

\appendix{}

\chapter{\label{Apend VarTransg}Variaci\'{o}n de la Forma de Transgresi\'{o}n}

En est\'{e} ap\'{e}ndice se deduce la variaci\'{o}n del Lagrangeano
transgresor [ec.~(\ref{Ec VarLagrangT})] bajo las variaciones infinitesimales%
\begin{align*}
\boldsymbol{A}  &  \rightarrow\boldsymbol{A}^{\prime}=\boldsymbol{A}%
+\delta\boldsymbol{A},\\
\bar{\boldsymbol{A}}  &  \rightarrow\bar{\boldsymbol{A}}^{\prime}%
=\bar{\boldsymbol{A}}+\delta\bar{\boldsymbol{A}}.
\end{align*}

Bajo estas variaciones, tenemos%
\begin{align*}
\delta\boldsymbol{\Theta}  &  =\delta\boldsymbol{A}-\delta\bar{\boldsymbol{A}%
},\\
\delta\boldsymbol{A}_{t}  &  =\delta\bar{\boldsymbol{A}}+t\delta
\boldsymbol{\Theta},
\end{align*}
y en t\'{e}rminos de ellas, el Lagrangeano transgresor
[ec.~(\ref{Ec LagrangT 2})] var\'{\i}a de la forma%
\begin{equation}
\delta L_{\mathrm{T}}^{\left(  2n+1\right)  }=kn\left(  n+1\right)  \int
_{0}^{1}\mathrm{d}t~\left\langle \boldsymbol{\Theta}\mathrm{D}_{t}%
\delta\boldsymbol{A}_{t}\boldsymbol{F}_{t}^{n-1}\right\rangle +k\left(
n+1\right)  \int_{0}^{1}\mathrm{d}t~\left\langle \delta\boldsymbol{\Theta
F}_{t}^{n}\right\rangle .\label{Ec EcMov Original}%
\end{equation}

Usando la regla de Leibniz para $\mathrm{D}_{t}$ y ec.~(\ref{EcCondInvConm}),
integramos por partes,%
\begin{equation}
\left\langle \boldsymbol{\Theta}\mathrm{D}_{t}\delta\boldsymbol{A}%
_{t}\boldsymbol{F}_{t}^{n-1}\right\rangle =\left\langle \mathrm{D}%
_{t}\boldsymbol{\Theta}\delta\boldsymbol{A}_{t}\boldsymbol{F}_{t}%
^{n-1}\right\rangle -\mathrm{d}\left\langle \boldsymbol{\Theta}\delta
\boldsymbol{A}_{t}\boldsymbol{F}_{t}^{n-1}\right\rangle
.\label{Ec EcMov IntgrPartes}%
\end{equation}

Usando las relaciones%
\begin{align*}
\frac{\mathrm{d}}{\mathrm{d}t}\boldsymbol{F}_{t}  &  =\mathrm{D}%
_{t}\boldsymbol{\Theta},\\
\frac{\mathrm{d}}{\mathrm{d}t}\delta\boldsymbol{A}_{t}  &  =\delta
\boldsymbol{\Theta},
\end{align*}
en ec.~(\ref{Ec EcMov IntgrPartes}) integramos por partes en $t$,%
\[
n\left\langle \boldsymbol{\Theta}\mathrm{D}_{t}\delta\boldsymbol{A}%
_{t}\boldsymbol{F}_{t}^{n-1}\right\rangle =\frac{\mathrm{d}}{\mathrm{d}%
t}\left\langle \delta\boldsymbol{A}_{t}\boldsymbol{F}_{t}^{n}\right\rangle
-\left\langle \delta\boldsymbol{\Theta F}_{t}^{n}\right\rangle -n\mathrm{d}%
\left\langle \boldsymbol{\Theta}\delta\boldsymbol{A}_{t}\boldsymbol{F}%
_{t}^{n-1}\right\rangle .
\]

Reemplazando en ec.~(\ref{Ec EcMov Original}), llegamos al resultado final
\[
\delta L_{\mathrm{T}}^{\left(  2n+1\right)  }=k\left(  n+1\right)  \left(
\left\langle \delta\boldsymbol{AF}^{n}\right\rangle -\left\langle \delta
\bar{\boldsymbol{A}}\bar{\boldsymbol{F}}^{n}\right\rangle \right)  +kn\left(
n+1\right)  \mathrm{d}\int_{0}^{1}\mathrm{d}t~\left\langle \delta
\boldsymbol{A}_{t}\boldsymbol{\Theta F}_{t}^{n-1}\right\rangle .
\]

Es interesante observar que el t\'{e}rmino correspondiente a las ecuaciones
del movimiento, $k\left(  n+1\right)  \left(  \left\langle \delta
\boldsymbol{AF}^{n}\right\rangle -\left\langle \delta\bar{\boldsymbol{A}}%
\bar{\boldsymbol{F}}^{n}\right\rangle \right)  ,$ es f\'{a}cilmente deducible
del Teorema de Chern--Weil, ec.~(\ref{EcTeoChern-WeilSobreM}).

\chapter{\label{Apend TeoNoether}Teorema de Noether}

En esta secci\'{o}n, escribiremos el teorema de Noether en general, sin romper
con la estructura de forma diferencial del lagrangeano.

Sea $M$ una variedad $d$-dimensional y sea $L^{\left(  d\right)  }\left(
\varphi^{A}\right)  $ una $d$-forma Lagrangeana para un conjunto de campos
$\varphi^{A}.$ Entonces, bajo la variaci\'{o}n $\varphi^{A}\rightarrow
\varphi^{A}+\delta\varphi^{A},$ el lagrangeano variar\'{a} de la forma%
\begin{equation}
\delta L^{\left(  d\right)  }=E_{A}\delta\varphi^{A}+\mathrm{d}\left(
B_{A}\delta\varphi^{A}\right) \label{Ec dL = E+dB}%
\end{equation}
en donde%
\[
E_{A}\left(  \varphi\right)  =0
\]
corresponde a las ecuaciones de movimiento y%
\[
\left.  B_{A}\left(  \varphi\right)  \delta\varphi^{A}\right\vert _{\partial
M}=0
\]
a las condiciones de contorno. Debemos recalcar que con $\delta$ estamos
denotando la variaci\'{o}n funcional, \textit{i.e.},%
\[
\delta\varphi^{A}=\varphi^{A\prime}\left(  x\right)  -\varphi^{A}\left(
x\right)  .
\]

Consideremos que el lagrangeano $L^{\left(  d\right)  }$ posee dos
simetr\'{\i}as: una ser\'{a} la de difeomorfismos, y la otra, una
simetr\'{\i}a de gauge.

\section{Corriente de Difeormofismos On-Shell}

Bajo un difeomorfismo infinitesimal, $x\rightarrow x+\xi,$ la
\textit{variaci\'{o}n funcional} de una $p$-forma diferencial arbitraria
$\alpha=\frac{1}{p!}\alpha_{\mu_{1}\mu_{2}\cdots\mu_{p}}\mathrm{d}x^{\mu_{1}%
}\cdots\mathrm{d}x^{\mu_{p}}$ viene dada por%
\[
\delta_{\mathrm{dif}}\alpha=-\pounds _{\xi}\alpha,
\]
en donde $\pounds _{\xi}$ corresponde a la derivada de Lie%
\[
\pounds _{\xi}=\mathrm{dI}_{\xi}+\mathrm{I}_{\xi}\mathrm{d},
\]
y $\mathrm{I}_{\xi}$ al operador de contracci\'{o}n%
\[
\mathrm{I}_{\xi}\alpha=\frac{1}{\left(  p-1\right)  !}\xi^{\mu_{1}}\alpha
_{\mu_{1}\mu_{2}\cdots\mu_{p}}\mathrm{d}x^{\mu_{2}}\cdots\mathrm{d}x^{\mu_{p}%
}.
\]

Por lo tanto, evaluando ec.~(\ref{Ec dL = E+dB}) para la variaci\'{o}n
funcional asociada a un difeomorfismo, tenemos
\[
-\pounds _{\xi}L^{\left(  d\right)  }=-E_{A}\pounds _{\xi}\varphi
^{A}-\mathrm{d}\left(  B_{A}\pounds _{\xi}\varphi^{A}\right)  .
\]

Dado que $L^{\left(  d\right)  }$ es una $d$-forma, $\pounds _{\xi}L^{\left(
d\right)  }=\mathrm{dI}_{\xi}L^{\left(  d\right)  },$ y por lo tanto, llegamos
a la identidad%
\[
\mathrm{d}\left(  B_{A}\pounds _{\xi}\varphi^{A}-\mathrm{I}_{\xi}L^{\left(
d\right)  }\right)  +E_{A}\pounds _{\xi}\varphi^{A}=0.
\]

Definimos%
\begin{equation}
\ast J^{\left(  \mathrm{dif-on}\right)  }=B_{A}\pounds _{\xi}\varphi
^{A}-\mathrm{I}_{\xi}L^{\left(  d\right)  },\label{Ec J Dif On-Shell}%
\end{equation}
y por lo tanto, tenemos%
\begin{equation}
\mathrm{d}\ast J^{\left(  \mathrm{dif-on}\right)  }+E_{A}\pounds _{\xi}%
\varphi^{A}=0.\label{Ec J dif Identidad}%
\end{equation}

Cuando $\varphi^{A}$ corresponde a una configuraci\'{o}n on-shell,
\textit{i.e.}, que satisface $E_{A}\left(  \varphi\right)  =0,$ $J^{\left(
\mathrm{dif-on}\right)  }$ es conservada,%
\[
\mathrm{d}\ast J^{\left(  \mathrm{dif-on}\right)  }=0.
\]

\section{Corriente de Gauge On-Shell}

Cuando el lagrangeano es invariante bajo una transformaci\'{o}n de
simetr\'{\i}a infinitesimal $\varphi^{A}\rightarrow\varphi^{A}+\epsilon^{A},$
tenemos que ec.~(\ref{Ec dL = E+dB}) corresponde a%
\[
E_{A}\epsilon^{A}+\mathrm{d}\left(  B_{A}\epsilon^{A}\right)  =0.
\]

As\'{\i}, cuando definimos $\ast J^{\left(  \mathrm{gauge-on}\right)  }%
=B_{A}\epsilon^{A},$ tenemos%
\begin{equation}
\mathrm{d}\ast J^{\left(  \mathrm{gauge-on}\right)  }+E_{A}\epsilon
^{A}=0,\label{Ec J gauge Identidad}%
\end{equation}
y as\'{\i}, cuando $\varphi^{A}$ corresponde a una configuraci\'{o}n on-shell,
$J^{\left(  \mathrm{gauge-on}\right)  }$ se conserva,%
\[
\mathrm{d}\ast J^{\left(  \mathrm{gauge-on}\right)  }=0.
\]

\section{Corrientes Conservadas Off-Shell}

Debe de observarse que cuando $E_{A}\pounds _{\xi}\varphi^{A}$ corresponde a
una forma exacta,%
\begin{equation}
E_{A}\pounds _{\xi}\varphi^{A}=\mathrm{d}X,\label{Ec CondOffShell Dif}%
\end{equation}
es posible definir%
\begin{equation}
\ast J^{\left(  \mathrm{dif-off}\right)  }=\ast J^{\left(  \mathrm{dif-on}%
\right)  }+X\label{Ec Dif: Joff=Jon + X}%
\end{equation}
la cual se conservar\'{a} [v\'{e}ase ec.~(\ref{Ec J dif Identidad})] sin
requerir que $\varphi^{A}$ satisfaga las ecuaciones del movimiento,%
\[
\mathrm{d}\ast J^{\left(  \mathrm{dif-off}\right)  }=0.
\]

De la misma forma, cuando $E_{A}\epsilon^{A}$ corresponde a una forma exacta,%
\begin{equation}
E_{A}\epsilon^{A}=\mathrm{d}Y,\label{Ec CondOffShell Gauge}%
\end{equation}
entonces%
\begin{equation}
\ast J^{\left(  \mathrm{gauge-off}\right)  }=\ast J^{\left(  \mathrm{gauge-on}%
\right)  }+Y,\label{Ec Gauge Joff=Jon + Y}%
\end{equation}
se conservar\'{a} [v\'{e}ase ec.~(\ref{Ec J gauge Identidad})] para cualquier
configuraci\'{o}n de $\varphi^{A}.$

\section{Cargas conservadas}

Sea $\Sigma_{0}$ una variedad $\left(  d-1\right)  $-dimensional, e impongamos
la condici\'{o}n de que $M$ tenga la topolog\'{\i}a $M=\mathbb{R}\times
\Sigma_{0}.$ Esto significa que siempre existe un embebimiento
\[
\Sigma:\Sigma_{0}\rightarrow\Sigma_{\text{{\tiny M}}}\subset M
\]
de $\Sigma_{0}$ en $M.$ Sea $J$ una corriente conservada en $M,$
$\mathrm{d}_{\text{{\tiny M}}}\ast J=0.$ Entonces, definimos la carga $Q$
asociada a $J$ como%
\[
Q=\int_{\Sigma_{0}}\Sigma^{\ast}\ast J
\]
en donde $\Sigma^{\ast}\ast J$ corresponde a la imagen rec\'{\i}proca de $\ast
J$ inducido por el embebimiento $\Sigma.$

Cuando $\ast J$ corresponde a una forma exacta sobre $M$, $\ast J=\mathrm{d}%
_{\text{{\tiny M}}}\sigma,$ entonces la carga puede escribirse como una
integral sobre $\partial\Sigma_{0},$%
\begin{align*}
Q  &  =\int_{\Sigma_{0}}\Sigma^{\ast}\left(  \mathrm{d}_{\text{{\tiny M}}%
}\sigma\right)  ,\\
&  =\int_{\Sigma_{0}}\mathrm{d}_{{\tiny \Sigma}}\Sigma^{\ast}\sigma,\\
&  =\int_{\partial\Sigma_{0}}\left(  \partial\Sigma_{0}\right)  ^{\ast}\sigma.
\end{align*}

\chapter{\label{Apend ForzRes Zamponha}Reducci\'{o}n Resonante cuando
$\mathfrak{g}$ satisface las condiciones de Weimar-Woods}

\chaptermark{Reducci\'{o}n Resonante/Weimar-Woods}

En este ap\'{e}ndice demostraremos que ec.~(\ref{Ec ForzResZamp Np's})
corresponde a la condici\'{o}n de reducci\'{o}n resonante
ec.~(\ref{Ec Cond Forz SpSq=nSr}) para la sub\'{a}lgebra resonante encontrada
en Sec.~\ref{Sec Weimar-Woods}.

En efecto, sea $\mathfrak{g}=\bigoplus_{p=0}^{n}V_{p}$ una descomposici\'{o}n
en subespacios de $\mathfrak{g}$ la cual satisface las condiciones de
Weimar-Woods, ec.~(\ref{Ec Weimar--Woods Condition}) y sea $S_{\mathrm{E}%
}^{\left(  N\right)  }=\bigcup_{p=0}^{n}S_{p}$ la correspondiente
descomposici\'{o}n resonante de $S_{\mathrm{E}}^{\left(  N\right)  },$ con%
\[
S_{p}=\left\{  \lambda_{\alpha_{p}}\text{, tal que }\alpha_{p}=p,\ldots
,N+1\right\}  \qquad N+1\geq n.
\]

Ahora, particionemos cada subconjunto $S_{p}=\check{S}_{p}\cup\hat{S}_{p}$
como en ecs.~(\ref{spdown})--(\ref{spup}),%
\begin{align}
\check{S}_{p}  &  =\left\{  \lambda_{\alpha_{p}}\text{, tal que }\alpha
_{p}=p,\ldots,N_{p}\right\}  ,\\
\hat{S}_{p}  &  =\left\{  \lambda_{\alpha_{p}}\text{, tal que }\alpha
_{p}=N_{p}+1,\ldots,N+1\right\}  .
\end{align}

Observemos que esta partici\'{o}n satisface autom\'{a}ticamente las siguientes
tres propiedades:
\begin{align}
N_{p}\geq N_{q}  &  \Leftrightarrow\hat{S}_{p}\subset\hat{S}_{q}%
,\label{Ec Np>Nq <--> SpUp C SqUp}\\
\check{S}_{0}\times\hat{S}_{q}  &  =\hat{S}_{q}%
,\label{Ec SoDown x SqUp = SqUp}\\
\check{S}_{p}\times\hat{S}_{q}  &  \subset\hat{S}_{x}\text{ tal que }N_{x}\leq
H_{N+1}\left(  p+N_{q}\right)  .\label{Ec SpDown x Sq Up = Sx Up}%
\end{align}

Dado que $\mathfrak{g}=\bigoplus_{p=0}^{n}V_{p}$ satisface las condiciones de
Weimar-Woods, entonces la condici\'{o}n de reducci\'{o}n resonante
ec.~(\ref{Ec Cond Forz SpSq=nSr}) sobre la partici\'{o}n $S_{p}=\check{S}%
_{p}\cup\hat{S}_{p}$ toma la forma%
\begin{equation}
\check{S}_{p}\times\hat{S}_{q}\subset\bigcap_{r=0}^{H_{n}\left(  p+q\right)
}\hat{S}_{r}.\label{Ec W-WForcedCondition SpSq=nSr}%
\end{equation}

Analizemos primero esta condici\'{o}n para el caso $p=0,$%
\[
\check{S}_{0}\times\hat{S}_{q}\subset\bigcap_{r=0}^{q}\hat{S}_{r}.
\]

Utilizando ec.~(\ref{Ec SoDown x SqUp = SqUp}), tenemos que esto es
equivalente a exigir%
\begin{equation}
\hat{S}_{q}\subset\bigcap_{r=0}^{q}\hat{S}_{r}.
\end{equation}

As\'{\i}, para todo $r$ tal que $0\leq r\leq q$ tenemos $\hat{S}_{q}%
\subset\hat{S}_{r}$, y usando ec.~(\ref{Ec Np>Nq <--> SpUp C SqUp}), obtenemos
la condici\'{o}n equivalente sobre los $N$'s,%
\begin{equation}
\forall r\leq q,N_{r}\leq N_{q}.\label{r<q-->Nr<Nq}%
\end{equation}

Dada esta condici\'{o}n, tenemos que entonces%
\[
\bigcap_{r=0}^{H_{n}\left(  p+q\right)  }\hat{S}_{r}=\hat{S}_{H_{n}\left(
p+q\right)  }.
\]
y por lo tanto, la condici\'{o}n de reducci\'{o}n resonante
ec.~(\ref{Ec W-WForcedCondition SpSq=nSr}) toma la forma%
\begin{equation}
\check{S}_{p}\times\hat{S}_{q}\subset\hat{S}_{H_{n}\left(  p+q\right)
}.\label{Ec SpDown x SqUp = SHn(p+q)Up}%
\end{equation}

Para resolver esta condici\'{o}n, basta con imponer que para cada $\hat{S}%
_{x}$ tal que $N_{x}\leq H_{N+1}\left(  N_{q}+p\right)  $, se cumpla que
$\hat{S}_{x}\subset\hat{S}_{H_{n}\left(  p+q\right)  }$ (ve\'{a}se propiedad
ec.~(\ref{Ec SpDown x Sq Up = Sx Up}) de la partici\'{o}n). Utilizando la
propiedad de la partici\'{o}n ec.~(\ref{Ec Np>Nq <--> SpUp C SqUp}), es
posible escribir en forma equivalente%
\begin{equation}
\forall N_{x}\leq H_{N+1}\left(  N_{q}+p\right)  ,\quad N_{H_{n}\left(
p+q\right)  }\leq N_{x}.
\end{equation}

Por lo tanto, se tiene tambi\'{e}n que%
\begin{equation}
N_{H_{n}\left(  p+q\right)  }\leq H_{N+1}\left(  N_{q}+p\right)  .
\end{equation}

Para el caso $p=1,$ se tiene%
\[
N_{H_{n}\left(  p+1\right)  }\leq H_{N+1}\left(  N_{q}+1\right)  ,
\]
y utilizando ec.~(\ref{r<q-->Nr<Nq}), llegamos a las desigualdades%
\[
N_{q}\leq N_{H_{n}\left(  q+1\right)  }\leq H_{N+1}\left(  N_{q}+1\right)  .
\]

Su soluci\'{o}n es simplemente%
\[
N_{q+1}=\left\{
\begin{array}
[c]{l}%
N_{q}\text{ \'{o}}\\
H_{N+1}\left(  N_{q}+1\right)
\end{array}
\right.  .
\]

Esto resuelve la condici\'{o}n ec.~(\ref{Ec W-WForcedCondition SpSq=nSr}) y
por lo tanto, tenemos que%
\begin{equation}
\left\vert \mathfrak{\check{G}}_{\text{$\mathrm{R}$}}\right\vert
=\bigoplus_{p=0}^{n}\check{S}_{p}\otimes V_{p},
\end{equation}
con $\check{S}_{p}=\left\{  \lambda_{\alpha_{p}}\text{, tal que }\alpha
_{p}=p,\ldots,N_{p}\right\}  $ y en donde
\begin{equation}
N_{p+1}=\left\{
\begin{array}
[c]{l}%
N_{p}\text{ or}\\
H_{N+1}\left(  N_{p}+1\right)
\end{array}
\right.  ,
\end{equation}
es un \'{a}lgebra de Lie resonantemente reducida con constantes de estructura%
\begin{equation}
C_{\left(  a_{p},\alpha_{p}\right)  \left(  b_{q},\beta_{q}\right)
}%
^{\phantom{\left( a_{p}, \alpha_{p} \right) \left( b_{q}, \beta_{q} \right)}\left(
c_{r},\gamma_{r}\right)  }=K_{\alpha_{p}\beta_{q}}%
^{\phantom{\alpha_{p} \beta_{q}}\gamma_{r}}C_{a_{p}b_{q}}%
^{\phantom{a_{p} b_{q}}c_{r}},\text{ with }\alpha_{p},\beta_{p},\gamma
_{p}=p,\ldots,N_{p}.
\end{equation}

\chapter{\label{Apend Dirac}Matrices de Dirac y Componentes del Tensor
Invariante de $\mathfrak{osp}\left(  \mathfrak{32}|\mathfrak{1}\right)  $}

\chaptermark{Matrices de Dirac/Tensor Invariante $\mathfrak{osp}\left(
\mathfrak{32}|\mathfrak{1}\right)  $}

\begin{center}
\textbf{\quotedblbase}\textit{Es gibt keinen Gott, und Dirac ist sein
Prophet\footnote{\textquotedblleft\textit{No hay ning\'{u}n Dios, y Dirac es
su Profeta\textquotedblright}}}\textbf{\textquotedblleft}

(Wolfgang Pauli)
\end{center}

En la presente tesis hemos utilizado la representaci\'{o}n en supermatrices de
$\mathfrak{osp}\left(  \mathfrak{32}|\mathfrak{1}\right)  $ para escribir su
\'{a}lgebra (Sec.~\ref{Sec S-Exp osp(32|1)}) y luego algunas de las
componentes de la supertraza simetrizada utilizadas para construir el tensor
invariante para el \'{A}lgebra~M (Sec.~\ref{Sec Lagrang M Alg}).

En el sector bos\'{o}nico del \'{a}lgebra [$\mathfrak{sp}\left(
\mathfrak{32}\right)  $] se ha escogido por conveniencia como base a las
matrices de Dirac\footnote{Esta no es la \'{u}nica elecci\'{o}n, tambi\'{e}n
es posible utilizar matrices de Dirac en $D=12$ para llevar a cabo la
construcci\'{o}n (Ve\'{a}se por ejemplo Ref. \cite{TesisTroncoso})} en $D=11$ ($32\times32$
componentes) como%
\begin{align*}
\boldsymbol{P}_{a}  &  =\frac{1}{2}\Gamma_{a},\\
\boldsymbol{J}_{ab}  &  =\frac{1}{2}\Gamma_{ab},\\
\boldsymbol{Z}_{abcde}  &  =\frac{1}{2}\Gamma_{abcde},
\end{align*}
en donde $\Gamma_{a_{1}\cdots a_{n}}=\frac{1}{n!}\Gamma_{\left[  a_{1}\right.
}\cdots\Gamma_{\left.  a_{n}\right]  }.$

Con la excepci\'{o}n de la identidad, las matrices de Dirac son de traza nula;
por lo tanto, para calcular la supertraza de un producto de ellas basta con
expresar \'{e}ste en la base de las matrices de Dirac y considerar el
t\'{e}rmino proporcional a la identidad.

As\'{\i}, tanto para calcular los conmutadores de $\mathfrak{sp}\left(
\mathfrak{32}\right)  $ como para calcular la supertraza de un producto de
matrices es necesaria una regla que permita expresar el producto de matrices
de Dirac en la base de Dirac. Resolver este problema no es tan trivial como
pudiera parecer, su soluci\'{o}n es tratada en Ref.~VanProeyen.

La soluci\'{o}n general de este problema en cualquier dimensi\'{o}n viene dada
en Ref.~\cite{VanPro99} la cual puede reescribirse en forma m\'{a}s
expl\'{\i}cita como (Ve\'{a}se Ref.~\cite{TesisEduardo})%
\[
\Gamma^{a_{1}\cdots a_{i}}\Gamma_{b_{1}\cdots b_{j}}=\sum_{s=0}^{\min\left(
i,j\right)  }\frac{1}{t!u!}\left(  -1\right)  ^{\frac{1}{2}s\left(
s-1\right)  }D_{\qquad b_{1}\cdots b_{j}}^{a_{1}\cdots a_{i}}\left(  s\right)
,
\]
en donde%
\[
D_{\qquad b_{1}\cdots b_{j}}^{a_{1}\cdots a_{i}}\left(  s\right)  =\left(
-1\right)  ^{\frac{1}{2}s\left(  s-1\right)  }\frac{1}{t!u!}\delta
_{d_{1}\cdots d_{t}b_{1}\cdots b_{j}}^{a_{1}\cdots a_{i}e_{1}\cdots e_{u}%
}\Gamma_{\qquad\;e_{1}\cdots e_{u}}^{d_{1}\cdots d_{t}},
\]
con%
\begin{align*}
t  &  =i-s,\\
u  &  =j-s,
\end{align*}
y en donde la matriz de Dirac sin ning\'{u}n \'{\i}ndice denota la identidad,
$\Gamma\equiv\boldsymbol{1}.$

Para manipular esta identidad, es de extrema utilidad descomponer la delta
antisimetrizada como%
\begin{align}
\delta_{b_{1}\cdots b_{m+n}}^{a_{1}\cdots a_{m+n}}  &  =\left(  -1\right)
^{\frac{1}{2}m\left(  m+1\right)  }\sum_{p_{1}=1}^{n+1}\sum_{p_{2}=p_{1}%
+1}^{n+2}\cdots\sum_{p_{m-1}=p_{m-2}+1}^{n+m-1}\sum_{p_{m}=p_{m-1}+1}%
^{n+m}\left(  -1\right)  ^{p_{1}+\cdots+p_{m}}\times\nonumber\\
&  \times\delta_{b_{p_{1}}\cdots b_{p_{m}}}^{a_{1}\cdots a_{m}}\delta
_{b_{1}\cdots\hat{b}_{p_{1}}\cdots\hat{b}_{p_{m}}\cdots b_{m+n}}%
^{a_{m+1}\cdots\cdots\cdots\cdots\cdots a_{m+n}}%
,\label{Ec DeltaKronecker Kaputt}%
\end{align}
en donde el s\'{\i}mbolo \textquotedblleft$^{\wedge}$\textquotedblright\ ha
sido utilizado para se\~{n}alar omisi\'{o}n de un \'{\i}ndice.

Por ejemplo, esta nos permite escribir
\begin{align*}
D_{a_{1}\cdots a_{i}b_{1}\cdots b_{j}}\left(  s\right)   &  =\left(
-1\right)  ^{s\left(  i-s\right)  +\frac{1}{2}s\left(  s-1\right)  }%
\sum_{p_{1}=1}^{1+j-s}\cdots\sum_{p_{s}=p_{s-1}+1}^{j}\sum_{q_{1}=1}%
^{1+i-s}\cdots\sum_{q_{s}=q_{s-1}+1}^{i}\left(  -1\right)  ^{p_{1}%
+\cdots+p_{s}+q_{1}+\cdots+q_{s}}\times\\
&  \times\eta_{\left[  a_{q_{1}}\cdots a_{q_{s}}\right]  \left[  b_{p_{1}%
}\cdots b_{p_{s}}\right]  }\Gamma_{a_{1}\cdots\hat{a}_{q_{1}}\cdots\hat
{a}_{q_{s}}\cdots a_{i}b_{1}\cdots\hat{b}_{p_{1}}\cdots\hat{b}_{p_{s}}\cdots
b_{j}}.
\end{align*}
en donde%
\[
\eta_{\left[  a_{1}\cdots a_{n}\right]  \left[  b_{1}\cdots b_{n}\right]
}\equiv\eta_{a_{1}c_{1}}\cdots\eta_{a_{n}c_{n}}\delta_{b_{1}\cdots b_{n}%
}^{c_{1}\cdots c_{n}},
\]
con lo que es posible escribir en forma directa $\Gamma_{a_{1}\cdots a_{i}%
}\Gamma_{b_{1}\cdots b_{j}}.$ A partir de estas identidades, es posible
expresar los conmutadores de $\mathfrak{sp}\left(  \mathfrak{32}\right)  $
como%
\[
A^{a_{1}\cdots a_{i}}B^{b_{1}\cdots b_{j}}\left[  \Gamma_{a_{1}\cdots a_{i}%
},\Gamma_{b_{1}\cdots b_{j}}\right]  =\sum_{s=0}^{\min\left(  i,j\right)
}\left[  1-\left(  -1\right)  ^{ij-s^{2}}\right]  A^{a_{1}\cdots a_{i}%
}B^{b_{1}\cdots b_{j}}D_{a_{1}\cdots a_{i}b_{1}\cdots b_{j}}\left(  s\right)
,
\]
o incluso anticonmutadores,%
\[
A^{a_{1}\cdots a_{i}}B^{b_{1}\cdots b_{j}}\left\{  \Gamma_{a_{1}\cdots a_{i}%
},\Gamma_{b_{1}\cdots b_{j}}\right\}  =\sum_{s=0}^{\min\left(  i,j\right)
}\left[  1+\left(  -1\right)  ^{ij-s^{2}}\right]  A^{a_{1}\cdots a_{i}%
}B^{b_{1}\cdots b_{j}}D_{a_{1}\cdots a_{i}b_{1}\cdots b_{j}}\left(  s\right)
,
\]
en donde%
\begin{align*}
A^{a_{1}\cdots a_{i}}B^{b_{1}\cdots b_{j}}D_{a_{1}\cdots a_{i}b_{1}\cdots
b_{j}}\left(  s\right)   &  =\binom{i}{s}\binom{j}{s}\left(  -1\right)
^{s\left(  s-1\right)  /2}\times\\
&  \times\eta_{\left[  b_{1}\cdots b_{s}\right]  \left[  c_{1}\cdots
c_{s}\right]  }A^{a_{1}\cdots a_{i-s}b_{1}\cdots b_{s}}B^{c_{1}\cdots
c_{s}a_{i-s+1}\cdots a_{i+j-2s}}\Gamma_{a_{1}\cdots a_{i+j-2s}}.
\end{align*}

El calcular los anticonmutadores entre matrices de Dirac resultar\'{a}
\'{u}til para calcular la traza simetrizada. En efecto, sean $X_{1}%
,\ldots,X_{n}$ $n$ matrices cuadradas y denotemos su producto simetrizado por
$\left\{  X_{1}\cdots X_{n}\right\}  $. Entonces, el producto simetrizado de
$n+1$ matrices puede ser escrito como%
\[
\left\{  X_{1}\cdots X_{n+1}\right\}  =\frac{1}{2\left(  n+1\right)  }%
\sum_{i=1}^{n+1}\left\{  X_{i},\left\{  X_{1}\cdots\hat{X}_{i}\cdots
X_{n+1}\right\}  \right\}  .
\]

As\'{\i} vemos que para obtener un producto simetrizado de matrices, basta con
conocer los anticonmutadores. Dada la estructura del tensor invariante para el
\'{A}lgebra~M, [ecs.~(\ref{itmalg1})-(\ref{itmalg5})] tenemos que un
anticonmutador particularmente importante corresponde a%
\[
\frac{1}{2}X^{a_{1}a_{2}}B^{b_{1}\cdots b_{j}}\left\{  \Gamma_{a_{1}a_{2}%
},\Gamma_{b_{1}\cdots b_{j}}\right\}  =X^{a_{1}a_{2}}B^{a_{3}\cdots a_{j+2}%
}\Gamma_{a_{1}\cdots a_{j+2}}-j\left(  j-1\right)  X_{a_{j-1}a_{j}}%
B^{a_{1}\cdots a_{j}}\Gamma_{a_{1}\cdots a_{j-2}},
\]
especialmente en el caso cuando $j=2n$ y
\[
B^{a_{1}\cdots a_{2n}}=\frac{1}{\left(  2n\right)  !}\delta_{b_{1}\cdots
b_{2n}}^{a_{1}\cdots a_{2n}}Y_{1}^{b_{1}b_{2}}\cdots Y_{n}^{b_{2n-1}b_{2n}}.
\]

En este punto, debe de utlizarse la identidad\footnote{El c\'{a}lculo de esta
expresi\'{o}n no es del todo trivial; debe de utilizarse en forma extensiva
ec.~(\ref{Ec DeltaKronecker Kaputt}), separar sumatorias y renombrar
\'{\i}ndices en forma cuidadosa.}%
\begin{align}
\left[  X\right]  _{a_{2n-1}a_{2n}}B^{a_{1}\cdots a_{2n}}  &  =-\frac
{2!}{\left(  2n\right)  !}\delta_{b_{1}\cdots b_{2n-2}}^{a_{1}\cdots a_{2n-2}%
}\sum_{\sigma\in S_{n}}\left(  \frac{1}{\left(  n-1\right)  !}%
\operatorname*{tr}\left(  XY_{\sigma\left(  1\right)  }\right)  Y_{\sigma
\left(  2\right)  }^{b_{1}b_{2}}\cdots Y_{\sigma\left(  n\right)  }%
^{b_{2n-3}b_{2n-2}}+\right. \nonumber\\
&  \left.  -\frac{2}{\left(  n-2\right)  !}\left[  Y_{\sigma\left(  1\right)
}\right]  _{\phantom{b_{1}}a}^{b_{1}}X_{\phantom{a}b}^{a}\left[
Y_{\sigma\left(  2\right)  }\right]  ^{bb_{2}}Y_{\sigma\left(  3\right)
}^{b_{3}b_{4}}\cdots Y_{\sigma\left(  n\right)  }^{b_{2n-3}b_{2n-2}}\right)
,\label{Ec Ident Permutante}%
\end{align}
para escribir%
\begin{multline*}
\frac{1}{2}X^{a_{1}a_{2}}Y_{1}^{b_{1}b_{2}}\cdots Y_{n}^{b_{2n-1}b_{2n}%
}\left\{  \Gamma_{a_{1}a_{2}},\Gamma_{b_{1}\cdots b_{2n}}\right\}  =\\
X^{a_{1}a_{2}}Y_{1}^{a_{3}a_{4}}\cdots Y_{n}^{a_{2n+1}a_{2n+2}}\Gamma
_{a_{1}\cdots a_{2n+2}}+\\
+2\sum_{i=1}^{n}\operatorname*{tr}\left(  XY_{i}\right)  \left[  Y_{1}\right]
^{a_{1}a_{2}}\cdots\hat{Y}_{i}\cdots\left[  Y_{n}\right]  ^{a_{2n-3}a_{2n-2}%
}\Gamma_{a_{1}\cdots a_{2n-2}}+\\
-2^{2}\sum_{\substack{i,j=1 \\i\neq j}}^{n}\left[  Y_{i}\right]
_{\phantom{b_{1}}a}^{a_{1}}X_{\phantom{a}b}^{a}\left[  Y_{j}\right]  ^{ba_{2}%
}\left[  Y_{1}\right]  ^{a_{3}a_{4}}\cdots\hat{Y}_{i}\cdots\hat{Y}_{j}%
\cdots\left[  Y_{n}\right]  ^{a_{2n-3}a_{2n-2}}\Gamma_{a_{1}\cdots a_{2n-2}}.
\end{multline*}

Con esta identidad, calcularemos el producto simetrizado de hasta 5 matrices
de la forma $\mathcal{X}=X^{ab}\Gamma_{ab},$%
\begin{align}
\left\{  \mathcal{X}_{1}\mathcal{X}_{2}\right\}   &  =X_{1}^{a_{1}a_{2}}%
X_{2}^{a_{3}a_{4}}\Gamma_{a_{1}\cdots a_{4}}+2\operatorname*{tr}\left(
X_{1}X_{2}\right)  \boldsymbol{1},\label{Ec Trazas Sim de 2}\\
\left\{  \mathcal{X}_{1}\mathcal{X}_{2}\mathcal{X}_{3}\right\}   &
=X_{1}^{a_{1}a_{2}}X_{2}^{a_{3}a_{4}}X_{3}^{a_{5}a_{6}}\Gamma_{a_{1}\cdots
a_{6}}+\nonumber\\
&  +\sum_{\sigma\in S_{3}}\left(  \operatorname*{tr}\left(  X_{\sigma\left(
1\right)  }X_{\sigma\left(  2\right)  }\right)  X_{\sigma\left(  3\right)
}^{a_{1}a_{2}}-\frac{4}{3}\left[  X_{\sigma\left(  1\right)  }\right]
^{a_{1}b}\left[  X_{\sigma\left(  2\right)  }\right]  _{bc}\left[
X_{\sigma\left(  3\right)  }\right]  ^{ca_{2}}\right)  \Gamma_{a_{1}a_{2}},
\end{align}%
\begin{align}
\left\{  \mathcal{X}_{1}\mathcal{X}_{2}\mathcal{X}_{3}\mathcal{X}_{4}\right\}
&  =X_{1}^{a_{1}a_{2}}X_{2}^{a_{3}a_{4}}X_{3}^{a_{5}a_{6}}X_{4}^{a_{7}a_{8}%
}\Gamma_{a_{1}\cdots a_{8}}+\nonumber\\
&  +\sum_{\sigma\in S_{4}}\left(  \frac{1}{2}\operatorname*{tr}\left(
X_{\sigma\left(  1\right)  }X_{\sigma\left(  2\right)  }\right)
X_{\sigma\left(  3\right)  }^{a_{1}a_{2}}X_{\sigma\left(  4\right)  }%
^{a_{3}a_{4}}+\right. \nonumber\\
&  \left.  -\frac{4}{3}\left[  X_{\sigma\left(  1\right)  }\right]
^{a_{1}a_{2}}\left[  X_{\sigma\left(  2\right)  }\right]  ^{a_{3}b}\left[
X_{\sigma\left(  3\right)  }\right]  _{bc}\left[  X_{\sigma\left(  4\right)
}\right]  ^{ca_{4}}\right)  \Gamma_{a_{1}\cdots a_{4}}+\nonumber\\
&  +\sum_{\sigma\in S_{4}}\left(  \frac{1}{2}\operatorname*{tr}\left(
X_{\sigma\left(  1\right)  }X_{\sigma\left(  2\right)  }\right)
\operatorname*{tr}\left(  X_{\sigma\left(  3\right)  }X_{\sigma\left(
4\right)  }\right)  +\right. \nonumber\\
&  \left.  -\frac{2}{3}\operatorname*{tr}\left(  X_{\sigma\left(  1\right)
}X_{\sigma\left(  2\right)  }X_{\sigma\left(  3\right)  }X_{\sigma\left(
4\right)  }\right)  \right)  \boldsymbol{1},
\end{align}%
\begin{align}
\left\{  \mathcal{X}_{1}\mathcal{X}_{2}\mathcal{X}_{3}\mathcal{X}%
_{4}\mathcal{X}_{5}\right\}   &  =X_{1}^{a_{1}a_{2}}X_{2}^{a_{3}a_{4}}%
X_{3}^{a_{5}a_{6}}X_{4}^{a_{7}a_{8}}X_{5}^{a_{9}a_{10}}\Gamma_{a_{1}\cdots
a_{10}}+\nonumber\\
&  +\sum_{\sigma\in S_{5}}\left(  \frac{1}{6}\operatorname*{tr}\left(
X_{\sigma\left(  1\right)  }X_{\sigma\left(  2\right)  }\right)
X_{\sigma\left(  3\right)  }^{a_{1}a_{2}}X_{\sigma\left(  4\right)  }%
^{a_{3}a_{4}}X_{\sigma\left(  5\right)  }^{a_{5}a_{6}}+\right. \nonumber\\
&  \left.  -\frac{2}{3}X_{\sigma\left(  1\right)  }^{a_{1}a_{2}}%
X_{\sigma\left(  2\right)  }^{a_{3}a_{4}}\left[  X_{\sigma\left(  3\right)
}\right]  ^{a_{5}b}\left[  X_{\sigma\left(  4\right)  }\right]  _{bc}\left[
X_{\sigma\left(  5\right)  }\right]  ^{ca_{6}}\right)  \Gamma_{a_{1}\cdots
a_{6}}+\nonumber\\
&  +\sum_{\sigma\in S_{5}}\left(  \frac{1}{2}\operatorname*{tr}\left(
X_{\sigma\left(  1\right)  }X_{\sigma\left(  2\right)  }\right)
\operatorname*{tr}\left(  X_{\sigma\left(  3\right)  }X_{\sigma\left(
4\right)  }\right)  X_{\sigma\left(  5\right)  }^{a_{1}a_{2}}+\right.
\nonumber\\
&  -\frac{4}{3}\operatorname*{tr}\left(  X_{\sigma\left(  1\right)  }%
X_{\sigma\left(  2\right)  }\right)  \left[  X_{\sigma\left(  3\right)
}\right]  ^{a_{1}b}\left[  X_{\sigma\left(  4\right)  }\right]  _{bc}\left[
X_{\sigma\left(  5\right)  }\right]  ^{ca_{2}}+\nonumber\\
&  -\frac{2}{3}\operatorname*{tr}\left(  X_{\sigma\left(  1\right)  }%
X_{\sigma\left(  2\right)  }X_{\sigma\left(  3\right)  }X_{\sigma\left(
4\right)  }\right)  \left[  X_{\sigma\left(  5\right)  }\right]  ^{a_{1}a_{2}%
}+\nonumber\\
&  \left.  +\frac{32}{15}\left[  X_{\sigma\left(  1\right)  }\right]
^{a_{1}b}\left[  X_{\sigma\left(  2\right)  }\right]  _{bc}\left[
X_{\sigma\left(  3\right)  }\right]  ^{cd}\left[  X_{\sigma\left(  4\right)
}\right]  _{de}\left[  X_{\sigma\left(  5\right)  }\right]  ^{ea_{2}}\right)
\Gamma_{a_{1}a_{2}}.\label{Ec Trazas Sim de 5}%
\end{align}

Para calcular supertrazas simetrizadas involucrando fermiones, se debe
proceder en forma an\'{a}loga, obteni\'{e}ndose (Ve\'{a}se
Ref.~\cite{TesisEduardo})%
\begin{align}
\operatorname*{Str}\left(  \boldsymbol{\bar{\chi}\zeta}\right)   &
=2\bar{\chi}\zeta,\label{Ec Prod Sim Fermion 0}\\
\operatorname*{Str}\left(  \boldsymbol{\bar{\chi}}\mathcal{X}_{1}%
\boldsymbol{\zeta}\right)   &  =0,\\
\operatorname*{Str}\left(  \boldsymbol{\bar{\chi}}\mathcal{X}_{1}%
\mathcal{X}_{2}\boldsymbol{\zeta}\right)   &  =\frac{2}{3}\bar{\chi}\left\{
\mathcal{X}_{1}\mathcal{X}_{2}\right\}  \zeta,\\
\operatorname*{Str}\left(  \boldsymbol{\bar{\chi}}\mathcal{X}_{1}%
\mathcal{X}_{2}\mathcal{X}_{3}\boldsymbol{\zeta}\right)   &  =0,\\
\operatorname*{Str}\left(  \boldsymbol{\bar{\chi}}\mathcal{X}_{1}%
\mathcal{X}_{2}\mathcal{X}_{3}\mathcal{X}_{4}\boldsymbol{\zeta}\right)   &
=\frac{2}{5}\bar{\chi}\left\{  \mathcal{X}_{1}\mathcal{X}_{2}\mathcal{X}%
_{3}\mathcal{X}_{4}\right\}  \zeta.\label{Ec Prod Sim Fermion 4}%
\end{align}
en donde $\boldsymbol{\bar{\chi}}=\bar{\chi}\boldsymbol{Q}$ y
$\boldsymbol{\zeta}=\boldsymbol{\bar{Q}}\zeta.$

En $D=11,$ s\'{o}lo la mitad de las matrices de Dirac son independientes, pues
$\Gamma_{a_{1}\cdots a_{n}}\propto\varepsilon_{a_{1}\cdots a_{n}a_{n+1}\cdots
a_{11}}\Gamma^{a_{n+1}\cdots a_{11}}.$ En particular, se tiene
que\footnote{Estamos usando la representaci\'{o}n donde $\Gamma_{1}\Gamma
_{2}\cdots\Gamma_{11}=+\boldsymbol{1};$ en el caso contrario, se tiene
$\Gamma_{a_{1}\cdots a_{11}}=-\varepsilon_{a_{1}\cdots a_{11}}\boldsymbol{1}%
.$}%
\[
\Gamma_{a_{1}\cdots a_{11}}=\varepsilon_{a_{1}\cdots a_{11}}\boldsymbol{1}.
\]

As\'{\i}, las \'{u}nicas matrices de Dirac con traza no nula son
$\boldsymbol{1}$ y $\Gamma_{a_{1}\cdots a_{11}}.$

Hasta el momento hemos considerado el producto simetrizado de hasta 5 matrices
$\Gamma_{ab};$ para obtener la traza simetrizada de seis matrices debemos
observar que dado que $\Gamma_{a_{1}\cdots a_{11}}$ es de traza no nula,
s\'{o}lo los siguientes productos de matrices de Dirac tienen traza no nula:%
\begin{align}
X_{1}^{a_{1}a_{2}}\cdots X_{5}^{a_{9}a_{10}}Y^{a_{11}}\operatorname*{Tr}%
\left(  \Gamma_{a_{1}\cdots a_{10}}\Gamma_{a_{11}}\right)   &
=\operatorname*{Tr}\left(  \boldsymbol{1}\right)  \varepsilon_{a_{1}\cdots
a_{11}}X_{1}^{a_{1}a_{2}}\cdots X_{5}^{a_{9}a_{10}}Y^{a_{11}}%
,\label{Ec Trazico 1}\\
X_{1}^{a_{1}a_{2}}X_{2}^{a_{3}a_{4}}\operatorname*{Tr}\left(  \Gamma
_{a_{1}a_{2}}\Gamma_{a_{3}a_{4}}\right)   &  =\operatorname*{Tr}\left(
\boldsymbol{1}\right)  2\operatorname*{tr}\left(  X_{1}X_{2}\right)  ,\\
X_{1}^{a_{1}a_{2}}\cdots X_{3}^{a_{5}a_{6}}Y^{a_{7}\cdots a_{11}%
}\operatorname*{Tr}\left(  \Gamma_{a_{1}\cdots a_{6}}\Gamma_{a_{7}\cdots
a_{11}}\right)   &  =\operatorname*{Tr}\left(  \boldsymbol{1}\right)
\varepsilon_{a_{1}\cdots a_{11}}X_{1}^{a_{1}a_{2}}\cdots X_{3}^{a_{5}a_{6}%
}Y^{a_{7}\cdots a_{11}},
\end{align}

\begin{multline}
X_{1}^{a_{1}a_{2}}\cdots X_{5}^{a_{9}a_{10}}Y^{b_{1}\cdots b_{5}%
}\operatorname*{Tr}\left(  \Gamma_{a_{1}\cdots a_{10}}\Gamma_{b_{1}\cdots
b_{5}}\right)  =\\
\operatorname*{Tr}\left(  \boldsymbol{1}\right)  \sum_{\sigma\in S_{5}}\left(
\frac{5}{6}X_{\sigma\left(  1\right)  }^{a_{1}a_{2}}\cdots X_{\sigma\left(
4\right)  }^{a_{7}a_{8}}\left[  X_{\sigma\left(  5\right)  }\right]
_{c_{2}c_{1}}Y^{c_{1}c_{2}a_{9}a_{10}a_{11}}+\right. \\
\left.  +\frac{20}{3}X_{\sigma\left(  1\right)  }^{a_{1}a_{2}}\cdots
X_{\sigma\left(  3\right)  }^{a_{5}a_{6}}\left[  X_{\sigma\left(  4\right)
}\right]  _{\phantom{a_{7}}b_{1}}^{a_{7}}\left[  X_{\sigma\left(  5\right)
}\right]  _{\phantom{a_{8}}b_{2}}^{a_{8}}Y^{b_{1}b_{2}a_{9}a_{10}a_{11}%
}\right)  \varepsilon_{a_{1}\cdots a_{11}}.\label{Ec Trazico 4}%
\end{multline}
La \'{u}ltima traza es particularmente d\'{\i}ficil de escribir, requiere el
uso cuidadoso de identidad ec.~(\ref{Ec Ident Permutante}).

Utilizando los productos simetrizados ecs.~(\ref{Ec Trazas Sim de 2}%
)-(\ref{Ec Trazas Sim de 5}), y las trazas ecs.~(\ref{Ec Trazico 1}%
)-(\ref{Ec Trazico 4}) junto con las supertrazas
ecs.~(\ref{Ec Prod Sim Fermion 0})-(\ref{Ec Prod Sim Fermion 4}), resulta
sencillo escribir las componentes de polinomio invariante, ecs.~(\ref{L5B1}%
)-(\ref{L4FF}), las cuales se utilizan para escribir el polinomio invariante
para el \'{A}lgebra~M.

Utilizando los productos simetrizados de 2, y 4 matrices, resulta directo
relajar las constantes de acoplamiento como se muestra en
Sec.~\ref{Sec Relajando Ctes Acopl}, provistos de la apropiada
simetrizaci\'{o}n de componentes del tensor invariante.

\end{document}